# QUANTUM THEORY OF UNAMBIGUOUS MEASUREMENTS[1]

**Michal Sedlák** [2]
*Research Center for Quantum Information, Institute of Physics, Slovak Academy of Sciences, Dúbravská cesta 9, 845 11 Bratislava, Slovakia*

In the present paper I formulate a framework that accommodates many unambiguous discrimination problems. I show that the prior information about any type of constituent (state, channel, or observable) allows us to reformulate the discrimination among finite number of alternatives as the discrimination among finite number of average constituents. Using this framework I solve several unambiguous tasks. I present a solution to optimal unambiguous comparison of two ensembles of unknown quantum states. I consider two cases: 1) The two unknown states are arbitrary pure states of qudits. 2) Alternatively, they are coherent states of single-mode optical fields. For this case I propose simple and optimal experimental setup composed of beam-splitters and a photodetector. As a second tasks I consider an unambiguous identification (UI) of coherent states. In this task identical quantum systems are prepared in coherent states and labeled as unknown and reference states, respectively. The promise is that one reference state is the same as the unknown state and the task is to find out unambiguously which one it is. The particular choice of the reference states is unknown to us, and only the probability distribution describing this choice is known. In a general case when multiple copies of unknown and reference states are available I propose a scheme consisting of beamsplitters and photodetectors that is optimal within linear optics. UI can be considered as a search in a quantum database, whose elements are the reference states and the query is represented by the unknown state. This perspective motivated me to show that reference states can be recovered after the measurement and might be used (with reduced success rate) in subsequent UI. Moreover, I analyze the influence of noise in preparation of coherent states on the performance of the proposed setup. Another problem I address is the unambiguous comparison of a pair of unknown qudit unitary channels. I characterize all solutions and identify the optimal ones. I prove that in optimal experiments for comparison of unitary channels the entanglement is necessary. The last task I studied is the unambiguous comparison of unknown non-degenerate projective measurements. I distinguish between measurement devices with apriori labeled and unlabeled outcomes. In both cases only the difference of the measurements can be concluded unambiguously. For the labeled case I derive the optimal strategy if each unknown measurement is used only once. However, if the apparatuses are not labeled, then each measurement device must be used (at least) twice. In particular, for qubit measurement apparatuses with unlabeled outcomes I derive the optimal test state in the two-shots scenario.

[1] Material in this article was presented as a PhD thesis of the author at the Institute of Physics, Slovak Academy of Sciences. The work was conducted under supervision of professor Vladimír Bužek.

[2]E-mail address: michal.sedlak@savba.sk

**Contents**

## 1 Introduction

In our everyday life we often see just consequences of some action and we wonder about its cause. Knowing how things around us work we have some candidates for the cause, which we call hypotheses. Usually we also have some expectations about the probability of those hypotheses to take place. Thus, based on the observed consequences we would like to make a conclusion. Most often we want to deduce the correct hypothesis i.e. the one, which took place. In this situation our logic tells us to select the hypotheses, which is "most probable". There are also situations, where the goal is to make a different type of conclusion. For example, imagine a policeman, who investigates a murder. As a first step he might want to narrow the number of suspects. Hence, based on the evidences he has, he should conclude, which of the hypothesis could not take place. In both cases our natural logic can be described using the probability theory and it relays on the conditional probabilities of having the observed consequences if one of the considered hypothesis is realized.

Our effort to make a conclusion can be formalized as follows. Imagine that we have a set of objects $\mathcal{A} \equiv \{A_1, \ldots, A_N\}$ and its element $A_i$, $i \in \{1, \ldots, N\}$, is randomly chosen with probability $\eta_i$. The action in which an element from $\mathcal{A}$ is involved, might be called a test and its possible observable consequences we denote as outcomes $\{w_1, \ldots, w_M\}$. Outcome of the test shall help us to infer something about the tested object and subsequently enables us to make the desired conclusion. The situation is very simple if the outcomes of the test have a direct relation to the tested object. If an outcome can occur for just one tested object (one hypotheses) then we may say that this outcome unambiguously indicates that object. If all the outcomes are either unambiguous or have zero probability to appear we may say that the test perfectly distinguishes among the objects $\{A_1, \ldots, A_N\}$. However, most often the relation of the outcomes to the tested objects will not be so direct and only some of the outcomes are unambiguous or even none of them is unambiguous. In such case we can either accept that our conclusions will be sometimes erroneous or we have to resign on conclusion we were originally aiming at and we shall make only weaker, but still error free statements. Until now, we were discussing the situation, when the test is fixed. In practice there are many situations in which we design the test together with the decision procedure that relates its outcomes with the conclusions we make. This is the case for example in communication, where the goal of the device on the receiver's side is to distinguish the signals coming out of the communication channel. In classical physics there is no principal restriction on the precision of the measurements we make. Thus, in theory arbitrary set of mutually different signals can be perfectly distinguished and the errors that occur are due to insufficient precision of the measurement apparatuses we have used. Nowadays, pulses of lights are used for communication through optical fibers. Permanent need for increase in the capacity of such communication leads to shortening of the pulses. Light is a quantum phenomenon and in certain situations exhibits its non-classical properties. Hence, it starts to be practically important to know how quantum mechanics restricts the distinguishability of quantum objects.

Quantum mechanics is a statistical theory, which describes how Nature behaves on an atomic scale. As each physical theory it has tools for describing state, evolution and predictions for measurable quantities of the considered system [1]. A specific feature of quantum mechanics is that probabilistic nature of its predictions can not be attributed to our insufficient knowledge about the state of the system and in general the state can not be inferred from a single measurement. In classical physics we are used to having a direct relation of the measured property of the physical system to its implicitly assumed preexisting value before the measurement. In quantum mechanics it follows from Bells inequalities that for some sets of observables assuming preexisting values before the measurement is forbidden. On the other hand, via experiments with quantum systems we either try to acquire information formerly encoded into the system or we are determining properties of the system itself. Usually we are not testing completely unknown quantum system, but we have some prior knowledge about it. For example there always exists a subset of orthogonal states that are perfectly distinguishable by some measurement, which for each of those states produces only one distinct outcome. Therefore, the prior knowledge of this subset from which the measured state originates, enables us to determine the state by a single measurement. However as soon as the possible states before the measurement are not mutually orthogonal, a perfect discrimination is impossible due to nonorthogonality of outcome probability distributions predicted by quantum mechanics.

At the first glance the impossibility to perfectly discriminate states of quantum objects might look as a disadvantage in the areas like computation and communication. Surprisingly, quite opposite seems to be true. The use of nonorthogonal states for quantum communication enables communicating parties to detect the eavesdropper who inevitably disturbs their measurement statistics. Moreover, it was recognized that an unambiguous discrimination can used to design quantum key distribution protocols (QKD) [2], but it can be also exploited by an eavesdropper to attack some QKD protocols [3]. In quantum computation (see e.g. [4]) encoding of a final result to nonorthogonal states can speed up the computation significantly even though some repetitions of the computation may be necessary to find the result. These quantum information processing applications currently motivate the investigation of state discrimination in situations with different prior knowledge on states being measured.

The discrimination of quantum states was first considered in 1970s in a pioneering work by Helstrom [5]. In his case the measured state is guaranteed to be in one of the two known states $\rho_1$, $\rho_2$, which appear with prior probabilities $\eta_1$, $\eta_2$, respectively. The task is to determine after each single measurement which of the two states we were given. If the states are not orthogonal we inevitably sometimes guess incorrectly and the goal is to minimize the probability of making an error. This approach is called *minimum error approach* and can be easily generalized also for more states $\rho_i$. Although we will determine the state correctly with the highest possible probability, we are never sure that our conclusion was correct.

The other (extreme) option to handle the inevitable errors in the discrimination is by ultimately increasing the reliability of some outcomes at the expense of one totally unreliable outcome (inconclusive result). This approach, called *unambiguous discrimination*, is adopted also in this paper and each of its conclusive outcomes unambiguously indicates one of the discriminated possibilities. Hence, the outcomes either imply error free conclusion or they are inconclusive and marked also as a failure of the measurement.

Unambiguous discrimination of a pair of pure states was solved in works of Ivanovic [6], Dieks [7] and Peres [8] in 1987. They found out that both states can be unambiguously determined, but the price to pay is the possibility of getting an inconclusive result, which means that the measurement failed to give an unambiguous answer. Naturally, we would like to maximize the probability of correct discrimination of the states, which is equivalent to minimization of the failure probability. The fact that a specific result of a (single shot) measurement allows us to determine with certainty which of the non-orthogonal states was prepared is astonishing and still attracts lot of attention. The original Ivanovic, Dieks and Peres's scenario with two pure states appearing with equal prior probability was generalized in several ways. The case with arbitrary prior probabilities was solved in 1995 by Jaeger and Shimony [9]. Chefles [10] showed that in contrast to minimum error approach only linearly independent states can be unambiguously discriminated. He also resolved unambiguous discrimination of $N$ symmetric pure states. Essentially no more pure state analytical solutions are known, however Sun [11] showed that the problem can be efficiently tackled numerically by the convex optimization. The transition to the *unambiguous discrimination of mixed states* yields very interesting and still open problem. The known results ranges from the upper and lower bounds on the failure probability, solutions for some special cases and some numerical approaches. The main focus is certainly on discrimination of a pair of mixed states. There, the important step on a way to the general solution are Raynal's reduction theorems [12, 13], which enable uniform derivation of nearly all previously solved special cases. They simplify the problem by reducing its dimension or splitting it into more pieces, which are often unambiguous discrimination of two pure states. In years 2007 and 2008 very general new results were obtained by Kleinmann, Kampermann, and Bruss. In particular, in [14] these authors found commutators, which reveal two dimensional block diagonal structure in the reduced states. Moreover, they succeeded [15, 16] to rewrite the necessary and sufficient optimality conditions by Eldar, Stojnic, and Hassibi [17] into an operational form, which enabled them to prove uniqueness of the optimal measurement within the set of proper USD measurements. Finally, they completely classify and derive the solutions in four dimensional Hilbert space. Unfortunately, there does not exist a closed formula for the probability of success and hence some people still consider the unsolved general problem as open also in four dimensional Hilbert space.

The unambiguous state discrimination started the investigation of tasks, in which the certain measurement can lead to unambiguous knowledge about some property of the system. It is clear that some kind of prior knowledge is needed for such tasks to be realizable. The closely related example to discrimination of states is *unambiguous discrimination of quantum channels*. The basic version of the task is to unambiguously distinguish among two fixed channels if we control the preparation of the initial states and the measurement after the channel. Obviously channels are distinguishable if there exist an input state, which is evolved to unambiguously distinguishable output states. The goal is thus to chose an input state for which the probability of unambiguously distinguishing final states is highest. The first results were obtained by G. Wang, and M. Ying [18] and tell us when the discrimination among $N$ quantum channels is possible if the tested channel is used multiple times.

Another intriguing unambiguous discrimination task, was proposed and solved by S.M. Barnett, A. Chefles, and I. Jex [19]. Imagine we are given two identical quantum systems, which are guaranteed to be in a product state. The task, called quantum state comparison, is to say unambiguously whether the systems are in the same or different state. No prior knowledge of the pure state of each of the systems implies that only symmetry with respect to exchange of the systems can be used. Perhaps a bit unexpectedly the analysis show that equivalence of states can not be concluded unambiguously, whereas the difference can be. A measurement outcome which would reveal the equality of the states will be always inconclusive, because a general pair of pure states can produce any outcome for a given measurement. On the other hand detection of the difference of the states is possible because a pair of identical pure states can not produce all measurement outcomes.

So far we have considered as a prior knowledge the information about the task we are solving and the structure of the expected states together with their corresponding prior probabilities. However the random choice of the quantum states in each run of the experiment was considered as a consequence of our insufficient prior knowledge. Another way how to look on this situation is to denote the prior knowledge we have about the problem as a *classical information* and to call the choice of quantum states as a *quantum information*. In this way the quantum state comparison can be also seen as a special kind of probabilistic quantum processor [20–23]. In such device one system serves as the data register and the second quantum system as the program register telling the machine what to do with the data (in our case with which state the data should be compared). The processor acts probabilistically, because we are asking for unambiguous discrimination of nonorthogonal possibilities.

The unambiguous discrimination task we intensively focus on in this work is denoted as *unambiguous identification (UI)*. Imagine that we are given $N$ identical quantum systems each of which is in different unknown pure state - the so called reference state. We are given one additional system, which is guaranteed to be in one of the reference states. Our task is to identify unambiguously with which reference state the additional system matches. This task was first proposed by J. Bergou and M. Hillery. They aptly named their solution ”Programmable Quantum State Discriminator”, because the reference states can be seen as a program, which tells the machine between which states to discriminate. Thus, this task seems to be a nice fusion of ideas from quantum processors and unambiguous discrimination of pure states. Its investigation can help to clarify how the prior knowledge differ if it is given as classical and quantum information, how those forms of information supplement each other and how they influence the solution of the discrimination tasks.

The paper is organized as follows. We begin by Chapter 2, which recalls some basics of quantum mechanics and provides some useful mathematical statements. In Chapter 3 we formulate a framework which accommodates unambiguous discrimination problems. We shall see that the prior information about any type of constituent (state, channel, observable) allow us to reformulate the discrimination among finite number of alternatives as discrimination among finite number of average constituents. In subsequent chapters 4, 5 and 6 we use this framework to investigate unambiguous tasks for states, channels and measurements, respectively. In each of these chapters we study discrimination and comparison in general setting. Moreover, for states we study also unambiguous identification. A more detailed treatment is given to unambiguous identification of coherent states in Section 4.3. Each chapter is supplemented by a brief review of recent results obtained in the specific topic that is covered in the chapter.

## 2 Mathematical tools

Quantum mechanics describes Nature and its behavior by mathematical objects called linear operators. The purpose of this chapter is to provide their definition together with their basic properties, we will often use. In order to do that we have to start with the first postulate of quantum mechanics:

*To any physical system there exist a complex separable Hilbert space $\mathcal{H}$, which provides sufficient ground for the complete description of the system.*

Hilbert space $\mathcal{H}$ is a complex vector space endowed with the inner product $\langle .|.\rangle$ and complete in the norm derived from the inner product. Separability of $\mathcal{H}$ guarantees that there exist $\{|i\rangle\}_{i=1}^{\dim\mathcal{H}}$ a countable orthonormal basis of $\mathcal{H}$ i.e. closure of the span of $\{|i\rangle\}$ equals $\mathcal{H}$ and $\langle i|j\rangle = \delta_{ij}$. Having a Hilbert space we can build various mathematical structures based on it. Let us note that for infinite dimensional Hilbert spaces many complications arise. We plan to work mostly in finite dimensional $\mathcal{H}$, thus we provide simpler definitions valid for such case. Complex linear functional $f_\psi : \mathcal{H} \mapsto \mathbb{C}$ is a mapping from Hilbert space $\mathcal{H}$ to complex numbers for which $\forall\alpha,\beta \in \mathbb{C}$, $\forall|\phi_1\rangle, |\phi_2\rangle \in \mathcal{H}$ $f_\psi(\alpha|\phi_1\rangle + \beta|\phi_2\rangle) = \alpha f_\psi(|\phi_1\rangle) + \beta f_\psi(|\phi_2\rangle)$ holds. Thanks to Riesz lemma, to any complex linear functional $f_\psi$ there exist a unique vector $|\psi\rangle \in \mathcal{H}$, such that $f_\psi(|\phi\rangle) = \langle\psi|\phi\rangle$ $\forall|\phi\rangle \in \mathcal{H}$. We denote such functional by symbol $\langle\psi|$ and the abbreviation for its action on vector $|\phi\rangle$ coincide with the resulting scalar product $\langle\psi|\phi\rangle$. The complex linear functionals from $\mathcal{H}$ also form a Hilbert space isomorphic to $\mathcal{H}$ and usually denoted $\mathcal{H}^*$. The mapping $A : \mathcal{H} \mapsto \mathcal{H}$ which is linear $(A(\alpha|\phi_1\rangle + \beta|\phi_2\rangle) = \alpha A|\phi_1\rangle + \beta A|\phi_2\rangle)$ and defined on the dense subset of $\mathcal{H}$, we denote as *linear operator*. If the norm of the linear operator $A$, defined as:

$$\|A\| := \sup_{|\phi\rangle} \frac{\sqrt{\langle A\phi|A\phi\rangle}}{\sqrt{\langle\phi|\phi\rangle}}, \quad A|\phi\rangle \equiv |A\phi\rangle, \tag{2.1}$$

is finite we call operator $A$ *bounded*. Bounded linear operators form a Banach space $\mathcal{L}(\mathcal{H})$ i.e. a normed vector space complete in its norm. Moreover, $\mathcal{L}(\mathcal{H})$ has also a structure of noncommutative algebra, because bounded linear operators are closed with respect to composition of operators.

In general any linear operator $A$ is completely determined by its action on the basis vectors. Therefore in $d$ dimensional Hilbert space $\mathcal{H}$ the action of $A$ can be efficiently specified by $d \times d$ matrix. If we use orthonormal basis the matrix elements are given by $A_{ij} = \langle i|A|j\rangle$. The elements $A_{ij}$ $i = 1, \ldots, d$ determine the transformation of vector $|j\rangle$ into vector $A|j\rangle = \sum_{i=1}^{d} A_{ij}|i\rangle$. The simple example of linear operator is $|i\rangle\langle j|$ - a complex linear functional $\langle j|$ whose result rescales the fixed vector $|i\rangle$ from $\mathcal{H}$. If indexes $i, j$ run through $1, \ldots, d$ the set of the operators we obtain form a basis of $\mathcal{L}(\mathcal{H})$ and each $A \in \mathcal{L}(\mathcal{H})$ can be written as $A = \sum_{i,j=1}^{d} A_{ij}|i\rangle\langle j|$.

Each linear operator $A \in \mathcal{L}(\mathcal{H})$ defines a linear operator $A^*$ on $\mathcal{H}^*$: $A^*(\langle\psi|) \mapsto \langle\chi|$ such that $\forall|\phi\rangle \in \mathcal{H}$ $\langle\chi|\phi\rangle = \langle\psi|A|\phi\rangle$. Thanks to isomorphism $T$ between $\mathcal{H}$ and $\mathcal{H}^*$ operator $A^*$ uniquely defines the so called *adjoint operator* $A^\dagger := T^{-1}A^*T$ on $\mathcal{H}$. The adjoint operator therefore obeys the following property $\langle\phi|A^\dagger|\psi\rangle = \langle A\phi|\psi\rangle$. Taking the adjoint of an operator is an antilinear operation i.e. $(A + \lambda B)^\dagger = A^\dagger + \overline{\lambda} B^\dagger$, where $\overline{\lambda}$ is a complex conjugate of $\lambda \in \mathbb{C}$. We will further focus only on *normal operators*. Linear operator $A$ is normal if and only if $AA^\dagger = A^\dagger A$. Different subsets of normal operators are important in quantum mechanics, so we recall them by listing their defining properties. *Selfadjoint operator* is equal to its adjoint ($A = A^\dagger$), *projector* is an operator obeying $P = P^\dagger = P^2$. A *unitary operator* $U$ fulfills the relation $UU^\dagger = U^\dagger U = I$, where $I$ is a unity operator acting trivially $I|\phi\rangle = |\phi\rangle$ on each vector $|\phi\rangle$ from $\mathcal{H}$. Operator $A$ is called *possitive* if it is selfadjoint and $\forall|\phi\rangle \in \mathcal{H}$ $\langle\phi|A|\phi\rangle \geq 0$. For characterization of the internal structure of operators we will need the following terms. Vector $|\phi\rangle \in \mathcal{H}$ is an eigenvector of operator $A$ and $\lambda_\phi$ is its corresponding eigenvalue if $A|\phi\rangle = \lambda_\phi|\phi\rangle$. A very useful *Spectral theorem* states that any normal operator has the following decomposition:

$$A = \sum_k \lambda_k P_k, \tag{2.2}$$

where $\lambda_k \in \mathbb{C}$ are the eigenvalues of operator $A$ and $P_k$ are projectors onto mutually orthogonal subspaces corresponding to those eigenvalues. Thus any vector from the subspace on which $P_k$ projects is an eigenvector with eigenvalue $\lambda_k$. The orthogonality of the subspaces is equivalent to $P_k P_j = P_j P_k = 0\ \forall k \neq j$ and implies that eigenvectors with different eigenvalues are orthogonal. Let us define a *kernel of an operator* as a subspace which is mapped into zero. For normal operators it is a subspace corresponding to eigenvalue zero. By a *support of an operator* we will understand an orthocomplement of operators kernel i.e. the biggest orthogonal subspace of the kernel. For normal operator the support is a subspace corresponding to all nonzero eigenvalues. The subsets of normal operators can be defined also through constraints on their eigenvalues. Selfadjoint operators must have only real eigenvalues, positive operators positive eigenvalues, unitary operators eigenvalues with modulus 1 and projectors only eigenvalues 0 and 1.

Description of compound quantum systems requires tensor product of Hilbert spaces corresponding to the parts of the system. Assume we are given two Hilbert spaces $\mathcal{H}_A$, $\mathcal{H}_B$ together with their orthonormal bases $\{|i\rangle_A\}_{i=1}^{\dim \mathcal{H}_A}$, $\{|k\rangle_B\}_{k=1}^{\dim \mathcal{H}_B}$. Hilbert space $\mathcal{H}_A \otimes \mathcal{H}_B$ is a vector space span by vectors $|i\rangle_A \otimes |k\rangle_B$ endowed with inner product, which is on these vectors defined as $({}_A\langle i| \otimes {}_B\langle k|)(|j\rangle_A \otimes |l\rangle_B) = ({}_A\langle i|j\rangle_A).({}_B\langle k|l\rangle_B)$ and linearly extended to the rest of the space. Completeness of $\mathcal{H}_A$, $\mathcal{H}_B$ in the norm derived from the scalar product assures completeness of $\mathcal{H}_A \otimes \mathcal{H}_B$ which is therefore also a Hilbert space. Usually we keep the same ordering of the subsystems and thus we often simplify the notation by omitting the subscripts denoting the parts of the system. Moreover for $\forall z \in \mathbb{C}$, $\forall|\varphi\rangle, |\psi\rangle \in \mathcal{H}_A$, $\forall|\phi\rangle, |\xi\rangle \in \mathcal{H}_B$, the following rules hold:

$$\begin{aligned} z|\varphi\rangle \otimes |\phi\rangle &= (z|\varphi\rangle) \otimes |\phi\rangle = |\varphi\rangle \otimes (z|\phi\rangle), \\ (|\varphi\rangle + |\psi\rangle) \otimes |\phi\rangle &= |\varphi\rangle \otimes |\phi\rangle + |\psi\rangle \otimes |\phi\rangle, \\ |\varphi\rangle \otimes (|\phi\rangle + |\xi\rangle) &= |\varphi\rangle \otimes |\phi\rangle + |\varphi\rangle \otimes |\xi\rangle. \end{aligned}$$

A pair of linear operators $A \in \mathcal{L}(\mathcal{H}_A)$, $B \in \mathcal{L}(\mathcal{H}_B)$ naturally defines a bounded linear operator $A \otimes B$ on $\mathcal{H}_A \otimes \mathcal{H}_B$. Its action is first defined on the *product states* (states of the form $|\varphi\rangle \otimes |\phi\rangle$) by:

$$(A \otimes B)|\varphi\rangle \otimes |\phi\rangle := (A|\varphi\rangle) \otimes (B|\phi\rangle),$$

and then linearly extended to the whole $\mathcal{H}_A \otimes \mathcal{H}_B$. Let us note that $\mathcal{L}(\mathcal{H}_A \otimes \mathcal{H}_B) = \mathcal{L}(\mathcal{H}_A) \otimes \mathcal{L}(\mathcal{H}_B)$ and algebras $\mathcal{L}(\mathcal{H}_A)$, $\mathcal{L}(\mathcal{H}_B)$ are contained as a subalgebras of operators acting by unit on one of the subsystems.

We should also define the *trace* which connects linear operators with numbers. Thus, it is a linear mapping from $\mathcal{L}(\mathcal{H})$ to $\mathbb{C}$ defined as: $Tr(A) = \sum_{i=1}^{d} \langle i|A|i\rangle$, where $A \in \mathcal{L}(\mathcal{H})$ and $\{|i\rangle\}_{i=1}^{d}$ is any basis of $\mathcal{H}_d$. In practice, use of orthonormal basis for calculation of the trace is usually more convenient. The trace has the following properties ($A, B, C, U \in \mathcal{L}(\mathcal{H})$, $\lambda \in \mathbb{C}$):

$$\begin{aligned} Tr(A+\lambda B) &= Tr(A) + \lambda Tr(B), \\ Tr(ABC) &= Tr(BCA) = Tr(CAB), \\ Tr(U^{-1}AU) &= Tr(A), \quad U - unitary, \\ Tr(A \otimes B) &= Tr(A)Tr(B). \\ Tr(A|\psi\rangle\langle\psi|) &= \langle\psi|A|\psi\rangle \end{aligned}$$

Working with compound quantum systems we often need to focus our attention only on one of the subsystem. In such situations we frequently use a mapping called *partial trace* which transforms operators acting on the whole Hilbert space $\mathcal{H}_A \otimes \mathcal{H}_B$ onto operators on Hilbert space of one of the subsystems. Partial trace over subsystem B is defined as:

$$Tr_B(C) = \sum_{i,j=1}^{\dim \mathcal{H}_A} \left( \sum_{k=1}^{\dim \mathcal{H}_B} {}_A\langle i| \otimes {}_B\langle k|C|j\rangle_A \otimes |k\rangle_B \right) |i\rangle_A \, {}_A\langle j|,$$

where $C \in \mathcal{L}(\mathcal{H}_A \otimes \mathcal{H}_B)$, $Tr_B(C) \in \mathcal{L}(\mathcal{H}_A)$ and $\{|i\rangle_A\}_{i=1}^{\dim \mathcal{H}_A}$, $\{|k\rangle_B\}_{k=1}^{\dim \mathcal{H}_B}$ are orthonormal bases of $\mathcal{H}_A$, $\mathcal{H}_B$ respectively. Partial trace over subsystem A is defined analogously and result of both of them does not depend on the choice of the basis.

Before discussing how all these mathematical notions are used in quantum physics, let us prove two lemmas and present one recipe, which will become useful later in chapter about unambiguous identification.

**Lemma 1** *Let A, B be positive operators acting on $\mathcal{H}$, such that $Tr(AB) = 0$. Then the support of A is orthogonal to support of B.*

**Proof 1** *Since operators A, B are positive they have spectral decomposition:*

$$A = \sum_{i=1}^{rankA} \lambda_i |\phi_i\rangle\langle\phi_i|, \quad \lambda_i > 0, \quad B = \sum_{j=1}^{rankB} \kappa_j |\chi_j\rangle\langle\chi_j|, \quad \kappa_j > 0 \tag{2.3}$$

*Now let's evaluate the trace:*

$$0 = Tr(AB) = \sum_{i=1}^{rankA} \sum_{j=1}^{rankB} \lambda_i \kappa_j |\langle\phi_i|\chi_j\rangle|^2 \tag{2.4}$$

*Eigenvalues $\lambda_i$, $\kappa_j$ are positive and $|\langle\phi_i|\rangle\chi_j|^2$ is nonnegative, therefore we have a sum of nonnegative terms equal to zero. This is possible if and only if each term in the sum is equal to zero. Thus we have:*

$$\langle\phi_i|\chi_j\rangle = 0 \quad \forall i, \forall j, \tag{2.5}$$

*which means that eigenvectors of A are orthogonal to all eigenvectors of B and so concludes the proof.*

**Lemma 2** *Let $E$, $\rho(\psi) = |\psi\rangle\langle\psi|$ be positive operators acting on $\mathcal{H}$. If for set $S$, an integral $\Omega = \int_S |\psi\rangle\langle\psi| d\psi$ exists the following conditions on operator $E$ are equivalent:*

$$\forall|\psi\rangle \in S \quad Tr(E\rho(\psi)) = 0 \Longleftrightarrow Tr(E\Omega) = 0. \tag{2.6}$$

*In the integral defining $\Omega$ we use the Haar unitary invariant integration measure as well as in all other integrals in this paper.*

**Proof 2** *Integration of equation from left hand side (LHS) of equivalence together with linearity of trace obviously implies the right hand side (RHS). To prove the oposite implication we first rewrite the equation from RHS.*

$$0 = Tr(E\Omega) = Tr(E \int_S |\psi\rangle\langle\psi| d\psi) = \int_S \underbrace{\langle\psi|E|\psi\rangle}_{\geq 0} d\psi \tag{2.7}$$

*Integral of continuous nonnegative function is equal to zero only if the function is zero in the integration region. Therefore we have:*

$$\forall|\psi\rangle \in S \quad \langle\psi|E|\psi\rangle = Tr(E\rho(\psi)) = 0, \tag{2.8}$$

*which completes the proof.*

**Recipe 1 Construction of Jordan basis for two subspaces**

*Suppose we are given two orthonormal bases $\{|a'_i\rangle\}_{i=1}^{\dim V_1}$, $\{|b'_j\rangle\}_{j=1}^{\dim V_2}$ of subspaces $V_1$, $V_2$ respectively. We would like to rotate those bases in such a way that the rotated bases $\{|a_i\rangle\}_{i=1}^{\dim V_1}$, $\{|b_i\rangle\}_{i=1}^{\dim V_2}$ of $V_1$, $V_2$ will obey the following property:*

$$\forall i = 1, \dots, \dim V_1, \quad \forall j = 1, \dots, \dim V_2, \quad \langle a_i|b_j\rangle = \delta_{ij} \cos\theta_i \geq 0. \tag{2.9}$$

*Such a bases are called Jordan basis. They can always be relabeled so that $\cos\theta_i \geq \cos\theta_j$ for $i < j$. In order to construct Jordan basis from $\{|a'_i\rangle\}$, $\{|b'_i\rangle\}$ we first create $(\dim V_1) \times (\dim V_2)$ matrix of overlaps $H$ with elements $H_{ij} = \langle a'_i|b'_j\rangle$. Next we have to find its Singular value decomposition:*

$$H = U_1.D.U_2^\dagger, \tag{2.10}$$

*where $U_i$ ($i = 1, 2$) is $\dim V_i \times dimV_i$ unitary matrix and $\dim V_1$ times $\dim V_2$ matrix $D$ has non vanishing only diagonal elements $D_{ii} = \cos(\theta_i)$. Choosing $\{|a_i\rangle\}$, $\{|b_i\rangle\}$ to be:*

$$|a_i\rangle = \sum_{k=1}^{\dim V_1} (U_1)_{ki}|a'_k\rangle \qquad |b_j\rangle = \sum_{l=1}^{\dim V_2} (U_2)_{lj}|b'_l\rangle$$

*we obtain new orthonormal bases of $V_1$, $V_2$, since we have only unitarily rotated the former bases*[3]*, with the following mutual overlap:*

$$\begin{aligned}
\langle a_i|b_j\rangle &= \sum_{k=1}^{\dim V_1} \sum_{l=1}^{\dim V_2} (U_1)^*_{ki} \langle a'_k|b'_l\rangle (U_2)_{lj} = \\
&= \sum_{k,m=1}^{\dim V_1} \sum_{n,l=1}^{\dim V_2} (U_1^\dagger)_{ik}(U_1)_{km} D_{mn} (U_2^\dagger)_{nl} (U_2)_{lj} = \\
&= D_{ij} = \delta_{ij} \cos\theta_i
\end{aligned}$$

[3]Transpose of a unitary matrix is also unitary

*Thus we fulfilled the definition (2.9) i.e. we found the Jordan basis of subspaces* $V_1$*,* $V_2$*. Let us note that the basis of the common subspace of* $V_1$ *and* $V_2$ *is given by vectors* $|a_k\rangle = |b_k\rangle$ *such that* $\cos\theta_k = 1$*. On the other hand vectors* $|a_m\rangle$ *such that* $\cos\theta_m = 0$ *or* $m > \dim(V_2)$ *form a basis of a subspace of* $V_1$*, which is orthogonal to subspace* $V_2$*. Analogously, vectors* $|b_n\rangle$ *such that* $\cos\theta_n = 0$ *or* $n > \dim(V_1)$ *define a subspace of* $V_2$ *orthogonal to* $V_1$*.*

After the brief summary of mathematics we will use, we should describe how those mathematical structures are used in quantum mechanics. The entity of a quantum system determines the Hilbert space $\mathcal{H}$ appropriate for its description. Our knowledge about the state of a quantum system is expressed by a *density matrix* $\rho$ - a positive linear operator on $\mathcal{H}$ with trace one. The set of all possible density matrixes we denoted as $\mathcal{S}(\mathcal{H})$ and its elements are often also called *mixed states*. If the density matrix $\rho$ is a projector or equivalently $Tr(\rho^2) = 1$ we call the state *pure*, because $\rho = |\psi\rangle\langle\psi|$ for some vector $|\psi\rangle \in \mathcal{H}$ with norm one. Vectors $e^{i\varphi}|\psi\rangle$, $\varphi \in \mathbb{R}$ represent the same pure state. Pure states can therefore be identified with elements of projective Hilbert space, which is formed by cosets of the type $\lambda|\psi\rangle$, $\lambda \in \mathbb{C}$. We will denote the set of all pure states of a $d$-dimensional quantum system (qudit) $S_{pure}$.

Physical quantities, which can be measured, are called *observables* and correspond to self-adjoint operators. If the observable $A$ is measured, only the eigenvalues of operator $A$ can appear as measurement outcomes. The experience from experiments tells us that the immediate repetition of a measurement always gives the same outcome. This property of measurement is guaranteed by the projection postulate. It claims that the measurement of observable $A$ collapses the state of the system into an eigenstate of $A$, corresponding to the observed eigenvalue. The statistics of measurement outcomes for the observable $A$ can be collected by repeating the whole experiment many times. Quantum mechanics predicts that the mean value of these outcomes will be $\langle A\rangle = \sum_k \lambda_k p(\lambda_k|\rho) = \sum_k \lambda_k Tr(\Pi_k\rho) = Tr(A\rho)$, where $\rho$ is the state of the system immediately before the measurement, $p(\lambda_k|\rho)$ is the probability of obtaining the outcome $\lambda_k$ and $\Pi_k$ projects onto a subspace corresponding to $\lambda_k$. The measurement of an observable $A$ can be therefore characterized as a *projective measurement* and might be specified by a set of orthogonal projectors $\{\Pi_k\}$ and the corresponding measurement outcomes.

However, quantum mechanics permits a broader class of measurements to be performed. The most general measurement is described by a *Positive Operator Valued Measure* (POVM). From a mathematical point of view POVM is a mapping $\mathcal{A}$ from the set of outcomes $\{\omega_1, \ldots, \omega_n\}$ into the set of effects $\mathcal{E}(\mathcal{H})$, i.e. a set of positive operators $E$ on a Hilbert space $\mathcal{H}$ such that $O \leq E \leq I$, where $O$ is the zero operator and $I$ is the identity operator. Moreover, the POVM is normalized to identity i.e. $\mathcal{A}_1 + \cdots + \mathcal{A}_n = I$, where $\mathcal{A}_i \equiv \mathcal{A}(\omega_i)$.

We say that an observable or a measurement is *sharp* or equivalently *projective* if each effect composing the corresponding POVM is a projection, i.e., $\mathcal{A}_j = \mathcal{A}_j^2$ for all $j$. If, moreover, $\mathcal{A}_j\mathcal{H}$ is a one-dimensional subspace of $\mathcal{H}$ for each $j$, then the observable is *non-degenerate*. In such a case we can write $\mathcal{A}_j = |\psi_j\rangle\langle\psi_j| \equiv \psi_j$ and $\langle\psi_j|\psi_k\rangle = \delta_{jk}$. In fact, each orthonormal basis of the Hilbert space defines a sharp non-degenerate POVM. We denote by $\mathcal{M}$ the set of all non-degenerate sharp observables. From the point of view of a physicist a POVM is specified by a set of positive operators $\{E_i \equiv \mathcal{A}_i\}$, which sum up to identity operator $\sum_i E_i = I$. The probability of obtaining an outcome corresponding to the measurement operator $E_i$ is $p(E_i|\rho) = Tr(E_i\rho)$. The change of state associated with the measurement is not specified in this concept and in general depends on the particular realization of the POVM. Neumark [30] showed that each POVM can be realized as a projective measurement on the quantum system supplemented with an auxiliary quantum system. The auxiliary system is usually called the ancilla and its Hilbert space can have as many dimensions as the number of the POVM elements.

The description of discrimination tasks we consider does not require characterization of the dynamics of quantum systems, thus the aforementioned terms should suffice for our discussion.

## 3 Model of quantum experiment

The purpose of this chapter is to define a perspective through which we will look on quantum experiments. For us, an experiment is a sequence of three consecutive events: preparation, evolution and measurement of the considered quantum system. In practice executing any of these events involves several steps and takes a finite amount of time. However, for the purposes of this paper we will not need the details of the execution procedure, but rather its final effects. Thus, in our view the preparation part of an experiment is sufficiently described by the quantum state $\rho$ to which the system was set. The overall effect of the evolution part of an experiment is characterized by a *channel* [4], i.e. a completely positive trace preserving map on the state space of the quantum system. A measurement of a quantum system has two important aspects. One is the classical information about the obtained outcome of the measurement and the other is the state change inevitably induced by the measurement. The more information we gain about the state of the system before the measurement the more disturbed is the state of the quantum system by the measurement. In principle, we can further evolve the system after the measurement and then measure it again. However, the additional amount of information we acquire about the state of the system before the first measurement goes to zero as we iterate this process. Moreover, the whole development of the system starting by the first measurement can be always perceived as the realization of one big complicated measurement (see figure 3.1). We adopt this point of view, because our goal will be to exploit the information of a particular measurement outcome to maximize our knowledge about the constituents of an experiment. Hence, we do not need to work with the quantum system after the measurement and we are only interested in the probabilities that quantum mechanics predicts for the possible measurement outcomes. This is the reason why we describe the measurement part of an experiment by a positive operator valued measure (POVM).

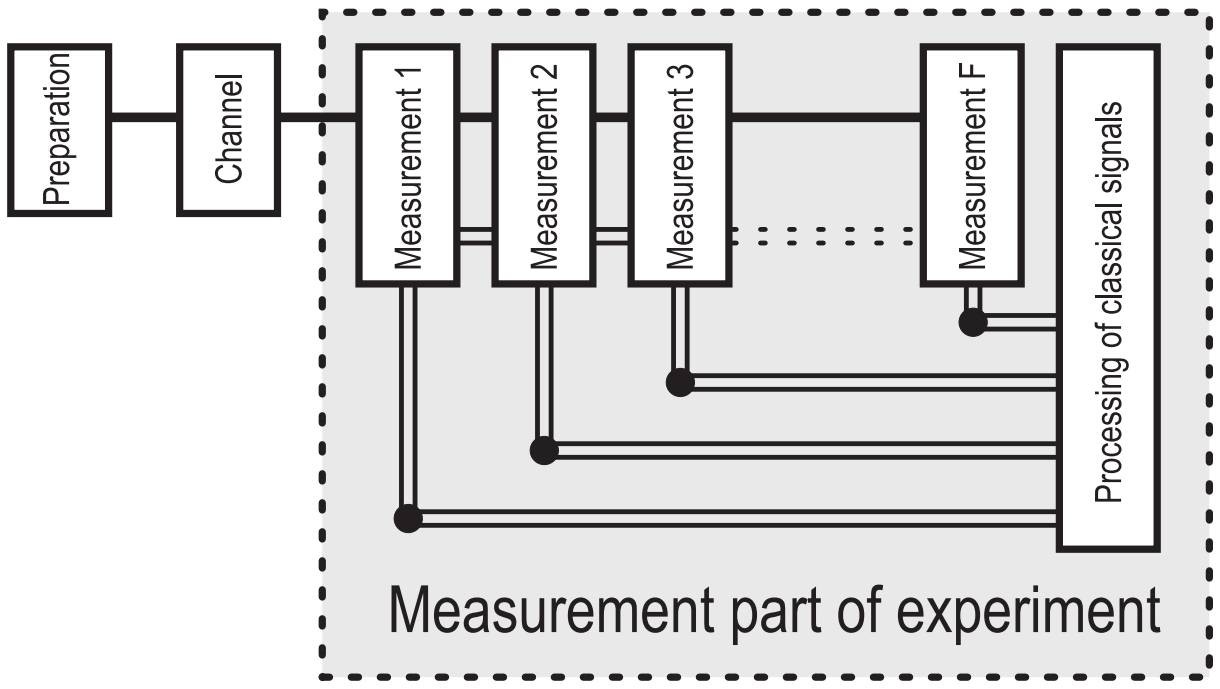

Figure 3.1. The schematical model of quantum experiment.

[4]In each run of the experiment the state after the evolution will be measured, so post selection can not be involved in the evolution part of the experiment.

### 3.1 Unambiguous discrimination problem

Having established all three constituents of a quantum experiment (preparation, evolution, and measurement) let us now define the tasks we want to study. Imagine we are given one of these constituents i.e. either a preparator or a quantum channel or a measurement apparatus. The considered device is equipped with an incomplete description of its function. We are given a question about the operation of the device which we should answer by using the device only once. Our answers are required to be unambiguous, that is error free, and also an inconclusive answer indicating failure is allowed. The goal is to design an experiment, which gives a correct conclusive answer as often as possible.

More formally, this task called the *unambiguous discrimination problem* (UDP), can be defined in the following way. Let the labels $\mathsf{P}$, $\mathsf{C}$, and $\mathsf{M}$ indicate the preparator, channel, and measurement, respectively. We will call $S_{\mathsf{X}}(\mathcal{H})$ for $\mathsf{X} \in \{\mathsf{P}, \mathsf{C}, \mathsf{M}\}$ the set of all possible constituents of type $\mathsf{X}$ for the Hilbert space $\mathcal{H}$. The incomplete knowledge about the constituent we investigate should be reflected by a probability measure $dA$ defined on $S_{\mathsf{X}}(\mathcal{H})$. Thus, $\int_{S_{\mathsf{X}}(\mathcal{H})} dA = 1$ and the measure $dA$ represents the probability density that the tested constituent is actually $A$. In the studied problem a valid question is such that its conceivable answers divide all possibly appearing elements of $S_{\mathsf{X}}(\mathcal{H})$ (i.e. support of $dA$) into a finite number $M$ of disjoint subsets $S_i$. Hence, it must hold that $S_i \bigcap S_j = \emptyset$ for $i \neq j$ and $\sum_{i=1}^{M} \int_{S_i} dA = 1$. Let $p_j(A)$ be the probability of concluding answer $j$ if constituent $A \in S_{\mathsf{X}}(\mathcal{H})$ is tested in the experiment we have designed. The probability of obtaining an inconclusive answer, denoted as $j = 0$, will be $p_0(A)$. The unambiguity of conclusive answers is mathematically stated by the following no error conditions:

$$\forall i, j \in \{1, \ldots, M\}, \quad i \neq j, \quad \forall A \in S_i; \quad p_j(A) = 0. \tag{3.1}$$

The objective is to design an experiment that maximizes

$$P_{succ} = \sum_{i=1}^{M} \int_{S_i} p_i(A) dA \ , \tag{3.2}$$

the probability of successfully answering the investigated question about the constituent.

One might ask what is the motivation for defining this slightly artificially looking UDP. Although it might not be obvious at first sight, but the definition of UDP covers all the unambiguous tasks discussed in the introduction. Thus, understanding some aspects of the unambiguous discrimination problem might give us some insight to a wide variety of unambiguous discrimination tasks. In the next section we will see that in finite dimensional Hilbert spaces UDP can be reformulated as unambiguous discrimination of $M$ known constituents of a given type. This suggests that investigation of discrimination of known states, channels, and measurements is certainly an important thing to study. Unfortunately, regardless of effort made by many authors, even the discrimination of two mixed states is still not completely solved. Much less is known for discrimination of more than two states or for discrimination of channels and observables. The investigation of state comparison and identification has lead to progress in discrimination of certain types of states and in a similar way the study of comparison and identification could help in discrimination of certain types of channels or observables. UDP allows us to see these tasks in a unified way and hopefully observe similarities and differences between the tasks for states, channels and observables more clearly.

### 3.2 Reformulation of UDP

Let us now consider a finite dimensional Hilbert space $\mathcal{H}$. We define the average constituents[5]

$$A_i = \frac{1}{\eta_i} \int_{S_i} A \, dA, \tag{3.3}$$

where the factor $\eta_i = \int_{S_i} dA$ guarantees that $A_i$ is a convex combination of constituents from $S_i$. Hence, $A_i$ is a valid state, channel or measurement. Probabilities predicted by quantum mechanics are linear with respect to states, channels and measurements, which allow us to rewrite $P_{succ}$ in terms of average constituents

$$P_{succ} = \sum_{i=1}^{M} \int_{S_i} p_i(A) dA = \sum_{i=1}^{M} p_i \left( \int_{S_i} A \, dA \right) = \sum_{i=1}^{M} \eta_i p_i(A_i). \tag{3.4}$$

Let us now integrate the original no error conditions from Eq. (3.1)

$$\forall i \neq j, \quad \forall A \in S_i : \quad p_j(A) = 0 \quad \Rightarrow \quad p_j(A_i) = \frac{1}{\eta_i} \int_{S_i} p_j(A) \, dA = 0. \tag{3.5}$$

We see that the no error conditions from Eq. (3.1) imply the no error conditions for unambiguous discrimination of the constituents $A_i$. The converse is also true, because if integral of non negative continuous function is zero then the integrated function must be zero in the integrated region. Hence UDP can be equivalently reformulated as unambiguous discrimination of average constituents $A_i$ appearing with prior probabilities $\eta_i$.

### 3.3 Mathematical apparatus for discrimination

Let us now sketch a general experiment for discrimination of a given type of constituent and briefly discuss its mathematical description. An experiment for discrimination of states inevitably starts by an action of the investigated preparator, which sets the system into one of the possible states $\rho_i$. Anything that happens afterwards (e.g. use of ancillary quantum systems, any kind of evolution, processing of measurement results) is incorporated into the measurement part of the experiment described by a positive operator valued measure. The most general POVM $\{E_i\}$ for unambiguous discrimination of $M$ states consists of $M+1$ elements. Without loss of generality we require that the observation of outcome $i \in \{1, \ldots, M\}$ indicates the preparation of state $\rho_i$ and the outcome 0 is inconclusive. The optimal measurement maximizes the probability of success $P_{succ} = \sum_{i=1}^{M} \eta_i p_i(\rho_i) = \sum_{i=1}^{M} \eta_i Tr(\rho_i E_i)$, while preserving normalization $\sum_{i=0}^{M} E_i = I$ and positivity ($E_i \geq 0$) of the POVM elements.

A general experiment for the discrimination of quantum channels is more complicated. Except for the principal quantum system exposed to the tested channel we have to consider also an ancillary system, whose evolution is incorporated into the preparation or measurement of the compound system. The class of such experiments is very broad, because the Hilbert space of an ancilla can be arbitrary. Fortunately, as Ziman [31] showed one can restrict the ancilla to have a Hilbert space isomorphic with the principal quantum system. Moreover, he showed that the experiment is uniquely described by a *process positive operator valued measure (PPOVM)* $\{M_i\}$. The tested channel $\mathcal{E}$ is equivalently represented via the Choi-Jamiolkowski isomorphism as a *process state* $\omega_{\mathcal{E}}$ and consequently the probability of observing the result $i$ reads $p_i(\mathcal{E}) = Tr(\omega_{\mathcal{E}} M_i)$. The most general PPOVM $\{M_i\}$ for unambiguous discrimination of $M$ channels consists of $M+1$ elements. We associate the observation of result $i \in \{1, \ldots, M\}$ with the use of channel $\mathcal{E}_i$ and declare the outcome 0 as inconclusive. The optimal measurement should maximize the probability of success $P_{succ} = \sum_{i=1}^{M} \eta_i p_i(\mathcal{E}_i) = \sum_{i=1}^{M} \eta_i Tr(\omega_i M_i)$, while preserving positivity of PPOVM elements $M_i$ and normalization $\sum_{i=0}^{M} M_i = \xi^T \otimes I$ with $\xi$ being a state of the principal quantum system. Thus, from the mathematical point of view, the optimization is similar to the discrimination of states except for the normalization of the operator measure. In particular, the choice of this normalization $\xi$ outlines the additional freedom, which complicates the optimization.

[5]Under some strange circumstances it may happen that $\int_{S_i} dA = 0$ for some $i$. We can remove these answers from the definition of UDP, because they are forbidden to appear ($\forall A \notin S_i \; p_i(A)dA = 0 \bigwedge \int_{S_i} dA = 0 \Rightarrow \int_{S_{X(\mathcal{H})}} p_i(A)dA = 0$) and the elements $A$ from $S_i$ can not spoil the unambiguity ($\int_{S_i} p_j(A)dA = 0$). Hence such sets $S_i$ do not influence the UDP.

The testing of measurements is a bit different from the experiments for channels and states, because the outcomes of the investigated measurement apparatus may not be directly linked to the results of the test. For example, imagine the discrimination of $M$ POVMs each having $N$ possible outcomes. For $N < M$, a single outcome of the tested measurement could not indicate each of the possibly used POVMs. Hence, the most general strategy uses the principal system measured by the tested POVM as well as an ancillary quantum system, whose measurement depends on the outcome of the tested POVM. One can show that it suffices to consider ancilla with the same Hilbert space as the principal quantum system. Unfortunately, a suitable mathematical framework for describing these type of experiments is not yet developed. Moreover, sometimes it may happen that we are not allowed to use any other measurement then the tested one. In such a situation the possible experiment consists of the preparation that we control and a tested measurement, whose outcomes can be linked to test results in many ways.

The following three chapters are devoted to states, channels and measurements. Each of them first summarizes the crucial known results on discrimination and then investigates the comparison and the identification of the given constituents. For states, these tasks are also studied for coherent states, which in this case do not allow the average constituent approach to the problem. Nevertheless, we are able to find and compare several solutions, which are also easily optically realizable.

## 4 Unambiguous tasks for states

### 4.1 Unambiguous discrimination of two mixed states

The aim of this section is to collect in one place the most useful constructive procedures proposed for unambiguous discrimination of a pair of general mixed states. The material we present further in this section is mainly adopted from the PhD thesis of P. Raynal [13], which provides a thorough review on unambiguous discrimination of two mixed states until year 2006. In years 2007 and 2008 other very general results were obtained by Matthias Kleinmann et. al. and we summarize them at the end of the section.

The notation we use is the following: The quantum system we are given is guaranteed to be either in the mixed state $\rho_1$ or in the mixed state $\rho_2$. This two possibilities appear with a priori probabilities $\eta_1,\, \eta_2 = 1 - \eta_1$, respectively. The POVM elements $E_1, E_2$ correctly identify states $\rho_1, \rho_2$, respectively and element $E_0$ correspond to an inconclusive result. The unambiguity of the measurement is ensured by fulfilling a pair of the no-error conditions: $0 = Tr(E_1\rho_2) = Tr(E_2\rho_1)$. We use the superscript *opt* to indicate that the *Unambiguous State Discrimination*(USD) measurement maximizes the probability of discrimination $P_D \equiv P_{succ} = \eta_1 Tr(E_1\rho_1) + \eta_2 Tr(E_2\rho_2)$. Due to validity of the no-error conditions this is equivalent to minimization of the probability of failure $Q = \eta_1 Tr(E_0\rho_1) + \eta_2 Tr(E_0\rho_2) = 1 - P_D$.

Let us note that it suffice to focus on a subspace $\mathcal{S}$ given by the span of the supports of the density matrices $\rho_1,\, \rho_2$. If we denote by $\Pi_{\mathcal{S}}$ the projector onto $\mathcal{S}$ then we have $Tr(E_k\rho_i) = Tr(E_k\Pi_{\mathcal{S}}\rho_i\Pi_{\mathcal{S}}) = Tr(E'_k\rho_i)$, where $E'_k = \Pi_{\mathcal{S}}E_k\Pi_{\mathcal{S}}$ is the part of the operator $E_k$ acting only on $\mathcal{S}$. Due to normalization of POVM $\{E_k\}$ we have $\sum_{k=0}^{2} E'_k = \Pi_{\mathcal{S}} \sum_{k=0}^{2} E_k\Pi_{\mathcal{S}} = \Pi_{\mathcal{S}}$, so $\{E'_k\}$ forms a POVM on a subspace $\mathcal{S}$ of the Hilbert space $\mathcal{H}$ ($E'_k$ are positive operators supported on $\mathcal{S}$ and sum up to identity on subspace $\mathcal{S}$). Hence, the no error conditions hold for $E'_k$ and the probability of discrimination stays the same as for the measurement $\{E_k\}$. This means that the search for the optimal measurement can be done in the smaller Hilbert space specified by the subspace $\mathcal{S}$ and any choice of POVM $\{E_k\}$ leading to $\{E'_k\}$ performs the unambiguous discrimination equally well. For $\mathcal{S} \neq \mathcal{H}$ there are infinitely many POVMs $\{E_k\}$ leading to $\{E'_k\}$. For example $E_k = E'_k + F_k$ with positive operators $F_k$ supported in $\mathcal{S}^{\perp}$ and summing to $I - \Pi_{\mathcal{S}}$ defines one such class. An USD measurement corresponding to $F_1 = F_2 = 0,\; F_0 = I - \Pi_{\mathcal{S}}$ is called *proper USD measurement*.

P. Raynal is the author of the following three reduction theorems, which can be used to simplify the problem by contracting its effective Hilbert space $\mathcal{H}$.

The first theorem tells us that the common subspace of the supports of the mixed states $\rho_1$, $\rho_2$ cannot be used for unambiguous discrimination and thus can be split off from the problem.

**Theorem 1** *Reduction Theorem for a Common Subspace*

*Suppose supports $S_{\rho_1}$ and $S_{\rho_2}$ have a non-empty common subspace $\mathcal{H}_{\bigcap}$. We denote by $\mathcal{H}'$ the orthogonal complement of $\mathcal{H}_{\bigcap}$ in $\mathcal{H}$ while $\Pi_{\mathcal{H}_{\bigcap}}$ and $\Pi_{\mathcal{H}'}$ denote respectively the projector onto $\mathcal{H}_{\bigcap}$ and $\mathcal{H}'$. Then the optimal USD measurement is characterized by POVM elements of*

*the form*

$$\begin{aligned} E_1^{opt} &= E_1'^{opt} \\ E_2^{opt} &= E_2'^{opt} \\ E_0^{opt} &= E_0'^{opt} + \Pi_{\mathcal{H}_\cap} \end{aligned} \tag{4.1}$$

*where the operators* $E_0'^{opt}, E_1'^{opt}, E_2'^{opt}$ *form a POVM* $E_k'^{opt}$ *with support on* $\mathcal{H}'$ *describing the optimal USD measurement of a reduced problem defined by*

$$\begin{aligned} \rho_1' &= \frac{1}{N_1}\Pi_{\mathcal{H}'}\rho_1\Pi_{\mathcal{H}'}, \quad \eta_1' = \frac{N_1\eta_1}{N}, \quad N_1 = Tr(\rho_1\Pi_{\mathcal{H}'}) \\ \rho_2' &= \frac{1}{N_2}\Pi_{\mathcal{H}'}\rho_2\Pi_{\mathcal{H}'}, \quad \eta_2' = \frac{N_2\eta_2}{N}, \quad N_2 = Tr(\rho_2\Pi_{\mathcal{H}'}) \\ & \qquad N = N_1\eta_1 + N_2\eta_2 \end{aligned} \tag{4.2}$$

*And finally, the optimal failure probability* $Q^{opt}$ *can be written in terms of* $Q'^{opt}$*, the optimal failure probability of the reduced problem, as*

$$Q^{opt} = 1 - N + NQ'^{opt}. \tag{4.3}$$

The second reduction theorem proves that $E_1^{opt}$, the conclusive element of the optimal measurement, is a projector on the part of the support of $\rho_1$, which is orthogonal to the support of $\rho_2$ and analogously for $E_2^{opt}$. Thus these parts of the supports can be eliminated and it suffices to vary the POVM elements on a smaller Hilbert space, which corresponds to USD of two mixed states with reduced rank.

**Theorem 2** *Reduction Theorem for Orthogonal Subspaces*

*Let us assume that supports* $S_{\rho_1}$ *and* $S_{\rho_2}$ *have no common subspace. Then one can construct a decomposition*

$$\mathcal{H} = \mathcal{H}' \oplus \mathcal{H}'^{\perp} \tag{4.4}$$

*with* $\mathcal{H}'^{\perp} = S_1^{\perp} + S_2^{\perp}$, $S_1^{\perp} = K_{\rho_1} \bigcap S_{\rho_2}$ *and* $S_2^{\perp} = K_{\rho_2} \bigcap S_{\rho_1}$*. The solution of the optimal USD measurement problem can be given, with help of* $\Pi_{S_1^{\perp}}$ *and* $\Pi_{S_2^{\perp}}$*, the projection onto* $S_1^{\perp}$ *and* $S_2^{\perp}$ *, respectively, in* $\mathcal{H} = \mathcal{H}' \oplus \mathcal{H}'^{\perp}$*, by*

$$\begin{aligned} E_1^{opt} &= E_1'^{opt} + \Pi_{S_2^{\perp}} \\ E_2^{opt} &= E_2'^{opt} + \Pi_{S_1^{\perp}} \\ E_0^{opt} &= E_0'^{opt} \end{aligned} \tag{4.5}$$

*the operators* $E_0'^{opt}, E_1'^{opt}, E_2'^{opt}$ *form a POVM* $E_k'^{opt}$ *with support on* $\mathcal{H}'$ *describing the optimal USD measurement of a reduced problem defined by*

$$\begin{aligned} \rho_1' &= \frac{1}{N_1}\Pi_{\mathcal{H}'}\rho_1\Pi_{\mathcal{H}'}, \quad \eta_1' = \frac{N_1\eta_1}{N}, \quad N_1 = Tr(\rho_1\Pi_{\mathcal{H}'}) \\ \rho_2' &= \frac{1}{N_2}\Pi_{\mathcal{H}'}\rho_2\Pi_{\mathcal{H}'}, \quad \eta_2' = \frac{N_2\eta_2}{N}, \quad N_2 = Tr(\rho_2\Pi_{\mathcal{H}'}) \\ & \qquad N = N_1\eta_1 + N_2\eta_2 \end{aligned} \tag{4.6}$$

*And finally, the optimal failure probability $Q^{opt}$ can be written in terms of $Q'^{opt}$, the optimal failure probability of the reduced problem, as*

$$Q^{opt} = NQ'^{opt}. \tag{4.7}$$

The third reduction theorem tells us that if we have an orthonormal basis in which matrices of $\rho_1$, $\rho_2$ are simultaneously block diagonal then it suffice to optimally solve the USD of two mixed states for each subblock.

**Theorem 3** *Reduction Theorem for two block diagonal density matrices*

*Suppose that $\rho_1$ and $\rho_2$ are block diagonal (in other words, there exists a set of orthogonal projectors $\Pi_k$ such that $\sum_{k=1}^n \Pi_k = I$ and $\rho_i = \sum_{k=1}^n \Pi_k \rho_i \Pi_k, i = 1, 2$. Then the optimal USD measurement can be chosen block diagonal where each block is optimal onto its restricted subspace.*

*More precisely, the optimal USD measurement is characterized by POVM elements of the form*

$$E_i^{opt} = \sum_{k=1}^{n} E_i^{k\,opt} \tag{4.8}$$

*For $k = 1, \ldots, n$, the operators $E_0^{k\,opt}, E_1^{k\,opt}, E_2^{k\,opt}$ form a POVM $E_j^{k\,opt}$ with support on $S_{\Pi_k}$ describing the optimal USD measurement of the reduced problem defined by*

$$\begin{aligned}
\rho_1^k &= \frac{1}{N_1^k}\Pi_k \rho_1 \Pi_k, \quad \eta_1^k = \frac{N_1^k \eta_1}{N^k}, \quad N_1^k = Tr(\rho_1 \Pi_k) \\
\rho_2^k &= \frac{1}{N_2^k}\Pi_k \rho_2 \Pi_k, \quad \eta_2^k = \frac{N_2^k \eta_2}{N^k}, \quad N_2^k = Tr(\rho_2 \Pi_k) \\
& \qquad\qquad\qquad N^k = N_1^k \eta_1 + N_2^k \eta_2
\end{aligned} \tag{4.9}$$

*And finally, the optimal failure probability can be written in terms of $Q^{k\,opt}$, the failure probability of the reduced problems, as*

$$Q^{opt} = \sum_{k=1}^{n} N_k Q_k^{opt}. \tag{4.10}$$

The problem one obtains after application of the above three theorem is called a Standard form of USD of two mixed states. The three reduction theorems are powerful in a sense that all previously (prior to P. Raynal thesis) solved cases of USD of two mixed states can be using these theorems reduced to unambiguous discrimination of two known pure states. Therefore, we remind the form of optimal measurement for this basic unambiguous discrimination problem.

### 4.1.1 Unambiguous discrimination of two known pure states

We are given one instance of a quantum system guaranteed to be either in a pure state $|\psi_1\rangle$ or in a pure state $|\psi_2\rangle$. The states $|\psi_1\rangle$, $|\psi_2\rangle$ are known to us and appear with a priori probabilities $\eta_1$, $\eta_2 = 1 - \eta_1$. The goal is to design a measurement, which *unambiguously* distinguishes the two possibilities with the highest possible probability. The existence of the inconclusive measurement element $E_0$ is implied by the fact that in quantum mechanics nonorthogonal quantum states cannot be perfectly distinguished, when only finite number of copies is provided. Thus the presence of inconclusive results is the price we are paying for the unambiguity of the measurement.

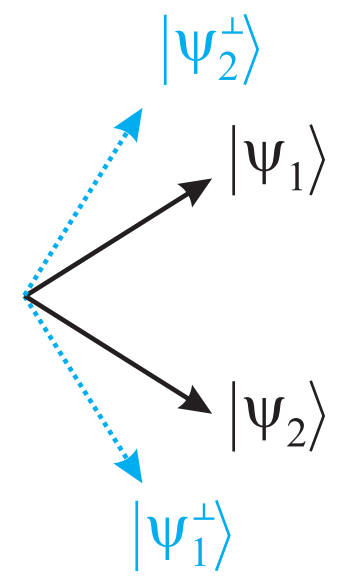

Figure 4.1. Measurement directions for the unambiguous discrimination of two known pure states.

This task was first formulated and solved for equal prior probabilities in works of Ivanovic, Dieks and Peres [6–8] in 1987. However the solution for arbitrary prior probabilities was obtained 8 years later by Jaeger and Shimony [9] in 1995.

The construction of the optimal measurement is relatively simple. First, one observes that the relevant part of the Hilbert space is only a plane in which states $|\psi_1\rangle$, $|\psi_2\rangle$ lay. This is because the support of the measurement also in the orthocomplement of the plane does not affect the overlap with $|\psi_i\rangle$ and implies the same positivity conditions for the POVM elements. Secondly, the requirement of unambiguity implies (see figure 4.1) $E_1 = c_1|\psi_2^\perp\rangle\langle\psi_2^\perp|$, $E_2 = c_2|\psi_1^\perp\rangle\langle\psi_1^\perp|$, where $c_i \in \mathbb{R}$, $|\psi_i^\perp\rangle$ is orthogonal to $|\psi_i\rangle$ and lays in the span of $|\psi_1\rangle$, $|\psi_2\rangle$. Finally, the maximization of the probability of discrimination $P_D$ can be done explicitly, because it is constrained only by the positivity of elements $E_k$ (k=0,1,2), which have rank at most two.

The final form of the optimal measurement depends on the relation of the overlap $\lambda = |\langle\psi_1|\psi_2\rangle|$ to the prior probability $\eta_1 = 1 - \eta_2$. For a given overlap of the states $\lambda$, there always exist three regimes:

- If $\eta_1 \in [0, \frac{\lambda^2}{1+\lambda^2}]$ the optimal measurement is a projective measurement, which either unambiguously identify state $|\psi_2\rangle$ or produce an inconclusive result:

$$E_1^{opt} = 0, \quad E_2^{opt} = |\psi_1^\perp\rangle\langle\psi_1^\perp|, \quad E_0^{opt} = |\psi_1\rangle\langle\psi_1|, \tag{4.11}$$

  The corresponding probability of discrimination is:

$$P_D = \eta_2(1-\lambda^2) = (1-\eta_1)(1-|\langle\psi_1|\psi_2\rangle|^2) \tag{4.12}$$

- If $\eta_1 \in [\frac{\lambda^2}{1+\lambda^2}, \frac{1}{1+\lambda^2}]$ the optimal measurement is a true POVM measurement, which con-

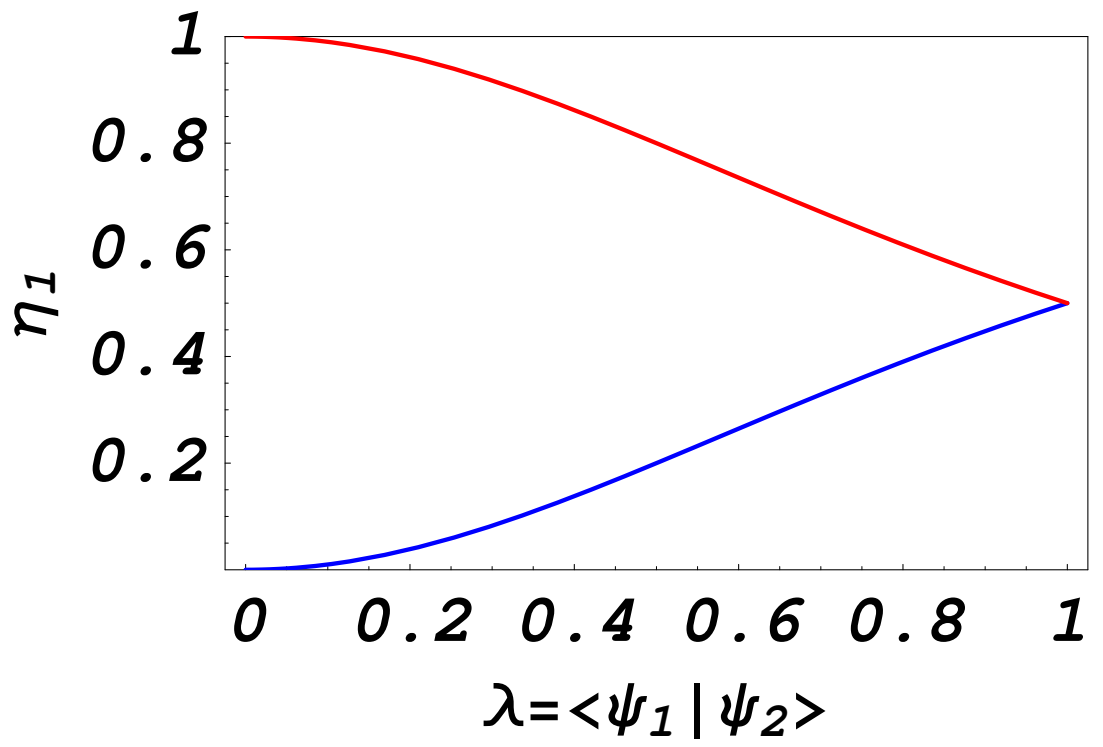

Figure 4.2. Dependence of prior probability $\eta_1$ transitions between the three regimes of the optimal POVM on the overlap of discriminated pure states $\lambda$.

clusively identify both states $\psi_1, \psi_2$:

$$E_1^{opt} = \frac{1-\sqrt{\frac{\eta_2}{\eta_1}}\lambda}{1-\lambda^2}|\psi_2^\perp\rangle\langle\psi_2^\perp|, \quad E_2^{opt} = \frac{1-\sqrt{\frac{\eta_1}{\eta_2}}\lambda}{1-\lambda^2}|\psi_1^\perp\rangle\langle\psi_1^\perp|, \qquad (4.13)$$

$$E_0^{opt} = I - E_1^{opt} - E_2^{opt}.$$

The resulting probability of discrimination is:

$$P_D = 1 - 2\sqrt{\eta_1\eta_2}\lambda = 1 - 2\sqrt{\eta_1\eta_2}|\langle\psi_1|\psi_2\rangle| \qquad (4.14)$$

- If $\eta_1 \in [\frac{1}{1+\lambda^2}, 1]$ the optimal measurement is a projective measurement, which either unambiguously identify state $|\psi_1\rangle$ or produce an inconclusive result

$$E_1^{opt} = |\psi_2^\perp\rangle\langle\psi_2^\perp|, \quad E_2^{opt} = 0, \quad E_0^{opt} = |\psi_2\rangle\langle\psi_2|, \qquad (4.15)$$

and gives the probability of discrimination:

$$P_D = \eta_1(1-\lambda^2) = \eta_1(1 - |\langle\psi_1|\psi_2\rangle|^2) \qquad (4.16)$$

The position of the borders between the three regimes depending on the overlap of the states $\lambda$ is depicted on figure 4.2.

Let us now come back to the unambiguous discrimination of two general mixed states and summarize some of the very general results found in the last two years by Kleinmann, Kampermann, and Bruss. We will use notion of the weighted density operators $\gamma_\mu = \eta_\mu\rho_\mu$ $\mu = 1, 2$. As we already illustrated the subspace orthogonal to supports of discriminated states $\rho_1, \rho_2$ provides freedom in the choice of the optimal measurement. However, one might ask whether at least the part of the measurement operators ($E'_k$) acting only on $\mathcal{S}$, the span of the supports of $\rho_1, \rho_2$, is uniquely determined. For this purpose we can work without loss of generality with the proper USD measurements. This is because any USD measurement $\{E_k\}$ can be turned into proper USD measurement with the same operators $E'_k$ and the same probability of success. Kleinmann [15] shows that a POVM $\{E_k\}$ is a proper USD measurement if and only if $E_0$ acts as identity on $\mathcal{S}^\perp$, $E_0 \geq 0, I - E_0 \geq 0$, and $\gamma_1(I - E_0)\gamma_2 = 0$. Hence, the inconclusive element $E_0$ uniquely determines the proper USD measurement.

For proper USD measurements the first two Raynal's reduction theorems are show [15] to commute and to be idempotent. Hence, only one application of each of them is needed. Moreover, the application of this two theorems can be done using Jordan basis of the supports of $\gamma_1, \gamma_2$. The supports of the reduced states (denoted $S_{\rho_1}, S_{\rho_2}$) are *strictly skew*, i.e. $S_{\rho_1} \bigcap S_{\rho_2}^\perp = \{0\}$, $S_{\rho_2} \bigcap S_{\rho_1}^\perp = \{0\}$, and do not have common subspace $S_{\rho_1} \bigcap S_{\rho_2} = \emptyset$.

In general it is not known how to check whether the reduced states have some common block diagonal structure. Nevertheless, in [14] Kleinmann et. al. shows that for operators without common part of support two dimensional block diagonal structure exists if and only if $[\gamma_1, \gamma_1\gamma_2\gamma_1] = 0$, $[\gamma_2, \gamma_2(\gamma_1)^2\gamma_2] = 0$, and $[\gamma_1, \gamma_1(\gamma_2)^2\gamma_1] = 0$. If these commutators vanish then they provide a method to construct Jordan basis, which using the third Raynal's reduction theorem splits the problem into several independent discriminations of two known pure states. Each of these subproblems has a unique solution (see Section 4.1.1) as well as the whole optimal USD measurement that it forms.

However, proof of the uniqueness in general situation requires a different approach. Kleinmann et. al. succeeded to rewrite the necessary and sufficient conditions on the optimality of USD measurement by Eldar, Stojnic, and Hassibi [17] into operational form. If we denote by $\Lambda_1$ the projector onto $ker\gamma_2 \bigcap \mathcal{S}$ and by $\Lambda_2$ the projector onto $ker\gamma_1 \bigcap \mathcal{S}$ then the rewritten optimality conditions for proper USD measurement read:

$$(\Lambda_1 - \Lambda_2)E_0(\gamma_2 - \gamma_1)(\Lambda_1 + \Lambda_2) \geq 0 \tag{4.17}$$

$$(\Lambda_1 - \Lambda_2)E_0(\gamma_2 - \gamma_1)(I - E_0) = 0 \tag{4.18}$$

It turns out that Eq. (4.18) in a compact way expresses the following three conditions:

$$\Lambda_1 E_0(\gamma_2 - \gamma_1)E_0\Lambda_1 \geq 0 \tag{4.19}$$

$$\Lambda_2 E_0(\gamma_1 - \gamma_2)E_0\Lambda_2 \geq 0 \tag{4.20}$$

$$\Lambda_1 E_0(\gamma_2 - \gamma_1)E_0\Lambda_2 = 0 \tag{4.21}$$

An important consequence of the optimality conditions is the fact that optimal $E_0$ and hence also the whole optimal and proper USD measurement is completely determined by $\Pi_0$ the projector onto the support of $E_0$. Another thing needed in the proof of the uniqueness is the knowledge of the rank of the POVM element $E_0$. For optimal proper USD measurements this can be shown to be $\mathrm{rank}E_0 = \mathrm{rank}\gamma_1\gamma_2 + \dim\ker(\gamma_1 + \gamma_2)$. Let us consider two optimal proper USD measurements $\{E_k\}$, $\{\overline{E}_k\}$. Due to linearity of the probability rule and convexity of measurements POVM $\{\frac{1}{2}(E_k + \overline{E}_k)\}$ is also optimal. However, all these three measurements must have the same rank of the inconclusive POVM element. Positivity of operators $E_0$, $\overline{E}_0$ implies that this is possible only if all the three inconclusive elements have the same support given by projector $\Pi_0$. Since $\Pi_0$ completely determines the POVM one concludes that the optimal and proper USD measurement is unique.

Rank of the conclusive POVM elements $E_1, E_2$ can be used to classify the optimal USD measurement. For operators $\gamma_1, \gamma_2$ with strictly skew supports the ranks $e_1 = \text{rank} E_1$, $e_2 = \text{rank} E_2$ obey (see [15]) inequalities $e_1 \leq r$, $e_2 \leq r$, $r \leq e_1 + e_2 \leq 2r$, where $r = \text{rank}\gamma_1 = \text{rank}\gamma_2$. Among these $(r+1)(r+2)/2$ possibilities for $(e_1, e_2)$ three are better understood. Measurement types $(0, r)$ and $(r, 0)$ correspond to the so called *single state detection*, where only state $\rho_2$ or state $\rho_1$, respectively, is unambiguously detected by projective measurement. For fixed $\rho_1, \rho_2$ single state detection is always optimal for a prior probability $\eta_1$ or $\eta_2$ sufficiently small[6].

Measurement type $(r, r)$ corresponds to the so called *fidelity form measurement*. The name comes from the Uhlmann's fidelity

$$F(\rho_1, \rho_2) = Tr(|\sqrt{\rho_1}\sqrt{\rho_2}|) = Tr(\sqrt{\sqrt{\rho_1}\rho_2\sqrt{\rho_1}}),$$

which is closely related to this measurement. An upper bound on the probability of success $P_D \leq 1 - 2\sqrt{\eta_1\eta_2}|\langle\varphi_1|\varphi_2\rangle|$ follows from considering USD measurement among maximally overlapping purifications $|\varphi_1\rangle, |\varphi_2\rangle$ of states $\rho_1, \rho_2$. Uhlmann's fidelity is equal to the overlap $|\langle\varphi_1|\varphi_2\rangle|$, so we have $P_D \leq 1 - 2\sqrt{\eta_1\eta_2}\, F(\rho_1, \rho_2) = 1 - 2\, Tr(\sqrt{\sqrt{\gamma_1}\gamma_2\sqrt{\gamma_1}})$. It can be shown [15], [32], [33] that this bound is tight[7] if and only if $\gamma_1 \geq \sqrt{\sqrt{\gamma_1}\gamma_2\sqrt{\gamma_1}}$ and $\gamma_2 \geq \sqrt{\sqrt{\gamma_2}\gamma_1\sqrt{\gamma_2}}$. In such case the inconclusive POVM element of optimal and proper USD measurement has the following form:

$$E_0 = I - (\gamma_1 + \gamma_2)^{-}\{\sqrt{\gamma_1}(\gamma_1 - F_1)\sqrt{\gamma_1} + \sqrt{\gamma_2}(\gamma_2 - F_2)\sqrt{\gamma_2}\}(\gamma_1 + \gamma_2)^{-},$$

where $(\gamma_1 + \gamma_2)^{-}$ denotes the inverse of $\gamma_1 + \gamma_2$ on its support and $F_1 = \sqrt{\sqrt{\gamma_1}\gamma_2\sqrt{\gamma_1}}$, $F_2 = \sqrt{\sqrt{\gamma_2}\gamma_1\sqrt{\gamma_2}}$. Let us note that for fixed $\rho_1, \rho_2$ the region of $\eta_1$, in which the fidelity form measurement is optimal, might be empty. For USD of two known pure states this region covers the whole interval between the two single state detections. Hence, the aforementioned types of USD measurements describe the whole solution from Section 4.1.1 originally found by Jaeger and Shimony. On the other hand, already for $r = \text{rank}\gamma_1 = \text{rank}\gamma_2 = 2$ there are weighted density operators $\gamma_1, \gamma_2$ with strictly skew supports for which none of the aforementioned types of USD measurement is optimal. Kleinmann et. al. applied their optimality criteria and for each measurement type belonging to $r = 2$ reduced potentially optimal POVMs to a finite number of candidates. Because of the uniqueness of the optimal POVM only one of the candidates really exists and forms a valid USD measurement. Construction of the candidates is quite technical and involves solving of high degree polynomials. For $\text{rank}\ \gamma_1 = r > 2$ the measurement types were not yet investigated. Unfortunately, they are not expected to give some clear insight into the problem.

### 4.2 Quantum state comparison

In the classical world it is relatively easy to compare (quantitatively, as well as qualitatively) features of physical systems and to conclude with certainty whether two systems exhibit the same properties, or not. On the other hand, the statistical nature of the quantum theory restricts our ability to provide deterministic conclusions/predictions even in the simplest experimental situations. Therefore comparison of quantum states is different compared to classical situation. To be specific, let us consider that we are given two independently prepared quantum systems of the same physical origin (e.g., two photons). Our task is to determine unambiguously whether these two photons have been prepared in the same polarization state. That is, we want to compare the two states and we want to know whether they are identical or not. If we have just a single copy of each state and no further information then the scenario according to which we first measure each state does not work. For that we would need an infinite ensemble of identically prepared systems. In this case also all other strategies would fail, because our knowledge about the states is insufficient [34]. Simply if each photon can also be in any mixed state, then it is impossible to test equality of states of the two photons. However, there are often situations in which we know something more about the states we need to compare. For example, we might know that each photon was prepared in the pure state. Barnett, Chefles, and Jex [19] were the first who considered this kind of scenario for two qudits. We shall call it *comparison of unknown states*. Here unknown means that there is still a continuum of possibilities for each of the compared states. In the same paper Barnett *et al.* discuss also different kind of prior information we might have about the compared states. We call the scenario *comparison of states chosen from a finite set*, because the compared states are known to be randomly chosen from a finite set of possible states $|\varphi_i\rangle$ with probability $q_i$. These two scenarios have distinct features, therefore we discuss them separately.

[6]Region in which single state detection is optimal can be calculated exactly (see [15]) thanks to the optimality conditions (4.17),(4.18)

[7]Here we still assume strictly skew supports of $\rho_1, \rho_2$

### 4.2.1 Comparison of states chosen from a finite set

The simplest version of this problem is a comparison of two quantum systems, each of them guaranteed to be prepared either in state $|\varphi_1\rangle$ or in state $|\varphi_2\rangle$ with equal prior probabilities $q_1 = q_2 = 1/2$. In this case two USD measurements can be used to determine the state of each individual system. This strategy fails if one or both outcomes are inconclusive. Otherwise, we can unambiguously conclude equality as well as difference of the compared states. Barnett *et al.* proposed and proved optimality of the strategy that measures simultaneously both systems and succeeds more often. It unambiguously detects both equality and difference of the states, but in contrast to the former strategy it does not tell us in which states the systems were. However, for this problem the emergence of inconclusive results is unavoidable and it is the price we are paying for the unambiguity of the conclusions. Rudolph, Spekkens, and Turner reformulated this comparison problem as discrimination the following two mixed states

$$\begin{aligned}\rho_1 &= \frac{(q_1)^2}{(q_1)^2+(q_2)^2}|\varphi_1,\varphi_1\rangle\langle\varphi_1,\varphi_1| + \frac{(q_2)^2}{(q_1)^2+(q_2)^2}|\varphi_2,\varphi_2\rangle\langle\varphi_2,\varphi_2| \\ \rho_2 &= \frac{1}{2}|\varphi_1,\varphi_2\rangle\langle\varphi_1,\varphi_2| + \frac{1}{2}|\varphi_2,\varphi_1\rangle\langle\varphi_2,\varphi_1| \end{aligned} \tag{4.22}$$

appearing with prior probabilities $\eta_1 = (q_1)^2 + (q_2)^2$, $\eta_2 = 2q_1q_2$ respectively. They considered $q_1 = q_2$ and using their upper and lower bounds on USD of mixed states confirmed the optimality of the solution by Barnett *et al.*. Later, Kleinmann *et. al.* [35] for arbitrary $q_1, q_2$ used Raynal's reduction theorems to reduce the problem to the discrimination of two known pure states. Applications of this comparison problem in quantum information processing are e.g. quantum fingerprinting [36] and quantum digital signatures [37].

The problem was generalized in two ways. First, the extension to more quantum systems in pure states was studied by Chefles et. al. [38]. While for two systems their states can be either equal or different for more systems there are more possibilities. For example, one might ask whether are all systems in the same state, called *identicality confirmation*, or there is at least one different state. Similarly, we might want to know whether all the states of the $N$ compared systems are different from each other or at least one pair of states is the same.

Chefles *et al.* showed that identicality confirmation is possible[8] if and only if the set of possible states $\{|\varphi_i\rangle\}_{i=1}^{M}$ is linearly independent. Let us note that linear independence of the possible states allows *incoherent* strategy, which determines each of the compared states by USD measurement. Certainly, if no inconclusive outcome arise then the individual measurement outcomes fix the conclusion. For two possible pure states ($M = 2$) and arbitrary $N$ Kleinmann *et. al.* [35] found the optimal solution for arbitrary prior probabilities $q_1, q_2$. They have used Raynal's reduction theorems to turn the problem into USD of two mixed states of rank 2. Unfortunately, except for $q_1 = q_2$, where the problems splits into two USD of two known pure states the solution is cumbersome and we will not discuss it here.

Surprisingly, Chefles *et al.* in [38] found that linear independence restriction does not apply if we want to unambiguously confirm that not all states are the same. This is exactly the point that is used in comparison of unknown states. There we will present a strategy proposed by Chefles, who also proved its optimality for wide range of circumstances.

Similar requirements as those above apply also to confirmation that all the compared systems are in different states. Let us note that this task is meaningful only if $M \geq N$ i.e. there are at least as many possible states as there are quantum systems. As Chefles *et.al.* in [38] shows unambiguous confirmation of all states of the compared systems being different is possible[9] if and only if any $N$ element subset of the set of possible states $\{|\varphi_i\rangle\}_{i=1}^{M}$ is linearly independent. Also universal strategy exists for this task, which means that the measurement has nonzero probability to detect that all systems are in different states for any $N$ tuple of linearly independent states. If any product state can enter the measurement then it is optimal (for details see [38]) to project onto totally antisymmetric subspace $\mathcal{S}_{anti}$ of the Hilbert space of $N$ compared quantum systems. If the projection succeeds then we conclude unambiguously that the states were linearly independent and hence all mutually different. This happens with probability equal to $\frac{1}{N!}\det\Gamma$, where $\Gamma$ is Gram matrix of the overlaps of the $N$ compared states. On the other hand, projection onto $\mathcal{S}_{anti}^{\perp}$ is inconclusive.

Except for looking on more than two quantum systems one can generalize comparison of states chosen from a finite set also by allowing finite sets of mixed states $\varphi_i$. Kleinmann *et.al.* considered this task in [35]. They showed that the identicality of states can be confirmed if and only if $\mathcal{S}_{\varphi_i} \nsubseteq \sum_{k\neq i}\mathcal{S}_{\varphi_k}$, where $\mathcal{S}_{\varphi_i}$ denotes the support of mixed state $\varphi_i$.

The *unambiguous* state comparison as introduced by Barnett *et al.* is a positive-operator-value-measure (POVM) measurement that has two possible outcomes associated with the two answers: the two states are different, or outcome of the measurement corresponds to an inconclusive answer.

[8] Here all N-tupples of states $|\varphi_i\rangle^{\otimes N}$ are required to have nonzero probability of identicality confirmation.

[9] Here all such N-tupples of mutually different states are expected to be detectable with nonzero probability.

### 4.2.2 Comparison of unknown states

In this section we consider comparison of states of quantum systems, which are known to be pure. Apart from that we have no other information that would further restrict the set of possible states. Hence, there is a continuum of possible states for each of the compared systems. It is important to know that there is a nonzero prior probability $\eta_{same}, \eta_{diff}$ of compared systems being in the same or different states respectively. Otherwise, the conclusion of the comparison would be clear just from our prior knowledge and no measurement would be needed. The first who formulated and solved this task were Barnett, Chefles, and Jex [19]. They considered two qudits each of them prepared in the arbitrary pure state. The task was to determine unambiguously whether the qudits are in same or different states.

We shall discuss slightly more general version of the task solved by Chefles *et al.* in [38], where $N$ instead of two systems are compared and we have to decide whether are all systems in the same state or not. More precisely we want to either unambiguously confirm identicality of the compared states or prove existence of at least one difference among them. We will rephrase this task in the spirit of unambiguous discrimination problem defined in Chapter 3. In this case state $\rho$ of the $N$ compared systems plays the role of the tested constituent $A$. Our prior knowledge about the possible constituents/states is the following: any pure product state $\rho$ can emerge and with probability $\eta_{same}$ state $\rho = |\psi\rangle\langle\psi|^{\otimes N}$ for some $|\psi\rangle \in \mathcal{H}$. Hence, the set of possibly emerging states splits into two subsets $S_1$, $S_2$ that correspond to all states being identical and some states being different, respectively. Moreover, the probability measure $dA$ describing the occurrence of the compared states should be uniform on the set $S_i$ $(i = 1, 2)$, because we have no further information, which would make some states from the set $S_i$ more favoured. This determines $dA$ to be $\eta_{same} d\psi$ on $S_1$ and $\eta_{diff} d\psi_1 \dots d\psi_N$ on $S_2$. Consequently, we can calculate the average states/constituents to be

$$A_1 \equiv \rho_1 = \frac{1}{\eta_{same}} \int_{S_1} A dA = \int |\psi\rangle\langle\psi|^{\otimes N} d\psi \tag{4.23}$$

$$A_2 \equiv \rho_2 = \frac{1}{\eta_{diff}} \int_{S_2} A dA = \frac{1}{\eta_{diff}} \int \psi_1 \otimes \dots \otimes \psi_N \; \eta_{diff} d\psi_1 \dots d\psi_N$$

$$= \left( \int |\psi\rangle\langle\psi| d\psi \right)^{\otimes N} = \frac{1}{d^N} I \tag{4.24}$$

where we used $\eta_{same} = \int_{S_1} dA$, $\eta_{diff} = \int_{S_2} dA$ and abbreviated $\psi_k \equiv |\psi_k\rangle\langle\psi_k|$. Furthermore, as it is shown for example in [39] the integral in Eq. 4.23 is equal to $\frac{1}{\binom{d+N-1}{d-1}} P^{sym}_{1\dots N}$, where $P^{sym}_{1\dots N}$ is the projector onto the totally symmetric subspace of $\mathcal{H}^{\otimes N}$. The considered version of comparison problem is thus equivalent to USD of mixed states $\rho_1 = \frac{N!(d-1)!}{(d+N-1)!} P^{sym}_{1\dots N}$, $\rho_2 = \frac{1}{d^N} I$ appearing with prior probabilities $\eta_{same}, \eta_{diff}$, respectively. Kernel of $\rho_2$ is a trivial subspace $\{0\}$, which implies that state $\rho_1$ can not be unambiguously detected. This means that it is impossible to unambiguously confirm identicality of the compared states. This impossibility is suggested also by the linear dependence of the possible states that form a continuum of all pure states in $\mathcal{H}$. Hence, $E_1 = 0$ and the Raynal reduction theorems quickly tell us that for $\eta_{same} \neq 0$ optimal POVM has $E_2 = I - P^{sym}_{1\dots N}$, $E_0 = P^{sym}_{1\dots N}$. Simply the totaly symmetric subspace is common support of $\rho_1, \rho_2$ and its orthocomplement is due to the second Raynal's reduction theorem optimally used to detect state $\rho_2$ i.e. to indicate existence of at least one difference among

the compared states. On average the success probability reads (see Eq. 3.4)

$$P_{succ} = \eta_{same} Tr(E_1\rho_1) + \eta_{diff} Tr(E_2\rho_2) = \eta_{diff}(1 - \frac{1}{d^N} Tr(P^{sym}_{1\ldots N})). \quad (4.25)$$

It depends on the ratio of dimensions of the totaly symmetric subspace $\binom{d+N-1}{d-1}$ and the whole Hilbert space $\mathcal{H}^{\otimes N}$. We may also ask what is the probability of detecting at least one difference for particular states $|\psi_1\rangle, \ldots |\psi_N\rangle$ entering the comparison measurement. Chefles *et. al.* showed that this conditional probability is given by the permanent of $\Gamma$ the Gram matrix of overlaps of the compared states

$$P(|\psi_1\rangle, \ldots |\psi_N\rangle) = Tr(E_2 \psi_1 \otimes \ldots \otimes \psi_N) = 1 - \frac{1}{N!} \mathrm{per}\Gamma, \quad (4.26)$$

where $\mathrm{per}\,\Gamma = \sum_{\sigma \in S(N)} \Gamma_{1\sigma(1)} \ldots \Gamma_{N\sigma(N)} = \sum_{\sigma \in S(N)} \langle\psi_1|\psi_{\sigma(1)}\rangle \ldots \langle\psi_N|\psi_{\sigma(N)}\rangle$. Moreover, Chefles *et.al.* in [38] showed that the above measurement is optimal for detection of at least one difference whenever all states $|\psi\rangle\langle\psi|^{\otimes N}$ $|\psi\rangle \in \mathcal{H}$ can appear. This can be seen also through average states/constituents. If all systems can be identically prepared in any pure state $|\psi\rangle$ then state $\rho_1$ has the same kernel whatever is the measure $dA$ on the set $S_1$. Only the kernel of $\rho_1$ can be used for detection of at least one difference[10], hence $E_2 = I - P^{sym}_{1\ldots N}$ obviously maximizes $Tr(E_2\rho_2)$ for any $\rho_2$.

### 4.2.3 Comparison of two ensembles of pure states

The aim of this part of the paper is to find the optimal unambiguous state comparison procedure in the case we have more copies of the two quantum states which we need to compare. The compared states are guaranteed to be pure and to belong to a $d$-dimensional Hilbert space $\mathcal{H}$. The dimensionality of the Hilbert space is known, otherwise the only information about the states is the probability of them being the same $\eta_{same} \neq 0$. In more physical terms this means that we have two preparators $A$ and $B$. First preparator produce the state $|\psi_1\rangle$, while the second produces state $|\psi_2\rangle$. Suppose we are given $k$ copies of states produced by the preparator $A$ and $l$ copies of states originated from the second preparator $B$. We want to decide whether the states prepared by the preparators $A$ and $B$ are the same, or different. Thus, want to distinguish between two sets of states $S_1 = \{|\psi\rangle_A^{\otimes k} \otimes |\psi\rangle_B^{\otimes l} : \psi \in \mathcal{H}\}$ and $S_2 = \{|\psi_1\rangle_A^{\otimes k} \otimes |\psi_2\rangle_B^{\otimes l} : |\psi_1\rangle \neq |\psi_2\rangle\}$. The first set corresponds to the situation when the two preparators prepare the same (though unknown) states, while the second set corresponds to the situation when the prepared states are different.

[10]The requirement can be mathematically stated as $E_2 \leq I - P^{sym}_{1\ldots N}$ and the claim follows using any pure state decomposition of $\rho_2$

In the same way as in the previous section we can calculate the average states to be:

$$\begin{aligned}A_1 \equiv \rho_1 &= \frac{1}{\eta_{same}}\int_{S_1} A dA = \int |\psi\rangle\langle\psi|^{\otimes k+l} d\psi \\ &= \frac{1}{d_{k+l}} P^{sym}_{1\ldots k+l} \qquad (4.27)\\ A_2 \equiv \rho_2 &= \frac{1}{\eta_{diff}}\int_{S_2} A dA = \frac{1}{\eta_{diff}}\int \psi_1^{\otimes k}\otimes\psi_2^{\otimes l}\ \eta_{diff} d\psi_1 d\psi_2 \\ &= \int |\psi\rangle\langle\psi|^{\otimes k} d\psi \otimes \int |\psi\rangle\langle\psi|^{\otimes l} d\psi \\ &= \frac{1}{d_k d_l} P^{sym}_{1\ldots k}\otimes P^{sym}_{k+1\ldots k+l}\ , \qquad (4.28)\end{aligned}$$

where we abbreviated by $d_k \equiv \binom{d+k-1}{d-1}$ the dimension of the symmetric subspace of $k$ systems. Also for more copies of the two compared states it is not possible to unambiguously conclude that the compared states are the same, because support of $\rho_1$ is included in the support of $\rho_2$ i.e. $P^{sym}_{1\ldots k+l} \leq P^{sym}_{1\ldots k}\otimes P^{sym}_{k+1\ldots k+l}$. Thus, $E_1 = 0$ and the comparison procedure succeeds with the probability of detecting the difference of the compared states $P_{succ} = \eta_{diff} Tr(E_2\rho_2)$. The measurement for detecting of at least one difference, discussed in the previous section, is optimal also for this task, because it optimally reveals any dissimilarity from the totally symmetric states of $N = k+l$ systems. Raynal's reduction theorems lead to the same proper USD measurement[11]

$$E_1 = 0, \quad E_2 = I - P^{sym}_{1\ldots N}, \quad E_0 = P^{sym}_{1\ldots N}. \qquad (4.29)$$

The optimal measurement can be derived also directly by optimizing the probability of success under the constraint of unambiguity and preservation of positivity and normalization of the POVM elements. We present this derivation, because it will guide us in the case when the average constituent approach can not be used. For simplicity we will omit the indexes $A, B$ in the rest of the section.

In order to construct the optimal POVM for detecting the difference of the compared states we first use the unambiguity requirement expressed by the (no-error) condition

$$\forall|\psi\rangle \in \mathcal{H}, \quad Tr[E_2(|\psi\rangle\langle\psi|)^{\otimes k+l}] = 0\,. \qquad (4.30)$$

that guarantees that whenever we obtain the result $E_2$ we can conclude that the states were indeed different. Integrating uniformly over all pure states $S_{pure} = \{|\psi\rangle \in \mathcal{H}\}$ we obtain an equivalent no-error condition that reads

$$0 = \int_{S_{pure}} d\psi \mathrm{Tr}\Big[E_2(|\psi\rangle\langle\psi|)^{\otimes k+l}\Big] = \mathrm{Tr}[E_2\Delta]\,, \qquad (4.31)$$

where

$$\Delta = \int_{S_{pure}} d\psi (|\psi\rangle\langle\psi|)^{\otimes k+l} = \frac{1}{\binom{k+l+d-1}{d-1}} P^{sym}_{1\ldots N}, \qquad (4.32)$$

[11]There is freedom provided by the trivial subspace and amounts to division of $I - P^{sym}_{1\ldots k}\otimes P^{sym}_{k+1\ldots k+l}$ among $E_i$.

and $P^{sym}_{1...N}$ is the projector onto a completely symmetric subspace of $\mathcal{H}^{\otimes(k+l)}$. Because of the positivity of the operators $E_2$ and $\Delta$ the equation (4.31) implies that these two operators have orthogonal supports (see Lemma 1 in Chapter 2). Hence, the biggest support operator $E_2$ can have is the orthogonal complement to the support of $\Delta$. POVM element $E_2$ is an effect ($0 \leq E_2 \leq I$) so the latter requirement can be written as $E_2 \leq I - P^{sym}_{1...N}$. The average success probability of detecting the difference between the compared states can be written as

$$\begin{aligned} P_{succ} \equiv P_{succ}(k,l) &= \eta_{same}.0 + \eta_{diff} \int_{S_{pure}} \int_{S_{pure}} d\psi_1 d\psi_2 P(|\psi_1\rangle, |\psi_2\rangle), \\ P(|\psi_1\rangle, |\psi_2\rangle) &= \langle\psi_1|^{\otimes k} \otimes \langle\psi_2|^{\otimes l} E_2 |\psi_1\rangle^{\otimes k} \otimes |\psi_2\rangle^{\otimes l}, \end{aligned} \tag{4.33}$$

The requirement on the support of $E_2$ implies inequality $\langle\Psi|E_2|\Psi\rangle \leq \langle\Psi|I - P^{sym}_{1...N}|\Psi\rangle$ $\forall|\Psi\rangle \in \mathcal{H}^{\otimes k+l}$, which is saturated if $E_2 = I - P^{sym}_{1...N}$. This choice obviously maximizes $P(|\psi_1\rangle, |\psi_2\rangle)$ $\forall|\psi_1\rangle, |\psi_2\rangle \in \mathcal{H}$ and consequently also $P_{succ}(k,l)$. Thus, the optimal comparison of $k$ and $l$ copies of unknown pure states prepared by the two preparators is accomplished by the projective measurement given in Eq. (4.29).

In what follows we calculate the probability of revealing the difference of the states $|\psi_1\rangle$, $|\psi_2\rangle$ measured by the optimal comparator, i.e.

$$\begin{aligned} P(|\psi_1\rangle, |\psi_2\rangle) &= \mathrm{Tr}[(I - P^{sym}_{1...N})|\Psi\rangle\langle\Psi|] \\ &= 1 - \langle\Psi|\Psi_S\rangle , \end{aligned} \tag{4.34}$$

where $|\Psi\rangle \equiv |\psi_1\rangle^{\otimes k} \otimes |\psi_2\rangle^{\otimes l}$ and

$$|\Psi_S\rangle \equiv P^{sym}_{1...N}|\Psi\rangle = \frac{1}{(k+l)!} \sum_{\sigma \in S(k+l)} \sigma(|\Psi\rangle) . \tag{4.35}$$

In the above formulas we denoted by $S(N)$ a group of permutations of $N$ elements and $\sigma(|\Psi\rangle)$ denotes the state $|\Psi\rangle$ in which subsystems have been permuted via the permutation $\sigma$. For example, a permutation $\nu_k$ exchanging only the $k$-th and the $(k+1)$-th position acts as

$$\nu_k(|\Psi\rangle) = |\psi_1\rangle^{\otimes k-1}|\psi_2\rangle|\psi_1\rangle|\psi_2\rangle^{\otimes l-1} . \tag{4.36}$$

The state $|\Psi\rangle$ has $N$ subsystems defining $N$ positions, which are interchanged by the permutation $\sigma$. Let us denote by $N_1$ the subset of the first $k$ positions (originally copies of $|\psi_1\rangle$) and by $N_2$ the remaining $l$ positions (originally occupied by systems in the state $|\psi_2\rangle$). For our purposes it will be useful to characterize each permutation $\sigma \in S(k+l)$ by the number of positions $m$ in the subset $N_1$ occupied by subsystems originated from the subset $N_2$. Literally, $m(\sigma)$ is the number of states $|\psi_2\rangle$ moved into the first $k$ subsystems ($N_1$) by the permutation $\sigma$ acting on the state $|\Psi\rangle$. Using this number we can write

$$\langle\Psi|\sigma(\Psi)\rangle = |\langle\psi_1|\psi_2\rangle|^{2m(\sigma)} . \tag{4.37}$$

For instance,

$$\begin{aligned} \langle\Psi|\nu_k(|\Psi\rangle) &= \langle\psi_1|^{\otimes k}\langle\psi_2|^{\otimes l}|\psi_1\rangle^{\otimes k-1}|\psi_2\rangle|\psi_1\rangle|\psi_2\rangle^{\otimes l-1} \\ &= |\langle\psi_1|\psi_2\rangle|^{2m(\nu_k)} = |\langle\psi_1|\psi_2\rangle|^2 . \end{aligned}$$

In order to evaluate the scalar product

$$\langle\Psi|\Psi_S\rangle = \frac{1}{(k+l)!}\sum_{\sigma\in S(k+l)}\langle\Psi|\sigma(|\Psi\rangle). \tag{4.38}$$

we need to calculate the number of permutations $C_m$ with the same value $m = m(\sigma)$. For each permutation $\sigma$ there are exactly $k!l!$ permutations leading to the same state $\sigma(|\Psi\rangle)$. The number of different quantum states $\sigma_1(|\Psi\rangle), \sigma_2(|\Psi\rangle), \ldots$ having the same overlap $|\langle\psi_1|\psi_2\rangle|^{2m}$ with the state $|\Psi\rangle$ (i.e. the same $m$) is $\binom{k}{m}\binom{l}{m}$. This is because each such state is fully specified by enumerating $m$ from the first $k$ subsystems to which $|\psi_2\rangle$ states were permuted and by enumerating $m$ from the last $l$ subsystems to which $|\psi_1\rangle$ states were moved. To sum up our derivation, we have $C_m = k!l!\binom{k}{m}\binom{l}{m}$, and consequently Eq. (4.38) can be rewritten as

$$\langle\Psi|\Psi_S\rangle = \sum_{m=0}^{\min(k,l)}\frac{\binom{k}{m}\binom{l}{m}}{\binom{k+l}{k}}|\langle\psi_1|\psi_2\rangle|^{2m}\,. \tag{4.39}$$

The optimal probability reads

$$P(|\psi_1\rangle,|\psi_2\rangle) = 1 - \sum_{m=0}^{\min(k,l)}\frac{\binom{k}{m}\binom{l}{m}}{\binom{k+l}{k}}|\langle\psi_1|\psi_2\rangle|^{2m}\,. \tag{4.40}$$

Before calculating the average probability of success $P_{succ}(k,l)$ it is useful to evaluate the mean values of the overlaps

$$\begin{aligned}
&\overline{|\langle\psi_1|\psi_2\rangle|^{2m}} = \int_{S_{pure}}\int_{S_{pure}} d\psi_1 d\psi_2 \langle\psi_1|\psi_2\rangle^m\langle\psi_2|\psi_1\rangle^m \\
&= \int_{S_{pure}} d\psi_1 \langle\psi_1|^{\otimes m}\left(\int_{S_{pure}} d\psi_2 |\psi_2\rangle\langle\psi_2|^{\otimes m}\right)|\psi_1\rangle^{\otimes m} \\
&= \frac{1}{\binom{m+d-1}{d-1}}\int_{S_{pure}} d\psi_1\langle\psi_1|^{\otimes m}P^{sym}_{1\ldots N}|\psi_1\rangle^{\otimes m} \\
&= \frac{1}{\binom{m+d-1}{d-1}}\,,
\end{aligned} \tag{4.41}$$

where we exploited the identity in Eq. (4.32).

We will insert Eqs. (4.40) and (4.41) into the definition (4.33) and utilize the Vandermonde's identity

$$\binom{a+b}{r} = \sum_{m=0}^{r}\binom{a}{m}\binom{b}{r-m}$$

to evaluate the summation to obtain

$$\begin{aligned}
\frac{1}{\eta_{diff}} P_{succ}(k,l) &= 1 - \frac{1}{\binom{k+l}{k}} \sum_{m=0}^{\min(k,l)} \frac{\binom{k}{m}\binom{l}{m}}{\binom{m+d-1}{d-1}} \\
&= 1 - \frac{k!(d-1)!}{(k+d-1)!} \frac{1}{\binom{k+l}{k}} \sum_{m=0}^{k} \binom{k+d-1}{k-m} \binom{l}{m} \\
&= 1 - \frac{k!(d-1)!}{(k+d-1)!} \frac{\binom{k+l+d-1}{k}}{\binom{k+l}{k}} \\
&= 1 - \frac{\binom{k+l+d-1}{k+l}}{\binom{k+d-1}{k}\binom{l+d-1}{l}} .
\end{aligned}$$

The previous steps are valid for $k < l$, however we can perform analogous calculation for $l \leq k$ and obtain the same final result, which can be nicely written as

$$P_{succ}(k,l) = \eta_{diff} \left( 1 - \frac{\dim(\mathcal{H}_{sym}^{\otimes k+l})}{\dim(\mathcal{H}_{sym}^{\otimes k}) \dim(\mathcal{H}_{sym}^{\otimes l})} \right), \tag{4.42}$$

where $\mathcal{H}_{sym}^{\otimes k}$ stands for a completely symmetric subspace of $\mathcal{H}^{\otimes k}$. Thus, we see that the success rate is essentially given by one minus the ratio of dimensionality of the failure subspace to the dimension of the potentially occupied space.

*Additional copy of an unknown state*

Next we will analyze properties of $P(|\psi_1\rangle, |\psi_2\rangle)$ with respect to $k, l$. In particular, we will study how it behaves as a function of the number $k, l$ of available copies of the two compared states. We are going to confirm a very natural expectation that any additional copy of one of the compared states always increases the probability of success. Stated mathematically, it suffices to prove that

$$P(|\psi_1\rangle, |\psi_2\rangle, k+1, l) \geq P(|\psi_1\rangle, |\psi_2\rangle, k, l), \tag{4.43}$$

since $P(|\psi_1\rangle, |\psi_2\rangle, k, l)$ is symmetric with respect to $k, l$. For $k \geq l$

$$\begin{aligned}
\delta &\equiv P(|\psi_1\rangle, |\psi_2\rangle, k+1, l) - P(|\psi_1\rangle, |\psi_2\rangle, k, l) \\
&= \frac{1}{\binom{k+l}{k}} \sum_{m=0}^{\min(k,l)} \left( 1 - \frac{(k+1)^2}{(k+1-m)(k+l+1)} \right) \\
&\quad \times \binom{k}{m} \binom{l}{m} |\langle\psi_1|\psi_2\rangle|^{2m} .
\end{aligned} \tag{4.44}$$

For $k < l$ the additional term $-|\langle\psi_1|\psi_2\rangle|^{2k+2} \binom{k+l+1}{k+1} / \binom{l}{k+1}$ appears in the expression for $\delta$, however it is possible to proceed in the same way in both cases. We can think of $\delta$ as being a polynomial in $x \equiv |\langle\psi_1|\psi_2\rangle|^2$, which vanishes for $x = 1$, because $P(|\psi\rangle, |\psi\rangle) = 0$. The coefficients $a_m$ of the polynomial $\delta = \sum_m a_m x^m$ are nonnegative for $m \leq (k+1)l/(k+l+1)$ and negative otherwise. Therefore, we can apply the Lemma from Appendix A.1 to conclude that $\delta(x) \geq 0$ for $x \in [0, 1]$, which is equivalent to Eq. (4.43). We have proved that for any pair of compared states the additional copies of the states improve the probability of success, so the statement holds also for the average success probabilities, i.e.

$$P_{succ}(k+1, l) \geq P_{succ}(k, l) \,. \tag{4.45}$$

*Optimal choice of resources*

Now we consider the situation when the total number $N$ of copies of the two states is fixed, i.e. $N = k + l$. Our aim is to maximize the success probability with respect to the splitting of the $N$ systems into $k$ copies of the state $|\psi_1\rangle$ and $l$ copies of the state $|\psi_2\rangle$. In order to find the solution to this problem we prove the following inequality

$$\begin{aligned} \Lambda &\equiv P(|\psi_1\rangle, |\psi_2\rangle, k+1, N-k-1) - P(|\psi_1\rangle, |\psi_2\rangle, k, N-k) \\ &\geq 0 \quad \text{for} \quad k \leq \lfloor N/2 \rfloor \,, \end{aligned} \tag{4.46}$$

where $\lfloor a \rfloor$ indicates the floor function, i.e. the integer part of the number. The previous inequality automatically implies $\Lambda \leq 0$ for $k > \lfloor N/2 \rfloor$, because $P(|\psi_1\rangle, |\psi_2\rangle, k, l)$ is symmetric in $k$ and $l$. Therefore, this would mean that the optimal value is $k = \lfloor N/2 \rfloor$.

Thus, to complete the proof it is sufficient to confirm the validity of Eq. (4.46). This is done in the same way as for Eq. (4.43) i.e. by looking on $\Lambda$ as on a polynomial in $x \equiv |\langle\psi_1|\psi_2\rangle|^2$ and showing that the assumptions of the Lemma from Appendix A.1 hold.

Hence, given the total number $N$ of copies it is most optimal to have half of them in the state $|\psi_1\rangle$ and the other half in the state $|\psi_2\rangle$. In this case the average probability of success

$$P_{succ}(\lfloor N/2 \rfloor, N - \lfloor N/2 \rfloor) = \max_k P_{succ}(k, N-k) \tag{4.47}$$

is maximized.

More quantitative insight into the behavior of $P(|\psi_1\rangle, |\psi_2\rangle)$ and $P_{succ}(k, k)$ is presented in figures (4.3) and (4.4). The figure (4.3) illustrates that the more copies of the compared states we have and the smaller is their overlap, the higher is the probability of revealing the difference between the states. The overlap of a pair of randomly chosen states decreases with the dimension of $\mathcal{H}$. Therefore the mean probability $P_{succ}(k, k)$ for a fixed number of copies $k$ grows with the dimension $d$. This fact is documented in Fig. 4.4.

*Comparison with large number of copies*

Let us now study the situation when $k = 1$ and $l \to \infty$. In this case the sum in Eq. (4.40) has only two terms, which can be easily evaluated to obtain

$$\begin{aligned} P(|\psi_1\rangle, |\psi_2\rangle) &= \lim_{l\to\infty} \left(1 - \frac{1 + l|\langle\psi_1|\psi_2\rangle|^2}{l+1}\right) = \\ &= 1 - |\langle\psi_1|\psi_2\rangle|^2 \,. \end{aligned} \tag{4.48}$$

In this limit the same probability of success can be reached also by a different comparison strategy. We can first use the state reconstruction techniques to precisely determine the state $|\psi_2\rangle$ and then by projecting the remaining $|\psi_1\rangle$ state onto $I - |\psi_2\rangle\langle\psi_2|$ reveal the difference between the states.

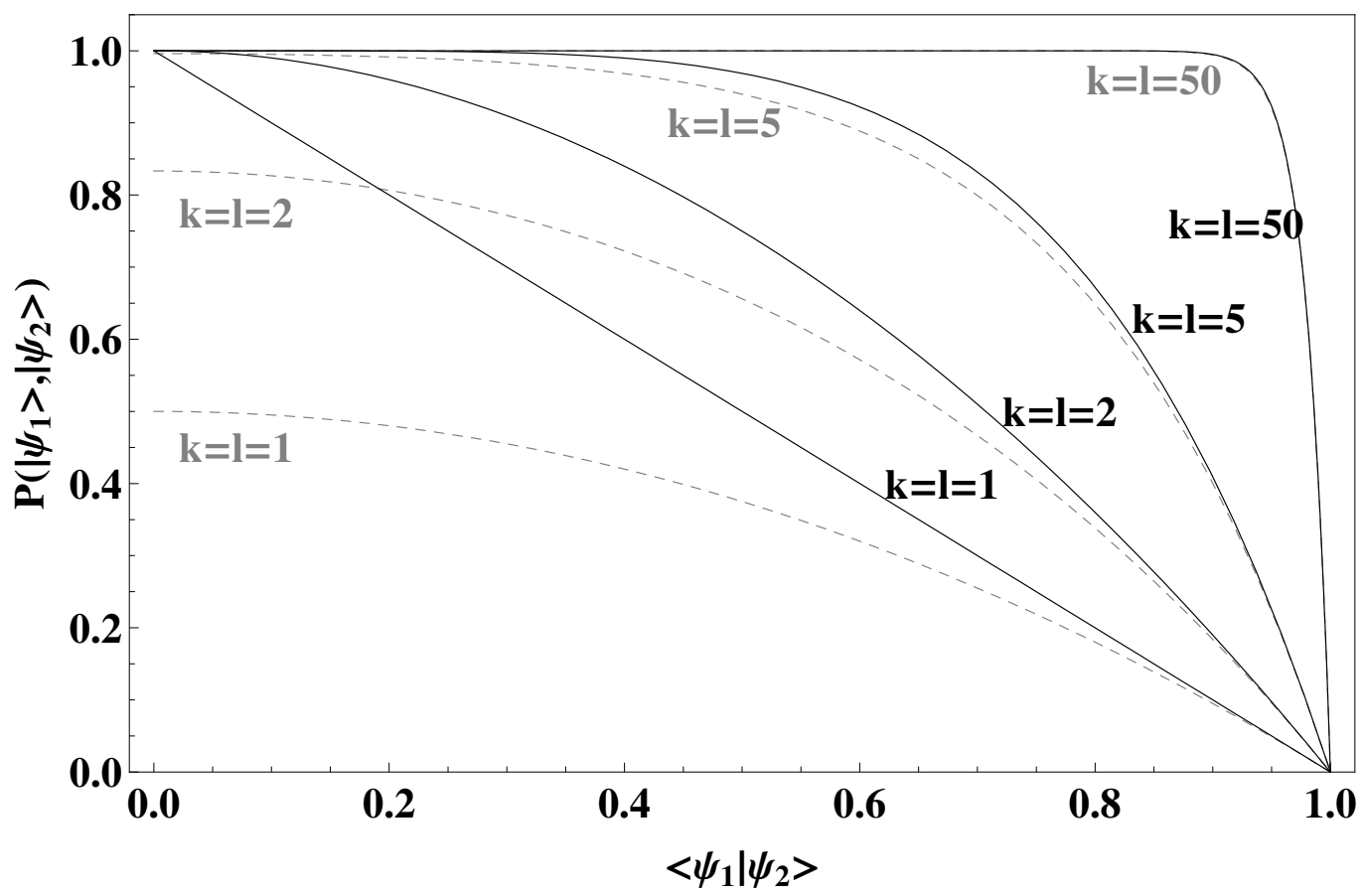

Figure 4.3. The probability of revealing the difference between the compared states $|\psi_1\rangle$, $|\psi_2\rangle$. The gray dashed lines are valid for the optimal state comparison among all pure states. Each line corresponds to a different number of copies of the compared states. The solid black lines indicate the performance of the optimal comparison if we are restricted to coherent states only.

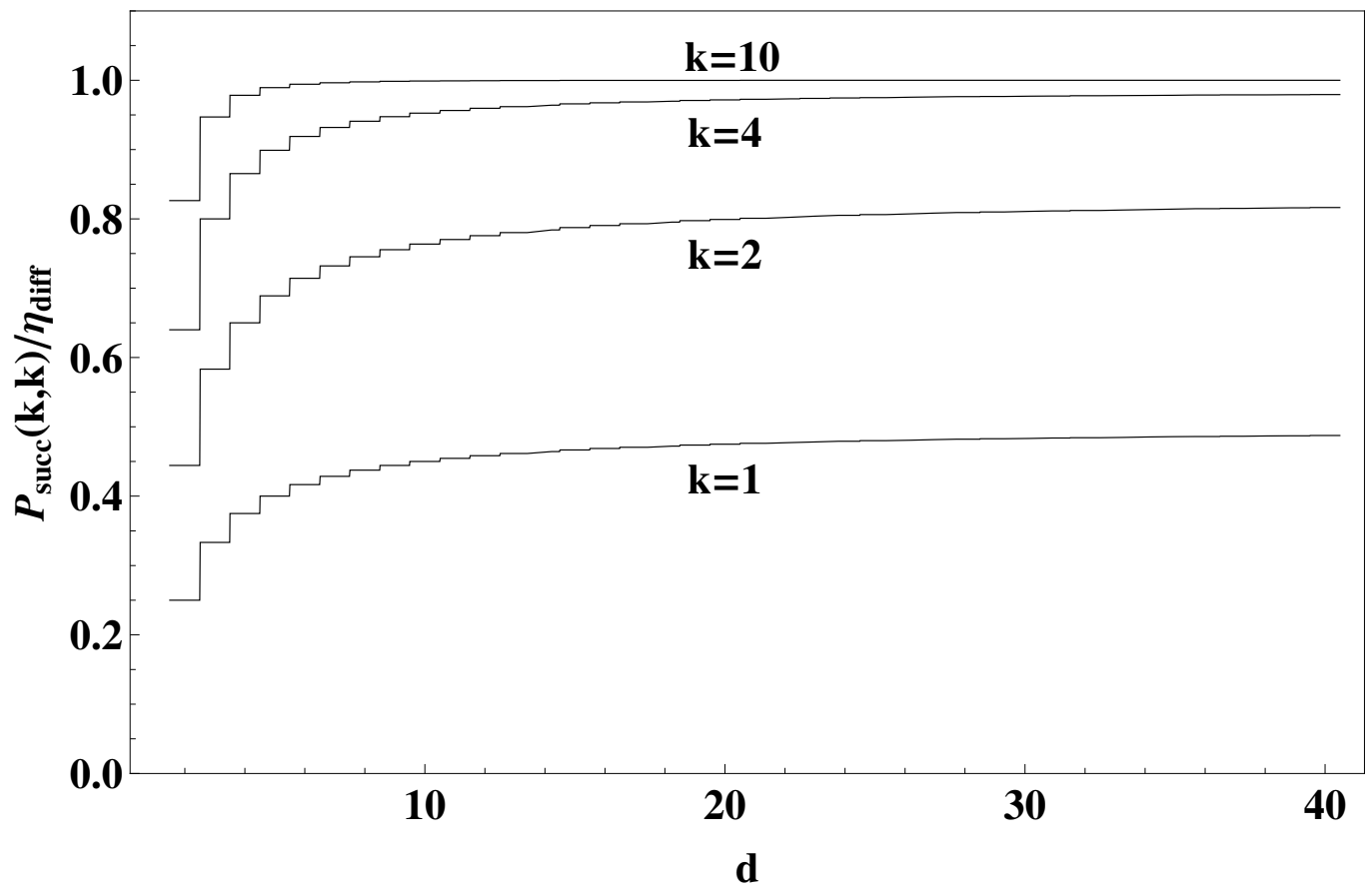

Figure 4.4. The mean probability of the detection of a difference between the compared states $|\psi_1\rangle$, $|\psi_2\rangle$ as a function of the dimension of the Hilbert space of the compared systems.

For the limit, where the number of both compared states goes to infinity simultaneously ($k = l \to \infty$), from Eq. (4.42) we recover for any finite $d$ the classical behavior i.e.

$$\lim_{k\to\infty} P_{succ}(k,k) = 1\,. \tag{4.49}$$

Therefore we can conclude that larger the number of the copies $k$ and $l$ of the two states higher the probability to determine that the two states are different is. In the limit $k = l \to \infty$ we essentially obtain a classical comparison problem.

#### 4.2.4 Comparison of two ensembles of coherent states

In any quantum information processing the prior knowledge about the system in which information is encoded plays an important role. The most explicit example one can name is the state estimation when the prior knowledge about the state is crucial. In what follows we will analyze the quantum state comparison and instead of assuming that the two compared states are totally arbitrary we will restrict a class of possible states. To be more specific, we will consider a harmonic oscillator and we focus our attention on comparison of two ensembles of coherent states.

Coherent states [40] are defined as eigenstates of the annihilation operator $a$ (acting on $\mathcal{H}_\infty$) associated with eigenvalues taking arbitrary value in the complex plane. The set of coherent states is defined as

$$S_{\rm coh} = \{|\alpha\rangle \in \mathcal{H}_\infty : \quad \alpha \in \mathbb{C}\,, \quad a|\alpha\rangle = \alpha|\alpha\rangle\}\,. \tag{4.50}$$

The first, who considered unambiguous quantum state comparison of coherent states were E. Andersson, M. Curty and I. Jex [41]. They proposed a simple optical setup realizing the comparison of a pair of coherent states, which consisted of a beamsplitter and a photodetector. The optimality of the setup was an open question, hence in [25] we proved optimality of the setup and derived its POVM description. The aforementioned results are a special case ($k = l = 1$) of the comparison of two ensembles of coherent states, which we shall investigate now. Our next task is two-fold: Firstly we introduce an optimal protocol for comparison of two ensembles of coherent states. Secondly we propose an experimental realization of the optimal coherent states comparator.

Following the same line of reasoning as in the previous section the measurement operator $E_1^{\rm coh}$ unambiguously revealing that the coherent states ($k$ copies of state $|\alpha_1\rangle$ and $l$ copies of the state $|\alpha_2\rangle$) are different must obey the following "no-error" conditions

$$Tr\left(E_1^{\rm coh}(|\alpha\rangle\langle\alpha|)^{\otimes k+l}\right) = 0 \quad \forall\, |\alpha\rangle \in S_{\rm coh}\,, \tag{4.51}$$

or equivalently

$$0 \;=\; \int_{S_{\rm coh}} d\alpha\, Tr\left(E_1^{\rm coh}|\alpha\rangle\langle\alpha|^{\otimes k+l}\right) = Tr(E_1^{\rm coh}\Delta)\,, \tag{4.52}$$

where $d\alpha$ is an arbitrary positive measure such that its support contains all coherent states[12].

Since the operators $E_1^{\rm coh}$ and $\Delta$ are positive, the identity $Tr(E_1^{\rm coh}\Delta) = 0$ implies that their supports are orthogonal. As before (in the case of all pure states) it is optimal to choose $E_1^{\rm coh}$ to be a projector onto the orthocomplement of the support of $\Delta$. Denoting by $\Delta_{\rm coh}^N$ the projector onto the support of $\Delta$ we can write $E_1^{\rm coh} = I - \Delta_{\rm coh}^N$. As it is shown in Appendix A.2 using a properly normalized Lebesgue measure on a complex plane we can write

$$\Delta = \frac{N}{\pi}\int_{\mathbb{C}} d\alpha|\alpha\rangle\langle\alpha|^{\otimes N} = \Delta_{\rm coh}^N\,. \tag{4.53}$$

[12] Under $\int_{\mathbb{C}} d\beta f(\beta)$ we mean $\int_{\mathbb{R}^2} dxdyf(x+iy)$.

Consider $|\Psi\rangle = |\alpha_1\rangle^{\otimes k} \otimes |\alpha_2\rangle^{\otimes l}$ to be a general input state of the coherent-state comparison machine. Using the Eq.(4.53) we obtain the following expression for the success probability $P(|\alpha_1\rangle, |\alpha_2\rangle)$

$$\begin{aligned} P(|\alpha_1\rangle, |\alpha_2\rangle) &= Tr\left(\Pi_1^{\mathrm{coh}} |\Psi\rangle\langle\Psi|\right) = 1 - \langle\Psi|\Delta_{\mathrm{coh}}^{k+l}|\Psi\rangle \\ &= 1 - \frac{k+l}{\pi}\int_{\mathbb{C}} d\beta |\langle\alpha_1|\beta\rangle|^{2k} |\langle\alpha_2|\beta\rangle|^{2l} \\ &= 1 - \frac{k+l}{\pi}\int_{\mathbb{C}} d\beta e^{-k|\alpha_1-\beta|^2 - l|\alpha_2-\beta|^2} \\ &= 1 - \frac{k+l}{\pi} e^{-\frac{kl}{k+l}|\alpha_1-\alpha_2|^2} \int_{\mathbb{C}} d\beta e^{-\left|\sqrt{k+l}\beta - \frac{1}{\sqrt{k+l}}(k\alpha_1 + l\alpha_2)\right|^2} \\ &= 1 - e^{-\frac{kl}{k+l}|\alpha_1-\alpha_2|^2}, \end{aligned} \tag{4.54}$$

where we used the following modification of the rectangular identity

$$\begin{aligned} k \quad & |\alpha_1 - \beta|^2 + l|\beta - \alpha_2|^2 \\ = \quad & \left|\sqrt{k+l}\beta - \frac{k\alpha_1 + l\alpha_2}{\sqrt{k+l}}\right|^2 + \frac{kl}{k+l}|\alpha_1 - \alpha_2|^2. \end{aligned}$$

*Optical setup for unambiguous comparison of coherent states*

In this subsection we will describe an optical realization of an unambiguous coherent-states comparator that achieves the optimal value of the success probability (see above). The experimental setup we are going to propose will consist of several beam-splitters and only a single photodetector. A beam-splitter acts on a pair of coherent states in a very convenient way, in particular, the output beams remain unentangled and coherent, i.e.

$$|\alpha\rangle \otimes |\beta\rangle \mapsto |\sqrt{T}\alpha + \sqrt{R}\beta\rangle \otimes |-\sqrt{R}\alpha + \sqrt{T}\beta\rangle\,, \tag{4.55}$$

where $T, R$ stand for transmissivity and reflectivity, respectively, and $T + R = 1$. The aforementioned property of the beam-splitter transformation enables us to consider each of its outputs separately.

Our setup is composed of $k + l - 1$ beam-splitters and one photodetector. The $k - 1$ beam-splitters are used to "concentrate" (focus) the information encoded in $k$ copies of the first state. Namely, they are arranged according to Fig. 4.5 and they perform the unitary transformation $|\alpha_1\rangle^{\otimes k} \mapsto |\sqrt{k}\alpha_1\rangle \otimes |0\rangle^{\otimes k-1}$. To do this the transmissivities of the beam-splitters must be set as follows

$$T_j = \frac{j}{j+1} \qquad R_j = \frac{1}{j+1}\,.$$

Similarly, $l - 1$ beam-splitters are used to "concentrate" the $l$ copies of the second state. The "concentrated" states $|\sqrt{k}\alpha_1\rangle$, $|\sqrt{l}\alpha_2\rangle$ are then launched into the last beam-splitter in which the comparison of input coherent states is performed. It performs the following unitary transformation

$$\begin{aligned} |\sqrt{k}\alpha_1\rangle \otimes |\sqrt{l}\alpha_2\rangle \quad \mapsto \quad & |\sqrt{T_f k}\alpha_1 + \sqrt{R_f l}\alpha_2\rangle \\ & \otimes|\sqrt{T_f l}\alpha_2 - \sqrt{R_f k}\alpha_1\rangle\,, \end{aligned} \tag{4.56}$$

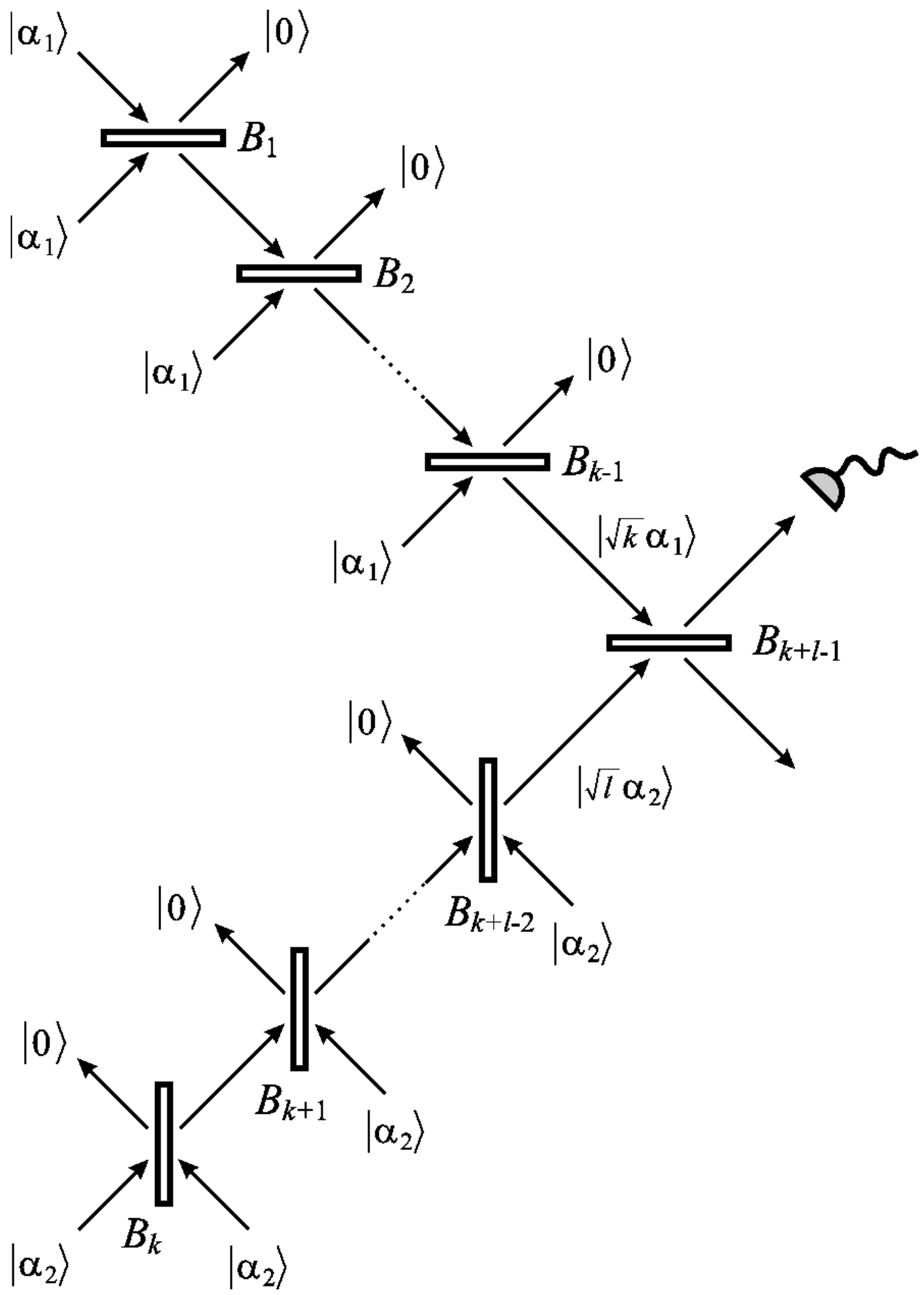

Figure 4.5. The beam-splitter setup for the comparison of two finite-size ensembles composed of $k$ copies of the coherent state $|\alpha_1\rangle$ and $l$ copies of the coherent state $|\alpha_2\rangle$, respectively.

where $R_f, T_f$ is the reflectivity and transmissivity of the last beam-splitter. To obtain the vacuum in the upper output (see Fig. 4.5) we need to adjust the values of reflectivity and transmissivity so that the identity $kR_f = lT_f$ holds, i.e.

$$T_f = \frac{k}{k+l}, \qquad R_f = \frac{l}{k+l}.$$

Finally, a photodetector will measure the presence of photons in the upper output port of the last beam-splitter (see Fig. 4.5). If the two compared states are identical, in the output port we have zero photons - that is this port is in the vacuum state. Therefore a detection of at least one photon unambiguously indicates the difference between the compared states. On the other hand the observation of no photons is inconclusive, since each coherent state has a nonzero overlap with the vacuum. As a result we obtain the success probability

$$\begin{aligned} P(|\alpha_1\rangle, |\alpha_2\rangle) &= 1 - |\langle 0|\sqrt{\frac{kl}{k+l}}(\alpha_2 - \alpha_1)\rangle|^2 \\ &= 1 - e^{-\frac{kl}{k+l}|\alpha_1-\alpha_2|^2}, \end{aligned} \tag{4.57}$$

which is the optimal one. Analyzing the last equation we find out that $P(|\alpha_1\rangle, |\alpha_2\rangle, m, n) > P(|\alpha_1\rangle, |\alpha_2\rangle, k, l)$ if and only if $\frac{mn}{m+n} > \frac{kl}{k+l}$. This equivalence implies that $P(|\alpha_1\rangle, |\alpha_2\rangle, k+1, l) > P(|\alpha_1\rangle, |\alpha_2\rangle, k, l)$. Thus, also in the case of coherent states the additional copy of one of the compared states helps to increase the mean success of the state comparison. For a fixed number of copies of both compared states $N$ the fraction $k(N-k)/N$ is maximized for $k = N/2$. Therefore, the probability of revealing the difference of the states is maximized if $k = l$.

### 4.2.5 Summary

Let me summarize my original results on the comparison of ensembles of quantum states derived in this part of the chapter. The difference of arbitrary unknown pure states $|\psi_1\rangle, |\psi_2\rangle$ can be unambiguously detected with the probability

$$P(|\psi_1\rangle, |\psi_2\rangle) = 1 - \sum_{m=0}^{\min(k,l)} \frac{\binom{k}{m}\binom{l}{m}}{\binom{k+l}{k}} |\langle\psi_1|\psi_2\rangle|^{2m}, \tag{4.58}$$

providing that we have $k$ copies of state prepared by the first preparator and $l$ copies produced by the second preparator. This result does not depend on the dimension of the system in contrast to the average success probability, which reads

$$P_{succ}(k,l) = \eta_{diff}\left(1 - \frac{\dim(\mathcal{H}_{sym}^{\otimes k+l})}{\dim(\mathcal{H}_{sym}^{\otimes k})\dim(\mathcal{H}_{sym}^{\otimes l})}\right). \tag{4.59}$$

Given the a priori knowledge that the states are coherent one can increase the probability (see Fig. 4.3) to

$$P(|\alpha_1\rangle, |\alpha_2\rangle) = 1 - e^{-\frac{kl}{k+l}|\alpha_1-\alpha_2|^2}. \tag{4.60}$$

The improvement is significant (Fig. 4.3) for small number of copies.

We also addressed the problem of maximizing the success probability providing that the total number of available copies is fixed. We have shown that it is optimal if the number of copies is the same, i.e. $k = l = N/2$. In the limit of the large number of copies the comparison approach reduces to "classical" comparison based on the quantum-state estimation.

We have proposed an optical implementation of the optimal quantum-state comparator of two finite ensembles of coherent states. This proposal is relatively easy to implement, since it consists only of $N-1$ beam-splitters and a single photodetector. Unfortunately, the success of unambiguous state comparison is very fragile with respect to small imperfections. The reason is that the device can be only used for pure states. Therefore our device can be used only in the situation when sources of a noise $\mathcal{N}$ can be modeled as quantum channels preserving the validity of the no-error conditions $\mathrm{Tr}(E_1^{\mathrm{coh}}\mathcal{N}[\Delta_{\mathrm{coh}}^N]) = 0$. An example of such noise is an application of random unitary channel (simultaneously on all copies) transforming coherent states into coherent states.

### 4.3 Unambiguous identification

Unambiguous identification is a discrimination task in which the description of the quantum states we should distinguish is partly given by the classical information and partly by the states of additional quantum systems. In order to unify the notation of the problems which fit into this concept we first define a sufficiently wide framework. Next, we summarize the previously known results. The aim is to sketch the scenarios, which were solved and name the techniques that were used. Afterwards we explain the relation of UI to discrimination of mixed states, which can be used to re-derive many of the known results in a uniform fashion. The remaining part of the chapter is devoted to investigation of influence of prior knowledge on unambiguous identification. This means we analyze how the distribution of the prior information between quantum states and the classical description affects the probability of success and the form of optimal measurement.

#### 4.3.1 Definition of the framework

The problems considered further in this chapter can be described within the following framework. Suppose we are given identical quantum systems, each of them living in $d$ dimensional Hilbert space $\mathcal{H}$. The systems are divided into $M+1$ groups: $A, B, C, \ldots$, containing $n_A$, $n_B$, $n_C$, ... systems respectively. Systems in one group are prepared in the same unknown pure state. Furthermore state of systems in A is guaranteed to be the same as the state of systems in one of the other groups. The task further denoted as *Unambiguous Identification(UI)* is to unambiguously recognize which group systems in A match. Strategy working for any choice of pure states we denote as *universal UI*. From mathematical point of view we should discriminate among the following $M$ types of states:

$$|\Psi_i\rangle_{ABC\ldots} \equiv |\psi_i\rangle_A^{\otimes n_A} \otimes |\psi_1\rangle_B^{\otimes n_B} \otimes |\psi_2\rangle_C^{\otimes n_C} \otimes \ldots \qquad i = 1, 2, \ldots, M \tag{4.61}$$

via a positive operator value measure (POVM) consisting of $M+1$ elements. Element $\mathrm{E}_i$ will correspond to correct identification of $|\Psi_i\rangle$ type of state and $\mathrm{E}_0$ corresponds to an inconclusive result. These elements must obey no error conditions (equation (4.62)) and constitute a valid POVM (equations (4.63)):

$$\forall i \neq j \quad Tr[\mathrm{E}_i \Psi_j] = 0, \quad \Psi_j \equiv |\Psi_j\rangle\langle\Psi_j| \tag{4.62}$$

$$\mathrm{E}_i \geq 0, \mathrm{E}_0 \geq 0, \quad \mathrm{E}_0 + \sum_{i=1}^{M} \mathrm{E}_i = I. \tag{4.63}$$

We assume that states of the type $|\Psi_i\rangle$ appear with a prior probability $\eta_i$. We will refer to states $|\psi_i\rangle$ as being *reference states* and denote the state of individual quantum systems in group A as an *unknown state*. The performance of the UI measurement can be quantified by a probability of

identification for a particular choice of reference states

$$P(|\psi_1\rangle,\ldots,|\psi_M\rangle) = \sum_{i=1}^{M} \eta_i Tr[\mathrm{E}_i \Psi_i] \tag{4.64}$$

However, more adequate figure of merit is its average value

$$\int P(|\psi_1\rangle,\ldots,|\psi_M\rangle)\ \chi(|\psi_1\rangle,\ldots,|\psi_M\rangle)\ d\psi_1\ldots d\psi_M,$$

where $\chi(|\psi_1\rangle,\ldots,|\psi_M\rangle)$ is the probability distribution describing our knowledge about the choice of reference states. Usually, there is no reason to expect any correlations in the choice of reference states and hence our classical information leads to an assumption that the reference states are independently and uniformly chosen from subset of pure states, further denoted by $S$. Consequently, the average probability of identification reads

$$P_{succ} \equiv \overline{P(S)} = \underbrace{\int_S \cdots \int_S}_{M} P(|\psi_1\rangle,\ldots,|\psi_M\rangle)\ d\psi_1\ldots d\psi_M \tag{4.65}$$

and the optimality of UI measurement is defined with respect to it. However, we will see that optimization of $P(|\psi_1\rangle,\ldots,|\psi_M\rangle)$ is in some situations closely related to maximization of the average probability of identification. We denote the set of all pure states of a $d$-dimensional quantum system $S_{pure}$ and the subscript of $P$ will indicate the used UI measurement.

### 4.3.2 Previous work

Quantum information processing most often deals with systems of qubits. Qubits are two-dimensional quantum systems, which implies that also Hilbert space for systems of small number of qubits is relatively simple and allow different tasks to be solved explicitly. Thus, many tasks are first formulated and solved for qubits and afterwards the solution is generalized to qudits. This was the case also for unambiguous identification.

*Qubits*

*One copy*

J. Bergou and M.Hillery [42] first formulated and solved the basic version of the UI problem. In this case we have only one copy of unknown and two reference states and all of them are qubits ($M = 2$, $n_A = n_B = n_C = 1$, $d = 2$). Thus two types of states:

$$|\Psi_1\rangle_{ABC} \equiv |\psi_1\rangle \otimes |\psi_1\rangle \otimes |\psi_2\rangle \qquad |\Psi_2\rangle_{ABC} \equiv |\psi_2\rangle \otimes |\psi_1\rangle \otimes |\psi_2\rangle. \tag{4.66}$$

enter the UI measurement, which distinguishes whether state of qubit A matches the state of qubit B or qubit C. The optimal measurement should maximize the mean probability of identification $\overline{P(S_{pure})}$. Bergou and Hillery first derived anzatz for the measurement from the symmetry considerations. The result of it's parameter optimization depends on the prior probabilities $\eta_1$, $\eta_2$. There are three different regimes in which the optimal measurement is either two-valued projective or a true POVM measurement with three outcomes:

$$\begin{aligned} 0 \leq \eta_1 < 1/5 \quad & \mathrm{E}_1 = 0, & \mathrm{E}_2 &= \mathrm{P}^{asym}_{AB} \otimes I_C, \\ 1/5 \leq \eta_1 \leq 4/5 \quad & \mathrm{E}_1 = \lambda \mathrm{P}^{asym}_{AC} \otimes I_B, & \mathrm{E}_2 &= \frac{4-4\lambda}{4-3\lambda} \mathrm{P}^{asym}_{AB} \otimes I_C, \\ 4/5 < \eta_1 \leq 1 \quad & \mathrm{E}_1 = \mathrm{P}^{asym}_{AC} \otimes I_B, & \mathrm{E}_2 &= 0, \end{aligned} \tag{4.67}$$

where $\mathrm{P}^{asym} \equiv |\psi^-\rangle\langle\psi^-|$ is a projector onto the antisymmetric subspace of the two qubit Hilbert space $\mathcal{H}^{\otimes 2}$, $|\psi^\pm\rangle = 1/\sqrt{2}(|01\rangle \pm |10\rangle)$ and $\lambda = \frac{2}{3}(2 - \sqrt{\frac{\eta_2}{\eta_1}})$. The inconclusive result is associated with the POVM element $\mathrm{E}_0 = I - \mathrm{E}_1 - \mathrm{E}_2$. The optimal UI measurement is closely related to quantum state comparison as we illustrate in next few lines. This relation will later serve us also as a motivation for the proposition of the UI measurement for coherent states.

*Relation to Quantum state comparison*

Let us examine how the optimal UI works for a prior probability $\eta_1 < \frac{1}{5}$. The measurement (4.67) has effectively two outcomes, which are either unambiguously identifying the more probable $|\Psi_2\rangle_{ABC}$ type of state or signaling inconclusive result. The unambiguous decision is not based on testing the equality of the unknown and the second reference state, but rather on revealing the difference between unknown and the first reference state – stored in qubits AB. Hence, qubit C is not used and the measurement distinguishes antisymmetric and symmetric states of the subsystem AB. States of the type $|\Psi_1\rangle$ are symmetric in qubits AB, whereas $|\Psi_2\rangle$ type of states are not. Therefore, the projection onto antisymmetric subspace of qubits AB unambiguously identifies $|\Psi_2\rangle$ type of state and the projection onto symmetric subspace is inconclusive, because of $|\Psi_2\rangle$ having overlap with it. This is exactly what happens in quantum state comparison of two unknown pure states. Two equal states are in symmetric subspace, therefore projection onto antisymmetric subspace unambiguously indicate that the compared states were different. On the other hand, projection onto symmetric subspace is inconclusive, because each pair of pure states has nonzero overlap with it. Thus, for $\eta_1 < \frac{1}{5}$ the optimal UI measurement is a quantum state comparison of unknown and the less probable reference state. The corresponding mean probability of identification is $\frac{1}{4}(1-\eta_1)$. Analogous consideration hold for $\eta_1 > \frac{4}{5}$ for which $\overline{P(S_{pure})}$ is $\frac{1}{4}(1-\eta_2) = \frac{1}{4}\eta_1$. For equal prior probabilities the optimal POVM elements $\mathrm{E}_1$, $\mathrm{E}_2$ are $2/3$ multiples of the above mentioned quantum state comparison measurement elements $\mathrm{P}^{asym}_{AC}$, $\mathrm{P}^{asym}_{AB}$. In this case the mean probability of identification equals $1/6$.

*More copies of unknown states*

J. Bergou et. al. in [43] investigated also the situation with more copies of an unknown state and with different prior knowledge of the two reference states. They used a technique based on the relation of UI to discrimination of known mixed states, which we explain in detail in next sections. For the already discussed case of only one copy of unknown state ($M = 2$, $n_A = n_B = n_C = 1$) the corresponding mixed states are:

$$\rho_1 = \frac{1}{3} P^{sym}_{AB} \otimes \frac{1}{2} I_C, \quad \rho_2 = \frac{1}{2} I_A \otimes \frac{1}{3} P^{sym}_{BC} \tag{4.68}$$

In this case Bergou et. al. explicitly used the no-error conditions to determine the most general form of the UI measurement. Its parameters were optimized to yield the maximum probability of identification, while keeping the POVM conditions satisfied. This optimization is quite technical, but finally yields the measurement (4.67) from [42]. Although this technique is essentially equivalent to that from the first paper of Bergou and Hillery [42] it can be more easily modified to consider different prior knowledge on the reference states. If one of the reference states is known (this corresponds to $n_B \to \infty$) one expects that the probability of UI should be higher and Bergou et. al. confirmed it. In this case the UI can be reformulated as an unambiguous discrimination of two-qubit mixed states:

$$\rho_1 = |0\rangle\langle 0|_A \otimes \frac{1}{2} I_C, \quad \rho_2 = \frac{1}{3} P_{AC}^{sym}, \tag{4.69}$$

where the known reference state (from group B) is for convenience denoted as a basis state $|0\rangle$. For the generalizations of this two scenarios, which differ by having $m$ copies of unknown state instead of one, a more sophisticated technique was used. It is essentially a solution to discrimination of a pair of certain mixed states, and it is called unambiguous subspace discrimination according to task it originally solves. Thanks to that technique the aforementioned scenarios ($M = 2$, $n_A = m$, $n_B = n_C = 1$), ($M = 2$, $n_A = m$,$n_B \mapsto \infty, n_C = 1$) were solved for arbitrary prior probabilities $\eta_1, \eta_2$. From the results we see that the more copies of unknown state we have the more quantum information is provided and thus the probability of identification is higher.

*More copies of unknown and reference states*

Bing He and J. Bergou [44] managed to apply the same technique also to a scenario where the number of copies for each of the two reference states was also varied ($M = 2$, $n_A = m$, $n_B = n_C = n$). However it is not easy to express the solution explicitly for all prior probabilities $\eta_1$, $\eta_2$. Thus, for arbitrary $n, m$ we have an optimum identification probability and the corresponding measurement only for prior probability $\eta_1$ in a small interval around $1/2$. In general work of J. Bergou et. al. in [43] shows that the more prior knowledge (more copies of the states or their more detailed classical description) we have about the states to be identified the higher is the possible probability of unambiguous identification.

*Qudits*

After the first results for unambiguous identification of qubits were obtained it was natural to investigate also more general situations. These certainly include the use of $d$ dimensional quantum systems instead of qubits and varying both the number of different reference states and the number of copies per reference state.

*Many copies of two reference states*

The scenario with one copy of unknown state and $n$ copies per each of the two reference states ($M = 2$, $n_A = 1$, $n_B = n_C = n$) was considered by A. Hayashi, M. Horibe, and T. Hashimoto [45]. They found analytical solution for the case of equal prior probabilities $\eta_1 = \eta_2 = \frac{1}{2}$ and any dimension $d$ of the used quantum systems. The situation offers symmetries whose exploitation substantially simplifies the form of the general UI measurement to such extend that for equal prior probabilities it was possible to solve the problem completely. The relevant symmetries are given by representations of permutation group $S(n)$ and unitary group $U(n)$. They arise from the possibility of exchanging the copies without changing the state of the system and from uniformity of the averaging in the probability of identification. Furthermore, the equality of the prior probabilities permits restriction to measurements, where the conclusive elements are the same except for acting on different subsystems. The final form of the optimal UI measurement is most conveniently written as:

$$\mathrm{E}_1 = e.I_B \otimes \mathrm{P}^{asym}_{AC}, \quad \mathrm{E}_2 = e.I_C \otimes \mathrm{P}^{asym}_{AB}, \quad e = \sum_{\lambda} e_\lambda \Gamma_\lambda, \tag{4.70}$$

where $\lambda$ specifies both $U(d)$ and $S(n)$ type of the irreducible representation, $\Gamma_\lambda$ is the projector onto that invariant subspace of $\mathcal{H}^{\otimes n}$, and $e_\lambda$ is non-negative real number depending on $n$ and $\lambda$. The authors managed to express the average probability of identification as a finite sum depending only on $n$ and $d$. The sum can be evaluated numerically, but in large $n$ limit it can be also rewritten as an integral using Stirling formula and then explicitly evaluated. As $n$ increases we can extract more information about the reference states and in the large $n$ limit the reference states become known to us. Thus, as expected, the integral yields the same result as an average of the probability of discrimination of two known pure states over all their possible choices. The authors finally employed numerics to produce graphs, which show how for a chosen dimensionality $d$ of the systems the average probability of identification grow with the increasing $n$ till it reaches its aforementioned asymptotic value.

*One copy per $M$ reference states*

Unambiguous identification of $M$ types of reference states presents qualitatively different type of generalization to basic UI problem. In this direction the first results were obtained by C. Zhang and M. Ying [46]. They considered the situation with one copy per each reference state and one copy of the unknown state ($n_A = 1$, $n_B = n_C = \cdots = 1$). The no-error conditions (4.62), which guarantee the correctness of the conclusive results, can be combined to express the constraints on the measurement in more compact and easily testable way. Precisely this was done by Zhang and Ying, who derived the following necessary and sufficient criterion for judging whether the chosen measurement performs unambiguous identification:

*POVM $\{E_i\}_{i=0}^M$ performs UI of $M$ reference states if and only if $\forall i = 1, \ldots, M$ $supp(Tr_i(E_i))$ is in the totally antisymmetric subspace of $\mathcal{H}^{\otimes M}$.*

For the special case, when the dimension of the quantum systems $d$ is equal the number of reference states $M$ the problem simplifies significantly. It is because the totally antisymmetric subspace of $\mathcal{H}^{\otimes M}$ is span only by one vector $|\phi\rangle$. Zhang and Ying finally optimize the measurement within the MiniMax approach, which rates the measurement via its worst case performance. Due to our framework we would optimize the average probability of identification. For equal prior probabilities $\eta_i = 1/M$ this gives the same UI measurement as MiniMax approach. This is because the convexity of the set of UI measurement permits restriction of the optimization to the same one parameter class and the maximization of the free parameter is in both cases restricted by the positivity of the measurement elements. The resulting optimal measurement has simple structure and reads:

$$\begin{aligned} \mathrm{E}_1 &= \frac{1}{M} I_A \otimes (|\phi\rangle\langle\phi|)_{BCD\ldots}, \quad \mathrm{E}_2 = \frac{1}{M} I_B \otimes (|\phi\rangle\langle\phi|)_{ACD\ldots}, \quad \ldots, \\ \mathrm{E}_0 &= I - \sum_{i=1}^{M} E_i \end{aligned} \tag{4.71}$$

The measurement element $E_i$ clicks only if the unknown state equals $i$-th reference state, because otherwise unknown and some other reference state are equal, which implies zero overlap with totally antisymmetric $|\phi\rangle$ in $E_i$.

Zhang and Ying propose the measurement with the same structure as (4.71) also for the case $d > M$, however its optimality is unknown. Recently, the optimal measurement for the special case $d = M$ was rederived by J. Bergou and U. Herzog in [47]. They identified block diagonal structure in the corresponding mixed states and employed the result of A.Chefles [10] for unambiguous discrimination of $M$ known symmetric pure states in each of these blocks. Unfortunately, also their result does not clarify the situation for $d > M$. If the dimension of the systems is smaller than the number of reference states ($d < M$) the totally antisymmetric subspace of $\mathcal{H}^{\otimes M}$ does not exist. Thus, according to the aforementioned criterion the support of $E_i$, $i = 1, \ldots, M$ must be empty, which implies that no useful UI measurement exist in this case. This explains why the one copy unambiguous identification of qubits was investigated only for two reference states.

### 4.3.3 UI as discrimination of known mixed states

The aim of this section is to explicitly show that the UI can be rephrased as unambiguous discrimination among $M$ multipartite mixed states. To see this we proceed according to general recipe introduced in section 3.2. Hence, we first rewrite the mean probability of identification in a suitable form, which will suggest the definition of the mixed states $\rho_i$, which play the role of the average constituents $A_i$. Secondly, we reformulate the no-error conditions (4.62) in terms of states $\rho_i$, prove their equivalence with (4.62) and finally discuss the optimality of a measurement for both tasks. The mean probability of identification (4.65) can be rewritten using equation (4.64) as:

$$\begin{aligned} \overline{P(S)} &= \underbrace{\int_S \ldots \int_S}_{M} \sum_{i=1}^{M} \eta_i Tr[\mathrm{E}_i \Psi_i] d\psi_1 \ldots d\psi_M = \\ &= \sum_{i=1}^{M} \eta_i Tr[\mathrm{E}_i \underbrace{\int_S \ldots \int_S}_{M} \Psi_i d\psi_1 \ldots d\psi_M] = \\ &= \sum_{i=1}^{M} \eta_i Tr[\mathrm{E}_i A_i] = \sum_{i=1}^{M} \eta_i Tr[\mathrm{E}_i \rho_i] = P_D, \end{aligned} \tag{4.72}$$

where we have defined the average constituent $A_i$:

$$A_i \equiv \rho_i = \underbrace{\int_S \dots \int_S}_{M} |\Psi_i\rangle\langle\Psi_i| d\psi_1 \dots d\psi_M, \quad Tr(A_i) = Tr(\rho_i) = 1.$$

In equation (4.72) we finally wrote the average probability of UI as the probability of discrimination among $M$ mixed states $\rho_i$ appearing with prior probability $\eta_i$.

The integration of the no-error conditions (4.62) over all expected reference states $|\psi_i\rangle$ gives us the following no-error conditions for mixed states $\rho_i$:

$$0 = \underbrace{\int_S \dots \int_S}_{M} Tr[\mathrm{E}_i \Psi_j] d\psi_1 \dots d\psi_M = Tr[E_i \rho_j] \quad \forall i \neq j. \tag{4.73}$$

We can directly apply Lemma 2 from Chapter 2 to conclude that these conditions (4.73) are equivalent to no-error conditions (4.62). Thus, any measurement unambiguously discriminating among mixed states $\rho_i$ is a valid UI measurement and vice versa. Because of the equation (4.72) such a measurement is optimal for both tasks at the same time. Hence, solution of unambiguous discrimination of general mixed states automatically gives solution to the UI problem. However, unambiguous discrimination of mixed states is a very complicated problem, which has drawn a lot of attention during last decade, and is still not completely solved.

*Typical structure of $\rho_i$*

In unambiguous identification we typically consider situations when the corresponding mixed states $\rho_i$ have a very simple structure. Namely, if $S$ the set of possible reference states is defined by the whole Hilbert space $\mathcal{H}$ or by its nontrivial subspace then the corresponding average mixed states $\rho_i$ are rescaled tensor products of projectors onto the symmetric subspaces. To see this, we use a result derived by A. Hayashi et. al. in [39]:

$$\int_S |\varphi\rangle\langle\varphi|^{\otimes k} d\varphi = \frac{1}{\binom{k+d_S-1}{d_S-1}} P^{sym}(S), \tag{4.74}$$

where $d_S$ is the dimension of the subspace $\mathcal{H}_S \subset \mathcal{H}$ determining the set $S$ and $P^{sym}(S)$ is the projector onto symmetric subspace of $\mathcal{H}_S{}^{\otimes k} \subset \mathcal{H}^{\otimes k}$. Therefore:

$$\begin{aligned} \rho_1 &= \underbrace{\int_S \dots \int_S}_{M} |\psi_1\rangle\langle\psi_1|_A^{\otimes n_A} \otimes |\psi_1\rangle\langle\psi_1|_B^{\otimes n_B} \otimes |\psi_2\rangle\langle\psi_2|_C^{\otimes n_C} \otimes \cdots d\psi_1 \dots d\psi_M = \\ &= \frac{1}{\binom{n_A+n_B+d_S-1}{d_S-1}\binom{n_C+d_S-1}{d_S-1}\cdots} P^{sym}_{AB}(S) \otimes P^{sym}_{C}(S) \otimes \dots, \\ \rho_2 &= \frac{1}{\binom{n_A+n_C+d_S-1}{d_S-1}\binom{n_B+d_S-1}{d_S-1}\cdots} P^{sym}_{AC}(S) \otimes P^{sym}_{B}(S) \otimes \dots, \end{aligned} \tag{4.75}$$

and analogously for the rest of $\rho_i$'s ($i = 1, \dots, M$). In Eq. (4.75) we have used subscript of $P^{sym}(S)$ to indicate subsystems on which the projection to symmetric subspace is performed.

Tensor product of projectors is a projector, so $\rho_i$'s are projectors scaled to have trace one. This simple structure of mixed states $\rho_i$ simplifies their unambiguous discrimination significantly as we will see below.

### 4.3.4 General approach to UI with two types of reference states

Unambiguous identification with two types of reference states is intuitively expected to be less complicated than the case with more types of reference states. The restriction to two types of reference states allow us to use the known results from unambiguous discrimination of two mixed states, which are for the above mentioned states $\rho_i$ sufficient to solve the UI for any number of copies of unknown and reference states and for any dimension of the occupied subspace $\mathcal{H}_S$. We will now formulate the task called unambiguous subspace discrimination and present its solution found by J. Bergou et. al. [48]. This will teach us how to optimally unambiguously discriminate among a specific type of two mixed states. Afterwards, we shall recognize that this type of mixed states emerge also in UI and hence we solve it in the same way.

*Unambiguous subspace discrimination*

Imagine we have a Hilbert space $\mathcal{H}$ and a description of two of its subspaces $V_1, V_2$. Someone will with prior probability $\eta_1$ respectively $\eta_2$ choose subspace $V_1$ respectively $V_2$ and prepare a quantum system in a state, which is chosen uniformly at random from that subspace. Our task is to determine *unambiguously* from which subspace was the state chosen.

This task was formulated and solved by J. Bergou, E. Feldman, and M. Hillery [48] and it was motivated by the basic version of the UI problem. We first explain how is this problem connected to unambiguous discrimination of two mixed states and then we rederive the result of Bergou et.al. via Raynal's reduction theorems for general position of the subspaces $V_1, V_2$.

Let us establish the notation. Without loss of generality we can assume that $\dim V_1 \geq \dim V_2$. The most general measurement one can perform is a POVM. For our task it should have three measurement elements: $E_1$, $E_2$ correctly identifying the use of subspace $V_1$, $V_2$ and the failure measurement operator $E_0$. The inconclusive outcome is necessary, since unambiguous discrimination of two pure states is a special case of subspace discrimination. The probability of correctly determining the used subspace $P_D$ is defined as:

$$P_D = \eta_1 \int_{S_{V_1}} \langle\psi_1|E_1|\psi_1\rangle d\psi_1 + \eta_2 \int_{S_{V_2}} \langle\psi_2|E_2|\psi_2\rangle d\psi_2, \tag{4.76}$$

where $\langle\psi_i|E_i|\psi_i\rangle$ is the conditional probability of correctly concluding that subspace $V_i$ was used if the state $|\psi_i\rangle \in V_i$ was chosen. The set of states in subspace $V_i$ is denoted $S_{V_i}$. The requirement of the unambiguity of the measurement can be mathematically formulated as:

$$\begin{aligned} &\forall|\psi_2\rangle \in V_2 \quad Tr(E_1|\psi_2\rangle\langle\psi_2|) = 0 \\ &\forall|\psi_1\rangle \in V_1 \quad Tr(E_2|\psi_1\rangle\langle\psi_1|) = 0 \end{aligned} \tag{4.77}$$

The success probability $P_D$ defined by Eq.(4.76) and the no error conditions from Eq. (4.77) can be equivalently rewritten using linearity and Lemma 2 from Chapter 2 as:

$$P_D = \eta_1 Tr(E_1\rho_1) + \eta_2 Tr(E_2\rho_2), \tag{4.78}$$

$$0 = Tr(E_1\rho_2) = Tr(E_2\rho_1), \tag{4.79}$$

where we have defined

$$\rho_i = \int_{S_{V_i}} |\psi_i\rangle\langle\psi_i| d\psi_i. \tag{4.80}$$

Thus, we can argue in the same way as in Subsection 4.3.3 to conclude that unambiguous subspace discrimination can be reformulated as unambiguous discrimination of two mixed states.

Next, we show that $\rho_i = \frac{1}{\dim V_i} P_{V_i}$, where $P_{V_i}$ is the projector onto the subspace $V_i$. We choose some state $|\varphi\rangle$ from the subspace $V_i$. To obtain any state in $S_{V_i}$ we transform state $|\varphi\rangle$ using group of operators, which acts on subspace $V_i$ as identical representation of the special unitary group $SU(\dim V_i)$ and trivially on the orthocomplement of $V_i$. If we use unitary invariant measure on the $SU(\dim V_i)$ we can rewrite the definition of $\rho_i$ as follows:

$$\begin{aligned}
\rho_i &= \int_{S_{V_i}} |\psi_i\rangle\langle\psi_i| d\psi_i = \int_{SU(\dim V_i)} U|\varphi\rangle\langle\varphi|U^\dagger dU = \\
&= \frac{1}{\dim V_i} \sum_{k=1}^{\dim V_i} \int_{SU(\dim V_i)} U|\varphi_k\rangle\langle\varphi_k|U^\dagger dU = \\
&= \frac{1}{\dim V_i} \int_{SU(\dim V_i)} U(\sum_{k=1}^{\dim V_i} |\varphi_k\rangle\langle\varphi_k|)U^\dagger dU
\end{aligned} \tag{4.81}$$

The choice of states $|\varphi_k\rangle \in V_i$ is completely arbitrary, so we use this freedom and chose them to be the orthonormal basis of subspace $V_i$. Then $\sum_{k=1}^{\dim V_i} |\varphi_k\rangle\langle\varphi_k|$ equals unity on the subspace $V_i$ i.e. $\sum_{k=1}^{\dim V_i} |\varphi_k\rangle\langle\varphi_k| = P_{V_i}$ and since $U$ acts unitarily on $V_i$ we have:

$$U(\sum_{k=1}^{\dim V_i} |\varphi_k\rangle\langle\varphi_k|)U^\dagger = U.P_{V_i}.U^\dagger = P_{V_i} \ . \tag{4.82}$$

If the Haar measure $dU$ is standardly normalized then equation (4.81) yields $\rho_i = \frac{1}{\dim V_i} P_{V_i}$.

We proceed further by solving the USD of $\rho_1$, $\rho_2$. It suffice to work in the Hilbert space $\mathcal{H}' = V_1 \oplus V_2 \subset \mathcal{H}$, because the supports of $\rho_1$, $\rho_2$ are contained there. The key step in the solution is the use of appropriate basis. We will construct orthonormal basis $\{|e_i\rangle\}$ of $\mathcal{H}'$ from Jordan basis $\{|a_i\rangle\}$, $\{|b_j\rangle\}$ of subspaces $V_1$, $V_2$.

It is always possible to construct the Jordan basis of a pair of subspaces with the following properties, which are discussed together with the basis construction in recipe 1 from Chapter 2:

$$\begin{aligned}
\langle a_i|a_k\rangle &= \delta_{ik} \quad \forall i,k = 1,\ldots,\dim V_1, \\
\langle b_j|b_l\rangle &= \delta_{jl} \quad \forall j,l = 1,\ldots,\dim V_2, \\
\langle a_i|b_j\rangle &= \delta_{ij}\cos\theta_i \geq 0 \ .
\end{aligned}$$

We shall use the following notation:

| | |
|---|---|
| $n_1$ | dimension of the subspace $V_1$ |
| $n_2$ | dimension of the subspace $V_2$ |
| $n_c$ | $\dim (V_1 \bigcap V_2)$ = number of i's such that $\cos\theta_i = 1$ |
| $n_0$ | number of i's such that $\cos\theta_i = 0$ |

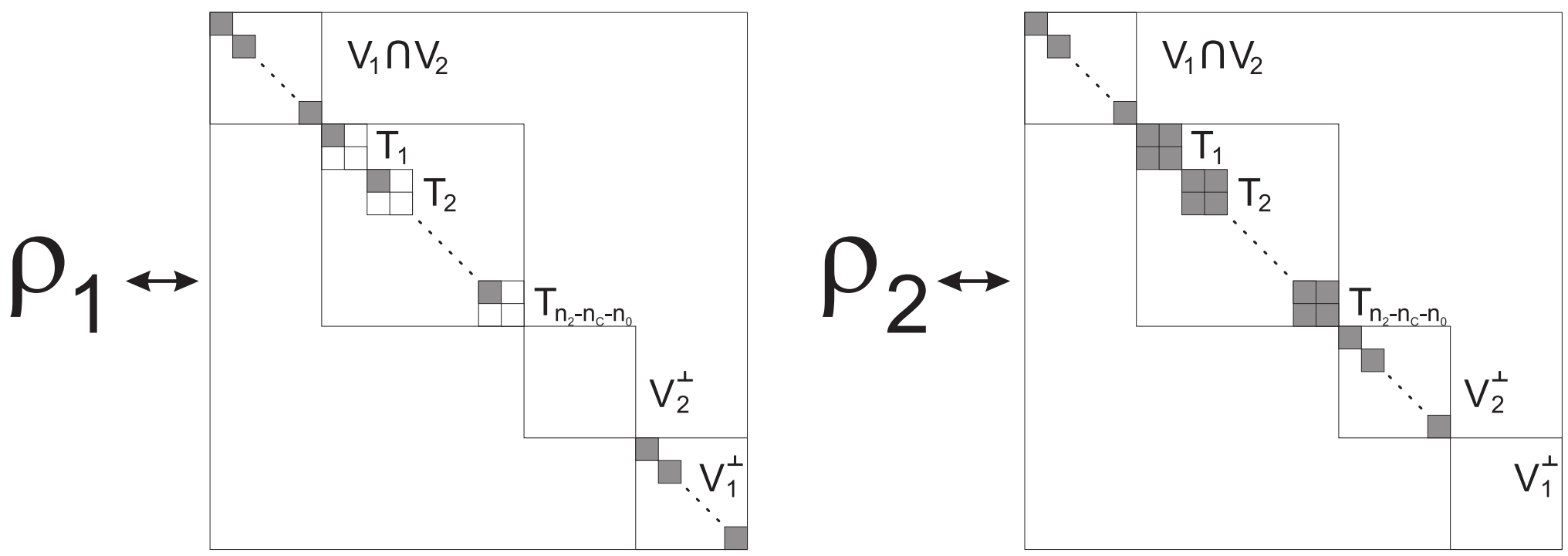

Figure 4.6. Structure of $\rho_1$, $\rho_2$ matrices in the basis $\{|e_i\rangle\}$.

| | |
|---|---|
| $T_k$ | subspace spanned by $\lvert a_{n_c+k}\rangle$, $\lvert b_{n_c+k}\rangle$, $k = 1, \ldots, n_2 - n_c - n_0$ |
| $\lvert a^{\perp}_{n_c+k}\rangle$ | state from subspace $T_k$ orthogonal to $\lvert a_{n_c+k}\rangle$ |
| $\lvert b^{\perp}_{n_c+k}\rangle$ | state from subspace $T_k$ orthogonal to $\lvert b_{n_c+k}\rangle$ |
| $V_1^{\perp}$ | Subspace of $V_2$ orthogonal to subspace $V_1$ |
| $V_2^{\perp}$ | Subspace of $V_1$ orthogonal to subspace $V_2$ |

We begin the fabrication of the basis $\{|e_i\rangle\}$ by prescribing first elements $|e_i\rangle = |a_i\rangle = |b_i\rangle$, $i = 1, \ldots, n_c$ to span the common subspace $V_1 \bigcap V_2$. We append them by pairs of vectors $|a_i\rangle$, $|a_i^{\perp}\rangle$, $i = n_c + 1, \ldots, n_2 - n_0$. We complete the basis by appending the unused Jordan basis vectors $|a_i\rangle$, $i = n_2 - n_0 + 1, \ldots, n_1$ from $V_1$ and $|b_j\rangle$, $j = n_2 - n_0 + 1, \ldots, n_2$ from $V_2$. The last mentioned vectors span the subspaces $V_2^{\perp}$, $V_1^{\perp}$ respectively. Each element $|a_i\rangle$ is contained directly in the basis $\{|e_i\rangle\}$ and each element $|b_j\rangle$ is either contained directly or can be obtained as a linear combination of vectors $|a_j\rangle$ and $|a_j^{\perp}\rangle$ from the subspace $T_{j-n_c}$. Thus, $\{|e_i\rangle\}$ is the basis of $\mathcal{H}' = V_1 \oplus V_2$, which is by construction orthonormal[13].

The matrixes of mixed states $\rho_1 = \frac{1}{n_1}\sum_{i=1}^{n_1} |a_i\rangle\langle a_i|$, $\rho_2 = \frac{1}{n_2}\sum_{j=1}^{n_2} |b_j\rangle\langle b_j|$ have in the basis $\{|e_i\rangle\}$ simple structure, which is depicted on the picture 4.6. The common part of the support of $\rho_1$, $\rho_2$ is $V_1 \bigcap V_2$ and is span by $|e_i\rangle$, $i = 1, \ldots, n_c$. Therefore, we can use the first Raynal's reduction theorem to split it off from $\rho$'s. We can then split off subspaces $V_1^{\perp}$, $V_2^{\perp}$ via the second reduction theorem. Finally, the problem can be reduced via the third reduction theorem to $n_2 - n_c - n_0$ unambiguous discriminations of pairs of pure states $|a_i\rangle$, $|b_i\rangle$. Application of the reduction theorems is not complicated, it only involves tedious calculations. Thus, we present

[13]The orthonormality follows from the properties of Jordan basis

directly its results. The final form of the POVM is the following:

$$\begin{aligned} E_1 &= P_2^{\perp} + \sum_{k=1}^{n_2-n_c-n_0} c_1^k |b_{n_c+k}^{\perp}\rangle\langle b_{n_c+k}^{\perp}| \\ E_2 &= P_1^{\perp} + \sum_{k=1}^{n_2-n_c-n_0} c_2^k |a_{n_c+k}^{\perp}\rangle\langle a_{n_c+k}^{\perp}| \\ E_0 &= I - E_1 - E_2, \end{aligned} \tag{4.83}$$

where $P_i^{\perp}$ is projector onto $V_i^{\perp}$ $(i = 1, 2)$ and $c_i^k$ are given by the solution of the USD in subspace $T_k$. The overall probability of discrimination of mixed states $\rho_1$, $\rho_2$ and also the probability of discrimination of subspaces $V_1$, $V_2$ is given by:

$$P_D = 1 - \left(\frac{\eta_1}{n_1} + \frac{\eta_2}{n_2}\right)\left(n_c + \sum_{k=1}^{n_2-n_c-n_0} (1 - P_D^k)\right), \tag{4.84}$$

where $P_D^k$ is the probability of discrimination of pure states $|a_{n_c+k}\rangle$, $|b_{n_c+k}\rangle$ $(k = 1, \ldots, n_2 - n_c - n_0)$ appearing with prior probabity $\eta_1^k$, $\eta_2^k$ respectively. It is worth to note that all these prior probabilities are equal and read:

$$\eta_1^k = \frac{\frac{\eta_1}{n_1}}{\frac{\eta_1}{n_1} + \frac{\eta_2}{n_2}} \equiv \eta_1', \quad \eta_2^k = \frac{\frac{\eta_2}{n_2}}{\frac{\eta_1}{n_1} + \frac{\eta_2}{n_2}} \equiv \eta_2' \tag{4.85}$$

On the other hand, due to varying overlap $\lambda = \langle a_{n_c+k}|b_{n_c+k}\rangle = \cos\theta_{n_c+k}$ of the discriminated states from the subspace $T_k$, the borders between the regimes of the USD measurement can be different for each $k$. The intermediate regime with the measurement being a true POVM is legitimate for prior probability $\eta_1^k$ from interval $B_k = [\frac{\cos^2\theta_{n_c+k}}{1+\cos^2\theta_{n_c+k}}, \frac{1}{1+\cos^2\theta_{n_c+k}}]$. The intervals are successively included one into another: $B_k \subset B_m$ for $k \leq m$. Therefore, the intersection of all the intervals is always not empty and is equal to $B_1$. Suppose $\eta_1' \in B_1$, then the intermediate regime is legitimate for each $k$, so:

$$\begin{aligned} \eta_1 &\in [\frac{n_1\cos^2\theta_{n_c+1}}{n_1\cos^2\theta_{n_c+1} + n_2}, \frac{n_1}{n_1 + n_2\cos^2\theta_{n_c+1}}] \equiv I_1 \\ c_1^k &= \frac{1 - \sqrt{\frac{\eta_2}{\eta_1}\frac{n_1}{n_2}}\cos\theta_{n_c+k}}{1 - \cos^2\theta_{n_c+k}}, \quad c_2^k = \frac{1 - \sqrt{\frac{\eta_1}{\eta_2}\frac{n_2}{n_1}}\cos\theta_{n_c+k}}{1 - \cos^2\theta_{n_c+k}} \\ P_D^k &= 1 - \frac{2\cos\theta_{n_c+k}}{\frac{\eta_1}{n_1} + \frac{\eta_2}{n_2}}\sqrt{\frac{\eta_1\eta_2}{n_1 n_2}} \end{aligned} \tag{4.86}$$

which enables the overall probability of discrimination to be written explicitly as:

$$P_D = 1 - \left(\frac{\eta_1}{n_1} + \frac{\eta_2}{n_2}\right) n_c - 2\sqrt{\frac{\eta_1\eta_2}{n_1 n_2}} \sum_{k=1}^{n_2-n_c-n_0} \cos\theta_{n_c+k} \tag{4.87}$$

However, if we move $\eta_1$ away from $I_1$, the number of intervals in which $\eta_1'$ is contained would decrease successively. Thus, in more and more $k$'s the regime with projective measurement ($c_i^k$ equal 0 or 1) is optimal. Let us denote by $R$ the number of different values of $\cos\theta_i$ such that $0 < \cos\theta_i < 1$. The interval [0,1] is split into $2R+1$ subintervals each of them prescribing different form of the optimal measurement and different formula for the overall probability of discrimination. Therefore, as $R$ increases it becomes very complicated to specify the optimal measurement and the probability of subspace discrimination in the whole range $\eta_1 \in [0,1]$. On the other hand, if all Jordan angles from $(0,1)$ are equal, then $R = 1$ and unambiguous subspace discrimination has only three regimes as well as USD of two known pure states.

*Application of unambiguous subspace discrimination on UI*

We saw in the previous subsection that unambiguous subspace discrimination can be reformulated and solved as discrimination of a pair of mixed states, which are multiples of projectors on those subspaces. Conversely, we can think of unambiguous discrimination of a pair of mixed states, which are multiples of projectors as about discrimination of subspaces given by the projectors. In Subsection 4.3.3 we showed that UI can be reformulated as discrimination of mixed states, which are multiples of projectors if $S$, the set of possible reference states, is derived from a subspace $\mathcal{H}_S$ of $\mathcal{H}$. Thus, for two types of reference states UI can be viewed and explicitly solved via unambiguous subspace discrimination.

If any choice of the reference states from $\mathcal{H}$ is expected then the subspaces to be discriminated are completely determined by the exchange symmetry that comes from equality of the unknown state and the given reference state. This means that UI measurement is distinguishing two types of symmetry the input states $|\Psi_1\rangle$, $|\Psi_2\rangle$ have.

In the rest of this subsection we would like to explain the known results on UI of two types of reference states in the spirit and terms of unambiguous subspace discrimination.

*Qubits*

As we already mentioned in the beginning of the chapter qubit scenarios with more than one copy of unknown or reference states were solved via unambiguous subspace discrimination. The key property that strongly influence the structure of the optimal measurement is $R$, the number of different values of $\cos\theta_i \in (0,1)$ in the Jordan basis we need to construct for the corresponding pair of subspaces. UI for $m$ copies of unknown state and one copy per each of the two reference states ($M = 2$, $n_A = m$, $n_B = n_C = 1$) was the first scenario in which the unambiguous subspace discrimination was used. Surprisingly, all the Jordan angles are in this case either zero ($\cos\theta_i = 1$) or $\cos\theta_i = 1/(m+1)$, thus $R = 1$ and the optimal measurement has only three regimes depending on the prior probability $\eta_1$. Another consequence of such Jordan angles is that the measurement elements $E_1$, $E_2$ are multiples of projectors[14]. One could expect the same situation also for the scenario with one copy of unknown state and $n$ copies for each of the two reference states ($M = 2$, $n_A = 1$, $n_B = n_C = n$). However, quite opposite is true and the situation is complicated also if we increase the number of copies of unknown states ($M = 2$, $n_A = m$, $n_B = n_C = n$). The Jordan angles are quite distinct resulting in $R = n$. Thus, it is complicated to work out probability of identification explicitly for any prior probability $\eta_1$. For $\eta_1 = 1/2$ the formula can be easily written, because the true POVM regime is optimal for all involved unambiguous pure state discriminations. The resulting probability of identification is a finite sum of fractions depending only on $n$ and $m$. Unfortunately, it can not be summed up easily and the expression covers few lines.

[14] The corresponding subspaces are given by vectors $\{|b^{\perp}_{n_c+k}\rangle\}_{k=1}^{n_2-n_c}$, $\{|a^{\perp}_{n_c+k}\rangle\}_{k=1}^{n_2-n_c}$ respectively.

*Qudits*

The results for qubits were obtained by unambiguous subspace discrimination, therefore the previous paragraph only point out some specific features of the known results. However, the results on qudit UI with two reference states were obtained by Hayashi et.al. [45] in a different way. The view on the problem through unambiguous subspace discrimination will enable as to extend the results also to unequal prior probabilities $\eta_1$, $\eta_2$. Let us first focus on the scenario with one copy of unknown state and one copy of each of the two reference states ($M = 2$, $n_A = 1$, $n_B = n_C = 1, d > 2$). The solution of the problem must be the same irrespective of the method we have used. Thus, we can use Hayashi's solution to infer the Jordan angles that we would obtain by constructing the Jordan basis corresponding to the problem. There is no ambiguity in such inference as we will see. The corresponding mixed states $\rho_1$,$\rho_2$ read:

$$\rho_1 = \frac{2}{d(d+1)} P_{AB}^{sym} \otimes I_C, \quad \rho_2 = \frac{2}{d(d+1)} P_{AC}^{sym} \otimes I_B, \tag{4.88}$$

Therefore, dimensions $n_1$, $n_2$ of the discriminated subspaces are equal. On the subspace spanned by the supports of the mixed states $\rho_1$,$\rho_2$ Hay-ashi's optimal measurement elements $E_1$, $E_2$ are $2/3$ multiples of projectors[15]. From the general form of POVM elements of unambiguous subspace discrimination given in equations (4.83) we can certainly conclude that:

- $P_1^\perp = P_2^\perp = 0$, because $E_1$, $E_2$ do not have part on which they precisely project, thus Jordan angles $\pi/2$ ($\cos\theta_i = 0$) will not occur i.e. $n_0 = 0$.
- all the coefficients $c_i^k = 2/3$, which together with $n_1 = n_2$, $\eta_1 = \eta_2$ and equations (4.86) implies $\cos\theta_{n_c+k} = 1/2$, $k = 1, \ldots, (d+1)d(d-1)/3$, where the range of $k$ can be derived from the explicit form of Hayashi's measurement, similarly as the dimension of common subspace $n_c = d(d+1)(d+2)/6$

This suffice to recover the Hyashi's average probability of identification (valid for equal prior probabilities) from the equation (4.87):

$$\overline{P_{opt}(S_{pure})} = \frac{1}{3}\frac{d-1}{d} \tag{4.89}$$

However, the knowledge of the Jordan basis and Jordan angles we obtained suffice to give the optimal measurement for arbitrary prior probabilities. Equality of all Jordan angles (different from zero) implies that only the multiples in front of the projectors in optimal $E_1$, $E_2$ will vary as we change the prior probabilities $\eta_1$, $\eta_2$. In other words $R = 1$ and the optimal measurement has only three different regimes. Thus, the unambiguous subspace discrimination together with the known solution for the special case $\eta_1 = \eta_2 = 1/2$ enabled as to find the solution in the whole range of prior probabilities. The same reasoning can be used also for the scenario with $n$ copies of the reference states ($M = 2$, $n_A = 1$, $n_B = n_C = n, d > 2$). However, the Jordan angles are distinct in this case and although we formally obtain the optimal measurement for any $\eta_1$, it is hard to write the measurement explicitly.

[15] More details about the structure of the optimal Hayashi's measurement for this case will be presented in the next section.

### 4.3.5 Influence of prior knowledge on UI

In this part of the chapter we illustrate how a prior knowledge of $S$, the subset of expected reference states, influence the optimal UI measurement and its performance with respect to universal UI measurement ($S = S_{pure}$). We illustrate this on two examples. The first one are equatorial qubits. They are a variation of the basic UI problem ($M = 2$, $n_A = n_B = n_C = 1$, $d = 2$) with the additional prior information that the reference states are chosen only from the equator of the qubit Bloch sphere. The second example is UI of coherent states i.e. $M = 2$, $n_A = n_B = n_C = 1$, $d = \infty$ with the additional prior information that the reference states are coherent states.

*Equatorial qubits*

We denote the state of a qubit $|e_\varphi\rangle = 1/\sqrt{2}(|0\rangle + e^{\imath\varphi}|1\rangle)$ with $\varphi \in [0, 2\pi]$ as an equatorial state. Let us denote the subset of all equatorial states $S_{eq}$. We shall find optimal UI measurement in the case we have one copy both from the unknown state and from the two reference states ($M = 2$, $n_A = n_B = n_C = 1$). Thus, the aim is to optimize $\overline{P(S_{eq})}$ the probability of identification averaged over the set $S_{eq}$. We shall first calculate the corresponding mixed states $\rho_1, \rho_2$:

$$\begin{aligned}\rho_i &= \int_{S_{eq}}\int_{S_{eq}} \Psi_i d\psi_1 d\psi_2 = \\ &= \frac{1}{(2\pi)^2}\int_0^{2\pi}\int_0^{2\pi} |e_{\varphi_i}\rangle\langle e_{\varphi_i}| \otimes |e_{\varphi_1}\rangle\langle e_{\varphi_1}| \otimes |e_{\varphi_2}\rangle\langle e_{\varphi_2}|\, d\varphi_1 d\varphi_2.\end{aligned}$$

The integration yields:

$$\begin{aligned}\rho_1 &= \frac{1}{8} I_C \otimes (|00\rangle\langle 00| + |11\rangle\langle 11| + 2|\psi^+\rangle\langle\psi^+|)_{AB} \\ \rho_2 &= \frac{1}{8} I_B \otimes (|00\rangle\langle 00| + |11\rangle\langle 11| + 2|\psi^+\rangle\langle\psi^+|)_{AC}.\end{aligned} \tag{4.90}$$

Consequently, we will solve USD of $\rho_1, \rho_2$ in the same way as J. Bergou et. al. in [43]. Hence, we calculate zero eigenvectors of $\rho_1$, $\rho_2$, because they determine the subspaces in which POVM elements $E_1$, $E_2$ can operate:

$$\begin{aligned} \rho_2|a_i\rangle &= 0, & \rho_1|b_i\rangle &= 0 \\ |a_1\rangle &= |0\rangle_B \otimes |\psi^-_{AC}\rangle & |b_1\rangle &= |0\rangle_C \otimes |\psi^-_{AB}\rangle \\ |a_2\rangle &= |1\rangle_B \otimes |\psi^-_{AC}\rangle & |b_2\rangle &= |1\rangle_C \otimes |\psi^-_{AB}\rangle \end{aligned} \tag{4.91}$$

$$E_1 = \sum_{i,j=1}^{2} \alpha_{ij}|a_i\rangle\langle a_j| \qquad E_2 = \sum_{i,j=1}^{2} \beta_{ij}|b_i\rangle\langle b_j| \tag{4.92}$$

Our goal is to maximize $\overline{P(S_{eq})}$, while keeping the POVM elements positive. We use equations (4.90) and (4.92) to express $\overline{P(S_{eq})}$ via coefficients $\alpha_{ij}$, $\beta_{ij}$:

$$\begin{aligned}\overline{P(S_{eq})} &= \eta_1 Tr(E_1\rho_1) + \eta_2 Tr(E_2\rho_2) \\ &= \frac{\eta_1}{8}(\alpha_{11} + \alpha_{22}) + \frac{\eta_2}{8}(\beta_{11} + \beta_{22})\end{aligned} \tag{4.93}$$

Accidentally, the states $|a_i\rangle$, $|b_j\rangle$ are the same as in the paper of J. Bergou et. al. [43] and also the expression for $\overline{P(S_{eq})}$ coincides with $\overline{P(S_{pure})}$ optimized there (see equations 3.19-3.22). Therefore, the optimization task and the resulting measurement is in our case exactly the same as for universal UI of qubits. Hence, the optimal UI measurement for equatorial qubits is also given by equation (4.67).

*Unambiguous Identification of coherent states*

Under unambiguous identification of coherent states we will mainly think of the UI task from our framework with $M = 2$, $n_A = n_B = n_C = 1$, $d = \infty$ with the additional prior information that the reference states are coherent states. Instead of dealing directly with infinite dimensional quantum systems and their coherent states we first show how intuitive universal UI measurement for arbitrary dimension can be constructed and compare it to the optimal universal UI measurement found by Hayashi et.al. [45]. This will enable us to find (in large $d$ limit) the corresponding universal UI measurement for quantum systems described in infinite dimensional $\mathcal{H}_\infty$ and to calculate the probability of identification for particular choice of reference states $P(|\psi_1\rangle, |\psi_2\rangle)$.

*The Swap Based approach for qudits*

The optimal POVM elements $\mathrm{E}_1, \mathrm{E}_2$ for universal UI of qubits ($M = 2$, $n_A = n_B = n_C = 1$, $d = 2$) are proportional to the projectors onto the antisymmetric subspace of the two qubit subsystem AC, respectively AB. The simple generalization of the aforementioned universal UI measurement to the case of qudits is the following POVM, which we abbreviate by sb (stands for the "swap based"):

$$\begin{aligned}
\mathrm{E}_1^{sb} &= c_1 I_B \otimes \mathrm{P}_{AC}^{asym} = c_1 I_B \otimes \frac{1}{2}(1 - \mathrm{Swap}_{\mathrm{AC}}), \\
\mathrm{E}_2^{sb} &= c_2 I_C \otimes \mathrm{P}_{AB}^{asym} = c_2 I_C \otimes \frac{1}{2}(1 - \mathrm{Swap}_{\mathrm{AB}}), \\
\mathrm{E}_0^{sb} &= I - \mathrm{E}_1^{sb} - \mathrm{E}_2^{sb},
\end{aligned} \tag{4.94}$$

where $\mathrm{P}_{XY}^{asym}$ denotes the projector onto the antisymmetric subspace of subsystems $X$ and $Y$, and $c_1$, $c_2$ are so far unspecified real numbers. Elements $\mathrm{E}_0^{sb}, \mathrm{E}_1^{sb}, \mathrm{E}_2^{sb}$ have to form a valid POVM, therefore certain conditions for $c_1, c_2$ must hold. Indeed, positivity of $\mathrm{E}_1^{sb}$ and $\mathrm{E}_2^{sb}$ implies $c_1 \geq 0$ and $c_2 \geq 0$, whereas the inequality imposed by the positivity of $\mathrm{E}_0^{sb}$ is not so apparent and we have to calculate the eigenvalues of $\mathrm{E}_0^{sb}$ explicitly. Let $\{|\,i\,\rangle\}_{i=1}^d$ denote a basis of the qudit Hilbert space $\mathcal{H}$. Then $|ijk\rangle \equiv |i\rangle_A \otimes |j\rangle_B \otimes |k\rangle_C$ is the basis of the three-qudit Hilbert space $\mathcal{H}^{\otimes 3}$. The operator $E_0^{sb}$ can be expressed in terms of unit and $\mathrm{Swap}$ operators so $\langle ijk|\mathrm{E}_0^{sb}|lmn\rangle \equiv 0$ whenever $\{ijk\}$ is not a permutation of $\{lmn\}$. In other words if we properly reorder this basis $\mathrm{E}_0^{sb}$ is block diagonal matrix. The blocks are of three types depending on the number of equivalent indeces:

- the trivial $1 \times 1$ block $\langle iii|\mathrm{E}_0^{sb}|iii\rangle = 1$
- the $3 \times 3$ block with matrix $\langle \sigma_1(iij)|\mathrm{E}_0^{sb}|\sigma_2(iij)\rangle$
- the $6 \times 6$ block with matrix $\langle \sigma_1(ijk)|\mathrm{E}_0^{sb}|\sigma_2(ijk)\rangle$.

The dimensionalities of the blocks are given by the number of inequivalent permutations $\sigma$ of the three indexes.

For qubits only the blocks of first two types occur in the matrix $\mathrm{E}_0^{sb}$ whereas for qudits $(d > 2)$ blocks of all three types arise. Hence we reduced the problem of finding the eigenvalues of a rank $d^3$ operator $\mathrm{E}_0^{sb}$ to calculation of the eigenvalues of the small matrices mentioned above. The calculation of those eigenvalues is only technical and therefore treated in detail in the appendix B.1. The condition resulting from requiring them to be nonnegative is a particulary simple inequality:

$$c_1 + c_2 \leq 1. \tag{4.95}$$

A straightforward consequence of the block diagonality of $\mathrm{E}_0^{sb}$ is that the inequality $(4.95)$ assures positivity of $\mathrm{E}_0^{sb}$ regardless of the dimension $d$ (provided that $d > 2$). We want to choose $c_1, c_2$ so that POVM requirements are satisfied and the average probability of UI (4.65) is maximal. However, let us first look on the probability of identification for particular choice of reference states $(4.64)$

$$\begin{aligned} P_{sb}(|\psi_1\rangle, |\psi_2\rangle) &= \eta_1 \langle\Psi_1|\mathrm{E}_1^{sb}|\Psi_1\rangle_{ABC} + \eta_2 \langle\Psi_2|\mathrm{E}_2^{sb}|\Psi_2\rangle_{ABC} = \\ &= \frac{\eta_1 c_1 + \eta_2 c_2}{2}(1 - |\langle\psi_1|\psi_2\rangle|^2). \end{aligned} \tag{4.96}$$

We see that optimal choice of $c_1, c_2$ does not depend on $|\psi_1\rangle$, $|\psi_2\rangle$ but only on prior probabilities $\eta_1, \eta_2$. Hence, the values of $c_1, c_2$ simultaneously maximizing $P_{sb}(|\psi_1\rangle, |\psi_2\rangle)$ and the average probability of UI (4.65) read: $c_1 = 0$, $c_2 = 1$ for $\eta_1 < \eta_2$, $c_1 = 1$, $c_2 = 0$ for $\eta_1 > \eta_2$ and $c_1 + c_2 = 1$ for $\eta_1 = \eta_2$. Therefore, for equal prior probabilities the UI probability is independent on the particular choice of $c_1$ and $c_2$ and is given by:

$$P_{sb}(|\psi_1\rangle, |\psi_2\rangle) = \frac{1}{4}(1 - |\langle\psi_1|\psi_2\rangle|^2). \tag{4.97}$$

However, due to symmetry reasons we further consider $c_1 = c_2 = 1/2$ in case $\eta_1 = \eta_2$, which gives:

$$\mathrm{E}_1^{sb} = \frac{1}{2} I_B \otimes \mathrm{P}_{AC}^{asym}, \quad \mathrm{E}_2^{sb} = \frac{1}{2} I_C \otimes \mathrm{P}_{AB}^{asym} \tag{4.98}$$

Averaging over all pure states can be easily done by using the integral

$$\int\int_{S_{pure}} |\langle\psi_1|\psi_2\rangle|^2 d\psi_1 d\psi_2 = 1/d$$

from [49]. We obtain the average probability of UI for the swap-based measurement

$$\overline{P_{sb}(S_{pure})} = \frac{1}{4}\left(\frac{d-1}{d}\right). \tag{4.99}$$

Although the probability (4.97) itself is independent of the dimension the average value converges to $1/4$ in the limit of $d \to \infty$. This corresponds to an intuitive expectation that two randomly chosen unit vectors in $\mathcal{H}$ are more likely to be orthogonal for higher values of $d$.

*Optimal universal UI for qudits - Hayashi's result*

Although POVM elements $\mathrm{E}_1^{sb}, \mathrm{E}_2^{sb}$ proportional to projectors onto antisymmetric part of the subsystem AC respectively AB seem intuitively as the best universal UI measurement, actually results of Hayashi et.al. [45] imply they are not. We now present explicit form of Hayashi's result for our considered problem and explain how it actually differs from the intuitive expectation. First of all, his universal UI measurement maximizes for equal prior probabilities the mean probability of identification $\overline{P(S_{pure})}$. This allows symmetry to be fruitfully exploited via representations of the unitary group $U \in U(d) \mapsto U^{\otimes 3}$ and via the permutation group $S(3)$ permuting the subsystems $A$,$B$,$C$ of $\mathcal{H}^{\otimes 3}$. As we already mentioned earlier the optimal measurement has the following form:

$$\mathrm{E}_1^{opt} = e.I_B \otimes \mathrm{P}_{AC}^{asym}, \quad \mathrm{E}_2^{opt} = e.I_C \otimes \mathrm{P}_{AB}^{asym}, \quad e = \sum_\lambda e_\lambda \Gamma_\lambda, \tag{4.100}$$

where $\lambda$ specifies both $U(d)$ and $S(3)$ type of the irreducible representation, $\Gamma_\lambda$ is the projector onto that invariant subspace of $\mathcal{H}^{\otimes 3}$, and $e_\lambda$ is non-negative real number. In this case ($M = 2$, $n_A = 1$, $n_B = n_C = n$) only two irreducible $U(d)$ representations specified by $\lambda = (2, 1, 0, \ldots, 0)$, $\lambda = (1, 1, 1, \ldots, 0)$ are relevant. The corresponding $e_\lambda$'s are $2/3$ and $1/2$. Therefore we have:

$$e = \frac{2}{3}\Gamma_{(2,1,0,\ldots,0)} + \frac{1}{2}\Gamma_{(1,1,1,\ldots,0)}. \tag{4.101}$$

Projectors $\Gamma_{(2,1,0,\ldots,0)}$ and $\Gamma_{(1,1,1,\ldots,0)}$ project onto the subspaces $(V_S \oplus V_{AS})^\perp$ and $V_{AS}$, where $V_S$ (respectively $V_{AS}$) is the totaly symmetric (respectively antisymmetric) subspace of $\mathcal{H}^{\otimes 3}$. Operators $I_C \otimes \mathrm{P}_{AB}^{asym}$, $I_B \otimes \mathrm{P}_{AC}^{asym}$ do not mix subspaces $(V_S \oplus V_{AS})^\perp$ and $V_{AS}$ on which operator $e$ is only a multiple of identity. Therefore $\mathrm{E}_1^{opt}$ (analogously $\mathrm{E}_2^{opt}$) is essentially $\frac{2}{3} I_B \otimes \mathrm{P}_{AC}^{asym}$ except for $V_{AS}$, where it is $\frac{1}{2} I_B \otimes \mathrm{P}_{AC}^{asym}$. Furthermore, part of the POVM elements $\mathrm{E}_1^{opt}$, $\mathrm{E}_2^{opt}$ acting on the totaly antisymmetric subspace $V_{AS}$ does not contribute to $P_{opt}(|\psi_1\rangle, |\psi_2\rangle)$ and $\overline{P_{opt}(S_{pure})}$, because input states $|\Psi_i\rangle_{ABC}$ (equation (4.61)) are symmetric in a pair of subsystems. Thus, for calculation of probabilities of identification we can as well use $\mathrm{E}_1 = \frac{2}{3} I_B \otimes \mathrm{P}_{AC}^{asym} = \frac{4}{3}\mathrm{E}_1^{sb}$, $\mathrm{E}_2 = \frac{2}{3} I_C \otimes \mathrm{P}_{AB}^{asym} = \frac{4}{3}\mathrm{E}_2^{sb}$ to obtain:

$$P_{opt}(|\psi_1\rangle, |\psi_2\rangle) = \frac{1}{3}(1 - |\langle\psi_1|\psi_2\rangle|^2) \tag{4.102}$$

$$\overline{P_{opt}(S_{pure})} = \frac{1}{3}\frac{d-1}{d} \tag{4.103}$$

Unlike previous sections, where we have considered unambiguous identification of quantum states from finite dimensional Hilbert space $\mathcal{H}$, here we will work with infinite dimensional Hilbert space of linear harmonic oscillator $\mathcal{H}_\infty$, which models a single mode of electromagnetic field (EM). The two techniques for UI of qudits presented above work for any dimension $d$. The resulting POVM elements are expressed via constant multiples of projectors, which in large d limit define also projectors on $\mathcal{H}_\infty^{\otimes 3}$. Therefore, we have formally same looking universal UI measurement also for states from $\mathcal{H}_\infty$. This measurement should be optimal for universal UI for the case of equal prior probability $\eta_1 = \eta_2$.

Our goal in this subsection is to show that UI of coherent states can be done with much better probability of identification than universal UI of all pure states from $\mathcal{H}_\infty$. The basic intuition for this is that coherent states form a very small subset $S_{coh}$ of all pure states from $\mathcal{H}_\infty$ and there could be a better way to identify them. The more reasonable motivation is based on the following observation: As we showed in Section 4.3.2 in UI for qubits appearing with arbitrary prior probabilities the optimal POVM is constructed via quantum state comparison measurement. Hence, if there is a better quantum state comparison of coherent states, which can be employed in UI setup for coherent states then this setup could outperform the universal UI measurement identifying all states from $\mathcal{H}_\infty$. E. Andersson, M. Curty and I. Jex [41] proposed such a quantum state comparison setup, which is also simply realizable by beamsplitter and a photodetector. In what follows we shortly explain how their setup works and afterwards we show how it can be used for a proposition of UI setup for coherent states.

*Quantum comparison of coherent states*

In comparison of coherent states we want to unambiguously distinguish between $|\alpha\rangle = |\beta\rangle$ and $|\alpha\rangle \neq |\beta\rangle$. This is equivalent to distinguishing $\beta - \alpha = 0$ and $\beta - \alpha \neq 0$, which can be done by 50/50 beamsplitter ($T = R = 1/2$) in the following way. The operation of the beamsplitter on coherent states is particularly simple, because it does not entangle the output modes:

$$|\alpha\rangle \otimes |\beta\rangle \mapsto |\sqrt{T}\alpha + \sqrt{R}\beta\rangle \otimes | - \sqrt{R}\alpha + \sqrt{T}\beta\rangle. \tag{4.104}$$

The state of the second mode after passing the beamsplitter will be either vacuum $|0\rangle$ or $|\frac{1}{\sqrt{2}}(\beta - \alpha)\rangle$ when $\alpha \neq \beta$. Thus, if we detect at least one photon in the second mode, which happens with probability $1 - |\langle 0|\frac{1}{\sqrt{2}}(\beta - \alpha)\rangle|^2 = 1 - e^{-\frac{1}{2}|\alpha-\beta|^2}$, we are sure that the states were different. On the other hand the detection of no photons is inconclusive, because all coherent states have nonzero overlap with vacuum.

*UI with three beamsplitters*

Motivated by the UI measurement in the case of qubits (equation (4.67)), we want to design a measurement that in a sense for each single run simultaneously performs comparisons of coherent states of subsystems AC and AB. For two separate state comparisons we can use two beamsplitters, so it seems natural to employ a third one to perform them simultaneously. Therefore, we consider a setup consisting of three beamsplitters $B_1$, $B_2$, and $B_3$ depicted on figure 4.7. We keep the notation of subsystems from our framework (mode A contain unknown state, B and C reference states) except for the added fourth ancillary mode D initially prepared in vacuum. Thus, the whole product state we are in general given can be written as $|\alpha_?\rangle_A \otimes |\alpha_1\rangle_B \otimes |\alpha_2\rangle_C \otimes |0\rangle_D$, where $|\alpha_?\rangle$ is guaranteed to be either $|\alpha_1\rangle$ or $|\alpha_2\rangle$.

Our three beamsplitters act on it in the following way:

$$|\alpha_?\rangle_A|\alpha_1\rangle_B|\alpha_2\rangle_C|0\rangle_D \mapsto U_3(DC).U_2(BA).U_1(DA)|\alpha_?\rangle_A|\alpha_1\rangle_B|\alpha_2\rangle_C|0\rangle_D,$$

where $U_i(XY)$ is unitary transformation performed by the $i$-th beamsplitter on the modes X and Y. Let us fix the transmitivity $T_1$ of $B_1$ for a moment and calculate the output states of $B_1$ using (4.104):

$$|0\rangle_D \otimes |\alpha_?\rangle_A \mapsto |\sqrt{R_1}\alpha_?\rangle_D \otimes |\sqrt{T_1}\alpha_?\rangle_A. \tag{4.105}$$

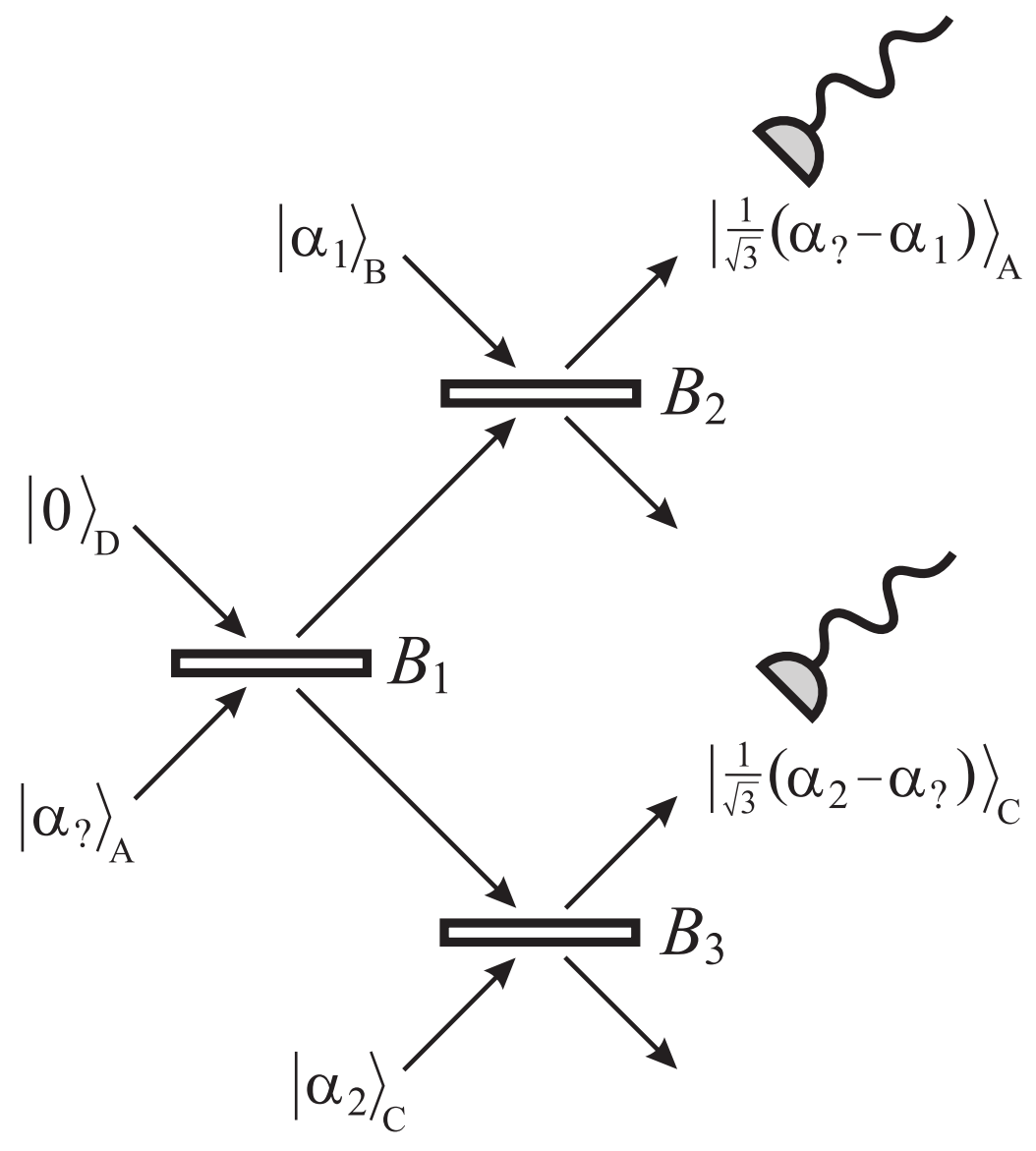

Figure 4.7. The beamsplitter setup designed for an unambiguous identification of coherent states.

The outputs of $B_1$ are in product state, so it suffice to analyze beamsplitters $B_2$ and $B_3$ separately. Beamsplitter $B_2$ transforms state of modes $A$, $B$ in the following way:

$$|\alpha_1\rangle_B \otimes |\sqrt{T_1}\alpha_?\rangle_A \mapsto |\sqrt{T_2}\alpha_1 + \sqrt{R_2 T_1}\alpha_?\rangle_B \otimes |-\sqrt{R_2}\alpha_1 + \sqrt{T_2 T_1}\alpha_?\rangle_A. \tag{4.106}$$

In case $\alpha_? = \alpha_1$ we want beamsplitter $B_2$ to behave as in comparison of identical states $|\alpha_1\rangle$, $|\alpha_1\rangle$. This means we want mode $A$ to be transformed into vacuum if $\alpha_? = \alpha_1$, which implies $\sqrt{T_2}\sqrt{T_1} - \sqrt{R_2} = 0$. This condition can be rewritten as:

$$T_2 = \frac{1}{1 + T_1}. \tag{4.107}$$

We proceed analogously for beamsplitter $B_3$:

$$\begin{aligned} &|\sqrt{R_1}\alpha_?\rangle_D \otimes |\alpha_2\rangle_C \mapsto \\ &\mapsto |\sqrt{T_3 R_1}\alpha_? + \sqrt{R_3}\alpha_2\rangle_D \otimes |-\sqrt{R_3 R_1}\alpha_? + \sqrt{T_3}\alpha_2\rangle_C. \end{aligned} \tag{4.108}$$

In case $\alpha_? = \alpha_2$ we want mode $C$ to be transformed into vacuum. This implies $\sqrt{T_3} - \sqrt{R_3 R_1} = 0$, which can be written as:

$$T_3 = \frac{1 - T_1}{2 - T_1}. \tag{4.109}$$

Equations (4.107), (4.109) can be met simultaneously, therefore we set the transmitivities $T_2, T_3$ according to them. The final state of our four modes after passing all three beamsplitters can be simply obtained from equations (4.106), (4.108) and reads:

$$|\sqrt{R_2}(\alpha_? - \alpha_1)\rangle_A \otimes |\sqrt{T_2}\alpha_1 + \sqrt{R_2 T_1}\alpha_?\rangle_B \otimes \\ \otimes |\sqrt{T_3}(\alpha_2 - \alpha_?)\rangle_C \otimes |\sqrt{T_3 R_1}\alpha_? + \sqrt{R_3}\alpha_2\rangle_D$$

The field modes are factorized, therefore as our State comparison motivation suggests we can focus only on state of modes $A$ and $C$. Depending on $\alpha_?$, modes $A$ and $C$ end up in state:

$$\begin{aligned} \alpha_? = \alpha_1 \quad &: \quad |0\rangle_A \otimes |\sqrt{T_3}(\alpha_2 - \alpha_1)\rangle_C \\ \alpha_? = \alpha_2 \quad &: \quad |\sqrt{R_2}(\alpha_2 - \alpha_1)\rangle_A \otimes |0\rangle_C \end{aligned} \tag{4.110}$$

We measure modes $A$ and $C$ by photodetectors $P_2$ and $P_1$ respectively. In each single run of experiment we can distinguish four situations: none of the detectors clicks, only $P_1$ clicks, only $P_2$ clicks, both detectors click. In our situation both detectors cannot click at the same time, because at least one of the modes is in vacuum. If only detector $P_1$ clicks from equations (4.110) we unambiguously conclude that $\alpha_? = \alpha_1$. Similarly if only detector $P_2$ clicks we unambiguously conclude that $\alpha_? = \alpha_2$. If none of the detectors click we cannot determine which mode was not in vacuum and therefore it is an inconclusive result.

In case $\alpha_? = \alpha_1$ the probability of correct identification follows from equations (4.110) and is given by the probability of detecting at least one photon in mode C:

$$1 - |\langle 0|\sqrt{T_3}(\alpha_2 - \alpha_1)\rangle|^2 = 1 - e^{-\frac{1-T_1}{2-T_1}|\alpha_1 - \alpha_2|^2}.$$

In case $\alpha_? = \alpha_2$ the probability of correct identification is given by the probability of detecting at least one photon in mode A:

$$1 - |\langle 0|\sqrt{R_2}(\alpha_2 - \alpha_1)\rangle|^2 = 1 - e^{-\frac{T_1}{1+T_1}|\alpha_1 - \alpha_2|^2}.$$

Thus, the probability of identification for reference states $|\alpha_1\rangle$, $|\alpha_2\rangle$ reads:

$$P_{bs}(|\alpha_1\rangle, |\alpha_2\rangle) = \eta_1(1 - e^{-\frac{1-T_1}{2-T_1}|\alpha_1 - \alpha_2|^2}) + \eta_2(1 - e^{-\frac{T_1}{1+T_1}|\alpha_1 - \alpha_2|^2}) \tag{4.111}$$

Next we want to optimize the performance of the setup by properly choosing transmitivity $T_1$. The definition of the uniform distribution on the set of coherent states is problematic, therefore we first focus on the probability of identification for a particular choice of reference states $|\alpha_1\rangle_B$, $|\alpha_2\rangle_C$ expressed by equation (4.111). In fact, this we later help us to draw more general conclusions. By plotting the $P_{bs}(|\alpha_1\rangle, |\alpha_2\rangle)$ for various ranges of $|\alpha_1 - \alpha_2|$,$\eta_1 \in [0, 1]$ and $T_1 \in [0, 1]$ one quickly finds that for the fixed values of $\eta_1$ and $|\alpha_1 - \alpha_2|$ the probability $P_{bs}(|\alpha_1\rangle, |\alpha_2\rangle)$ is maximal for the values of $T_1$ that depend on $\eta_1$ and $|\alpha_1 - \alpha_2|$. Thus, in general, for arbitrary prior probability optimal transmitivity $T_1$ depends on the reference states to be identified. However, we will show that in the special case of equal prior probabilities there is only one value of transmitivity $T_1$, which is optimal for all reference states. This value turns out to be $T_1 = 1/2$ as we for equal prior probabilities expect from symmetry reasons. In order to show this we calculate $\frac{\partial P_{bs}(|\alpha_1\rangle, |\alpha_2\rangle)}{\partial T_1}$ from Eq. (4.111) for $\eta_1 = \eta_2 = 1/2$ and the condition for critical points (vanishing the first derivative) yields:

$$1 = \frac{(1 + T_1)^2}{(2 - T_1)^2} e^{-|\alpha_1 - \alpha_2|^2 (\frac{1-T_1}{2-T_1} - \frac{T_1}{1+T_1})}. \tag{4.112}$$

For $0 \leqq T_1 < 1/2$ both terms on the right hand side (RHS) of (4.112) are greater than 1, for $1/2 < T_1 \leqq 1$ both terms are less than 1 and for $T_1 = 1/2$ both terms on the RHS are 1. Thus, $T_1 = 1/2$ is the only critical point for all reference states and because of the second derivative being negative it is the global maximum of $P_{bs}(|\alpha_1\rangle, |\alpha_2\rangle)$ for $T_1 \in [0, 1]$.

Further, we consider the UI of coherent states appearing with equal prior probabilities. In this case the optimal choice of transmitivities for our three beamsplitter setup is $T_1 = 1/2, T_2 = 2/3$ $T_3 = 1/3$, which enables the probability of identification (4.111) to be written as:

$$P_{bs}(|\alpha_1\rangle, |\alpha_2\rangle) = 1 - e^{-\frac{1}{3}|\alpha_1 - \alpha_2|^2} \tag{4.113}$$

*Comparison of UI strategies acting on coherent states*

In the previous paragraphs we have discussed three different UI measurements that can be used to identify coherent states:
*i)* the swap-based measurement,
*ii)* the optimal measurement,
*iii)* the beamsplitter setup.
The first two schemes unambiguously identify arbitrary states of qudits in arbitrary dimensions. The beamsplitter setup is designed to identify only coherent states. Although the comparison is usually understood in terms of average probabilities, we will adopt a different comparison method evaluating the performance directly in terms of probabilities $P(|\alpha_1\rangle, |\alpha_2\rangle)$ for all pairs of states. It turns out that for all the measurements these probabilities depend only on a scalar product of states under consideration.

As we mentioned in the beginning of this subsection qudit POVM elements $\mathrm{E}_i^{sb}$ and $\mathrm{E}_i^{opt}$ in large $d$ limit define also POVM elements in $\mathcal{H}_\infty^{\otimes 3}$. For simplicity we use the same notation for these operators. These two UI strategies are universal, so they work for any pure states from $\mathcal{H}_\infty$. If applied on coherent states the corresponding probabilities are given by Eqs. (4.97) and (4.102)

$$P_{sb}(|\alpha_1\rangle, |\alpha_2\rangle) = \frac{1}{4}(1 - |\langle\alpha_1|\alpha_2\rangle|^2) = \frac{1}{4}(1 - e^{-|\alpha_1 - \alpha_2|^2}) \tag{4.114}$$

$$P_{opt}(|\alpha_1\rangle, |\alpha_2\rangle) = \frac{1}{3}(1 - |\langle\alpha_1|\alpha_2\rangle|^2) = \frac{1}{3}(1 - e^{-|\alpha_1 - \alpha_2|^2}). \tag{4.115}$$

In what follows we will compare $P_{sb}(|\alpha_1\rangle, |\alpha_2\rangle)$, $P_{opt}(|\alpha_1\rangle, |\alpha_2\rangle)$, and $P_{bs}(|\alpha_1\rangle, |\alpha_2\rangle)$, which is a probability of identification for a beamsplitter setup designed especially for coherent states (see equation (4.113)). The following inequality holds for arbitrary coherent states $|\alpha_1\rangle$, $|\alpha_2\rangle$:

$$P_{sb} \leq P_{opt} \leq P_{bs} \tag{4.116}$$

Hence the same relation between the measurements holds also on average. The inequality can be derived as follows. We denote $e^{-|\alpha_1 - \alpha_2|^2}$ as $x$ ($x \in [0, 1]$). All the probabilities are zero for $\alpha_1 = \alpha_2$ ($x = 1$) as they should, because then the reference states are the same. The validity of the inequality can be proved by showing the reversed inequality for the first derivative of the probabilities with respect to $x$ (4.116), i.e.

$$\partial_x P_{sb} \geq \partial_x P_{opt} \geq \partial_x P_{bs} \quad \Leftrightarrow \quad -\frac{1}{4} \geq -\frac{1}{3} \geq -\frac{1}{3x^{\frac{2}{3}}} .$$

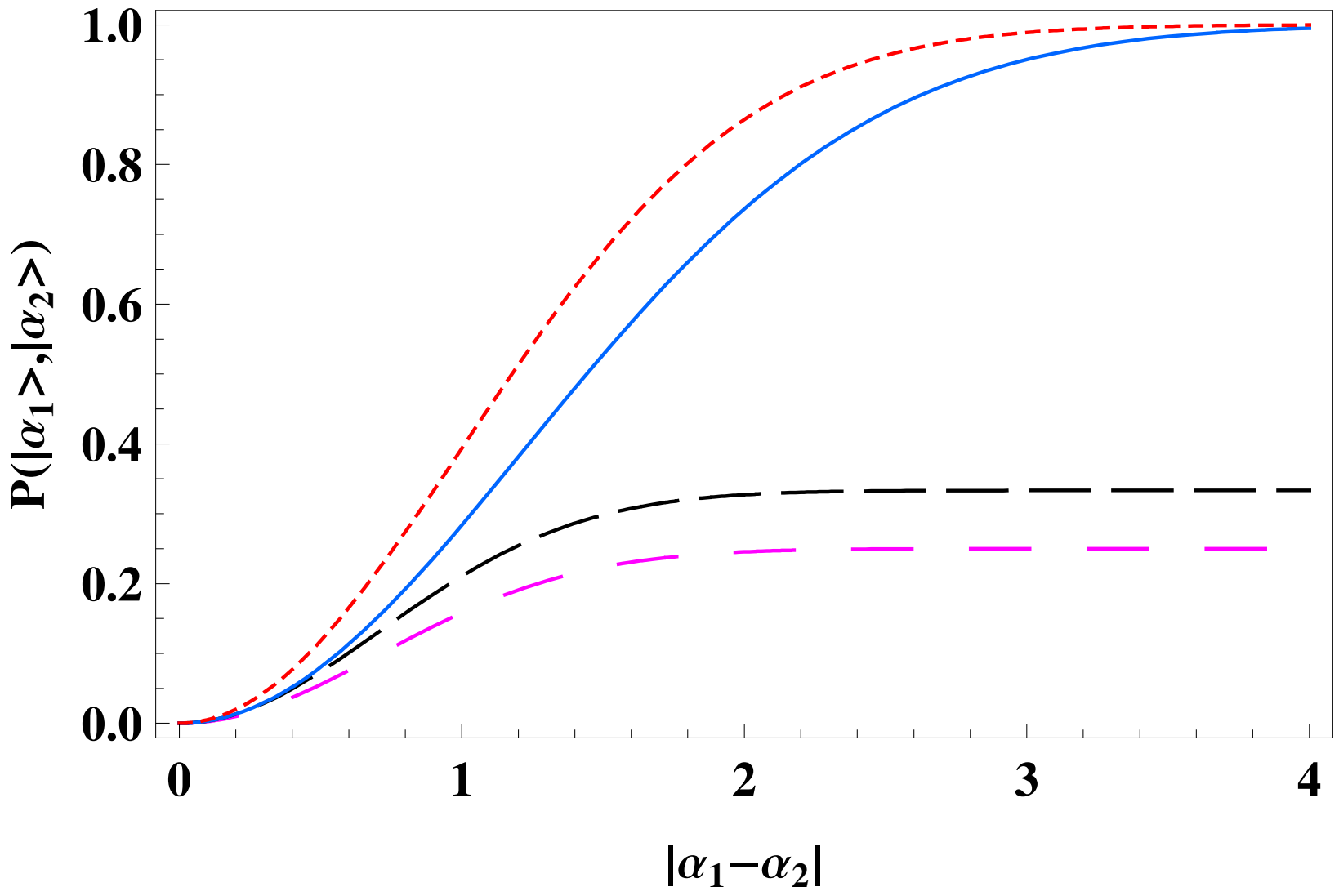

Figure 4.8. The probability of identification $P(|\alpha_1\rangle, |\alpha_2\rangle)$ as a function of the scalar product (given by $|\alpha_1 - \alpha_2|$) for three UI strategies applied on coherent states $|\alpha_1\rangle$, $|\alpha_2\rangle$. Starting from the bottom the two lowest lines correspond to universal UI measurements (the swap-based is in magenta and the optimal strategy is in black, respectively). The blue line is associated with the beam splitter setup that was designed especially for coherent states. The top (red) curve corresponds to the optimal discrimination probability among two known states.

The last row obviously holds in the interval $x \in [0, 1]$, so inequality (4.116) is proved. More quantitative insight is given in the figure (4.8) showing the dependence of the probability of identification for the considered UI strategies on the $|\alpha_1 - \alpha_2|$. As a result we can conclude that the beamsplitter setup designed for an unambiguous identification of coherent states performs better than the optimal universal UI measurement. Another remarkable feature is that the beamsplitter setup attains $P_{bs}(|\alpha_1\rangle, |\alpha_2\rangle) = 1$ for large values of $|\alpha_1 - \alpha_2|$, i.e. in the limit when two coherent states are orthogonal.

In this section we have addressed the problem of an unambiguous identification of unknown coherent states. We have explicitly designed UI measurement taking into account the a priori knowledge about a particular family of states and compared the proposed measurement with the universal unambiguous identification, i.e. the UI measurements (either the swap-based or the optimal one) that can be applied for all pure states. Our main goal was to design a simple experimental setup consisting of three beamsplitters (see Fig. 4.7) that performs best.

The proposed beamsplitters setup for unambiguous state identification can be compared with the measurement proposed in Ref. [50] discriminating optimally among two *known* coherent states. Both of them consists of three beamsplitters, but arranged differently. An interesting observation is that the differences between the probabilities are not very large (see Fig. 4.8) and even more surprising is the fact that two unknown nearly orthogonal coherent states can be identified almost perfectly. For the universal optimal UI measurement (see Fig. 4.8) there is a significant gap between the probabilities for state discrimination and state identification.

### 4.3.6 Optimal UI of coherent states with linear optics

In this part of the chapter we study possible generalization of an optical setup proposed in the previous section. More specifically, we show how the UI of coherent states can be performed in a general case when multiple copies of unknown and reference states are available. This investigation is motivated by the observation that with the increase of the number of identically prepared particles we can better identify the preparator. We shall also prove optimality of the discussed UI setups for multiple copies of unknown and reference states.

We start our investigation by stating the problem within our framework for unambiguous identification. We consider modes of a quantum electromagnetic field (linear harmonic oscillators) each described by a semi-infinite-dimensional Hilbert space $\mathcal{H}_\infty$ and prepared in a coherent state of a specific amplitude. We denote the complex amplitude of the unknown coherent state by $\alpha_?$ and similarly we denote the reference states as $|\alpha_1\rangle, |\alpha_2\rangle, \ldots, |\alpha_M\rangle$. Thus, in general, we should unambiguously distinguish the following $M$ possible types of states:

$$|\Psi_i\rangle_{ABC\ldots} \equiv |\alpha_i\rangle_A^{\otimes n_A} \otimes |\alpha_1\rangle_B^{\otimes n_B} \otimes |\alpha_2\rangle_C^{\otimes n_C} \otimes \ldots , \tag{4.117}$$

where $\alpha_? = \alpha_i$ and $i = 1, 2, \ldots, M$. We assume that the states of the type $|\Psi_i\rangle$ appear with an equal prior probability $\eta_i = 1/M$. The performance of the considered UI measurement will be most often quantified by the probability of identification for a particular choice of reference states $P(|\alpha_1\rangle, \ldots, |\alpha_M\rangle)$, because the optimal parameters of the measurement setup will not depend on specific reference states. Hence, most of the features that the averaged probability $\overline{P(S_{coh})}$ would have should be apparent already in $P(|\alpha_1\rangle, \ldots, |\alpha_M\rangle)$. Nevertheless, in accordance with our framework the aim is the maximization of the average value

$$\overline{P(S_{coh})} = \int_{\mathbb{C}^M} P(|\alpha_1\rangle, \ldots, |\alpha_M\rangle)\chi(\alpha_1, \ldots, \alpha_M)\, d\alpha_1 \ldots d\alpha_M, \tag{4.118}$$

where $\chi(\alpha_1, \ldots, \alpha_M)$ is the probability distribution describing our knowledge about the choice of reference states. In Eq. (4.118) we integrate over multiple infinite (complex) planes of complex amplitudes. Unfortunately, a uniform distribution on an infinite plane can not be properly defined. Thus, $\chi(\alpha_1, \ldots, \alpha_M)$ can not be uniform, but instead should be "regularized", i.e. it should satisfy some reasonable physical requirements. For example, the probability of having reference states with very big amplitudes, i.e. of very high energy, should be vanishing. This illustrates that UI of coherent states can not be so easily reformulated as unambiguous discrimination of $M$ mixed states. Even if there was a natural and mathematically allowed choice for $\chi(\alpha_1, \ldots, \alpha_M)$ we could not apply the tools for unambiguous discrimination of mixed states directly, because most of them were proved only for finite dimensional Hilbert spaces. Our approach to the problem is a bit more operational. We shall construct optical setups, which by construction perform UI for any coherent reference states and prove their optimality under the restriction that only linear optical elements and photodetectors are used in the measurement.

Naturally, coherent states encode complex numbers. From this point of view the state $|\frac{1}{\sqrt{2}}\alpha_?\rangle$ carries formally the whole information about the complex amplitude $\alpha_?$. This is due to the fact

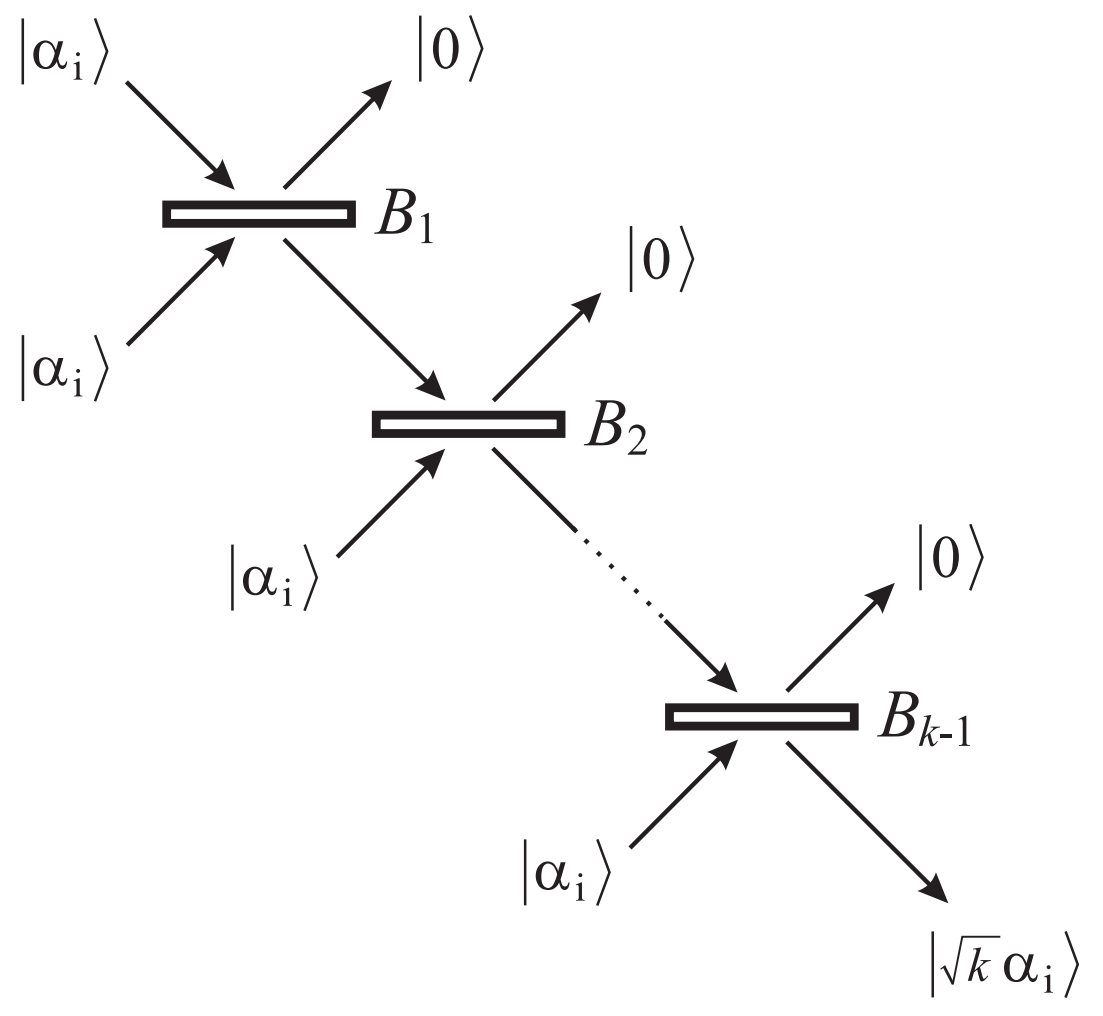

Figure 4.9. The beamsplitter setup designed for constructive interference of the same input coherent states. Out of $k$ copies of a coherent state $|\alpha_i\rangle$ we obtain at the output of a sequence of $k-1$ beamspitters one mode in the coherent state $|\sqrt{k}\alpha_i\rangle$ and $k-1$ modes in a vacuum state $|0\rangle^{\otimes k-1}$.

that we know the factor $\lambda = 1/\sqrt{2}$ by which $\alpha_?$ is rescaled. If the complex amplitude $\alpha_?$ is encoded in the state $|\lambda\alpha_?\rangle$ then for $0 \le \lambda < 1$ we will speak about a "diluted" unknown state while the case $\lambda > 1$ will be referred to as a "concentrated" unknown state $|\alpha_?\rangle$. These terms come from the fact that the "diluted" state can be obtained by superimposing a coherent state and a vacuum via a beam splitter. As a result of the beam splitter transformation two modes at the output of the beam splitter are in the diluted states. On the contrary, the "concentrated" state can be prepared by launching two copies of the same coherent state into the beam splitter. As a result we obtain one of the output modes in the "concentrated" state while the second mode in the vacuum state. Using a sequence of beamsplitters and corresponding resources one can prepare "diluted' or "concentrated" states with arbitrary value of the scaling factor $\lambda$. Actually, preparation of "concentrated" states is the main idea we will employ in our investigation of the UI measurement with multiple copies of unknown and reference states. At the beginning of the UI measurement we will, for each kind of state, concentrate the information encoded in its $k$ copies into a single quantum system. This can be done by a sequence of $k-1$ beam splitters (see Fig. 4.9) with transmitivities chosen so that the input state $|\alpha_i\rangle^{\otimes k}$ constructively interferes to produce the state $|\sqrt{k}\alpha_i\rangle \otimes |0\rangle^{\otimes k-1}$. More details about this transformation can be found in Section 4.2.4. The result of these preliminary transformations is a mapping of possible types of states $|\Psi_i\rangle$ into states $|\sqrt{n_A}\alpha_i\rangle_{A_1} \otimes |\sqrt{n_B}\alpha_1\rangle_{B_1} \otimes |\sqrt{n_C}\alpha_2\rangle_{C_1} \otimes \ldots \otimes |0\rangle^t$, where $t = n_A - 1 + n_B - 1 + \ldots$. As a next step we will use the setup proposed in previous section for a single copy of the unknown state and single copies of $M$ reference states. Of course, as we will see below the transmitivities of all beam splitters in the setup must be modified according to the number of copies of the unknown and the reference states we are given.

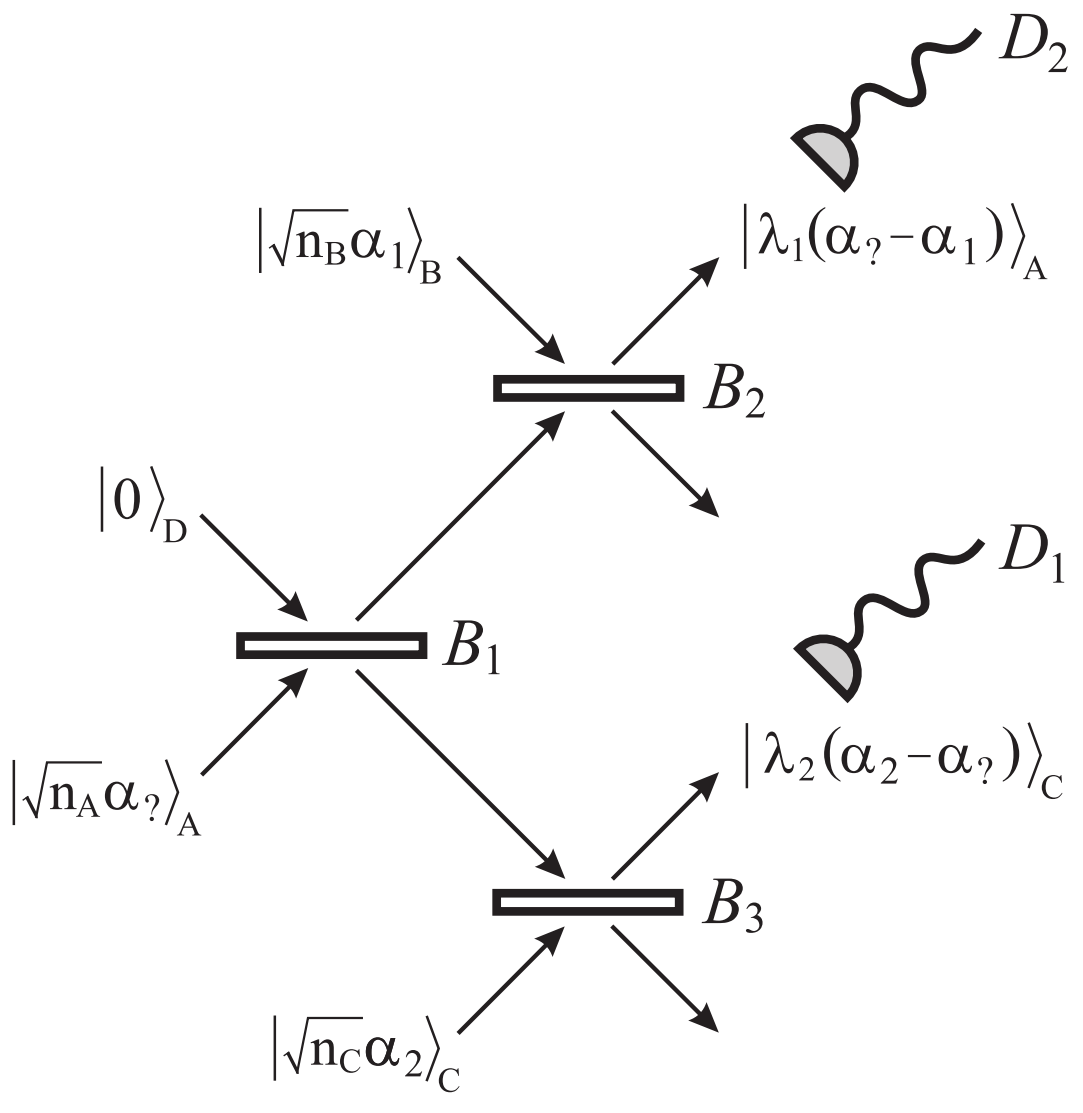

Figure 4.10. The beamsplitter setup designed for an unambiguous identification of multiple copies of two types of coherent states.

*Two types of reference states*

The unambiguous identification of two types of coherent reference states is the first natural step in generalizing the scenario with single copies of unknown and reference states investigated in previous section. In this section we consider $M = 2$ while $n_A, n_B, n_C$ are arbitrary. The above mentioned idea of "concentration" of quantum information implies that we first feed all provided copies of the unknown state into $n_A - 1$ beam splitters to obtain state $|\sqrt{n_A}\alpha_?\rangle$ in the mode $A_1$ (for brevity later called only $A$). The other modes $A_2 \ldots A_{n_A}$ end up in a vacuum state, therefore we will not consider them further. Similarly, $n_B - 1$ (respectively, $n_C - 1$) beam splitters are used to prepare the state $|\sqrt{n_B}\alpha_1\rangle$ (respectively, $|\sqrt{n_C}\alpha_2\rangle$) in the modes $B_1$ ($C_1$). Next, we feed these concentrated states into essentially the same scheme as proposed in the previous section (see Fig. 4.10). Thus, altogether we are going to use $n_A + n_B + n_C$ beam-splitters. The analysis of the setup presented in Fig. 4.10 is analogous to the one presented in the previous section, therefore we comment on it only briefly.

The beam splitter setup in Fig. 4.10 acts on input modes that are prepared in the state

$$|\Phi_{in}\rangle = |\sqrt{n_A}\alpha_?\rangle_A \otimes |\sqrt{n_B}\alpha_1\rangle_B \otimes |\sqrt{n_C}\alpha_2\rangle_C \otimes |0\rangle_D \,,$$

where $\alpha_?$ is guaranteed to be either $\alpha_1$ or $\alpha_2$. The action of the three beamsplitters in the setup is described by a unitary transformation

$$|\Phi_{in}\rangle \mapsto |\Phi_{out}\rangle = (U^{(2)}_{AB} \otimes U^{(3)}_{CD})(U^{(1)}_{AD} \otimes I_{BC})|\Phi_{in}\rangle \,,$$

where $U^{(j)}_{XY}$ is associated with the $j$-th beamsplitter $B_j$ acting on the modes $X$ and $Y$. Since beamsplitters do not entangle coherent states it follows that the output state $|\Phi_{out}\rangle$ remains fac-torized. In the first step the beamsplitter $B_1$ with transmittivity $T_1$ prepares two "diluted" copies of the state $|\sqrt{n_A}\alpha_?\rangle$, i.e.

$$|0\rangle_D \otimes |\sqrt{n_A}\alpha_?\rangle_A \mapsto |\sqrt{R_1 n_A}\alpha_?\rangle_D \otimes |\sqrt{T_1 n_A}\alpha_?\rangle_A \, . \tag{4.119}$$

In the second step the beamsplitters $B_2, B_3$ perform the transformation such that the output state reads

$$|\Phi_{out}\rangle = |out\rangle_A \otimes |out\rangle_B \otimes |out\rangle_C \otimes |out\rangle_D \, , \tag{4.120}$$

with

$$\begin{aligned} |out\rangle_A &= |-\sqrt{R_2 n_B}\alpha_1 + \sqrt{T_2 T_1 n_A}\alpha_?\rangle_A \, , \\ |out\rangle_B &= |\sqrt{T_2 n_B}\alpha_1 + \sqrt{R_2 T_1 n_A}\alpha_?\rangle_B \, , \\ |out\rangle_C &= |-\sqrt{R_3 R_1 n_A}\alpha_? + \sqrt{T_3 n_C}\alpha_2\rangle_C \, , \\ |out\rangle_D &= |\sqrt{T_3 R_1 n_A}\alpha_? + \sqrt{R_3 n_C}\alpha_2\rangle_D \, . \end{aligned}$$

A crucial observation is that the parameters $T_j, R_j = 1 - T_j$ can be adjusted so that either the mode $A$, or the mode $C$, ends up in a vacuum state providing that $\alpha_? = \alpha_1$, or $\alpha_? = \alpha_2$, respectively. In particular, setting the transmittivities to

$$T_2 = \frac{1}{1 + \frac{n_A}{n_B} T_1} \, ; \quad T_3 = \frac{1 - T_1}{\frac{n_C}{n_A} + 1 - T_1} \, , \tag{4.121}$$

we find

$$\begin{aligned} |out\rangle_A &= |\sqrt{R_2 n_B}(\alpha_? - \alpha_1)\rangle_A \, ; \\ |out\rangle_B &= |\sqrt{T_2 n_B}\alpha_1 + \sqrt{R_2 T_1 n_A}\alpha_?\rangle_B \, ; \\ |out\rangle_C &= |\sqrt{T_3 n_C}(\alpha_2 - \alpha_?)\rangle_C \, ; \\ |out\rangle_D &= |\sqrt{T_3 R_1 n_A}\alpha_? + \sqrt{R_3 n_C}\alpha_2\rangle_D \, . \end{aligned} \tag{4.122}$$

Finally, we perform a photodetection in the output modes $A$ and $C$ by the photodetectors $D_2$ and $D_1$, respectively. By detecting a photon in one of the two modes we can unambiguously identify the unknown state. In particular, for these two modes we have

$$\begin{aligned} &\alpha_? = \alpha_1 \leftrightarrow |0\rangle_A \otimes |\sqrt{T_3 n_C}(\alpha_2 - \alpha_1)\rangle_C \, ; \\ &\alpha_? = \alpha_2 \leftrightarrow |\sqrt{R_2 n_B}(\alpha_2 - \alpha_1)\rangle_A \otimes |0\rangle_C \, . \end{aligned} \tag{4.123}$$

We note that due to the fact that at least one of the modes is in a vacuum state both detectors cannot "click", i.e. cannot detect photons simultaneously. Therefore, in each single run of the experiment only three situations can happen:

*i)* none of the detectors click,

*ii)* only the detector $D_1$ clicks,

*iii)* only the detector $D_2$ clicks.

If only the detector $D_1$ clicks then following Eqs. (4.123) we unambiguously conclude that $\alpha_? = \alpha_1$. Similarly, if only the detector $D_2$ clicks we unambiguously conclude that $\alpha_? = \alpha_2$.

If none of the detectors click we cannot determine which mode was not in the vacuum state and therefore this situation represents an inconclusive result.

If $\alpha_? = \alpha_1$, then the probability of a correct identification is given as the probability of detecting at least one photon in the mode $C$

$$P_1 = 1 - |\langle 0|\sqrt{T_3 n_C}(\alpha_2 - \alpha_1)\rangle|^2 = 1 - e^{-\frac{n_C n_A(1-T_1)}{n_C + n_A(1-T_1)}|\alpha_1-\alpha_2|^2} . \tag{4.124}$$

Analogously, in the case $\alpha_? = \alpha_2$ the probability of a correct identification reads

$$P_2 = 1 - |\langle 0|\sqrt{R_2 n_B}(\alpha_2 - \alpha_1)\rangle|^2 = 1 - e^{-\frac{n_B n_A T_1}{n_B + n_A T_1}|\alpha_1-\alpha_2|^2} . \tag{4.125}$$

Thus the total probability of the identification of reference states $|\alpha_1\rangle$ and $|\alpha_2\rangle$ is equal to

$$P(|\alpha_1\rangle, |\alpha_2\rangle) = \eta_1 P_1 + \eta_2 P_2 = \frac{1}{2}(P_1 + P_2) . \tag{4.126}$$

Next we will optimize the performance of the setup by choosing an appropriate value of the transmittivity $T_1$. The definition of the uniform distribution on a set of coherent states is problematic, therefore we first focus our attention on the probability of identification for a particular choice of reference states $|\alpha_1\rangle$ and $|\alpha_2\rangle$ expressed by Eq. (4.126).

The investigation of the first derivative $\frac{\partial P(|\alpha_1\rangle,|\alpha_2\rangle)}{\partial T_1}$ reveals that the optimal choice of $T_1$ does not depend on the reference states $|\alpha_1\rangle$, $|\alpha_2\rangle$ only if $n_B = n_C$. As one expects, because of symmetry arguments, $T_1$ is optimally set to $1/2$ if $n_B = n_C$. In such a case, $P(|\alpha_1\rangle, |\alpha_2\rangle)$ can be simplified to take the following form:

$$P(|\alpha_1\rangle, |\alpha_2\rangle) = 1 - e^{-\frac{n_A n_B}{n_A + 2n_B}|\alpha_1-\alpha_2|^2} . \tag{4.127}$$

Let us note that if $n_B \neq n_C$ then there exists a prior probability $\eta_1 = 1 - \eta_2$ for which the optimal choice of $T_1$ does not depend on the reference states. However, as already mentioned, we focus on the $\eta_i = 1/M$ case and we will assume that we are given the same number of copies of each reference state.

*Trade-off between resources*

The number of copies of an unknown state or of a reference state we have can be seen as a measure of some resource. From this point of view an interesting question immediately arises: Which type of resource is more useful for an unambiguous identification of coherent states? Are unknown states more useful than reference states or vice versa? To answer these questions we consider the following situation. Imagine we will get altogether $N$ quantum systems (modes of electromagnetic field) but we have a liberty to specify whether the specific mode is prepared in the unknown state or in one of the two reference states. Thus, if we ask for $n_A$ copies of the unknown state we will obtain $n_B = n_C = (N - n_A)/2$ copies per a reference state. Let us for simplicity assume that $N$ and $n_A$ have the same parity. The probability of identification for a reference states $|\alpha_1\rangle$, $|\alpha_2\rangle$ then reads

$$P(|\alpha_1\rangle, |\alpha_2\rangle) = 1 - e^{-\frac{n_A(N-n_A)}{2N}|\alpha_1-\alpha_2|^2} \tag{4.128}$$

and it is maximized for $n_A = \lfloor N/2 \rfloor$, because the terms in the exponent are nonnegative. Hence, from the point of view of the resources, it is optimal to ask for a preparation of $\lfloor N/2 \rfloor$ unknown states and the equal number of copies per a reference state (specifically, $\lfloor N/4 \rfloor$).

*Infinite number of copies of reference states*

Unambiguous identification is a discrimination task in which we have very limited prior knowledge about the possible preparations of the quantum system. The amount of information about the possible preparations is essentially given by the number of copies of the reference states we obtain. In the limit of infinite number of copies the preparation of reference states becomes known (at least potentially via a tomographic measurement) and thus the UI is becoming equivalent to discrimination among *known* states. The unambiguous discrimination among pair of known pure states (for equal prior probabilities) was solved by Ivanovic, Dieks and Peres [6–8] in 1987. Their optimal measurement succeeds with a probability $1 - |\langle\varphi_1|\varphi_2\rangle|$, where $|\varphi_1\rangle, |\varphi_2\rangle$ are the known states in which the system can be prepared. In what follows we will show that in the aforementioned limit ($M = 2, n_B = n_C \to \infty$) our beam-splitter setup achieves the same optimal performance. In order to prove this we have to evaluate the limit of Eq. (4.127):

$$\begin{aligned} P(|\alpha_1\rangle, |\alpha_2\rangle, n_B &= n_C \to \infty) \\ &= \lim_{n_B \to \infty} 1 - e^{-\frac{n_A n_B}{n_A + 2n_B}|\alpha_1 - \alpha_2|^2} \\ &= 1 - e^{-\frac{n_A}{2}|\alpha_1 - \alpha_2|^2} \\ &= 1 - |\langle\alpha_1|\alpha_2\rangle|^{n_A}. \end{aligned} \tag{4.129}$$

In the last equality we have used the expression for the modulus of the overlap of the two coherent states $|\langle\alpha_1|\alpha_2\rangle|^2 = e^{-|\alpha_1 - \alpha_2|^2}$. In the limit $n_B = n_C \to \infty$ the two known states that could be unambiguously discriminated by the Ivanovic-Dieks-Peres measurement are $|\varphi_1\rangle = |\alpha_1\rangle^{\otimes n_A}$, $|\varphi_2\rangle = |\alpha_2\rangle^{\otimes n_A}$. Thus, we see that Eq. (4.129) is equal to $1 - |\langle\varphi_1|\varphi_2\rangle|$ and so our beam-splitter setup performs optimally in this limit. Let us note that for $n_A = 1$ our setup is in this limit equivalent to the setup proposed by K. Banaszek [50] for unambiguous discrimination between a pair of known coherent states. For $n_B = n_C \to \infty$ our $T_2 \to 1, T_3 \to 0$, i.e. the "concentrated" reference states are nearly reflected, which induces a displacement of the "diluted" unknown state $|\frac{1}{\sqrt{2}}\alpha_?\rangle$. In the same way K. Banaszek uses very unbalanced beam-splitters to cause the displacement of the outputs of the beam-splitter.

For the limiting case $M = 2$, $n_A = n_B = n_C \to \infty$ it is natural to expect a classical behavior, i.e. a unit probability of identification. For unequal reference states this result is easily obtained by taking the limit of Eq. (4.127).

*Weak implementation of UI measurement*

Let us consider a basic version of the UI of coherent states ($M = 2$, $n_A = n_B = n_C = 1$). We will describe a measurement, which in the case of success, leaves all the input states nearly unperturbed and achieves the probability of identification given by Eq. (4.127). The measurement procedure goes as follows: We first equally split each of our resource states into $N$ parts. Thus, we have $N$ copies of states $|\frac{1}{\sqrt{N}}\alpha_?\rangle, |\frac{1}{\sqrt{N}}\alpha_1\rangle, |\frac{1}{\sqrt{N}}\alpha_2\rangle$. We use the three beam-splitter setup ($M = 2$, $n_A = n_B = n_C = 1$) for each of these $N$ triples. The UI measurement performed on the first triple will succeed with the probability $1 - e^{-\frac{1}{3}|\frac{1}{\sqrt{N}}\alpha_1 - \frac{1}{\sqrt{N}}\alpha_2|^2} = 1 - e^{-\frac{1}{3N}|\alpha_1-\alpha_2|^2}$. If we find $\alpha_? = \alpha_1$ we can combine the unmeasured $3N-3$ modes into the states $|\sqrt{\frac{2N-2}{N}}\alpha_1\rangle, |\sqrt{\frac{N-1}{N}}\alpha_2\rangle$. For $\alpha_? = \alpha_2$ we operate analogously obtaining $|\sqrt{\frac{N-1}{N}}\alpha_1\rangle$, $|\sqrt{\frac{2N-2}{N}}\alpha_2\rangle$. If the UI measurement of the first triples fails we continue by measuring the other triples until we find a conclusive outcome or use all the triples. In the case of the $k$-th triple leading to the conclusive result we concentrate the remaining resources to obtain the states $|\sqrt{\frac{2(N-k)}{N}}\alpha_1\rangle, |\sqrt{\frac{N-k}{N}}\alpha_2\rangle$ or $|\sqrt{\frac{N-k}{N}}\alpha_1\rangle, |\sqrt{\frac{2(N-k)}{N}}\alpha_2\rangle$ depending on $\alpha_?$ being $\alpha_1$ or $\alpha_2$. We do not get a conclusive result only if the measurements of all $N$ triples yield inconclusive results. Hence, the overall probability of successful identification of the unknown state $1 - (e^{-\frac{1}{3N}|\alpha_1-\alpha_2|^2})^N = 1 - e^{-\frac{1}{3}|\alpha_1-\alpha_2|^2}$ is the same as in Eq. (4.127). However, in contrast to the three beam splitter setup proposed in previous section, if a conclusive result is obtained before measuring the $N$-th triple we still have "diluted" input states at our disposal.

*Optimality proof*

In this section we shall prove optimality of the proposed UI setups if only linear optical elements, photodetectors and sources of multimode coherent states are allowed to be used. Due to the fact that the linear optical transformations preserve the tensor product structure of coherent states it follows that in any measurement (using arbitrarily many photodetectors) the measured state is a factorized coherent state of $N$ modes of the form $|\beta_1 \otimes \cdots \otimes \beta_N\rangle = |\beta_1\rangle \otimes \cdots \otimes |\beta_N\rangle \equiv |\vec{\beta}\rangle$. In order to use an outcome of the measurement for the unambiguous conclusion the probabilities for all the other options must vanish. Let us note that for the considered family of states each photodetector measuring the individual mode has a non-vanishing probability to observe $n > 0$ photons unless this mode is in the vacuum state, i.e. if $|\beta_j\rangle \neq |0\rangle$, then $p_n(|\beta_j\rangle) = |\langle n|\beta_j\rangle|^2 > 0$ for all $n > 0$. Only for the vacuum state $p_n(|0\rangle) = 0$. Moreover, the probability to observe no photon is non-vanishing for all coherent states, i.e. this event cannot be used for unambiguous conclusion. Consequently, the unambiguous conclusions are necessarily associated with observation of the nonzero number of photons identifying the fact that the corresponding mode is not in the vacuum state.

In unambiguous identification of two types of reference states and equal number of copies per reference state ($M = 2, n_B = n_C$) our goal is to discriminate two families of states: either $|{\alpha_1}^{\otimes n_A} \otimes {\alpha_1}^{\otimes n_B} \otimes {\alpha_2}^{\otimes n_B}\rangle$, or $|{\alpha_2}^{\otimes n_A} \otimes {\alpha_1}^{\otimes n_B} \otimes {\alpha_2}^{\otimes n_B}\rangle$, where $|\alpha_1\rangle, |\alpha_2\rangle$ are arbitrary coherent states, but $\alpha_1 \neq \alpha_2$. In general, our (Gedanken) experiment starts with a preparation of a coherent state $|{\alpha_?}^{\otimes n_A} \otimes {\alpha_1}^{\otimes n_B} \otimes {\alpha_2}^{\otimes n_B} \otimes \beta_1 \otimes \cdots\rangle$, where $|\beta_j\rangle$ are fixed states of some ancillary modes. By linear optical elements this state is mapped into a state $|\Delta_1 \otimes \Delta_2 \otimes \Delta_3 \otimes \cdots\rangle$, where $\Delta_j$ are complex numbers depending on $\alpha_?, \alpha_1, \alpha_2$. Each of these modes is measured by a photodetector. In order to make an unambiguous conclusion $\alpha_? = \alpha_1$ based also on a click of the $j$th photodetector we need to guarantee for all values of $\alpha_1, \alpha_2$ that $\Delta_j = 0$ for $\alpha_? = \alpha_2$ and $|\Delta_j| > 0$ for $\alpha_? = \alpha_1$. Similarly, for the unambiguous conclusion $\alpha_? = \alpha_2$. As it was shown by He and Bergou in Ref. [51] the linear optical transformations of coherent states can be described by unitary matrices acting on vectors of amplitudes of individual modes, i.e.

$$G.(\underbrace{\alpha_?, \ldots, \alpha_?}_{n_A}, \underbrace{\alpha_1, \ldots, \alpha_1}_{n_B}, \underbrace{\alpha_2, \ldots, \alpha_2}_{n_B}, \beta_1, \ldots)^T = (\Delta_1, \ldots)^T,$$

where $G$ is a unitary matrix. Without the loss of generality we can write $G = W.Q$, where $W, Q$ are unitary and $Q$ performs the concentration operation. More precisely we set

$$Q = \begin{pmatrix} Q_1 & O & O & O \\ O & Q_2 & O & O \\ O & O & Q_2 & O \\ O & O & O & I \end{pmatrix}, \tag{4.130}$$

where $Q_1, Q_2$ are unitary matrices $n_A \times n_A$ (respectively $n_B \times n_B$) such that

$$\begin{aligned} Q_1 &: (d, \dots, d)^T \mapsto (\sqrt{n_A} d, 0, \dots, 0)^T ; \\ Q_2 &: (d, \dots, d)^T \mapsto (\sqrt{n_B} d, 0, \dots, 0)^T . \end{aligned}$$

The transformation described by the matrix $Q$ turns the vector of input coherent states amplitudes into

$$\overrightarrow{q} \equiv (\underbrace{\sqrt{n_A}\alpha_?, 0, \dots,}_{n_A} \underbrace{\sqrt{n_B}\alpha_1, 0, \dots,}_{n_B} \underbrace{\sqrt{n_B}\alpha_2, 0, \dots,}_{n_B} \beta_1, \dots)^T .$$

Hence, we can write the result of the overall transformation $G$ via the matrix $W$ acting on the above vector $\overrightarrow{q}$

$$W.\overrightarrow{q} = (\Delta_1, \Delta_2, \dots)^T \tag{4.131}$$

with

$$\Delta_j = W_{j,1}\sqrt{n_A}\alpha_? + W_{j,n_A+1}\sqrt{n_B}\alpha_1 + W_{j,n_A+n_B+1}\sqrt{n_B}\alpha_2 + \gamma_j \tag{4.132}$$

and $\gamma_j = \sum_k W_{j,k+n_A+2n_B}\beta_k$. The condition $\Delta_j = 0$ holding for all values $\alpha_1, \alpha_2$ if $\alpha_? = \alpha_2$ implies

$$\begin{aligned} 0 &= W_{j,n_A+1} = \gamma_j \\ \lambda_j &= W_{j,1}\sqrt{n_A} = -W_{j,n_A+2n_B+1}\sqrt{n_B} \\ |\Delta_j^{(1)}\rangle &= |\lambda_j(\alpha_? - \alpha_2)\rangle, \end{aligned}$$

where the upper index indicates the association of observation of photons in this mode with the conclusion $\alpha_? = \alpha_1$. Similarly, if the $j$th mode will be associated with the conclusion $\alpha_? = \alpha_2$, then the corresponding state has to be $|\Delta_j^{(2)}\rangle = |\lambda_j(\alpha_? - \alpha_1)\rangle$.

The detectors can be divided into three classes according to the type of states that are measured: i) $|\Delta_j^{(1)}\rangle$ (detecting $\alpha_? = \alpha_1$), ii) $|\Delta_j^{(2)}\rangle$ (detecting $\alpha_? = \alpha_2$), and, iii) different type of a state corresponding to an inconclusive result. The detectors from the third class can not be employed in making unambiguous decision and hence will not be considered further. An arbitrary click on the detector i) tells us that $\alpha_? = \alpha_1$ therefore we associate these clicks with the unambiguous result $\alpha_? = \alpha_1$. Analogously, clicks from the type ii) detector are associated with the unambiguous result $\alpha_? = \alpha_2$. In what follows we shall show that the events on detectors leading to the same conclusion can be replaced by a single detector while the success probability is preserved. In other words, an experiment in which $n_1$ detectors are used to conclude that $\alpha_? = \alpha_1$ and $n_2$ detectors to detect that $\alpha_? = \alpha_2$ can be replaced by an experiment with only two

photodetectors. In particular, by renaming the output ports the output vector can be rearranged into the form

$$\begin{pmatrix} \Delta_1^{(1)} \\ \vdots \\ \Delta_{n_1+1}^{(2)} \\ \vdots \\ \Delta_{n_1+n_2+1} \\ \vdots \end{pmatrix} = \begin{pmatrix} (\alpha_? - \alpha_2)\lambda_1 \\ \vdots \\ (\alpha_? - \alpha_1)\lambda_{n_1+1} \\ \vdots \\ \Delta_{n_1+n_2+1} \\ \vdots \end{pmatrix} \equiv \vec{\alpha}'_? \,. \tag{4.133}$$

In such case we denote $\Omega \equiv e^{-|\alpha_1-\alpha_2|^2}$ and the success probability reads

$$\begin{aligned} P_{\text{success}} &= \frac{1}{2}(1 - \prod_{j=1}^{n_1} e^{-|\lambda_j(\alpha_1-\alpha_2)|^2}) + \frac{1}{2}(1 - \prod_{j=n_1+1}^{n_1+n_2} e^{-|\lambda_j(\alpha_1-\alpha_2)|^2}) \\ &= 1 - \frac{1}{2}(\Omega^{\sum_{j=1}^{n_1}|\lambda_j|^2} + \Omega^{\sum_{j=n_1+1}^{n_1+n_2}|\lambda_j|^2})\,, \end{aligned} \tag{4.134}$$

because the UI measurement fails only if none of the conclusive detectors fire. However, there exist a unitary matrix of the block diagonal form

$$U = \begin{pmatrix} U_1 & O & O \\ O & U_2 & O \\ O & O & I \end{pmatrix}\,, \tag{4.135}$$

where $U_1, U_2$ are suitable unitary matrices $n_i \times n_i$ such that

$$\begin{aligned} U_1 &: (\lambda_1, \ldots, \lambda_{n_1})^T \mapsto (\kappa_1, 0, \ldots, 0)^T\,; \\ U_2 &: (\lambda_{n_1+1}, \ldots, \lambda_{n_1+n_2})^T \mapsto (\kappa_2, 0, \ldots, 0)^T \end{aligned}$$

with $\kappa_1 = \sqrt{\sum_{k=1}^{n_1}|\lambda_k|^2}$ and $\kappa_2 = \sqrt{\sum_{k=1}^{n_2}|\lambda_{n_1+k}|^2}$. This means that the overall product of coherent states transforms into

$$U : \vec{\alpha}'_? \quad \mapsto \quad \begin{pmatrix} \kappa_1(\alpha_? - \alpha_2) \\ 0 \\ \vdots \\ \kappa_2(\alpha_? - \alpha_1) \\ 0 \\ \vdots \\ \Delta_{n_1+n_2+1} \\ \vdots \end{pmatrix}\,. \tag{4.136}$$

Two detectors measuring the first and the $(n_1+1)^{\text{th}}$ output port are of the first respectively the second type and we see that the probability of success

$$\begin{aligned} P_{\text{success}} &= \frac{1}{2}(1 - e^{-|\kappa_1(\alpha_1-\alpha_2)|^2}) + \frac{1}{2}(1 - e^{-|\kappa_2(\alpha_2-\alpha_1)|^2}) \\ &= 1 - \frac{1}{2}(\Omega^{\sum_{j=1}^{n_1}|\lambda_j|^2} + \Omega^{\sum_{j=n_1+1}^{n_1+n_2}|\lambda_j|^2}) \end{aligned} \tag{4.137}$$

equals the multidetector case [see Eq.(4.134)]. This means we have shown that it suffice to consider one conclusive photodetector of the type one and one detector of the type two. We can now go back to Eq. (4.131) and require that the states measured by the photodetectors $D_1, D_2$ have the form $|\Delta_1\rangle = |\lambda_1(\alpha_? - \alpha_2)\rangle, |\Delta_2\rangle = |\lambda_2(\alpha_? - \alpha_1)\rangle$. This implies that first, $(n_A+1)^{th}$, and $(n_A+n_B+1)^{th}$ column of the matrix $W$ is constraint in the following way:

$$W = \begin{pmatrix} \sqrt{\frac{1}{n_A}}\lambda_1 & \dots & 0 & \dots & -\sqrt{\frac{1}{n_B}}\lambda_1 & \dots \\ \sqrt{\frac{1}{n_A}}\lambda_2 & \dots & -\sqrt{\frac{1}{n_B}}\lambda_2 & \dots & 0 & \dots \\ \vdots & \vdots & \ddots & & & \end{pmatrix}.$$

The unitarity of the matrix $W$ requires normalization of its rows i.e.

$$1 = (\frac{1}{n_A} + \frac{1}{n_B})|\lambda_1|^2 + a^2 = (\frac{1}{n_A} + \frac{1}{n_B})|\lambda_2|^2 + b^2 ,$$

where $a, b$ are norms of remaining parts of the first and the second row vectors, respectively. Their orthogonality and the Cauchy-Schwartz inequality give us the inequality $|\lambda_1\lambda_2|/n_A \leq ab$. With the help of the previous equation we find

$$\frac{1}{(n_A)^2}|\lambda_1|^2|\lambda_2|^2 \leq (1 - h|\lambda_1|^2)(1 - h|\lambda_2|^2), \tag{4.138}$$

where $h \equiv (\frac{1}{n_A} + \frac{1}{n_B})$. The probability of success in the UI for the scheme using linear optical elements described by the matrix $W$ is

$$P(|\alpha_1\rangle, |\alpha_2\rangle) = \frac{1}{2}\sum_{i=1}^{2}(1 - e^{-|\lambda_i|^2|\alpha_1-\alpha_2|^2}). \tag{4.139}$$

The higher the $|\lambda_i|$'s the higher $P(|\alpha_1\rangle, |\alpha_2\rangle)$ is. However, the values of $\lambda_1, \lambda_2$ must satisfy the inequality (4.138) and therefore the maximum is achieved [see Eq.(4.139)] only if the inequality (4.138) is saturated. Thus, we have to optimize $P(|\alpha_1\rangle, |\alpha_2\rangle)$ with respect to $|\lambda_1|$, while keeping

$$|\lambda_2|^2 = \frac{n_A n_B - (n_A + n_B)|\lambda_1|^2}{n_A + n_B - (2 + \frac{n_A}{n_B})|\lambda_1|^2}. \tag{4.140}$$

The optimal value of $|\lambda_1|$ for any value of $|\alpha_1 - \alpha_2|$ is $|\lambda_1|^2 = |\lambda_2|^2 = \frac{n_A n_B}{n_A + 2n_B}$, because at this point $\frac{\partial}{\partial|\lambda_1|}P(|\alpha_1\rangle, |\alpha_2\rangle) = 0$ and $P(|\alpha_1\rangle, |\alpha_2\rangle)$ is concave with respect to $|\lambda_1|$ in the allowed interval. The aforementioned choice of $|\lambda_1|$ corresponds to a performance of the setup we have proposed in section 4.3.6, and hence concludes the proof.

*More types of reference states*

In the previous section the optimal values of transmittivities in our beam-splitter setup were state-independent only in the case of equal number of copies per reference state. Thus, for more than two types of reference states we will discuss only cases with the same number of copies of each reference state. Unfortunately, we will see that even in this restricted scenario, the optimal choice of transmittivities in the setup we propose will depend on the reference states (which are supposed to be unknown).

The generalization of the beam-splitter unambiguous identification scheme from the previous subsection is straightforward: We start by preparing the "concentrated" states $|\sqrt{n_A}\alpha_?\rangle$, $|\sqrt{n_B}\alpha_1\rangle, |\sqrt{n_C}\alpha_2\rangle, \ldots$. We use $M-1$ beam-splitters to sequentially split the "concentrated" unknown state $|\sqrt{n_A}\alpha_?\rangle$ into $M$ states. Each of these $M$ states is then merged with one of the "concentrated" reference states $|\sqrt{n_B}\alpha_1\rangle, \ldots, |\sqrt{n_M}\alpha_M\rangle$ on beam-splitters $C_1, \ldots, C_M$. The transmittivity $T_k$ (of the beam-splitter $C_k$) is chosen so that the destructive interference yields the vacuum on the second output port of $C_k$ for $\alpha_? = \alpha_k$. These output ports are monitored by photodetectors $D_1, \ldots, D_M$. Detection of at least one photon by the photodetector $D_k$ unambiguously indicates $\alpha_? \neq \alpha_k$. If all photodetectors except the $k$-th fire, then we conclude that $\alpha_? = \alpha_k$. For $M = 2$ we had freedom in choosing the ratio $T_1$ with which the "concentrated" unknown state $|\sqrt{n_A}\alpha_?\rangle$ was split into two parts used for the two comparisons. In order to maximize the probability of identification we can tune $M-1$ transmittivities of the beam-splitters that govern the splitting of the "concentrated" unknown state. The optimal choice of these transmittivities even for equal prior probabilities $\eta_j = 1/M$ depends on the choice of the reference states. Once we consider $n_B = n_C = \ldots$ then let us consider equal splitting of the "concentrated" unknown state into $M$ parts, even though it is not necessarily the optimal choice. In such case the beam-splitters $C_1, \ldots, C_M$ are performing the following transformation:

$$\begin{aligned} C_k \quad : \quad & |\sqrt{\frac{n_A}{M}}\alpha_?\rangle \otimes |\sqrt{n_B}\alpha_k\rangle \mapsto |out1\rangle \otimes |out2\rangle; \\ & |out1\rangle = |\sqrt{\frac{T_k n_A}{M}}\alpha_? + \sqrt{R_k n_B}\alpha_k\rangle; \\ & |out2\rangle = |-\sqrt{\frac{R_k n_A}{M}}\alpha_? + \sqrt{T_k n_B}\alpha_k\rangle. \end{aligned} \tag{4.141}$$

The condition of $|out2\rangle$ being a vacuum for $\alpha_? = \alpha_k$ forces us to set the transmittivity to $T_k = n_A/(n_A + M n_B)$. The probability of observing at least one photon in $|out2\rangle$ if $\alpha_? = \alpha_j$ is $1 - e^{-\frac{n_A n_B}{n_A + M n_B}|\alpha_j - \alpha_k|^2}$. The corresponding probability of identification therefore reads:

$$P(|\alpha_1\rangle, \ldots, |\alpha_M\rangle) = \sum_{j=1}^{M} \frac{1}{M} \prod_{k \neq j} (1 - e^{-\frac{n_A n_B}{n_A + M n_B}|\alpha_j - \alpha_k|^2}) \,. \tag{4.142}$$

Let us note that for a single copy of an unknown state and a single copy of reference states ($n_A = n_B = n_C = \ldots = 1$) the preliminary part of the setup concentrating the input coherent states is not present and hence the setup is much simpler and it is depicted on Fig. 4.11.

The unambiguous identification of $M$ reference states described above can be considered as a search in a quantum database composed of $M$ elements, i.e. $M$ different though unknown coherent states $|\alpha_j\rangle$ that are encoded into $M$ modes of an electromagnetic field. We point out that we have only a single copy of each of the states $|\alpha_j\rangle$ so one can not acquire a complete classical knowledge about the state. This set of $M$ states corresponds to a quantum database. In addition we have the $(M+1)$-st mode of the light field in the state $|\alpha_?\rangle$. The search of the database corresponds to the task of matching of two modes such that $\alpha_? = \alpha_j$. So we can say that the two modes are in the same state without knowing what the state actually is.

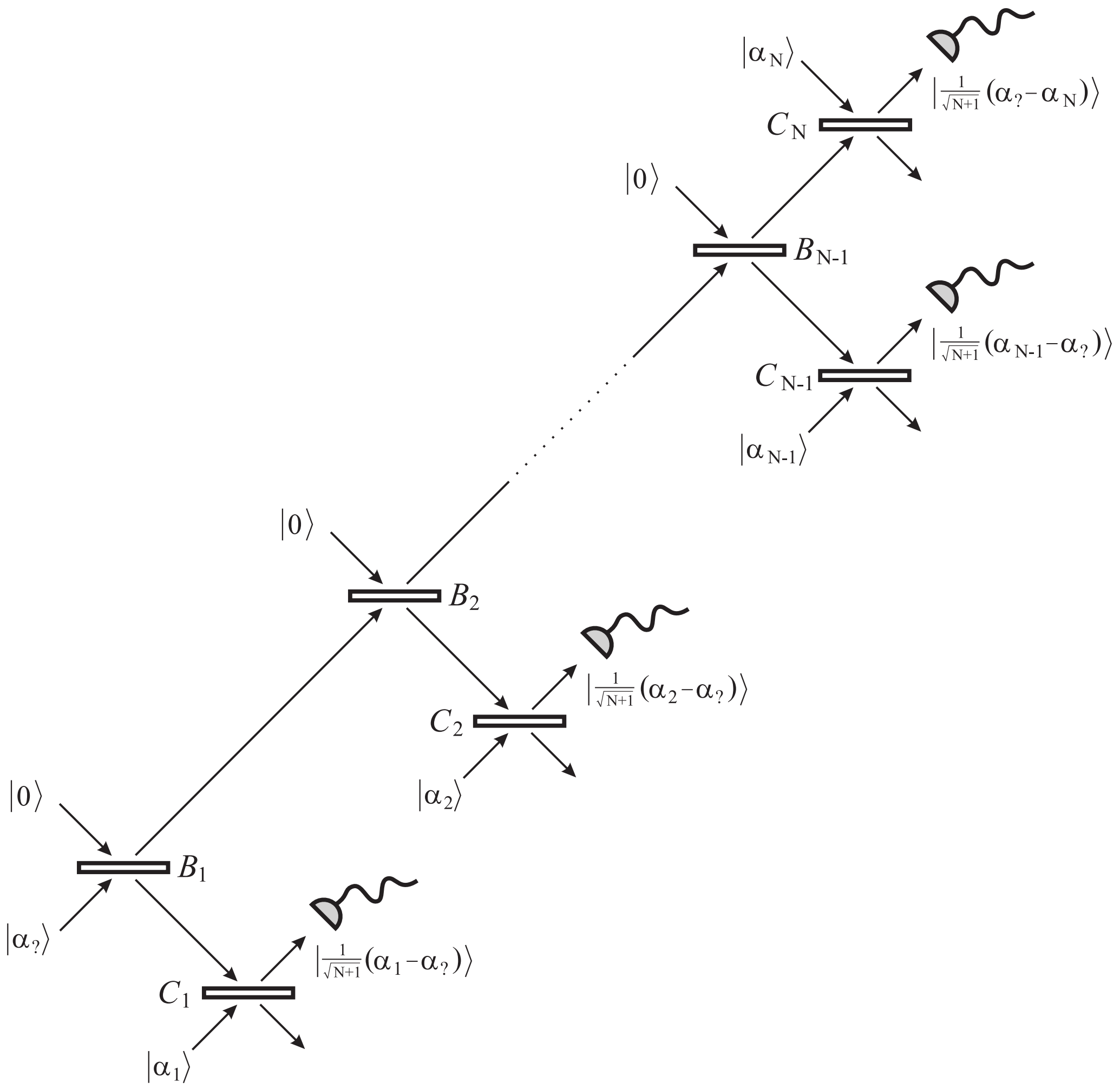

Figure 4.11. Unambiguous identification measurement setup identifying among $M$ coherent states.

### 4.3.7 Recovery of coherent reference states after UI

In this section we examine the information that remains in the unmeasured modes of our beam-splitter UI setups. In particular, we focus on a possibility of "recreating" the reference states after the unambiguous measurement. This recovery process might seem to be prohibited by the rules of quantum mechanics (due to irreversible disturbance of a quantum state by a measurement).

We show that in spite of the fact that the recovered reference states are "degraded" (disturbed), nevertheless they can be used in the subsequent round of the UI under the condition that new copy of an unknown state is provided. This can be useful for creating a quantum database, which would not be completely destroyed by the search performed on it. Instead, the data, i.e. the reference states, would degrade gradually with their repeated use.

First, we shall show that the coherent reference states can not be "recreated" without additional resources if the first unambiguous identification yields an inconclusive outcome. Although, this may seem disappointing, we show that the unmeasured states still can be used efficiently for the UI if the same unknown state is expected. Next, we examine the situation of the first UI producing a conclusive result known to us. In that case "diluted" reference states can be created, and they can be used for another independent unambiguous identification.

Let us consider the simplest version of unambiguous identification of coherent states ($M = 2, n_A = n_B = n_C = 1$). The beam-splitter setup for this scenario is depicted in Fig. 4.10. The modes $B$ and $D$ are not entangled with other modes, therefore their states do not depend on the measurement performed by the two photodetectors. The states of the modes $B, D$ are given by Eqs. (4.121) and (4.122), where $T_1$ is set to $1/2$ (for details see Section 4.3.6).

$$\begin{aligned} |out\rangle_B &= |\sqrt{\frac{2}{3}}\alpha_1 + \sqrt{\frac{1}{6}}\alpha_?\rangle_B \,; \\ |out\rangle_D &= |\sqrt{\frac{1}{6}}\alpha_? + \sqrt{\frac{2}{3}}\alpha_2\rangle_D \,. \end{aligned} \tag{4.143}$$

Using beam-splitters, phase shifters and known coherent states we can produce out of the states satisfying Eq. (4.143) a coherent state of the form

$$\left| a\left(\sqrt{\frac{2}{3}}\alpha_1 + \sqrt{\frac{1}{6}}\alpha_?\right) + b\left(\sqrt{\frac{1}{6}}\alpha_? + \sqrt{\frac{2}{3}}\alpha_2\right) + \gamma \right\rangle, \tag{4.144}$$

where $a, b, \gamma \in \mathbb{C}$. Imagine we want to recover the first reference state. Hence, we want the state from Eq. (4.144) to be $|\lambda\alpha_1\rangle$. Even though we know that either $\alpha_? = \alpha_1$, or $\alpha_? = \alpha_2$, a suitable choice of $a, b$ for one of these possibilities produces "junk" in the other case. Analogous reasoning works for the second reference state. For the inconclusive result of the UI measurement we do not know, which possibility took place, and thus the reference states can not be recovered.

*Repetition of UI for same unknown state*

Although the unmeasured modes of the beam-splitter setup seem to be useless nevertheless they can be exploited in the UI of the same unknown state $|\alpha_?\rangle$. Namely, we can feed them instead of the reference states into the beam-splitter scheme shown in Fig. 4.10. The concatenation is illustrated in Fig. 4.12. The transmittivity of the beamsplitter $B_2$ (respectively, $B_3$) can be set so that its measured output is in a vacuum if $\alpha_? = \alpha_1$ (respectively, if $\alpha_? = \alpha_2$). If we chose (for symmetry reasons) $T_1 = 1/2$ then the transmittivities $T_2$, $T_3$ should be set to $T_2 = 3/4$, $T_3 = 1/4$, respectively. This implies that the photodetectors measure the states $|(\alpha_? - \alpha_1)/\sqrt{6}\rangle$, $|(\alpha_2 - \alpha_?)/\sqrt{6}\rangle$. Thus, for both cases $\alpha_? = \alpha_1, \alpha_? = \alpha_2$ we can observe a photon in only one of the photodetectors and with the probability $1 - e^{-\frac{1}{6}|\alpha_1 - \alpha_2|^2}$ unambiguously conclude which possibility took place. Hence, the probability $1 - e^{-\frac{1}{6}|\alpha_1 - \alpha_2|^2}$ is a conditional UI probability after the first identification measurement returned an inconclusive result. The overall probability of an unambiguous identification for this two-round measurement is $1 - e^{-\frac{1}{2}|\alpha_1-\alpha_2|^2}$. This is due to the fact that the measurement fails only if both measurement rounds yield an inconclusive outcome.

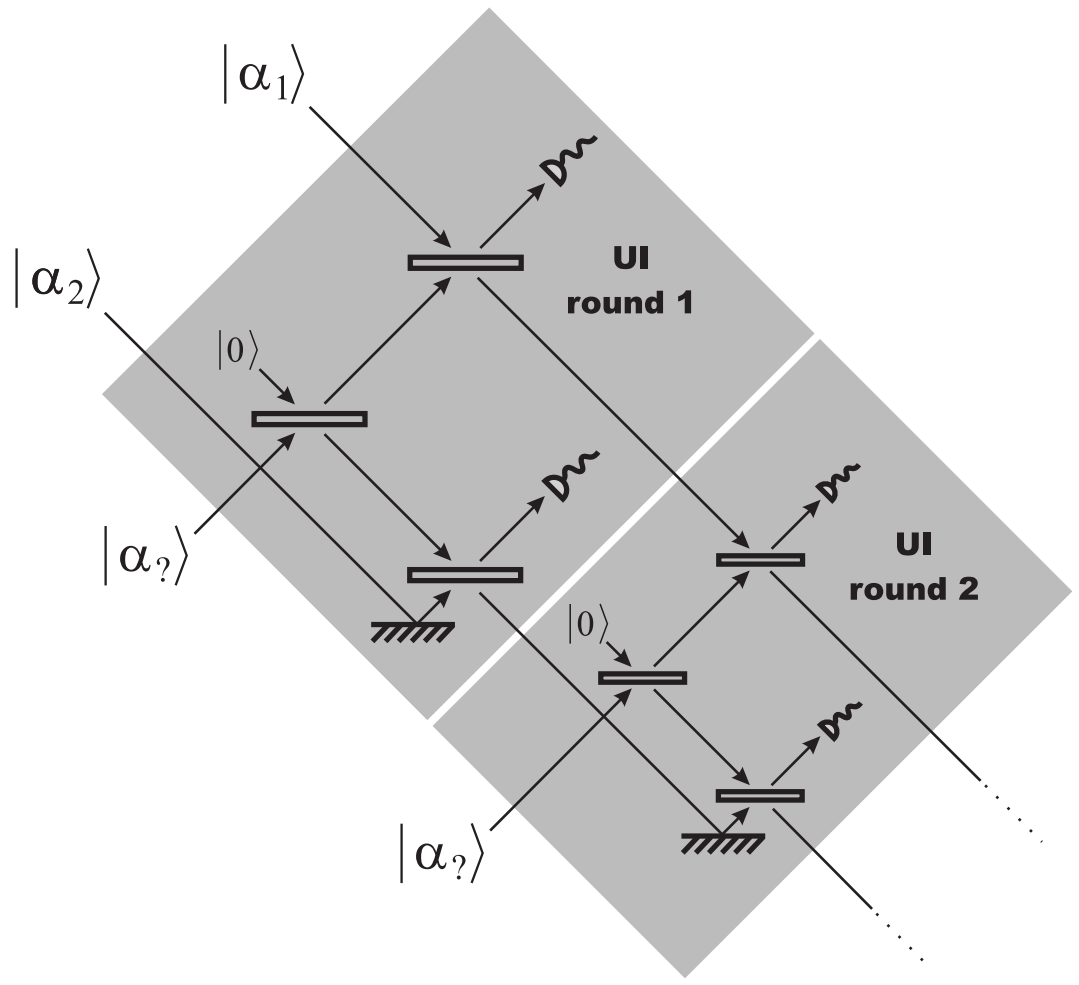

Figure 4.12. The beam-splitter setup designed for a subsequent unambiguous identification of multiple copies of an unknown coherent state.

The two-round measurement is essentially the UI scheme for $M = 2$, $n_A = 2$, $n_B = n_C = 1$, so we can compare its performance with the corresponding beam-splitter scheme analyzed in Section 4.3.6 [see Eq. (4.127)]. Indeed, the performance is the same, but the two round measurement has one possible advantage. If the first round gives a conclusive result then we still have an unmeasured copy of the unknown state (i.e. copy of $|\alpha_1\rangle$ or $|\alpha_2\rangle$) at our disposal. This is a similar advantage as in the case of weak implementation of the UI measurement discussed in Section 4.3.6.

*Repetition of UI with different unknown state*

As we illustrated in the beginning of Section 4.3.7 it is not possible to "recreate" the reference states by linear optics after an inconclusive result of an UI measurement is obtained. On the contrary, we will show that when a conclusive result is registered then both reference states can be "recreated". The proposed process will not perfect, because the recreated reference states will be a bit "diluted". Nevertheless, these states can be used as reference states for an UI with a different, independently prepared unknown state $|\beta_?\rangle$ (either $\beta_? = \alpha_1$ or $\beta_? = \alpha_2$).

When the result $\alpha_? = \alpha_1$ is found in the first round of the UI, the unmeasured modes $B, D$ are in states the $|\sqrt{\frac{3}{2}}\alpha_1\rangle_B$ and $|\sqrt{\frac{1}{6}}\alpha_1 + \sqrt{\frac{2}{3}}\alpha_2\rangle_D$, respectively. Thus, we have the "concentrated" first reference state $|\sqrt{\frac{3}{2}}\alpha_1\rangle$ in the mode B. Let us now examine whether the reference state $|\alpha_2\rangle$ can be "recreated" out of the modes $B$ and $D$. The obvious idea is to use the mode $B$ to shift the mode $D$ via a beam-splitter so that the $\alpha_1$ part of the amplitude in $|\sqrt{\frac{1}{6}}\alpha_1 + \sqrt{\frac{2}{3}}\alpha_2\rangle_D$ is canceled. This happens for the transmittivity of the beam-splitter equal to $9/10$:

$$|\sqrt{\frac{3}{2}}\alpha_1\rangle \otimes |\sqrt{\frac{1}{6}}\alpha_1 + \sqrt{\frac{2}{3}}\alpha_2\rangle \mapsto$$
$$\mapsto |\left(\sqrt{\frac{27}{20}} + \sqrt{\frac{1}{15}}\right)\alpha_1 + \sqrt{\frac{1}{60}}\alpha_2\rangle \otimes |\sqrt{\frac{3}{5}}\alpha_2\rangle\,. \tag{4.145}$$

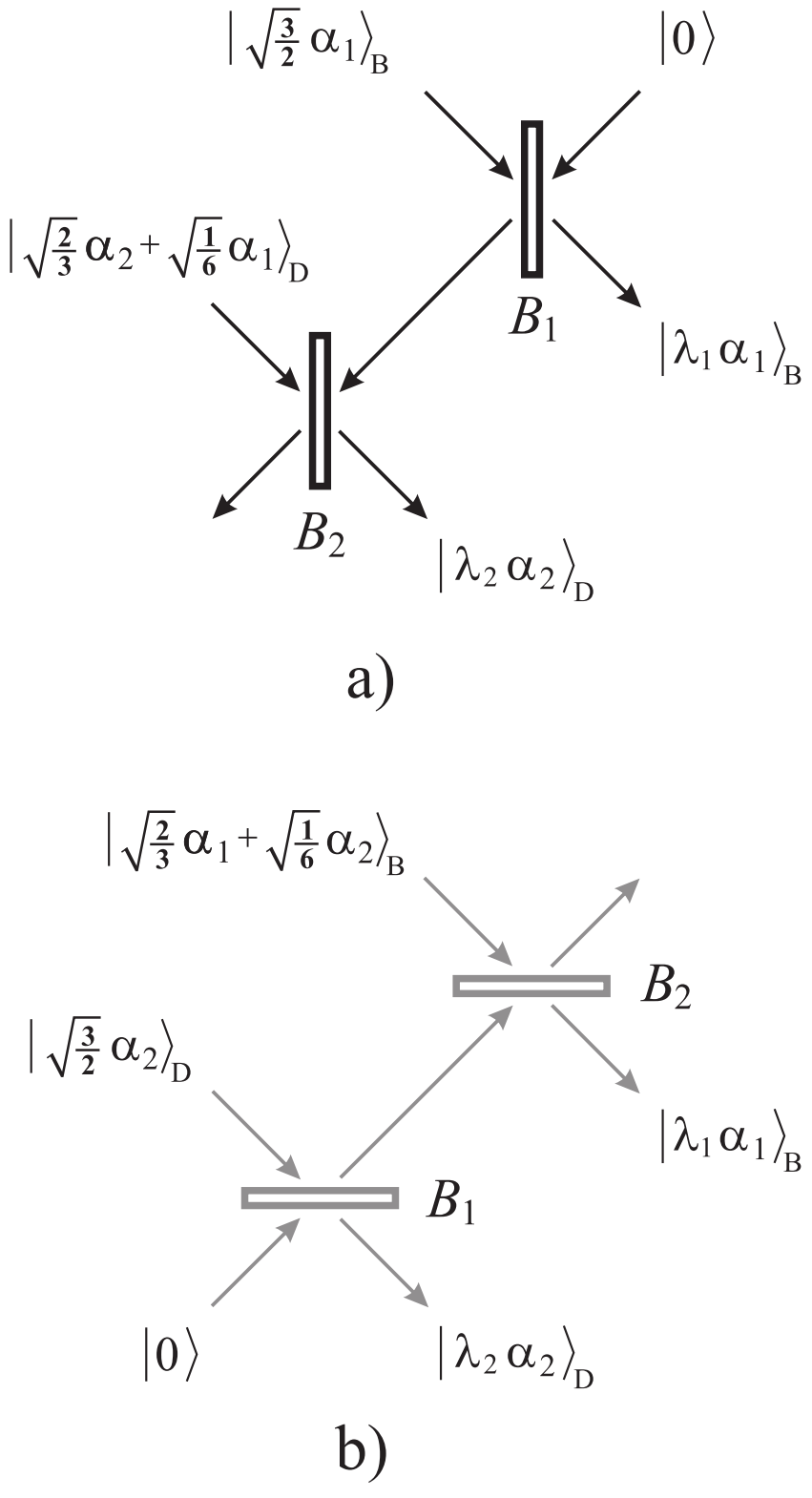

Figure 4.13. The beam-splitter setups designed for the recovery of unmeasured modes from Fig. 4.10. In the case $\alpha_? = \alpha_1$ the setup a) is used, however for $\alpha_? = \alpha_2$ the setup b) is used.

Hence, we know how to recover separately either the first or the second reference state. If solely such a single state is used in the subsequent UI measurement then the probability of success is bounded from above by $1/2$, because only one type of a reference state can be identified. Thus, we want to find a setup, which extracts both types of reference states simultaneously and allows for a subsequent round of the unambiguous identification of $|\beta_?\rangle$. Such a scheme is presented in Fig. 4.13a. The beam-splitter $B_1$ splits the "concentrated" first reference state into two parts. One part can be directly used for the next round of the UI, the second part cancels the $\alpha_1$ contribution in the amplitude of coherent state in mode $D$ via the beam-splitter $B_2$. If we set the transmittivity of the beam-splitter $B_1$ to be $T_1^R$, then the requirement of cancelation of the $\alpha_1$ contribution of the amplitude in the mode $D$ constrains the transmittivity of $B_2$ to be $T_2^R = (9 - 9T_1^R)/(10 - 9T_1^R)$. The corresponding "recreated" reference states then read

$$|\sqrt{\frac{3}{2}T_1^R}\alpha_1\rangle, \quad |\sqrt{\frac{6-6T_1^R}{10-9T_1^R}}\alpha_2\rangle\,. \tag{4.146}$$

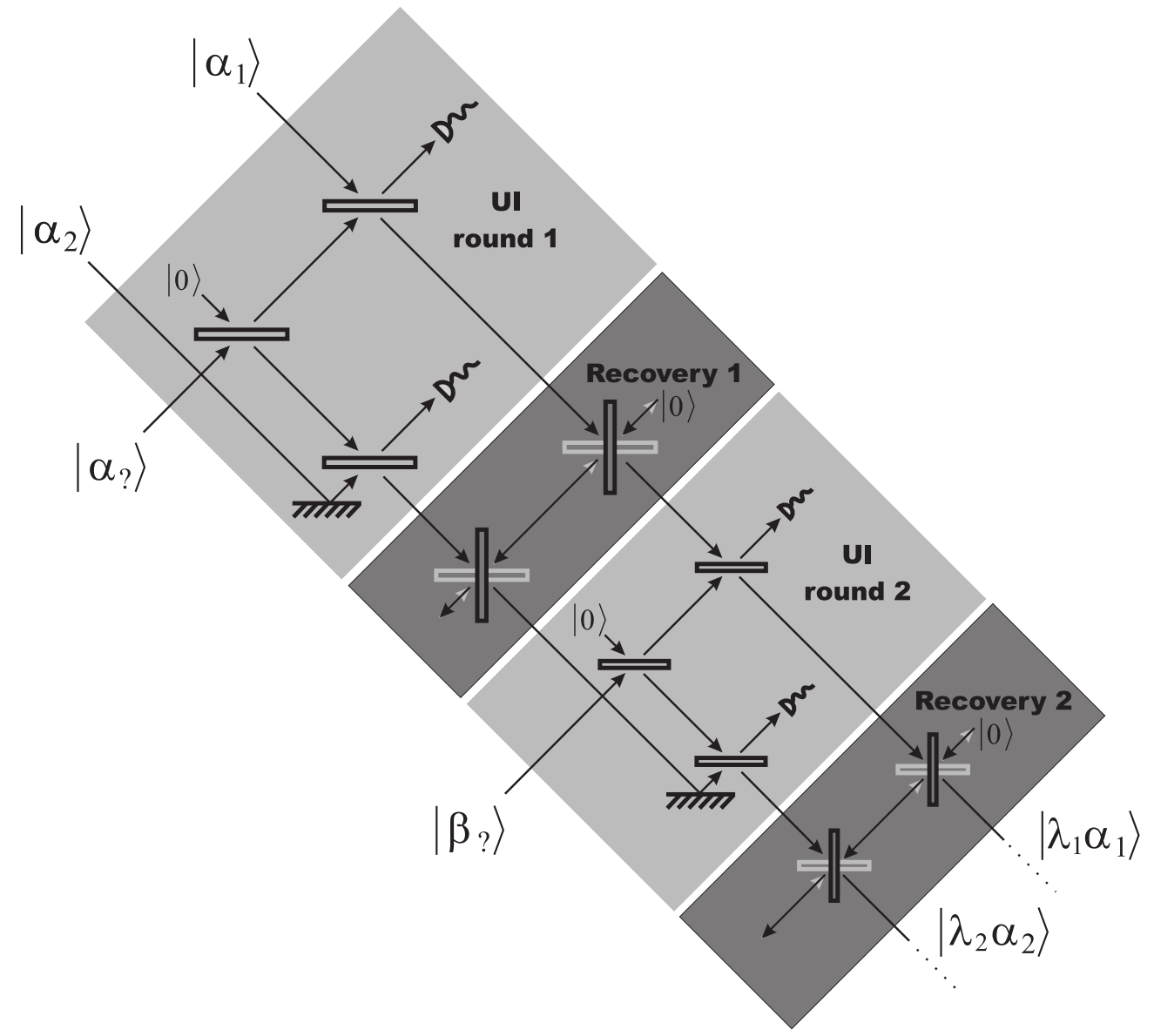

Figure 4.14. The beam-splitter setup designed for a repetition of the UI with different unknown state, which can be seen as a repeated search in a quantum database. The gray beam-splitters in recovery steps are used if the unknown state from previous round of the UI matches the second reference state otherwise the black beamsplitters are used.

We want to use these two states instead of the reference states $|\alpha_1\rangle$, $|\alpha_2\rangle$ in the next round of the UI. Both possible preparations $|\beta_?\rangle = |\alpha_1\rangle$, $|\beta_?\rangle = |\alpha_2\rangle$ will be equally likely, therefore we chose $T_1^R = (7 - \sqrt{13})/9$ so that equally diluted reference states

$$|\sqrt{\lambda_2}\alpha_1\rangle\,; \quad |\sqrt{\lambda_2}\alpha_2\rangle\,;\mathrm{with} \quad \lambda_2 \equiv \frac{7-\sqrt{13}}{6}\,, \tag{4.147}$$

enter the next round of the UI. If a conclusive result $\alpha_? = \alpha_2$ is obtained in the first round of the UI, then after exchanging the roles of the modes $B$ and $D$, analogous recovery setup (see Fig.4.13b) can be used to produce the "diluted" reference states given by Eq. (4.147). Thus, for both conclusive results from the first round of the UI, one type of the UI measurement using the recovered reference states can be used in the second round. Actually, the beam-splitter setup from Fig. 4.10 can be used (see Fig. 4.14) if we take into account that for our input states $n_A = 1$, $n_B = n_C = (7-\sqrt{13})/6$. Upon making this substitution the performance of the setup is the same as in Section 4.3.6, and all the formulas derived there remain valid. The aforementioned setup succeeds in the UI with the probability given by Eq. (4.127). However, the second round of the UI is possible only if the first UI succeeds, which implies the following probability of the UI in the second round

$$P^{(2)}(|\alpha_1\rangle, |\alpha_2\rangle) = (1 - e^{-\frac{1}{3}|\alpha_1-\alpha_2|^2})(1 - e^{-\frac{7-\sqrt{13}}{2(10-\sqrt{13})}|\alpha_1-\alpha_2|^2}). \tag{4.148}$$

It is interesting that the UI with nearly orthogonal reference states can be done also in the second round with a probability of success approaching unity.

Let us now see, whether further rounds of the UI are still possible. The first round of the UI can be seen as use of the beam-splitter setup from Fig. 4.10 with $n_A = n_B = n_C = 1$ followed by the setup from Fig. 4.13 recovering the reference states. In the second round we have used again the beam-splitter setup from Fig. 4.10 this time with $n_A = 1,\ n_B = n_C = (7-\sqrt{13})/6$. It turns out that we can perform infinitely many additional rounds of the UI, where in each round the unknown state is independently chosen to be either $|\alpha_1\rangle$ or $|\alpha_2\rangle$. It suffices to use the beam-splitter setup from Fig. 4.10 followed by the setup from Fig. 4.13 recovering the reference states in each round of the UI. However, the transmittivities of the beam-splitters used in those setups must be set as follows. Let us denote by $\sqrt{\lambda_k}$ the factor by which the reference states are suppressed at the beginning of the $k$-th round (e.g. $\lambda_1 = 1$). In $k$-th round of the UI we should set $T_1 = 1/2,\ T_2 = 2\lambda_k/(1+2\lambda_k),\ T_3 = 1/(1+2\lambda_k)$ in the scheme from Fig. 4.10 and

$$\begin{aligned} T_1^R &= 1 - \frac{2\lambda_k^2 + \sqrt{4\lambda_k^4 + (1+2\lambda_k)^2}}{(1+2\lambda_k)^2}; \\ T_2^R &= \frac{(1-T_1)(1+2\lambda_k)^2}{1+(1-T_1)(1+2\lambda_k)^2}, \end{aligned} \tag{4.149}$$

in the scheme presented in Fig. 4.13. The suppression of the amplitude of reference states is given by $\lambda_k \mapsto \lambda_{k+1} = f(\lambda_k)$, where

$$f(x) = \frac{(1+2x)^2 - 2x^2 - \sqrt{4x^4 + (1+2x)^2}}{2(1+2x)}. \tag{4.150}$$

The probability of successfully performing the UI in the $k$-th round is

$$P^{(k)}(|\alpha_1\rangle, |\alpha_2\rangle) = P^{(k-1)}(|\alpha_1\rangle, |\alpha_2\rangle)(1 - e^{-\frac{\lambda_k}{1+2\lambda_k}|\alpha_1-\alpha_2|^2}),$$

because the $k$-th round of the UI is possible only if all previous UIs succeeded[16]. The dependence of the probability of identification on the difference of the amplitudes of the reference states and on the number of measurement rounds is shown in Fig. 4.15.

Let us now discuss an alternative approach to the recovery of reference states. Imagine that our task is to identify $N$ independent unknown states with reference states. Instead of recovering reference states after identifying each of the unknown states we can first split the reference states into $N$ parts and then perform the identifications independently. We are going to illustrate that even though we know value $N$ ahead of time, the splitting strategy does not outperform the strategy based on recovery of reference states.

[16]We set $P^{(0)}(|\alpha_1\rangle, |\alpha_2\rangle) = 1$.

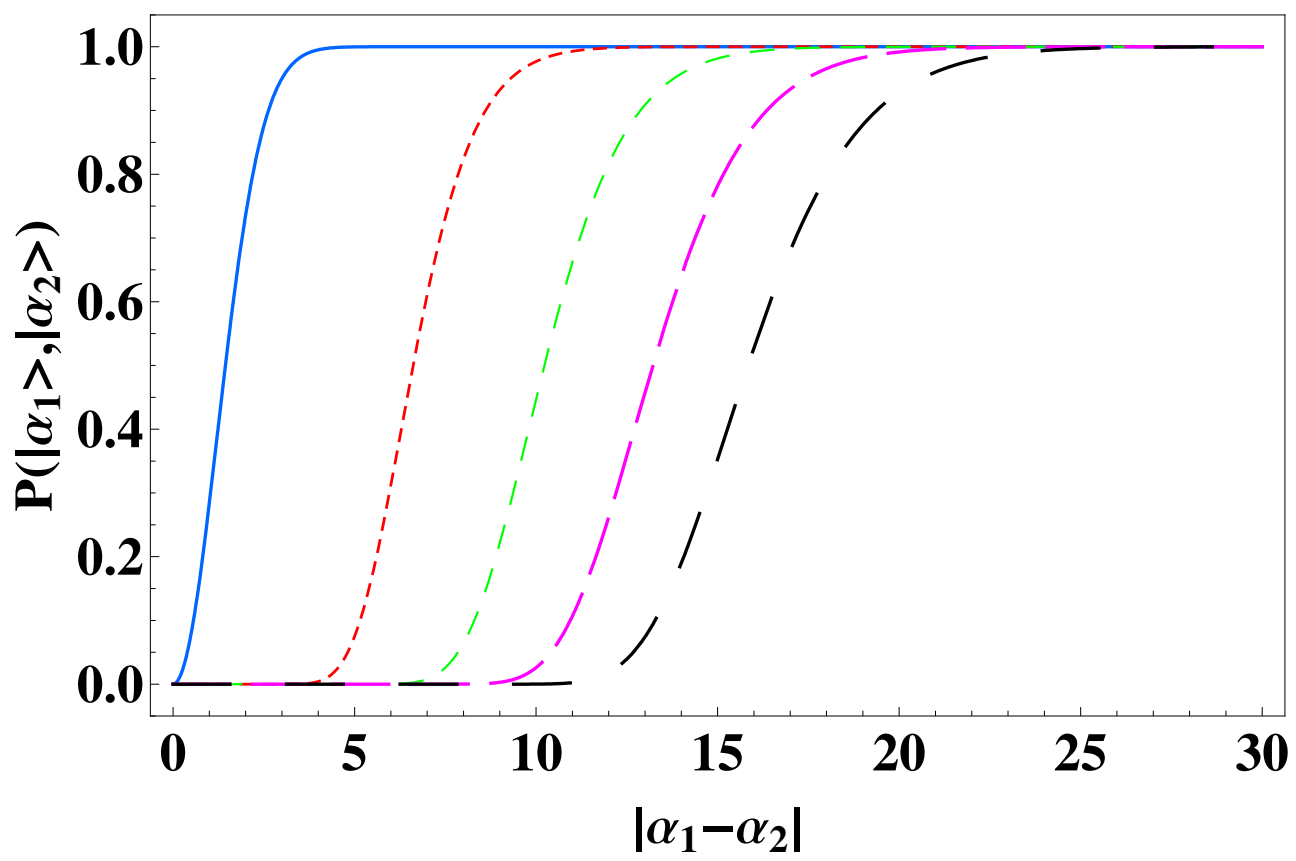

Figure 4.15. The performance of the recovery setup. The probability of identification $P(|\alpha_1\rangle, |\alpha_2\rangle)$ as a function of the scalar product (given by $|\alpha_1 - \alpha_2|$) depicted for various numbers of measurement rounds. Starting from the left the curves correspond to the probability of identification in the first, 20th, 40th, 60th, 80th round of the UI.

The splitting strategy begins by distributing the information in the two reference states into $N$ copies of the states $|\frac{1}{\sqrt{N}}\alpha_1\rangle, |\frac{1}{\sqrt{N}}\alpha_2\rangle$. These two states are then combined together with one of the unknown states and are unambiguously identified by the scheme for $M = 2, n_A = 1, n_B = n_C = 1/N$. The probability of a successful identification of the unknown state depends only on the reference states, hence for each of the $N$ UI measurements we have $P(|\alpha_1\rangle, |\alpha_2\rangle) = 1 - e^{-\frac{1}{N+2}|\alpha_1-\alpha_2|^2}$. The probability that all of them succeed is therefore $P_S^{(N)}(|\alpha_1\rangle, |\alpha_2\rangle) = (1 - e^{-\frac{1}{N+2}|\alpha_1-\alpha_2|^2})^N$. On the other hand in the scheme with the recovery of the reference states the $N$-th round can succeed only if all the previous identification rounds were successful. This means that the probability of success of the $N$-th round $P^{(N)}(|\alpha_1\rangle, |\alpha_2\rangle)$ is the same as the probability that all the $N$ rounds of the identification task were successful. The difference between the performance of the recovery and the splitting strategies for different $N$ is depicted in Fig. 4.16.

The investigated problem of finding a procedure for $N$ successful rounds of the unambiguous identification can be modified in several ways exhibiting the advantages of recovery or splitting strategies. For example, one may be interested to find a procedure, such that at least in $m \leq N$ out of $N$ rounds we find an unambiguous conclusion. In such formulation of the problem it is clear that there always exist $m$, for which the splitting strategy gives better results than the described recovery strategy adopted for exactly $m$ successes. However, as we have shown if $m = N$, then the recovery procedure outperforms the splitting strategy.

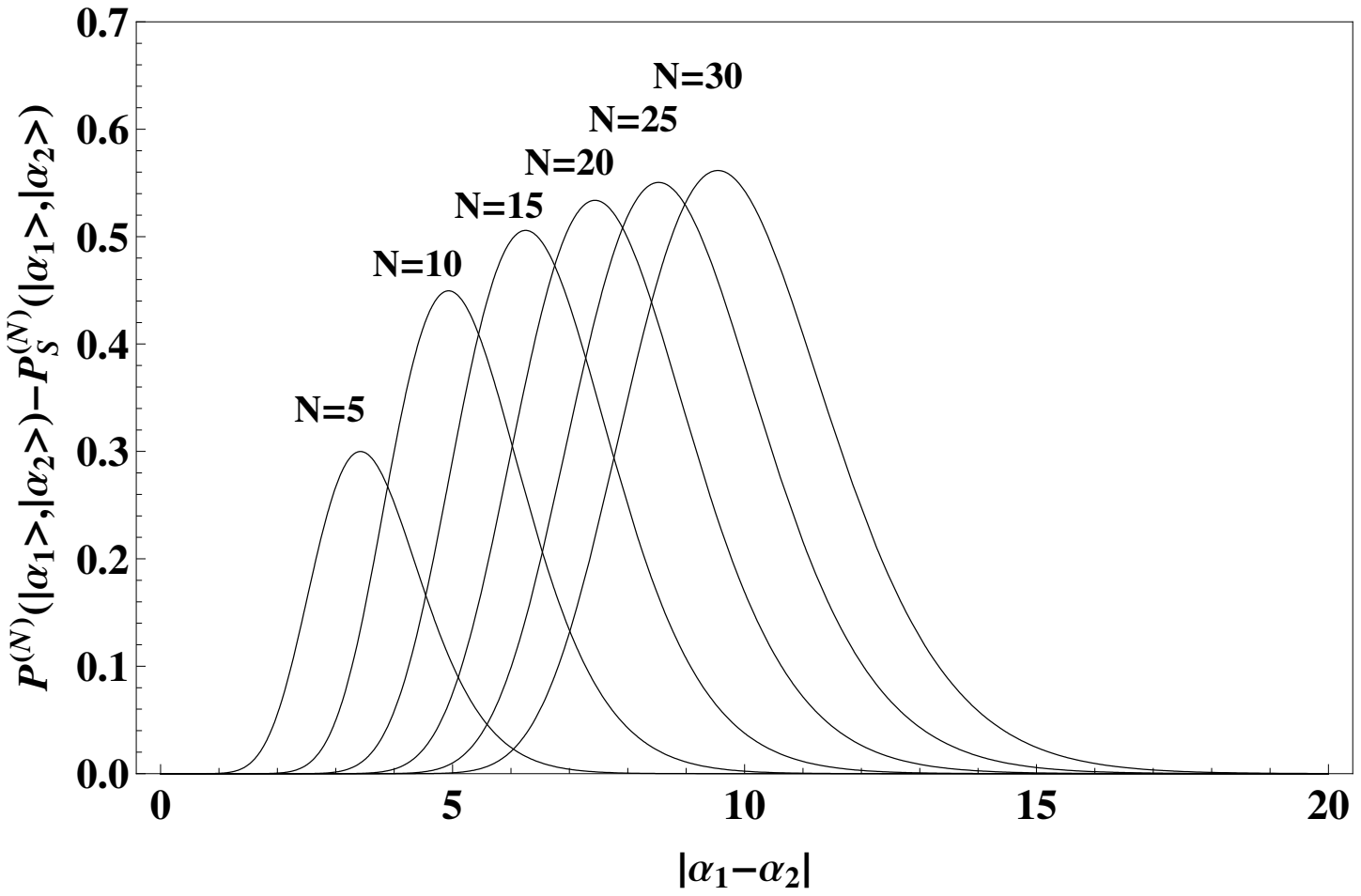

Figure 4.16. The difference between the performance of the recovery and the splitting strategies for different number of identification rounds $N$ as a function of the scalar product (given by $|\alpha_1 - \alpha_2|$).

### 4.3.8 Influence of noise on reliability of UI setups

In this section we shall investigate how noise (uncertainty) in the state preparation affects the reliability of the measurement results. The UI setups we have presented above are designed specifically for coherent states and ideally they are $100\%$ reliable, i.e. whenever we obtain a conclusive result $E_i$ then we are completely sure that the possibility $x_i$ (i.e., $\alpha_? = \alpha_i$) took place. However, it might be that the unknown and reference states are sent to us via a noisy channel or simply that their preparation is noisy. We assume that this disturbance has the form of a technical noise [52], and therefore the unknown and the reference states are not pure coherent states $|\alpha_i\rangle$, but rather their mixtures $\omega_i$:

$$\omega_i = \frac{1}{2\pi\sigma^2}\int_{\mathbb{C}} d\beta e^{-\frac{|\beta|^2}{2\sigma^2}}|\alpha_i+\beta\rangle\langle\alpha_i+\beta|\,; \quad (4.151)$$

$$\rho_i(\boldsymbol{\alpha}) = (\omega_i)^{\otimes n_A}\otimes(\omega_1)^{\otimes n_B}\otimes(\omega_2)^{\otimes n_C}\otimes\ldots, \quad (4.152)$$

with $\sigma$ defining the strength of the noise and $\boldsymbol{\alpha}$ indicating the dependence on the complex amplitudes $\alpha_i$. In such case conclusive results of our UI setups will no longer be unambiguous. More precisely, there will be a certain probability $P(x_i|E_i)$ with which the obtained outcome $E_i$ of the measurement is the consequence of the possibility $x_i$. This probability is called the *reliability* of the outcome $E_i$. The corresponding mathematical definition reads:

$$R(E_i) = P(x_i|E_i) = \frac{\eta_i P(E_i|x_i)}{\sum_{j=1}^{M}\eta_j P(E_i|x_j)}, \quad (4.153)$$

where $\eta_i$ is the a priori probability of the possibility $x_i$ and $P(E_i|x_j)$ is the probability that the measurement of the system prepared in the possibility $x_j$ will give a result $E_i$. Let us note that under the possibility $x_i$ we understand all situations in which the unknown state is the same as the $i$-th reference state. Thus $x_i$ stands for the whole set of situations, which differ by complex amplitudes $\alpha_k$ of the "centers" of the reference states $\omega_k$. How those "center points" of all reference states are chosen in $x_i$ is described by the probability distribution $\chi(\alpha_1, \ldots, \alpha_M)$. The support of $\chi$ is $\mathbb{C}^m$ corresponding to an infinite plane. Therefore a uniform probability distribution can not be defined on it. Nevertheless, we can express the reliability as:

$$R(E_i) = \frac{\eta_i \int_{\mathbb{C}^M} d\boldsymbol{\alpha} \chi(\boldsymbol{\alpha}) \mathrm{Tr}(E_i \rho_i(\boldsymbol{\alpha}))}{\sum_{j=1}^{M} \eta_j \int_{\mathbb{C}^M} d\boldsymbol{\alpha} \chi(\boldsymbol{\alpha}) \mathrm{Tr}(E_i \rho_j(\boldsymbol{\alpha}))}, \tag{4.154}$$

where $d\boldsymbol{\alpha} \equiv d\alpha_1 \ldots d\alpha_M$. In the limit $\sigma \to 0$ states $\omega_i$ become $|\alpha_i\rangle\langle\alpha_i|$. Because of the no-error conditions (4.62), which are for coherent states satisfied by our UI setups, only the $i$-th term of the sum in Eq. (4.153) survives. Thus, without noise the reliability is equal to unity. For $\sigma > 0$ also other terms in Eq. (4.153) will contribute and hence the reliability will be less than one. Moreover, the precise value of $R(E_i)$ will depend on the probability distributions $\chi(\boldsymbol{\alpha})$.

In the remaining part of this section we will investigate a scenario, which might be called as the *phase keying*. We assume that the two reference states ($M = 2$) have always opposite phases, i.e. if $\omega_1$ is centered around the amplitude $\alpha$ then $\omega_2$ is centered around the amplitude $-\alpha$. Values of $\alpha$ have a Gaussian distribution centered around 0 (the vacuum) with a dispersion $\xi$, so

$$\chi(\alpha_1, \alpha_2) = \delta(\alpha_1 + \alpha_2) \frac{1}{2\pi\xi^2} e^{-|\alpha_1|^2/(2\xi^2)}, \quad i = 1, 2. \tag{4.155}$$

In order to calculate the reliability we must first evaluate $\mathrm{Tr}[E_i \rho_j(\boldsymbol{\alpha})]$. This means we have to derive the probabilities with which detectors $D_1, D_2$ click if "fuzzy" states $\omega_?, \omega_1, \omega_2$ are fed into the UI setup instead of pure coherent states $|\alpha_?\rangle, |\alpha_1\rangle, |\alpha_2\rangle$. Our UI setup uses an additional mode $D$ that should be initially prepared in the vacuum. We assume that also this mode is noisy and initially in a state $\omega_i$ centered around 0 (the vacuum).

To present our calculations concisely, we first derive how the setup acts on displaced coherent input states (e.g. $|\alpha_i + \beta\rangle$) and then we integrate those partial results. Thus, for a single copy of the unknown and the reference states we derive how the UI setup acts on states $|\alpha_? + \nu\rangle, |\alpha_1 + \beta\rangle, |\alpha_2 + \gamma\rangle, |\varrho\rangle$ fed into the modes $A, B, C, D$ (see Fig. 4.10) and finally we perform integration over $\nu, \beta, \gamma, \varrho$.

For multiple copies of the unknown and the reference states we assume that the noise is acting independently on each of the copies, i.e. we analyze $n_B$ copies of the first reference state entering as states $|\alpha_1 + \beta_1\rangle, \ldots, |\alpha_1 + \beta_{n_B}\rangle$. The first part of the UI setup, which "concentrates" copies of the same species, generates the state $|\sqrt{n_B}\alpha_1 + \frac{1}{\sqrt{n_B}}(\beta_1 + \ldots + \beta_{n_B})\rangle$ and similarly, the state $|\sqrt{n_C}\alpha_2 + \frac{1}{\sqrt{n_C}}(\gamma_1 + \ldots + \gamma_{n_C})\rangle$ for the second reference state, and $|\sqrt{n_A}\alpha_? + \frac{1}{\sqrt{n_A}}(\nu_1 + \ldots + \nu_{n_A})\rangle$ for the unknown state. The beam splitter transformation for coherent input states does not entangle outputs, thus we can, in the same way as in Section 4.3.6, derive expressions for the states of the modes that the photodetectors $D_1, D_2$ measure. Consequently, the final states of the

modes $A$ and $C$ read:

$$\begin{aligned}\Big|\sqrt{\frac{n_A n_B}{n_A+2n_B}}\left[\alpha_? - \alpha_1 - \sqrt{\frac{1}{n_A}}\varrho + \frac{1}{n_A}\boldsymbol{\nu} - \frac{1}{n_B}\boldsymbol{\beta}\right]\Big\rangle_A &\equiv |\mu_1\rangle_A\,; \\ \Big|\sqrt{\frac{n_A n_C}{n_A+2n_C}}\left[\alpha_2 - \alpha_? - \sqrt{\frac{1}{n_A}}\varrho - \frac{1}{n_A}\boldsymbol{\nu} + \frac{1}{n_C}\boldsymbol{\gamma}\right]\Big\rangle_C &\equiv |\mu_2\rangle_C,\end{aligned} \tag{4.156}$$

where $\boldsymbol{\nu} \equiv \sum_{k=1}^{n_A}\nu_k$, $\boldsymbol{\beta} \equiv \sum_{k=1}^{n_B}\beta_k$, $\boldsymbol{\gamma} \equiv \sum_{k=1}^{n_C}\gamma_k$. Now we have to evaluate the probability of the projection of these states $|\mu_1\rangle_A, |\mu_2\rangle_C$ onto the vacuum. Subsequently, we will integrate this partial result to obtain the probability $P(D_k|\rho_i(\boldsymbol{\alpha}))$ that the photodetector $D_k$ ($k=1,2$) does not click. Probabilities $P(D_k|\rho_i(\boldsymbol{\alpha}))$ are related to $\mathrm{Tr}(E_i\rho_j(\boldsymbol{\alpha}))$ in the following way:

$$\begin{aligned}\mathrm{Tr}(E_1\rho_1) &= [1-P(D_1|\rho_1)].P(D_2|\rho_1)\,; \\ \mathrm{Tr}(E_1\rho_2) &= [1-P(D_1|\rho_2)].P(D_2|\rho_2)\,; \\ \mathrm{Tr}(E_2\rho_1) &= P(D_1|\rho_1).[1-P(D_2|\rho_1)]\,; \\ \mathrm{Tr}(E_2\rho_2) &= P(D_1|\rho_2).[1-P(D_2|\rho_2)]\,,\end{aligned} \tag{4.157}$$

where the argument of $\rho_i(\boldsymbol{\alpha})$ is omitted for brevity. Finally, we obtain the quantities $\mathrm{Tr}[E_i\rho_j(\boldsymbol{\alpha})]$ that we need for evaluating the reliability according to Eq. (4.154).

Using the formula $|\langle 0|\mu_i\rangle|^2 = e^{-|\mu_i|^2}$ for the modulus of the overlap of two coherent states we obtain:

$$P(D_1|\rho_i(\boldsymbol{\alpha})) = \int_{\mathbb{C}^m}\frac{d\varrho d\boldsymbol{\gamma} d\boldsymbol{\nu}}{(2\pi\sigma^2)^m}\exp\Big[-\frac{|\varrho|^2+\sum_{k=1}^{n_A}|\nu_k|^2+\sum_{k=1}^{n_B}|\gamma_k|^2}{2\sigma^2} - \frac{n_A n_C\Big|\alpha_2-\alpha_?-\sqrt{\frac{1}{n_A}}\varrho-\frac{1}{n_A}\boldsymbol{\nu}+\frac{1}{n_C}\boldsymbol{\gamma}\Big|^2}{n_A+2n_C}\Big]; \tag{4.158}$$

$$P(D_2|\rho_i(\boldsymbol{\alpha})) = \int_{\mathbb{C}^n}\frac{d\varrho d\boldsymbol{\beta} d\boldsymbol{\nu}}{(2\pi\sigma^2)^n}\exp\Big[-\frac{|\varrho|^2+\sum_{k=1}^{n_A}|\nu_k|^2+\sum_{k=1}^{n_B}|\gamma_k|^2}{2\sigma^2} - \frac{n_A n_B\Big|\alpha_?-\alpha_1-\sqrt{\frac{1}{n_A}}\varrho+\frac{1}{n_A}\boldsymbol{\nu}-\frac{1}{n_B}\boldsymbol{\beta}\Big|^2}{n_A+2n_B}\Big], \tag{4.159}$$

where $m = n_A + n_C + 1$, $n = n_A + n_B + 1$. The integrals in Eq. (4.158) and (4.159) can be performed using the relations derived in Appendix B.2. The results of the integration read:

$$P(D_1|\rho_i(\boldsymbol{\alpha})) = \frac{1}{1+2\sigma^2}e^{-\frac{1}{1+2\sigma^2}\frac{n_A n_C}{n_A+2n_C}|\alpha_i-\alpha_2|^2}\,; \tag{4.160}$$

$$P(D_2|\rho_i(\boldsymbol{\alpha})) = \frac{1}{1+2\sigma^2}e^{-\frac{1}{1+2\sigma^2}\frac{n_A n_B}{n_A+2n_B}|\alpha_i-\alpha_1|^2}\,, \tag{4.161}$$

where we have used the formulas for the case $x_i$, i.e. $\alpha_? = \alpha_i$. Consequently, using these results in Eq. (4.157) we obtain:

$$\begin{aligned}
\mathrm{Tr}(E_1\rho_1) &= \frac{1+2\sigma^2 - e^{-\frac{1}{1+2\sigma^2}\frac{n_A n_C}{n_A+2n_C}|\alpha_1-\alpha_2|^2}}{(1+2\sigma^2)^2};\\
\mathrm{Tr}(E_1\rho_2) &= \frac{2\sigma^2}{(1+2\sigma^2)^2}\, e^{-\frac{1}{1+2\sigma^2}\frac{n_A n_B}{n_A+2n_B}|\alpha_1-\alpha_2|^2};\\
\mathrm{Tr}(E_2\rho_1) &= \frac{2\sigma^2}{(1+2\sigma^2)^2}\, e^{-\frac{1}{1+2\sigma^2}\frac{n_A n_C}{n_A+2n_C}|\alpha_1-\alpha_2|^2};\\
\mathrm{Tr}(E_2\rho_2) &= \frac{1+2\sigma^2 - e^{-\frac{1}{1+2\sigma^2}\frac{n_A n_B}{n_A+2n_B}|\alpha_1-\alpha_2|^2}}{(1+2\sigma^2)^2}.
\end{aligned} \tag{4.162}$$

Now in order to obtain the reliability it remains to substitute Eqs. (4.155), (4.162) into Eq. (4.154) and to perform the remaining integrals. Those integrals can be performed in polar coordinates, where the angular dependence is trivial and the radial part can be simplified with the help of a substitution $t = e^{-r^2/2}$. After performing the integration we obtain the final result, which can be, for $n_B = n_C$, written in the compact form:

$$\begin{aligned}
R(E_1) = R(E_2) &= \frac{1+\theta}{1+2\theta};\\
\theta &= \frac{n_A+2n_B}{n_A n_B}\left(\frac{\sigma}{2\xi}\right)^2.
\end{aligned} \tag{4.163}$$

Let us note that $\lim_{\sigma\to 0} R(E_i) = 1$ as it should be. Moreover, the reliability depends only on the fuzziness $\sigma$ of the states entering the UI setup, the typical difference of the amplitudes of the reference states $2\xi$, and the number of copies that are available. If $\sigma \ll \xi$, i.e. the fuzziness of the states, is much smaller than the displacement used to encode the information, then $\theta \to 0$ and $R(E_i)$ approaches the unity. More quantitative insight in the case of a single copy of the unknown and the reference states is provided by Fig. 4.17. In order to see how the noise influences other relevant quantities we will calculate $\overline{P}, \overline{P_E}, \overline{P_F}$, which are called the averaged probability of success, the error, and the failure, respectively. Obviously, we either guess correctly, or incorrectly, or do not guest at all (inconclusive result/failure), therefore $\overline{P} + \overline{P_E} + \overline{P_F} = 1$ must hold. It is useful to rewrite the definition of these quantities in the following form:

$$\begin{aligned}
\overline{P} &= \frac{1}{2}\sum_{i=1}^{2}\int_{\mathbb{C}^2} d\boldsymbol{\alpha}\mathrm{Tr}(E_i\rho_i(\boldsymbol{\alpha}))\chi_i(\boldsymbol{\alpha})\,;\\
\overline{P_E} &= \frac{1}{2}\int_{\mathbb{C}^2} d\boldsymbol{\alpha}(\mathrm{Tr}(E_2\rho_1(\boldsymbol{\alpha})) + \mathrm{Tr}(E_1\rho_2(\boldsymbol{\alpha})))\chi_1(\boldsymbol{\alpha})\,;\\
\overline{P_F} &= 1 - \overline{P} - \overline{P_E}\,.
\end{aligned} \tag{4.164}$$

Now it suffice to substitute Eqs. (4.162) into the above equations and to perform the integration in polar coordinates in the same way as in the previous paragraph. The resulting expressions

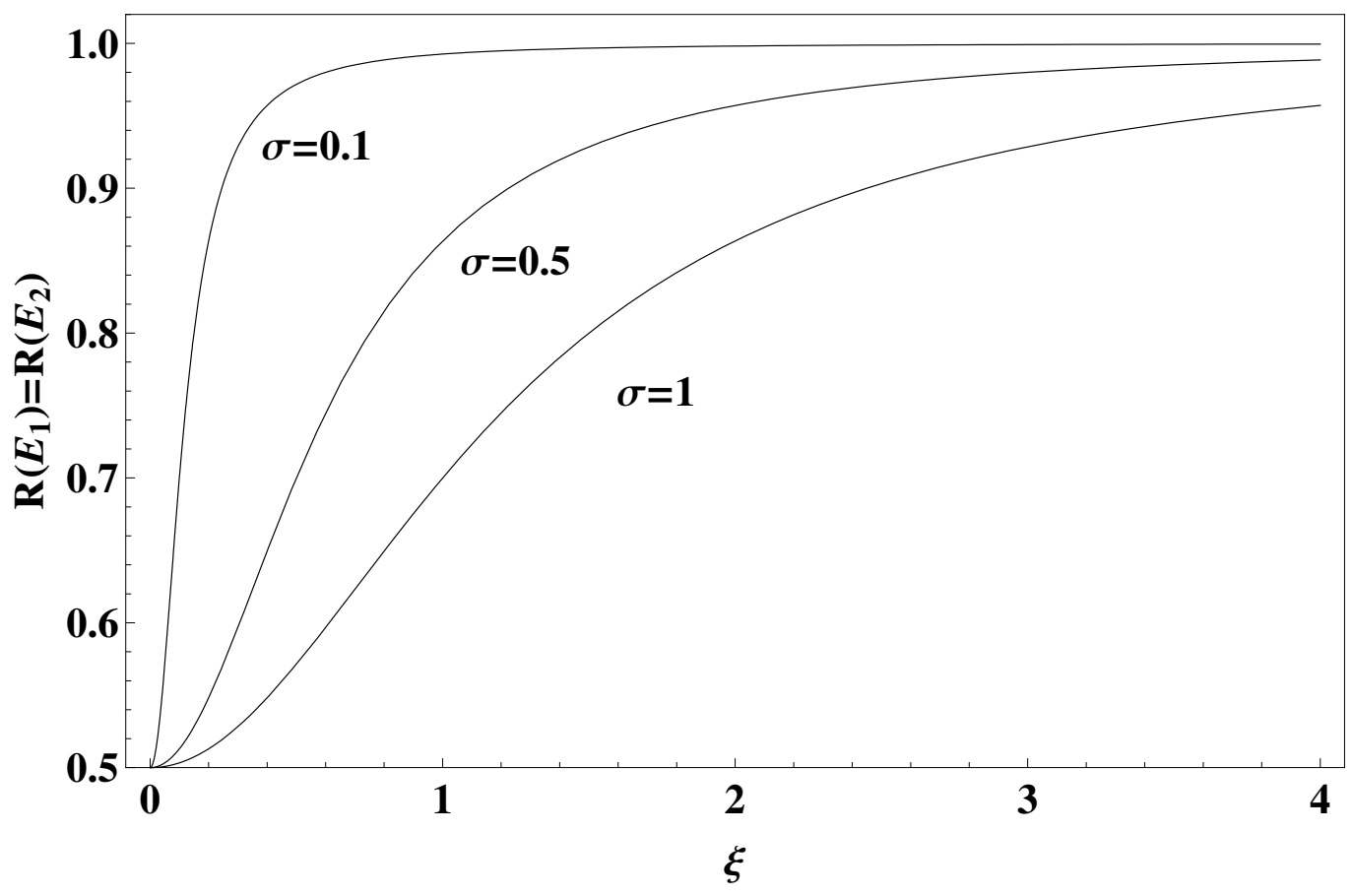

Figure 4.17. The reliability of the UI setup ($M = 2$, $n_A = n_B = n_C = 1$) as a function of the typical displacement $\xi$. Different curves correspond to different values of $\sigma$ i.e. to different fuzziness of the states. As is seen from the figure all curves in the limit of large $\xi$ are approaching the unity.

read:

$$\overline{P} = \frac{1}{1+2\sigma^2}(1 - \frac{1}{1+2\sigma^2+\frac{8n_A n_B}{n_A+2n_B}\xi^2}) ;$$

$$\overline{P_E} = \frac{1}{1+2\sigma^2}(\frac{2\sigma^2}{1+2\sigma^2+\frac{8n_A n_B}{n_A+2n_B}\xi^2}) ; \tag{4.165}$$

$$\overline{P_F} = \frac{2\sigma^2}{1+2\sigma^2} + \frac{1-2\sigma^2}{1+2\sigma^2}\frac{1}{1+2\sigma^2+\frac{8n_A n_B}{n_A+2n_B}\xi^2} . \tag{4.166}$$

More quantitative insight is presented in Fig. 4.18, which for the fixed $\sigma = 0.25$ presents the behavior of the calculated quantities $\overline{P}, \overline{P_E}, \overline{P_F}$ as a function of the typical displacement $\xi$. It is worth mentioning that for $\xi \to \infty$ the average probability of error goes to zero, but $\overline{P_F} > 0$, because the noise causes inconclusive results by firing both detectors simultaneously.

### 4.3.9 Summary

Let me summarize the unambiguous identification (UI) part of the chapter. In UI we are given a set of identical quantum systems prepared in pure states, which are labeled as unknown and reference states. The promise is that one type of reference state is the same as the unknown state and the task is to find out unambiguously which one it is. After stating precise definition of the problem and review of previous work we present the general approach to UI of two types of reference states. This approach is well suited for finite dimensional Hilbert spaces and it is based on reformulation of UI as discrimination of two known mixed states. When the probability distribution governing the choice of each reference state is uniform and supported on the whole Hilbert space $\mathcal{H}$ or on its subspace the problem can be solved completely due to simple structure of corresponding mixed states. However, my main focus was on the case where the set of possible reference states is formed by coherent states of an electromagnetic field. The relevance of this prior knowledge is illustrated in Section 4.3.5, where I show that the specialized measurement outperforms the universal unambiguous identification, i.e. the UI measurements that can be applied for all pure states. The difference between the measurements was quantified by the probability of identification for particular choice of reference states [see Eqs.(4.113),(4.115)] and is visualized by Figure 4.8. The interesting qualitative difference between the specialized and the universal measurement is in the probability of success for nearly orthogonal states. While our specialized measurement succeeds almost always the universal measurement produces conclusive result at most with probability $1/3$. Moreover, our specialized measurement can be easily experimentally realized, because it consists of three beam splitters and two photodetectors (see Figure 4.7). The setup was recently build and tested by L. Bartůšková et. al. [53] and we shortly summarize the experiment in Appendix C.

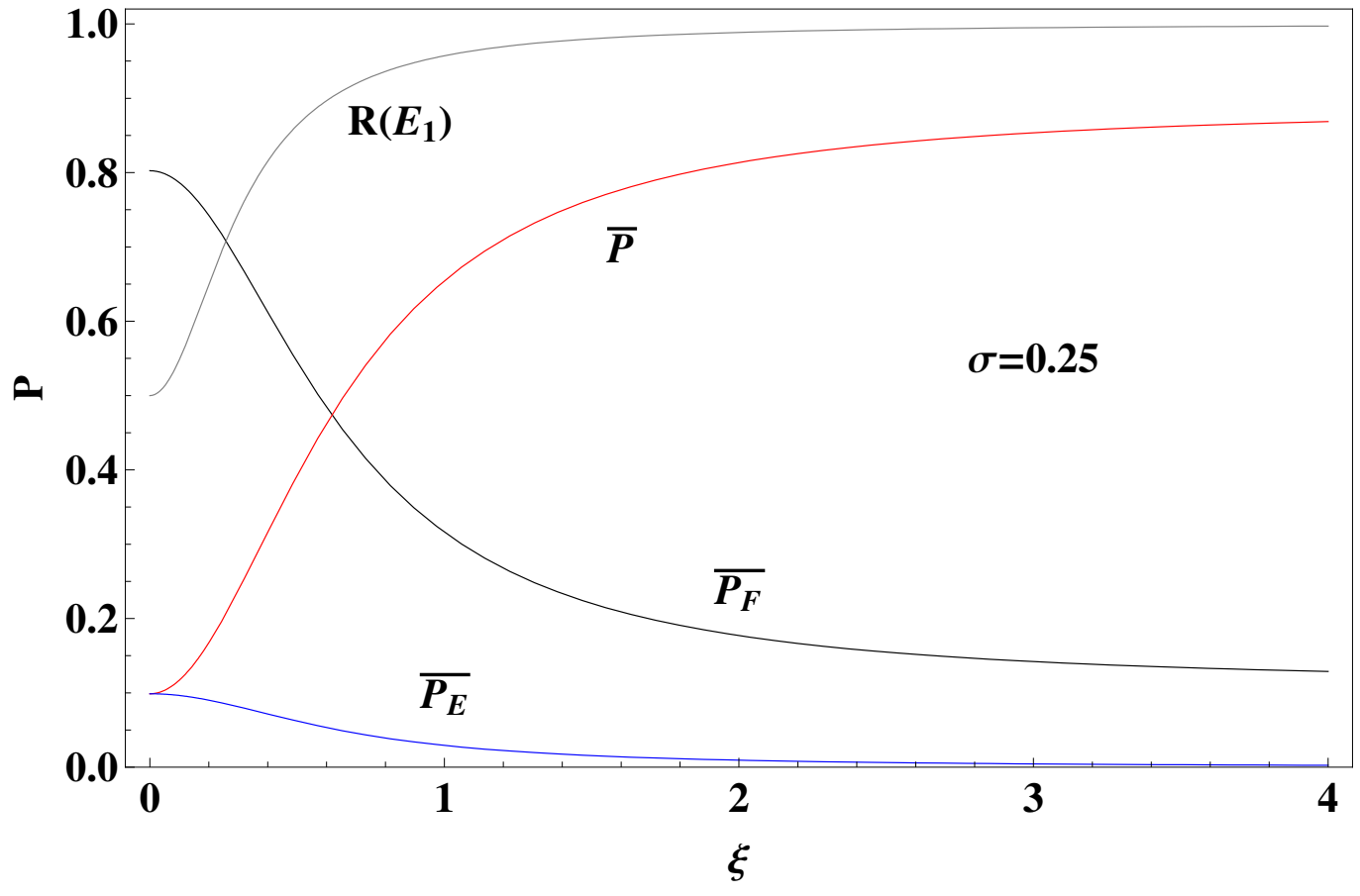

Figure 4.18. The reliability and the average probability of success ($\overline{P}$), the error ($\overline{P_E}$), and the failure ($\overline{P_F}$) for the "phase keying" scenario ($M = 2$, $n_A = n_B = n_C = 1$) with $\sigma = 0.25$ as a function of the typical displacement $\xi$ .

The beamsplitter setup was motivated by an intuitive reduction of the unambiguous identification problem into specific "distribution" task and an unambiguous state comparison. As a next step the generalization of this optical setup to situations with more copies of the unknown and the reference states was presented in Section 4.3.6. Our approach was based on an idea of the "concentration" of the same type of states into strong coherent states that were subsequently identified by setups for the single-copy scenario. In the UI task it is assumed that the particular choice of the reference states is unknown to us, and only the probability distribution $\chi$ describing this choice is known. Nevertheless, even without having $\chi$ it is possible to derive the optimal choice of transmittivities in the beam-splitter setup we proposed for two types of reference states and an equal number of copies of each of the reference states ($n_B = n_C$). In that case the probability of identification for the reference states $|\alpha_1\rangle, |\alpha_2\rangle$ reads:

$$P(|\alpha_1\rangle, |\alpha_2\rangle) = 1 - e^{-\frac{n_A n_B}{n_A + 2n_B}|\alpha_1 - \alpha_2|^2}. \tag{4.167}$$

Under the condition that the experimental setup consists only of linear optical elements and photodetectors we also proved the optimality of the setup. In the limit of $n_B = n_C \to \infty$ the two reference states become known. Therefore, one needs to unambiguously discriminate the unknown state between two known pure states. The probability of success of our setup in this case coincides with the optimal value achieved by the Ivanovic-Dieks-Peres measurement [6–8].

In Section 4.3.7 we addressed the question whether the coherent reference states can be recreated after our UI measurement. We showed that the reference states can be partially recovered only if the measurement yielded a conclusive outcome. The recovered reference states can be used in the next round of the UI if another unknown state is provided. This might be seen as a repeated search in a quantum database, where the data, i.e. the reference states, degrade with repeated use of the database.

Recently, a framework for transformations induced by linear optics on coherent states was proposed by B. He and J. Bergou in Ref. [51]. These authors illustrated their method on the three beam splitter setup proposed in Section 4.3.5 and suggested that the reference states can be always perfectly recovered. However, in their case the reference states are known, whereas in our case the complex amplitudes of all coherent states are not known in advance.

Finally, in Section 4.3.8 we investigated how a particular type of noise influence the reliability of the conclusions drawn by our UI setup. More precisely, we considered a communication scenario called the phase keying, with two coherent reference states of equal amplitude, but the opposite phases. We saw that the reliability of results, expressed by Eq. (4.163), depends only on the ratio of the amplitudes of the noise and the signal. However, for nonzero noise the unambiguity of the conclusions is lost.

## 5 Unambiguous tasks for channels

The aim of quantum channels is to describe the overall effect of a temporal evolution on the considered quantum system. A quantum channel prescribes the final state to any possible input state, but it does not characterize how the transformation is achieved. This is similar to having a black box evaluating function $f$ without knowing how the calculation proceeds inside. In this chapter we shall consider experiments in which the investigated quantum channel is used just once. Our aim will be to distinguish among the expected possibilities (i.e. expected channels) by a single run of the experiment. In order to do that we shall control both the state preparation and the final measurement surrounding the use of the investigated channel. General experiment of this kind (see figure 5.1) uses also an ancillary system that does not evolve, while the principal quantum system is exposed to the tested channel. Thus, the evolution of the compound system takes place in between the preparation and the measurement part of the experiment, which should be under our control. Class of such experiments is very broad, because the Hilbert space of ancilla can have arbitrary dimension and we can independently tune state preparation and the final measurement. Fortunately, many of those experiments are completely equivalent. For example, one can show that in general it suffices to consider pure state preparation and an ancilla with at most the dimension of the principal quantum system. Taking this into account some basic problems for quantum channels were studied. In particular, researchers investigated the minimum error discrimination for unitary channels [54, 55] and for some other specific channels [56–60]. In contrast to quantum states it was found that a finite number of uses of a unitary channel makes it possible to discriminate perfectly among discrete set of unitary channels, which are not distinguishable perfectly by a single use. For general quantum channels partial results were obtained also for the unambiguous discrimination. Wang and Ying [18] found the necessary and sufficient conditions for unambiguous channel discrimination in terms of Kraus operators, which characterize each channel. Chefles and Barnett in [61] investigated unambiguous discrimination of unitary operators. Focus on discrimination of unitary operators is naturally motivated by the area of quantum computation. Algorithms that provide speed up with respect to their classical counterparts are very often based on the use of coherent quantum superpositions, which are not degraded only by unitary channels. Many practically interesting problems (database search etc.) are theoretically formulated as an oracle identification problem. Here oracle is a black box, whose internal operating mechanism defines the solution of the problem. Unfortunately, we can not solve the problem by directly looking into the black box, instead we have to do it by testing the transformation that the oracle introduces on its inputs. Usually we know which finite set of transformations the oracle may implement and we want to discover which of them it actually is with as few uses (queries) of the oracle as possible. By convention the oracles in quantum computation are chosen to implement unitary channels. For unambiguous discrimination of such channels from their single use Chefles and Barnett proved the following statements: *Unitary operators $\{U_j\}_{j=1}^N$ acting on a Hilbert space $\mathcal{H}$ are unambiguously distinguishable if and only if they are linearly independent*. Moreover, set of linearly independent unitary operators can always be unambiguously discriminated using any entangled probe state with maximum Schmidt rank $d$, the dimensionality of $\mathcal{H}$. *For a set of commuting unitary operators $U_j$ any unambiguous discrimination experiment involving ancilla (see figure 5.1) is completely equivalent to an experiment without ancilla.* The above two statements restrict the number of commuting unitary operators that can be unambiguously discriminated to $d = \dim \mathcal{H}$. Chefles and Barnett in [61] were most concerned with the distinguishability of so called *standard oracle operators*, which are usually considered as (reversible) quantum realizations of classical oracles. These are unitary operators acting on two subsystems of the same dimension and modify the basis in the following way $|i\rangle|j\rangle \mapsto |i\rangle|j \oplus f(i)\rangle$, where $f$ is an integer valued function. The work of Chefles and Barnett laid foundations for unambiguous discrimination of quantum oracles and also reviewed the related previous work.

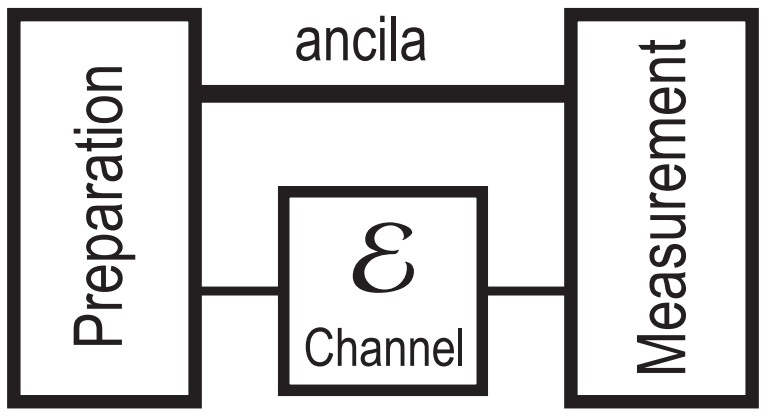

Figure 5.1. Scheme of general experiment for distinguishing quantum channels by their single use.

As we announced already in Chapter 3 for discussing experiments involving single use of the tested channel we will use framework of *process positive operator valued measure (PPOVM)*. This framework exploits a specific representation of channels defined via Choi-Jamiolkowski isomorphism [62, 63]. According to the theorem a channel on $D$ dimensional quantum system can be represented by a positive operator acting on a bipartite quantum system $D \times D$. In particular, a channel $\mathcal{E}$ is represented by an operator

$$\omega_{\mathcal{E}} = (\mathcal{I} \otimes \mathcal{E})[\Psi_D^+], \tag{5.1}$$

where $\Psi_D^+ = |\Psi_D^+\rangle\langle\Psi_D^+|$ and $|\Psi_D^+\rangle = \sum_{j=1}^D |j\rangle \otimes |j\rangle$. Let us note that $\Psi_D^+$ is not a projector, because it is not normalized and $Tr(\Psi_D^+) = D$. The operator $\frac{1}{D}\Psi_D^+$ is a one-dimensional projector onto the maximally entangled state $|\psi_+\rangle = \frac{1}{\sqrt{D}} \sum_j |j\rangle \otimes |j\rangle \in \mathcal{H} \otimes \mathcal{H}$.

Process POVM is defined [31, 64] as a collection of positive operators (effects) $M_1, \ldots, M_n$ such that $\sum_j M_j = \xi^T \otimes I$ for some state $\xi$ of the $D$ dimensional system. An event that we can observe in the experiment consists of a preparation of the test state $\varrho$ and an observation of the effect $E_j$ in the measurement $E$ of the output state. Let us note that in the experiment we are allowed to use an ancilla of arbitrary size, i.e. $\varrho$ and $E_j$ are operators defined on $d_{\rm anc} \times D$-dimensional Hilbert space. The conditioned probability to observe an event consisting of the state preparation $\varrho$ and the observation of an effect $E_j$ providing that channel $\mathcal{E}$ is tested equals

$$p(\varrho, E_j|\mathcal{E}) = Tr\left(E_j(\mathcal{I} \otimes \mathcal{E})[\varrho]\right). \tag{5.2}$$

Using the Choi-Jamiokowski relation $\varrho = (\mathcal{R}_\varrho \otimes \mathcal{I})[\Psi_D^+]$, where $\mathcal{R}_\varrho : \mathcal{L}(\mathcal{H}) \to \mathcal{L}(\mathcal{H}_{\rm anc})$ is a completely positive map, and the duality relation $Tr(Y\mathcal{F}[X]) = Tr(\mathcal{F}^*[Y]X)$ determining the dual channel $\mathcal{F}^*$ we can write

$$\begin{aligned} p(\varrho, E_j|\mathcal{E}) &= Tr((\mathcal{R}_\varrho^* \otimes \mathcal{I})[E_j](\mathcal{I} \otimes \mathcal{E})[\Psi_D^+]) \\ &= Tr(M_j\, \omega_{\mathcal{E}}), \end{aligned} \tag{5.3}$$

where $M_j$ is an element of PPOVM. By definition $M_j$ is positive and $\sum_j M_j = (\mathcal{R}^*_\varrho \otimes \mathcal{I})[I] = \xi^T \otimes I$, where $\xi = \mathrm{tr}_{\mathrm{anc}}[\varrho]$. Thanks to linearity of Eqs. (5.2),(5.3) the derivation of the PPOVM elements remains valid also for experiments in which the test state $\varrho_k$ and the measurement $\{E_j^{(k)}\}$ are together randomly chosen according to an ensemble $\varrho = \sum_k p_k \varrho_k$. Thus, any conceivable experiment in which the channel is used once can be formalized as a PPOVM and the converse also holds [31], i.e. any PPOVM can be experimentally implemented. Let us note that for a given PPOVM, i.e. a set of positive operators $M_1, \ldots, M_n$ such that $\sum_x M_x = \varrho^T \otimes I_D$ there exists many different experiments with different choices of test states and POVMs.

*Ancilla-free test state*

Consider a PPOVM such that $M_j = \varrho^T \otimes F_j$ for all $j$. Since the identity

$$Tr(\mathcal{E}[\varrho]F_j) = Tr((\mathcal{I} \otimes \mathcal{E})[\Psi_D^+](\varrho^T \otimes F_j)) = Tr(\omega_\mathcal{E} M_j)$$

holds for all qudit channels $\mathcal{E}$ and all qudit operators $\varrho, F$, it follows that this type of PPOVM can be realized by using a single ancilla-free test state $\varrho$ and performing the measurement described by POVM elements $F_j$.

*Maximally entangled probe*

On the other hand, let us consider that an unknown qudit channel is probed by a (normalized) maximally entangled state $|\psi_+\rangle = \frac{1}{\sqrt{D}} \sum_j |j\rangle \otimes |j\rangle \in \mathcal{H} \otimes \mathcal{H}$. In this case the mapping $\mathcal{R}_{\psi_+} = \frac{1}{D}\mathcal{I}$, i.e. $|\psi_+\rangle\langle\psi_+| = \frac{1}{D}\Psi_D^+$. That is, $M = (\mathcal{R}^*_{\psi_+} \otimes \mathcal{I})[F] = \frac{1}{D}F$, where $F$ is a two-qudit effect. Considering a POVM consisting of effects $F_1, \ldots, F_n$ the corresponding PPOVM is composed of positive operators $M_j = \frac{1}{D}F_j$.

In the following we shall apply the PPOVM framework to unambiguous tasks for quantum channels. We will first summarize the results on unambiguous discrimination of a pair of general channels and then restrict ourselves to unitary channels.

### 5.1 Unambiguous discrimination of two channels

PPOVM for unambiguous discrimination of two channels should consist of three elements $\{M_0, M_1, M_2\}$. We associate observation of result $i$ with use of channel $\mathcal{E}_i$ and declare outcome 0 as inconclusive. Unambiguity of measurement outcomes requires following two equations to be satisfied

$$Tr(M_1\omega_2) = 0, \quad Tr(M_2\omega_1) = 0. \tag{5.4}$$

The optimal measurement should maximize the probability of success

$$P_{succ} = \eta_1 Tr(\omega_1 M_1) + \eta_2 Tr(\omega_2 M_2), \tag{5.5}$$

while preserving positivity of PPOVM elements $M_i$ and normalization $M_0 + M_1 + M_2 = \xi^T \otimes I$ with $\xi$ being a state of the principal quantum system. Thus, the problem is similar to the unambiguous discrimination of states except for the normalization of the operator measure. This normalization $\xi$ provides additional freedom, which complicates the optimization. Nevertheless, some basic features of the problem remain unchanged. For example two mixed states can be unambiguously discriminated if and only if they have distinct supports. The same holds also for channels i.e. channels $\mathcal{E}_1, \mathcal{E}_2$ can be unambiguously distinguished by a single use of the channel if and only if their process states $\omega_i = (\mathcal{I} \otimes \mathcal{E}_i)[\Psi_D^+]$ have distinct supports. This can be easily seen after denoting by $\Pi_1, \Pi_2$ the projectors onto the supports of $\omega_1, \omega_2$, respectively. If $\Pi_1 = \Pi_2$ then Eq. (5.4) implies $0 = Tr(M_i \Pi_{3-i}) = Tr(M_i \Pi_i)$ and consequently $P_{succ} = 0$ i.e. the condition on supports is necessary. The sufficiency of the condition is proved by considering the PPOVM $M_1 = \frac{1}{2D}(I - \Pi_2)$, $M_2 = \frac{1}{2D}(I - \Pi_1)$, $M_0 = \frac{1}{D} I - M_1 - M_2$, which can be seen as an experiment with maximally entangled state $|\psi_+\rangle$. Let us stress that this PPOVM is not always optimal, but it succeeds with non-zero probability due to $\Pi_1 \leq \Pi_2$, $\Pi_2 \leq \Pi_1$ not holding simultaneously. The above condition judging the possibility of unambiguous channel discrimination via the supports of $\omega_1, \omega_2$ is an alternative to the condition found by Wang and Ying [18] and in some cases it may be more easily checked. Similarly as for unambiguous discrimination of states also for channels single channel detection measurements are optimal for very unbalanced prior probabilities. This, as we will see later, holds for a pair of unitary channels and seems to be a very plausible conjecture also for arbitrary two channels. For example, experiments with channels that contract everything into a fixed state are equivalent to experiments for state discrimination, because the state of the system that was affected by the channel is factorized from the potentially used ancilla. Hence, an attempt to unambiguously discriminate the possible final states for very unbalanced prior probabilities is an unambiguous state discrimination problem for which single state detection measurement is optimal.

In the language of PPOVM unambiguous detection of the channel $\mathcal{E}_1$ within the set $\{\mathcal{E}_1, \mathcal{E}_2\}$ corresponds to $M_2 = O$. In this case only the no-error condition $Tr(M_1\omega_2) = 0$ applies and the optimal probability of success contains just a single term $P_{succ} = \eta_1 Tr(M_1\omega_1)$.

Let us now return to a general case assuming arbitrary prior probabilities $\eta_1, \eta_2 = 1 - \eta_1$. In the following we present an upper bound on the probability of success proposed by Ziman et.al. [27]. This bound is an analog of the so called "Fidelity bound" known for the unambiguous discrimination of two mixed states (see Section 4.1 or for instance [65]).

**Proposition 1** *Let $\mathcal{E}_1, \mathcal{E}_2$ be channels (i.e. completely positive trace preserving linear maps) and $\eta_1, \eta_2$ be their prior probabilities. Then*

$$P_{succ} \leq 1 - 2\sqrt{\eta_1\eta_2} \min_{\xi} Tr|\sqrt{\omega_1}(\xi^T \otimes I)\sqrt{\omega_2}| \,, \tag{5.6}$$

*where $\omega_j = (\mathcal{I} \otimes \mathcal{E}_j)[\Psi_D^+]$.*

**Proof 3** *Proving the above proposition is equivalent to showing*

$$P_{fail} \geq 2\sqrt{\eta_1\eta_2} \min_{\xi} Tr|\sqrt{\omega_1}(\xi^T \otimes I)\sqrt{\omega_2}| \,, \tag{5.7}$$

*because probability of failure equals to $1 - P_{succ}$. We set $a = \eta_1 Tr(M_0\omega_1)$, $b = \eta_2 Tr(M_0\omega_2)$ and since for all numbers $a^2 + 2ab + b^2 \geq 4ab$ holds we get*

$$(P_{fail})^2 = (a+b)^2 \geq 4\eta_1\eta_2 Tr(M_0\Omega_1) Tr(M_0\omega_2) \,. \tag{5.8}$$

*Using the Cauchy-Schwartz inequality for arbitrary unitary operator $U$ we obtain*

$$\begin{aligned} Tr(M_0\omega_1)Tr(M_0\omega_2) &= Tr(U\sqrt{\omega_1}\sqrt{M_0}\sqrt{M_0}\sqrt{\omega_1}U^\dagger) \times \\ &\times \;\; Tr(\sqrt{\omega_2}\sqrt{M_0}\sqrt{M_0}\sqrt{\omega_2}) \\ &\geq (Tr(U\sqrt{\omega_1}M_0\sqrt{\omega_2}))^2 \,. \end{aligned}$$

*By definition* $M_0 = \xi^T \otimes I - M_1 - M_2$. *Due to no-error conditions* $\omega_1 M_2 = M_1 \omega_2 = O$ *holds and it follows that* $\sqrt{\omega_1} M_0 \sqrt{\omega_2} = \sqrt{\omega_1}(\xi^T \otimes I)\sqrt{\omega_2}$. *Thus,*

$$\overline{p}_{\rm failure} \geq 2\sqrt{\eta_1\eta_2}|Tr(U\sqrt{\omega_1}(\xi^T \otimes I)\sqrt{\omega_2})| \,. \tag{5.9}$$

*Using the identity* $\sup_U |Tr(XU)| = {\rm tr}|X|$ *holding for all operators* $X$ *the inequality reads*

$$\overline{p}_{\rm failure} \geq 2\sqrt{\eta_1\eta_2}{\rm tr}|\sqrt{\omega_1}(\xi^T \otimes I)\sqrt{\omega_2}| \,, \tag{5.10}$$

*which proves the proposition after the optimization over the PPOVM normalization is taken into account.*

The function $F(\omega_1, \omega_2) = \min_\xi {\rm tr}|\sqrt{\omega_1}(\xi \otimes I)\sqrt{\omega_2}|$ is called *completely bounded process fidelity* in analogy with the completely bounded norm $||\cdot||_{\rm cb}$.

### 5.2 Unambiguous discrimination of two unitary channels

Unitary channels are associated with Choi-Jamiokowski operators proportional to one-dimensional projectors. In particular, $\mathcal{E}_U = U \cdot U^\dagger$ is represented by $\omega_U = D|\psi_U\rangle\langle\psi_U|$, where $|\psi_U\rangle = (I \otimes U)|\psi_+\rangle$. Given a pair of unitary channels $U, V$, then the joint support of $\omega_U, \omega_V$ specifies a two-dimensional subspace $\mathcal{Q}$ of $\mathcal{H} \otimes \mathcal{H}$, which is relevant for discrimination.

Since supports of $\omega_U$ and $\omega_V$ are different, two unitaries can be always unambiguously distinguished. Let us denote by $Q$ a projector onto the linear subspace $\mathcal{Q}$ spanned by vectors $|\psi_U\rangle, |\psi_V\rangle$. The unambiguous no-error conditions require that on the relevant subspace $\mathcal{Q}$ the operators $M_U, M_V$ are rank-one and take the form

$$M_U^{\mathcal{Q}} = c_U(Q - |\psi_V\rangle\langle\psi_V|) \,, \tag{5.11}$$
$$M_V^{\mathcal{Q}} = c_V(Q - |\psi_U\rangle\langle\psi_U|) \,. \tag{5.12}$$

In addition, $M_U + M_V \leq \xi^T \otimes I$ for some state $\xi$. The optimal success probability reads

$$\begin{aligned} P_{succ} &= \max_{\rm PPOVM}(\eta_U Tr(M_U\omega_U) + \eta_V Tr(M_V\omega_V)) \\ &= \max_{\rm PPOVM}\left(\eta_U Tr(M_U^{\mathcal{Q}}\omega_U) + \eta_V Tr(M_V^{\mathcal{Q}}\omega_V)\right) \\ &= \max_{\varphi}\max_{\rm POVM}(\eta_U\langle\varphi_U|F_U|\varphi_U\rangle + \eta_V\langle\varphi_V|F_V|\varphi_V\rangle) \,. \end{aligned}$$

Here we used the fact that PPOVM can be always implemented using a pure test state (see [31]). This test state is associated with a suitable vector $|\varphi\rangle = (A \otimes I)|\Psi_D^+\rangle$ leading to $M_U = (A^\dagger \otimes I)F_U(A \otimes I)$, $M_V = (A^\dagger \otimes I)F_V(A \otimes I)$, where effects $F_U, F_V$ represent the conclusive outcomes of the performed POVM, i.e. $F_U + F_V \leq I \otimes I$. We have used the notation $|\varphi_U\rangle = (I \otimes U)|\varphi\rangle$ and $|\varphi_V\rangle = (I \otimes V)|\varphi\rangle$.

For a fixed test state $|\varphi\rangle\langle\varphi|$ the POVM maximizing the success probability $\eta_U\langle\varphi_U|F_U|\varphi_U\rangle + \eta_V\langle\varphi_V|F_V|\varphi_V\rangle$ is known from the problem of unambiguous discrimination of two pure states $|\varphi_U\rangle, |\varphi_V\rangle$ (see Section 4.1.1). Without loss of generality we can assume $\eta_U \geq \eta_V$. In such case the optimal POVM consists of effects

$$\begin{aligned} F_U &= \min\left\{\frac{1-\sqrt{\frac{\eta_V}{\eta_U}}|\langle\varphi_U|\varphi_V\rangle|}{1-|\langle\varphi_U|\varphi_V\rangle|^2},1\right\}(Q_\varphi-|\varphi_V\rangle\langle\varphi_V|)\,, \\ F_V &= \max\left\{\frac{1-\sqrt{\frac{\eta_U}{\eta_V}}|\langle\varphi_U|\varphi_V\rangle|}{1-|\langle\varphi_U|\varphi_V\rangle|^2},0\right\}(Q_\varphi-|\varphi_U\rangle\langle\varphi_U|)\,, \\ F_0 &= I-F_U-F_V\ , \end{aligned}$$

where $Q_\varphi$ is a projector onto the subspace spanned by vectors $|\varphi_U\rangle, |\varphi_V\rangle$. The success probability optimized also over choices of test state $|\varphi\rangle$ reads

$$P_{succ}=\begin{cases} 1-2\sqrt{\eta_U\eta_V}F(U,V) & \text{if } F(U,V)\le\sqrt{\frac{\eta_V}{\eta_U}}\le 1 \\ \eta_U(1-F(U,V)^2) & \text{if } F(U,V)\ge\sqrt{\frac{\eta_V}{\eta_U}}\le 1 \end{cases}\,, \tag{5.13}$$

where we denoted $F(U,V)=\min_\varphi|\langle\varphi_U|\varphi_V\rangle|$. As we will see $F(U,V)$ is the completely bounded process fidelity from the previous section, which for unitaries turns out be

$$F(U,V)=\min_{\xi\in S_{\mathrm{P}}(\mathcal{H})}|Tr(\xi U^\dagger V)|. \tag{5.14}$$

To see this we first consider PPOVM corresponding to experiments with pure test state $\varrho = |\varphi\rangle\langle\varphi|$. Any unit vector $|\varphi\rangle$ can be expressed as $|\varphi\rangle=\sqrt{D}(A\otimes I)|\psi_+\rangle$, thus $|\varphi\rangle\langle\varphi|=(\mathcal{R}_\varphi\otimes\mathcal{I})[\Psi_D^+]=(A\otimes I)\Psi_D^+(A^\dagger\otimes I)$ and the normalization requirement $Tr(|\varphi\rangle\langle\varphi|)=1$ corresponds to $Tr(A^\dagger A)=1$. The test state $|\varphi\rangle$ fixes the normalization of the related PPOVM to be $\xi^T\otimes I=\sum_j M_j=(\mathcal{R}^*_\varphi\otimes\mathcal{I})[I]=A^\dagger A\otimes I$ i.e. $A^\dagger A=\xi^T$. The minimum overlap of the final states $|\varphi_U\rangle, |\varphi_V\rangle$ reads

$$\begin{aligned} \min_\varphi|\langle\varphi_U|\varphi_V\rangle| &= D\min_{A:Tr(A^\dagger A)=1}|\langle(A\otimes U)\psi_+|(A\otimes V)\psi_+\rangle| \\ &= \min_A|Tr((A^\dagger A)^T U^\dagger V)| \\ &= \min_{\xi\in\mathcal{S}(\mathcal{H})}|Tr(\xi U^\dagger V)|\,, \end{aligned} \tag{5.15}$$

where we used the explicit form of state $|\psi_+\rangle=\frac{1}{\sqrt{D}}\sum_j|j\rangle\otimes|j\rangle$.

As a second step we evaluate the completely bounded process fidelity according to its definition from the previous section. For a pair of unitary channels $U$ and $V$ we have $\sqrt{\omega_U}=\sqrt{D}|\psi_U\rangle\langle\psi_U|$ and $\sqrt{\omega_V}=\sqrt{D}|\psi_V\rangle\langle\psi_V|$ and using the formula $Tr|X|=Tr\sqrt{X^\dagger X}$ the required expression follows:

$$\begin{aligned} F(U,V) &\equiv \min_\xi Tr|\sqrt{\omega_U}(\xi^T\otimes I)\sqrt{\omega_V}| \\ &= D\,Tr(|\psi_U\rangle\langle\psi_U|)\min_\xi|\langle\psi_U|(\xi^T\otimes I)\psi_V\rangle| \\ &= \min_\xi|Tr(\xi U^\dagger V)| \end{aligned} \tag{5.16}$$

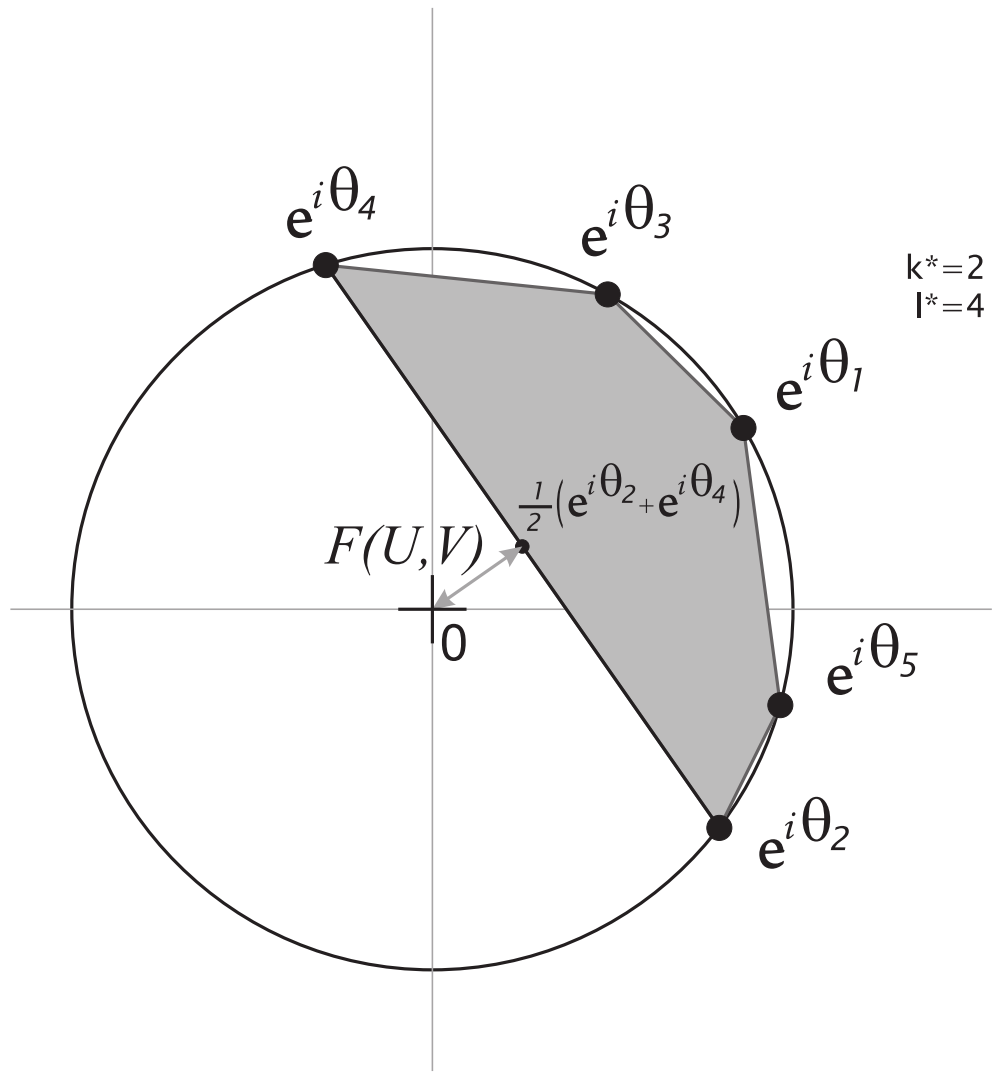

Figure 5.2. Geometrical interpretation of completely bounded process fidelity for two unitary channels.

From equations (5.13), (5.16) we see that the upper bound from Proposition 1 is saturated for $F(U,V) \leq \sqrt{\frac{\eta_V}{\eta_U}}$. However, for $F(U,V) \geq \sqrt{\frac{\eta_V}{\eta_U}}$ the optimal probability of success $P_{succ} = \eta_U(1 - F(U,V)^2)$ is smaller. We see that this bound is not achievable in general. In fact, the existence of the PPOVM giving the bound is not guaranteed in its derivation.

Let us now investigate equation $(5.14)$. The quantity on the right hand side was also analyzed in the study of perfect discrimination of unitary channels [54, 55] and we will repeat the analysis to illustrate its geometrical meaning. Let us denote by $\{\phi_k\}$ the eigenvectors of $U^\dagger V$ associated with eigenvalues $e^{i\theta_k}$. Then

$$F(U,V) = \min_{\xi \in S_{\mathrm{P}}(\mathcal{H})} |\sum_k e^{i\theta_k} \langle \phi_k | \xi | \phi_k \rangle| \tag{5.17}$$

The number on the right hand side is a convex combination of complex square roots of unity. Thus, it can be visualized as an element of the convex hull of points (eigenvalues of $U^\dagger V$) on the unit circle of the complex plane. Our aim is to find the complex number within this convex hull which is closest to zero. In particular, if 0 is not contained in the convex hull, then

$$F(U,V) = \frac{1}{2} \min_{k,l} |e^{i\theta_k} + e^{i\theta_l}|\,, \tag{5.18}$$

which means (see figure 5.2) that state $\xi$ minimizing the expression has only two nonanishing diagonal entries (equal to 1/2) in its matrix with respect to orthonormal basis $\{\phi_k\}$. The minimum in the Eq.(5.17) depends only on the diagonal entries of $\xi$, thus we can always choose optimal $\xi$ to be a pure state. That is, no ancilla is needed in order to realize an optimal experiment unambiguously discriminating two unitaries.

Since for two-dimensional Hilbert space the unitary operators have only two eigenvalues, the minimalization is trivial [54] and

$$F(U,V) = \frac{1}{2}|e^{i\theta_k} + e^{i\theta_l}| = \frac{1}{2}|Tr(U^\dagger V)|\,. \tag{5.19}$$

Hence in this case the orthogonality in the Hilbert-Schmidt sense is necessary and sufficient for perfect discrimination of $\mathcal{E}_U$ and $\mathcal{E}_V$. Moreover, the maximally entangled state (for which $\xi = \frac{1}{2}I$) is a universal test state, because it optimizes unambiguous discrimination of any two unitary channels.

Of course, the measurements depend on the particular task and the unitaries. Unfortunately, these properties do not hold in the higher dimensions. For example, CNOT and SWAP gate can be perfectly discriminated even without being orthogonal. In this particular case also the maximally entangled test state is not very usable.

### 5.3 Unambiguous comparison of unitary channels

Our goal in this section is to investigate a comparison of quantum devices implementing unknown unitary channels. Such universal comparator of unitary channels can be of use, for instance, in the calibration and testing of the quality of elementary quantum gates.

Quantum channels are tested in two steps. First we prepare a so-called test state and apply the channel. After that the output state is measured. Therefore, it seems natural to compare a pair of channels by comparing the states they produce out of the same initial state. Indeed state comparison is closely related to channel comparison, but there are also important differences concerning the optimal strategies as we shall see later. The first who considered unambigous comparison of unitary channels were Andersson, Jex, and Barnett [66]. They proposed several strategies and developed also their generalizations for comparing more than two unitaries. However, they did not investigate the optimality of the strategies, which is our aim here. In the following we reformulate the problem in the PPOVM framework and show existence of a solution. Consequently, the optimal solution shall be described together with its uniqueness.

#### 5.3.1 Formulation of the problem

Consider we are given two black boxes implementing unknown unitary channels $\mathcal{E}_U$ and $\mathcal{E}_V$ on qudit, i.e. $d$-dimensional quantum system. Our task is to unambiguously decide whether the black boxes perform the same *unknown* unitary channel, or not. More formally, whether a process implemented on $D = d \times d$ dimensional quantum system by the pair of devices is described by a channel $\mathcal{E}_U \otimes \mathcal{E}_V$ with $U \neq V$, or by a channel $\mathcal{E}_U \otimes \mathcal{E}_U$. As in any comparison problem we implicitly assume that the probability that the channels are the same is nonzero. Otherwise the problem would be senseless.

Let us note that unlike preparators (represented by states) the processes (associated with channels) can be used sequentially. In general, this is an important difference between the usage of preparators and processes providing us with a resource of a potential use. However, it does not give us any advantage in the case of the considered comparison problem. In particular, one cannot distinguish whether the product of two unknown unitary channels is $\mathcal{E}_U \circ \mathcal{E}_V$ (for $U \neq V$), or $\mathcal{E}_U \circ \mathcal{E}_U$, because for any unitary operator $W$ there exist unitary operators $U, V \neq W$ such that $W^2 = UV$.

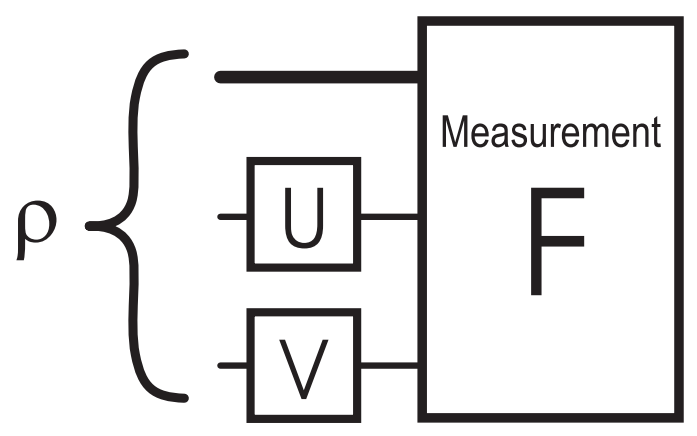

Figure 5.3. Experiment for comparison of two unitary channels $\mathcal{E}_U, \mathcal{E}_V$.

The experimental procedure for the comparison is illustrated in figure 5.3. Using each of the quantum boxes at most once the experiment will end by a measurement, whose outcome uniquely determines our conclusion. In particular, the experiment consists of three steps. At first, we prepare a so-called test state $\varrho$ on $\mathcal{H}_{\rm anc} \otimes \mathcal{H}_d \otimes \mathcal{H}_d$, where $\mathcal{H}_{\rm anc}$ is the Hilbert space of some ancilliary system. After that black boxes are applied and a measurement $F$ on the whole system including the ancilla is performed. Measurement outcomes are associated with effects $F_{\rm same}, F_{\rm diff}, F_0$ forming a three-valued POVM, i.e.

$$O \leq F_{\rm same}, F_{\rm diff}, F_0 \leq I\,; \qquad F_0 + F_{\rm same} + F_{\rm diff} = I\,.$$

As in any other unambiguous task the inconclusive outcome $F_0$ is needed in order to make the conclusive outcomes $F_{\rm same}, F_{\rm diff}$ unambiguous. In fact, we shall see explicitly that $F_0 \neq O$. An outcome $x \in \{\rm same,\ diff,\ 0\}$ is observed with the probability

$$p_x(U \otimes V) = Tr(F_x(\mathcal{I}_{\rm anc} \otimes \mathcal{E}_U \otimes \mathcal{E}_V)[\varrho])\,, \tag{5.20}$$

where $\mathcal{E}_U[\cdot] = U \cdot U^\dagger, \mathcal{E}_V[\cdot] = V \cdot V^\dagger$ are unitary channels implemented by the black boxes.

Our goal is to characterize all possible experiments (determined by pairs $\varrho, F$) performing the unambiguous comparison of unitary channels and identify the optimal strategy. The figure of merit for the optimization will be specified in details later.

### 5.3.2 Requirements on unambiguous comparators

Translating the comparison problem into PPOVM framework we set $D = d^2$ and associate the two black boxes acting on $d$-dimensional systems with operators

$$\omega_{U\otimes U} = (I_D \otimes U \otimes U)\Psi_D^+(I_D \otimes U^\dagger \otimes U^\dagger)\,, \tag{5.21}$$
$$\omega_{U\otimes V} = (I_D \otimes U \otimes V)\Psi_D^+(I_D \otimes U^\dagger \otimes V^\dagger)\,, \tag{5.22}$$

where $\Psi_D^+ = |\Psi_D^+\rangle\langle\Psi_D^+|$ and $|\Psi_D^+\rangle_{1234} = |\Psi_d^+\rangle_{13} \otimes |\Psi_d^+\rangle_{24} \in \mathcal{H}_d^{\otimes 4}$ with $|\Psi_d^+\rangle = \sum_{j=1}^d |j\rangle \otimes |j\rangle$. Operators $M_{\rm same}, M_{\rm diff}, M_0$ defining the PPOVM have to satisfy following no-error conditions ensuring the unambiguity of the corresponding conclusion:

$$p_{\rm diff}(U \otimes U) = Tr(\omega_{U\otimes U} M_{\rm diff}) = 0$$
$$p_{\rm same}(U \otimes V) = Tr(\omega_{U\otimes V} M_{\rm same}) = 0$$

for all $U, V \in U(d)$, where $U(d)$ denotes the group of unitary operators on $d$-dimensional Hilbert space.

Defining average channels as

$$\mathcal{A}[X] = \int_{U(d)} dU \;\; UXU^\dagger \,, \tag{5.23}$$

$$\mathcal{T}[Y] = \int_{U(d)} dU \; (U\otimes U)Y(U^\dagger \otimes U^\dagger) \,, \tag{5.24}$$

the above conditions can be equivalently rewritten as

$$0 = Tr((\mathcal{I}_{12}\otimes \mathcal{T}_{34})[\Psi_D^+]M_{\text{diff}}) \,, \tag{5.25}$$

$$0 = Tr((\mathcal{I}_{12}\otimes \mathcal{A}_3 \otimes \mathcal{A}_4)[\Psi_D^+]M_{\text{same}}) \,, \tag{5.26}$$

because all the relevant operators are positive. The actions of the twirling channel $\mathcal{T}$ and the average channel $\mathcal{A}$ are derived in Appendix D. In particular,

$$\mathcal{A}[X] = Tr(X)\,\frac{1}{d}\,I_d \,, \tag{5.27}$$

$$\mathcal{T}[Y] = \frac{Tr(YP^{sym})}{d_{sym}}P^{sym} + \frac{Tr(YP^{asym})}{d_{asym}}P^{asym} \,, \tag{5.28}$$

where $P^{sym}$, $P^{asym}$ are projectors onto symmetric and antisymmetric subspace of $\mathcal{H}_d\otimes\mathcal{H}_d$, respectively. Dimensions of these subspaces are denoted as $d_{sym}$, $d_{asym}$ and read

$$\begin{aligned} d_{sym} &= Tr(P^{sym}) = d(d+1)/2 \\ d_{asym} &= Tr(P^{asym}) = d(d-1)/2. \end{aligned} \tag{5.29}$$

Let us note that $P^{sym} = \frac{1}{2}(I + \text{Swap})$, $P^{asym} = \frac{1}{2}(I - \text{Swap})$, where the swap operator acts as $\text{Swap}|\psi\rangle\otimes|\varphi\rangle = |\varphi\rangle\otimes|\psi\rangle$ for all $\psi,\varphi\in\mathcal{H}_d$. Using these expressions we obtain

$$(\mathcal{I}_{12}\otimes \mathcal{A}_3 \otimes \mathcal{A}_4)[\Psi_D^+] = \frac{1}{d^2} I_d^{\otimes 4} \tag{5.30}$$

and since

$$\begin{aligned} \mathcal{T}[|jm\rangle\langle kn|] &= \frac{1}{d_{sym}}\langle kn|P^{sym}|jm\rangle\; P^{sym} + \frac{1}{d_{asym}}\langle kn|P^{asym}|jm\rangle\; P^{asym} \\ &= \frac{\delta_{jk}\delta_{mn}+\delta_{jn}\delta_{mk}}{2d_{sym}}\; P^{sym} + \frac{\delta_{jk}\delta_{mn}-\delta_{jn}\delta_{mk}}{2d_{asym}}\; P^{asym} \end{aligned}$$

we have

$$\begin{aligned}
\omega_{\mathcal{T}} &= (\mathcal{I}_{12}\otimes\mathcal{T}_{34})[(\Psi_d^+)_{13}\otimes(\Psi_d^+)_{24}] \\
&= \sum_{j,k,m,n} |jm\rangle_{12}\langle kn|\otimes\mathcal{T}_{34}[|jm\rangle_{34}\langle kn|] \\
&= \frac{1}{2d_{sym}}\sum_{j,m}\Big[|jm\rangle\langle jm| + |jm\rangle\langle mj|\Big]\otimes P^{sym} \\
&\quad + \frac{1}{2d_{asym}}\sum_{j,m}[|jm\rangle\langle jm| - |jm\rangle\langle mj|]\otimes P^{asym} \\
&= \frac{1}{4d_{sym}}\sum_{j,m}[(|jm\rangle+|mj\rangle)(\langle jm|+\langle mj|)]\otimes P^{sym} \\
&\quad + \frac{1}{4d_{asym}}\sum_{j,m}[(|jm\rangle-|mj\rangle)(\langle jm|-\langle mj|)]\otimes P^{asym} \\
&= \frac{1}{d_{sym}}P^{sym}\otimes P^{sym} + \frac{1}{d_{asym}}P^{asym}\otimes P^{asym}\,.
\end{aligned}$$

Putting all formulas together the conditions in Eqs.(5.25),(5.26) take the form

$$0 = Tr(\omega_{\mathcal{T}} M_{\rm diff})\,, \tag{5.31}$$

$$0 = \frac{1}{d^2}Tr(I_d^{\otimes 4}M_{\rm same}) = Tr(M_{\rm same})\,. \tag{5.32}$$

Since $M_{\rm same}, M_{\rm diff}$ are positive operators it follows that $M_{\rm same} = O$ and $M_{\rm diff}$ has support in the orthocomplement of $\omega_{\mathcal{T}}$. Consequently, we can unambiguously conclude only that the unitary channels are different. We can formulate the following proposition.

**Proposition 2** *If a PPOVM $M_{\rm same}, M_{\rm diff}, M_0$ describes an unambiguous comparison of arbitrary unitary channels, then necessarily*

$$\begin{aligned}
{\rm supp}M_{\rm diff}\perp{\rm supp}\omega_{\mathcal{T}}\,; \qquad M_{\rm same} = O\,; \\
M_0 = \xi^T\otimes I_D - M_{\rm diff}\,,
\end{aligned} \tag{5.33}$$

*for some state $\xi$ on $\mathcal{H}_D \equiv \mathcal{H}_d\otimes\mathcal{H}_d$.*

### 5.3.3 Optimal unambiguous comparator

Following the previous section as a figure of merit for unambiguous comparators of unitary channels we shall use the average conditioned probability of revealing their difference

$$\begin{aligned}
\overline{p}_{\rm diff} &= \int_{U(d)\times U(d)} dU dV\, p_{\rm diff}(U\otimes V) \\
&= Tr((\mathcal{I}_{12}\otimes\mathcal{A}_3\otimes\mathcal{A}_4)[\Psi_D^+]M_{\rm diff}) \\
&= \frac{1}{d^2}Tr(M_{\rm diff})\,.
\end{aligned} \tag{5.34}$$

The overall average success probability $P_{succ}$ equals $(1-\eta_{\rm same})\overline{p}_{\rm diff}$, where $\eta_{\rm same}\neq 0$ is the prior probability for channels being the same. This prior is independent of the particular PPOVM $\{M_{\rm diff}, M_0\}$ and therefore we shall use only the conditional average probability to evaluate the quality of the unambiguous comparison strategy. Hence, our task is to maximize the *conditional success probability* $\overline{p}_{\rm success}\equiv\overline{p}_{\rm diff}$ by finding a positive operator $M_{\rm diff}$ defined on $\mathcal{H}_D\otimes\mathcal{H}_D$ together with a state $\xi$ on $\mathcal{H}_D$ such that also the operator $M_0=\xi^T\otimes I_D-M_{\rm diff}$ is positive. Before specifying the optimal solution let us prove the following upper bound on the (conditional) success probability.

**Theorem 4** *If a process POVM consisting of positive operators $M_{\rm diff}, M_0$ with normalization $M_{\rm diff}+M_0=\xi^T\otimes I_D$ unambiguously compares an arbitrary pair of unitary channels, then*

$$\overline{p}_{\rm success}\leq\frac{d+1}{2d}\,. \tag{5.35}$$

**Proof 4** *The validity of the no-error condition $Tr(\omega_{\mathcal{T}}M_{\rm diff})=0$ implies that supports of $M_{\rm diff}$ and $\omega_{\mathcal{T}}$ are orthogonal. Let us denote by $|s_1\rangle,\ldots,|s_{d_{sym}}\rangle$, $|a_1\rangle,\ldots,|a_{d_{asym}}\rangle$ the vectors forming orthonormal bases of symmetric and antisymmetric subspaces of $\mathcal{H}_d\otimes\mathcal{H}_d$, respectively. Then* $\mathrm{supp}\,\omega_{\mathcal{T}}=\mathrm{span}\{|s_j\otimes s_k\rangle,|a_m\otimes a_{\cdot n}\rangle\}$*, where $j,k=1,\ldots,d_{sym}$ and $m,n=1,\ldots,d_{asym}$, and because of the mentioned orthogonality*

$$\mathrm{supp}M_{\rm diff}\subset\mathrm{span}\{|s_j\otimes a_n\rangle,|a_m\otimes s_k\rangle\}\,. \tag{5.36}$$

*It follows that in a spectral form*

$$M_{\rm diff}=\sum_\alpha\lambda_\alpha|\phi_\alpha\rangle\langle\phi_\alpha|\,, \tag{5.37}$$

*where $0\leq\lambda_\alpha\leq 1$ and*

$$|\phi_\alpha\rangle=\sum_{nj}c^\alpha_{nj}|a_n\otimes s_j\rangle+d^\alpha_{jn}|s_j\otimes a_n\rangle\,. \tag{5.38}$$

*Consequently,*

$$M_{\rm diff}=\sum_n|a_n\rangle\langle a_n|\otimes A_n+\sum_n B_n\otimes|a_n\rangle\langle a_n|+R\,,$$

*with*

$$
\begin{aligned}
A_n &= \sum_\alpha \lambda_\alpha \sum_{jl} c^\alpha_{nj} \overline{c^\alpha_{nl}} |s_j\rangle\langle s_l| \,; \\
B_n &= \sum_\alpha \lambda_\alpha \sum_{jl} d^\alpha_{nj} \overline{d^\alpha_{nl}} |s_j\rangle\langle s_l| \,; \\
R &= \sum_\alpha \lambda_\alpha \Bigg[ \sum_{m\neq n,j,l} c^\alpha_{mj}\, \overline{c^\alpha_{nl}} |a_m \otimes s_j\rangle\langle a_n \otimes s_l| + \\
&\quad + \sum_{m\neq n,j,l} d^\alpha_{jm}\, \overline{d^\alpha_{ln}} |s_j \otimes a_m\rangle\langle s_l \otimes a_n| + \\
&\quad + \sum_{m,n,j,l} c^\alpha_{mj}\, \overline{d^\alpha_{ln}} |a_m \otimes s_j\rangle\langle s_l \otimes a_n| + \\
&\quad + \sum_{m,n,j,l} d^\alpha_{jm}\, \overline{c^\alpha_{nl}} |s_j \otimes a_m\rangle\langle a_n \otimes s_l| \Bigg] \,.
\end{aligned}
$$

*Since $Tr(R) = 0$ we get for the average success probability*

$$
\overline{p}_{\text{success}} = \frac{1}{d^2} \sum_{n=1}^{d_{asym}} \left( Tr(A_n) + Tr(B_n) \right) . \tag{5.39}
$$

*The operators $A_n, B_n$ have the form of positive sum of one-dimensional projectors, hence they are positive.*

*Let us evaluate the mean value of operator $M_0 = \xi^T \otimes I - M_{\text{diff}}$ in a pure state associated with the vector $|s_j \otimes a_n\rangle$. Due to the required positivity of $M_0$ we get the inequality*

$$
0 \leq \langle s_j \otimes a_n | M_0 | s_j \otimes a_n \rangle = \langle s_j | \xi^T - B_n | s_j \rangle \,. \tag{5.40}
$$

*Similarly, also the inequality*

$$
\begin{aligned}
0 &\leq \langle a_n \otimes s_j | M_0 | a_n \otimes s_j \rangle \\
&\leq \langle a_n | \xi^T | a_n \rangle - \langle s_j | A_n | s_j \rangle
\end{aligned} \tag{5.41}
$$

*holds. These two inequalities can be used to bound the trace of the density operator $\xi^T$ as follows*

$$
\begin{aligned}
Tr(\xi^T) &= \sum_n \langle a_n | \xi^T | a_n \rangle + \sum_j \langle s_j | \xi^T | s_j \rangle \\
&\geq \sum_n \langle s_k | A_n | s_k \rangle + \sum_j \langle s_j | B_m | s_j \rangle \\
&\geq \langle s_k | \sum_n A_n | s_k \rangle + Tr(B_m) \,,
\end{aligned} \tag{5.42}
$$

*where we used the fact that by definition operators $B_m$ have support only on the symmetric subspace. The inequality holds for all choices of $k$ and $m$. Moreover, since $Tr(\xi^T) = 1$ and $B_m$* *is positive, i.e.* $Tr(B_m) \geq 0$*, we obtain that also*

$$\langle s_k | \sum_n A_n | s_k \rangle \leq 1 \,. \tag{5.43}$$

*for all* $k$*. Using these inequalities the success probability can be upper bounded as follows*

$$\begin{aligned}
\overline{p}_{\text{success}} &= \frac{1}{d^2}\left(\sum_{j=1}^{d_{sym}} \langle s_j | \sum_{n=1}^{d_{asym}} A_n | s_j \rangle + \sum_{m=1}^{d_{asym}} Tr(B_m)\right) \\
&= \frac{1}{d^2}\left[\sum_{m=1}^{d_{asym}} \left(\langle s_m | \sum_{n=1}^{d_{asym}} A_n | s_m \rangle + Tr(B_m)\right) + \right. \\
&\qquad \left. + \sum_{j=d_{asym}+1}^{d_{sym}} \langle s_j | \sum_{n=1}^{d_{asym}} A_n | s_j \rangle \right] \\
&\leq \frac{1}{d^2}(d_{asym} + d) = \frac{d_{sym}}{d^2} = \frac{d+1}{2d} \,,
\end{aligned} \tag{5.44}$$

*which proves the theorem.*

*Antisymmetric test states*

In what follows we shall design a process POVM saturating the upper bound on the success probability. In particular, for operators

$$M_{\text{diff}} = \varrho^T \otimes P^{sym} \,, \qquad M_0 = \varrho^T \otimes P^{asym} \,. \tag{5.45}$$

the success probability equals

$$\overline{p}_{\text{success}} = \frac{1}{d^2} Tr(M_{\text{diff}}) = \frac{1}{d^2} Tr(\varrho^T \otimes P^{sym}) = \frac{d_{sym}}{d^2} \,, \tag{5.46}$$

hence the upper bound is saturated. Let us note that the state $\varrho$ is not arbitrary, because the support of $M_{\text{diff}}$ must be orthogonal to support of $\omega_{\mathcal{T}}$ (see Eq.(5.36)). It implies that the state $\varrho$ has support only on antisymmetric subspace. We shall call such states antisymmetric. Similarly, if the support of a state is only in symmetric subspace we denote it as symmetric state.

The form of PPOVM in Eq. (5.45) suggests that one possible experimental realization consists of the folowing steps: i) prepare a two-qudit antisymmetric state $\varrho$; ii) insert each qudit into different black box; iii) measure a two-valued observable described by POVM $F_{\text{diff}} = P^{sym}$ and $F_0 = P^{asym}$, which identifies the exchange symmetry of the joint state of the two-qudit system.

The test state $\varrho$ is antisymmetric. If $U = V$ the action of the channels preserves the symmetry, i.e. the output state remains antisymmetric and in such case $F_0$ must be observed. For $U \neq V$ the measurement outcome cannot be predicted with certainty, so both outcomes $F_{\text{diff}}, F_0$ have nonvanishing probability of occurence. However, if an outcome $F_{\text{diff}}$ is observed, we can unambiguously conclude that $U$ and $V$ are different.

*Symmetric test states*

Alternatively, we can consider a process POVM

$$M_{\rm diff} = \varrho^T \otimes P^{asym} \,, \qquad M_0 = \varrho^T \otimes P^{sym} \tag{5.47}$$

satisfying all the constraints providing $\varrho$ has support in the symmetric subspace. For this choice the success probability reads

$$\overline{p}_{\rm success} = Tr(\varrho^T \otimes P^{asym}) = \frac{d_{asym}}{d^2} = \frac{d-1}{2d} \,, \tag{5.48}$$

which is not optimal. Such PPOVM describes an experiment in which a "symmetric" test state is used. The same measurement is carried out as in the antisymmetric case, but the role of conclusive and inconclusive results is exchanged, i.e. $F_{\rm diff} = P^{asym}$ and $F_0 = P^{sym}$.

As we have mentioned at the beginning of this section one possibility how to tackle the problem of unambiguous comparison of unitary channels is to adopt the universal comparison machines for states. Consider a pair of unitary channels applied on independent systems initially prepared in the same state. If $U = V$, then the resulting states are still described by the same state. However, if $U \neq V$, then the output states can be different. That is the state comparator can be used to find out whether the output states are different, which means that the unitary channels are different as well. In the language of channel comparison the described strategy can be interpreted as a strategy with pure, symmetric and factorized test state $\varrho = |\varphi \otimes \varphi\rangle\langle\varphi \otimes \varphi|$. Since the optimal state comparison is based on projective measurement described by projectors $P^{sym}, P^{asym}$, the value of the success probability is given in Eq.(5.48).

*Uniqueness of optimal solution*

In previous paragraphs we have shown that optimal strategy for comparison of unitary channels saturates the upper bound on probability of success imposed by Theorem 4. It means that PPOVM elements of each optimal strategy have to saturate all inequalities used in proof of this theorem. Analyzing this fact we can characterize all optimal strategies.

**Theorem 5** *If a process POVM $\{M_{\rm diff}, M_0\}$ with normalization $\xi^T \otimes I_D$ unambiguously compares arbitrary pair of unitary channels with $\overline{p}_{\rm success} = \frac{d+1}{2d}$, then*

$$M_{\rm diff} = \xi^T \otimes P^{sym} \,, \qquad M_0 = \xi^T \otimes P^{asym} \,, \tag{5.49}$$

*where $\xi$ is a state with a support belonging only to the antisymmetric subspace of $\mathcal{H}_d \otimes \mathcal{H}_d$.*

**Proof 5** *Saturation of inequality (5.43) for $k = d_{sym}$ together with inequality (5.42) implies that $Tr(B_n) = 0$ for all $n$. Consequently, positivity of operators $B_n$ implies $B_n = 0$ for all $n$ i.e. coefficients $d^{\alpha}_{jn}$ vanish. This in turn requires*

$$\langle s_k| \sum_n A_n |s_k\rangle = 1 \tag{5.50}$$

*for all $k$. Using Eq. (5.41) and Eq. (5.50) we get*

$$1 = \sum_n \langle s_k|A_n|s_k\rangle \leq \sum_n \langle a_n|\xi^T|a_n\rangle \leq Tr(\xi^T) = 1 \,,$$

*thus,* $\sum_n \langle a_n|\xi^T|a_n\rangle = 1$*. Due to positivity of* $\xi^T$ *we obtain* $\langle s_j|\xi^T|s_j\rangle = 0$ *for all* $j$*. This tells us that* $\xi^T$ *has support only on antisymmetric states. Since the used transposition is defined with respect to a product basis, the antisymmetric states preserve their antisymmetry, i.e. the state* $\xi$ *is antisymmetric as it is stated in the theorem.*

*Using the spectral form (5.37) and Eq. (5.38) we can rewrite* $M_{\rm diff}$ *as:*

$$M_{\rm diff} = \sum_j C_j \otimes |s_j\rangle\langle s_j| + H\,, \tag{5.51}$$

*with*

$$\begin{aligned} C_j &= \sum_\alpha \lambda_\alpha \sum_{nm} c^\alpha_{nj}\overline{c^\alpha_{mj}}|a_n\rangle\langle a_m|\,; \\ H &= \sum_\alpha \lambda_\alpha \sum_{j\neq l,m,n} c^\alpha_{mj}\,\overline{c^\alpha_{nl}}|a_m\otimes s_j\rangle\langle a_n\otimes s_l| \end{aligned}$$

*We rewrite also the probability of success [Eq. (5.34)] in terms of* $C_j$ *and because the operator* $H$ *is traceless we get*

$$\overline{p}_{\rm success} = \frac{1}{d^2}\sum_j Tr(C_j). \tag{5.52}$$

*Positivity of* $M_0 = \xi^T\otimes I - \sum_j C_j\otimes|s_j\rangle\langle s_j| - H$ *implies*

$$0 \leq \langle a\otimes s_j|M_0|a\otimes s_j\rangle = \langle a|\xi^T - C_j|a\rangle\,, \tag{5.53}$$

*where* $|a\rangle$ *is arbitrary vector from* $\mathcal{H}_D$*. Hence, we have that operator* $\xi^T - C_j$ *is positive for all* $j$ *and consequently that* $1 = Tr(\xi^T) \geq Tr(C_j)$*. Saturation of inequality (5.35) requires* $Tr(C_j) = 1$ *for all* $j$*, which in turn implies* $Tr(\xi^T - C_j) = 0$*. This together with the positivity of operator* $\xi^T - C_j$ *enables us to conclude that* $C_j = \xi^T$ *for all* $j$*. The operators* $M_{\rm diff}$*,*$M_0$ *therefore read*

$$\begin{aligned} M_{\rm diff} &= \xi^T\otimes P^{sym} + H\,, \\ M_0 &= \xi^T\otimes P^{asym} - H\,. \end{aligned}$$

*The support of the selfadjoint operator* $H$ *is orthogonal to the support of the operator* $\xi^T\otimes P^{asym}$*. Since* $H$ *is traceless it has both positive and negative eigenvalues unless* $H = O$*. However, positive eigenvalues of* $H$ *would spoil positivity of* $M_0$*, so the operator* $H$ *must vanish, which concludes the proof.*

### 5.3.4 Conclusions

The goal of this section was to find an optimal strategy for comparison of two unknown unitary channels. Exploiting the framework of process POVM we have shown that the optimal strategy achieves the average conditional success probability $\overline{p}_{\rm success} = (d+1)/(2d)$. An interesting observation is that the optimal strategy for comparison of unitary channels is very closely related to the comparison of pure states. In fact, the optimal state comparison is based on the implementation of the two-valued projective measurement measuring the exchange symmetry of the bipartite states. Outcomes are associated with projectors onto symmetric and antisymmetric subspaces of the joint Hilbert space. The optimal procedure for the comparison of unitary channels is exploiting the same measurement, but the outcomes are interpreted in the opposite way. Whereas for comparison of pure states the projector $P^{sym}$ corresponds to the inconclusive result, for unitaries this projector is associated with the unambiguous conclusion that the channels are different. Similarly, the projector $P^{asym}$ indicates the difference of compared pure states, but corresponds to no conclusion for unitaries. In both cases, the unambiguous conclusion that the states, or unitaries are the same, can not be made.

Devices implementing quantum channels are tested indirectly via their action on quantum states. In the experiment the unknown apparatuses are probed by some test states. We have shown that the optimal solution is achieved if and only if the test state is antisymmetric, i.e. its support is only in antisymmetric subspace. Let us note that if a state is separable, then necessarily its support contains product vectors. However, by definition there is no antisymmetric product vector, hence the support of each antisymmetric state does not contain any product vectors. Consequently, each antisymmetric state is necessarily entangled. In conclusion, the entanglement is the key ingredient for comparison of unitary channels. It enhances the success probability to reach the optimal value.

Let us note that the proposed optimal strategy is feasible in current quantum information experiments with photons and ions. In particular, in the qubit version the experiment consists of preparation of a singlet, application of the unknown single-qubit unitary channels on individual qubits and a projective measurement consisting of the projection onto a singlet, or arbitrary other maximally entangled state. As the measurement we can use, for instance, the Bell measurement, but it is not necessary. Moreover, for the optimal comparison of qudit unitary channels mixed test states are allowed.

## 6 Unambiguous tasks for measurements

The measurement part of an experiment correlates the state of the quantum system with some classically distinguishable signals. The number of possible signals is in practice finite and observation of such a signal is called an *event* or an *outcome*. The rules of quantum mechanics tell us that a quantum state is not an observable quantity. This means that there is always more than one quantum state in the state space that could trigger the observed event. In general this prevents us from deducing the (original) state precisely out of a single measurement outcome. In fact, quantum mechanics predicts the probability of a state $\varrho$ leading to the considered event to be $Tr(\varrho E)$, where $O \leq E \leq I$ is a positive operator i.e. an *effect* associated with the event. If the set of states before the measurement is restricted so that for only one of them $Tr(\varrho E) > 0$ then the observation of effect $E$ unambiguously implies that the measured state was $\varrho$. Also in other than unambiguous approaches the outcomes of the measurement are most often used to infer something about the measured state. Moreover, we often tune the measurements to optimize this inference.

On the other hand, there are also situations, where we want to design the experiment in such a way that it enables us to infer something about the measurement itself. Hence, the measurement is partly unknown and fixed and we want to deduce some of its properties out of its outcomes. This type of experiments is a bit different from the experiments probing channels or states, because the outcomes of the investigated measurement apparatus may not be directly linked to the conclusions we want to make. As an illustration consider a discrimination among $M$ measurement apparatuses, each having $N$ possible outcomes. For $N < M$ a single outcome of the tested measurement could not indicate each of the possibly used apparatuses. However, consider the following strategy for a single use of the unknown measurement. It would employ a principal system measured by the tested apparatus as well as an ancillary quantum system, whose measurement depends on the outcome of the tested POVM. The outcomes of the measurement on the ancilla could then directly correspond to the conclusions we want to make. One can show that it suffices to consider an ancilla with the same Hilbert space as the principal quantum system. Unfortunately, a suitable mathematical framework for description of these type of experiments is not yet developed. Moreover, sometimes it may happen that we are not allowed to use any other measurement than the tested one. In such a situation the possible experiments consist of the preparation that we control and a tested measurement, whose outcomes can be linked to the test results in many ways.

### 6.1 Labeled vs. unlabeled observables

Let us now look on the description of measurement apparatuses from a bit more operational point of view. Imagine we are given a Stern-Gerlach apparatus, whose outcomes are labeled by 1 and 2. Suppose outcome 1 emerges when the measured spin is along the measurement direction. Thus, if the measurement direction is along the $+z$ axes[17] the effects associated with the outcomes read: $E_1 = |+z\rangle\langle +z|$, $E_2 = |-z\rangle\langle -z|$. However, if the measurement direction is along the $-z$ axes the effects read: $E_1 = |-z\rangle\langle -z|$, $E_2 = |+z\rangle\langle +z|$. Although, these two different POVMs correspond to two different Stern-Gerlach apparatuses mutually rotated by 180 degrees, they provide us with the same information. More precisely, after exchanging the labels

[17] In our $3D$ space or in the Bloch sphere representation if we are considering other two level system than spin

on one of the apparatuses, quantum mechanics predicts the same probability distribution on their outcomes for any possible input state. This illustrates that by a suitable labeling (interpretation) of the outcomes a single physical apparatus can realize several different POVMs. Thus, all the measurements related in this way are equivalent after the desired labeling of the outcomes is performed. This motivates us to consider two types of equivalence among the POVMs. If the desired labeling of the apparatus outcomes is done we may always assume that the outcomes are labeled by numbers $1, \ldots, n$. If the outcomes were not properly labeled yet, we assign a number $j \in J_n \equiv \{1, \ldots, n\}$ to each of them. However, in such a case the ambiguity in the labeling must be taken into account and the equivalence of observables should be compatible with this freedom. Let us spell out the definitions of observable equivalence explicitly.

**Definition 1** *Observables $\mathcal{A} : J_n \to \mathcal{E}(\mathcal{H})$ and $\mathcal{B} : J_n \to \mathcal{E}(\mathcal{H})$ are* identical *if $\mathcal{A}_j = \mathcal{B}_j$ for all $j$.*

**Definition 2** *Observables $\mathcal{A} : J_n \to \mathcal{E}(\mathcal{H})$ and $\mathcal{B} : J_n \to \mathcal{E}(\mathcal{H})$ are* equivalent *(in the unlabeled sense) if there exists a permutation $\pi : J_n \to J_n$ such that $\mathcal{A}_j = \mathcal{B}_{\pi(j)}$ for all $j$.*

Thus, we shall use the word *identical* for equivalence in the strict labeled sense and the word *equivalent* for equivalence in the unlabeled sense. It follows from the definition that the equivalence class of an unlabeled observable consists of POVMs with the same range, i.e. the elements of the set of unlabeled measurements can be understood as unordered collections of effects summing up to identity.

Equivalence in the unlabeled sense of Def. (2) guarantees that the same property of the states is measured. An example are two Stern-Gerlach apparatuses measuring a spin along the same unoriented axes. However, in order to get the correct interpretation of the measured data one should calibrate the apparatus to have the labels correctly attached. On the other hand, when two measurements are identical, they have the same probability distributions without any relabeling of the outcomes. For example, an impetuous use of an equivalent measurement apparatus instead of the identical one in some quantum circuit may change the circuit behavior dramatically.

### 6.2 Perfect discrimination of two single qubit observables

Little is known about the discrimination of measurement apparatuses. The first paper focusing on unambiguous tasks was published in 2006 by Ji, Feng, Duan, and Ying [67]. These authors investigate the perfect discrimination of projective measurements, whose outcomes are numbered. Their motivation was the following. It is known (see e.g. [68]) that perfect discrimination between two nonorthogonal states is impossible unless we have access to an infinite number of copies of the tested state. However, only a finite number of uses suffices to discriminate between any two unitary channels. Thus, it was very interesting to investigate, whether a finite number of uses of the measurement apparatus suffice to distinguish between any two projective measurements. The authors of [67] showed that a finite number of uses of the apparatus always suffice, although the testing scheme depends on the particular projective measurements. They define three schemes. The *simple scheme* consists of a preparation of a test state, whose parts are then measured by the tested measurement apparatus. The second scheme they define is *M-M scheme*. In this scheme, additional known measurements can be used to measure the test state. In their third *M-U-M scheme* (depicted in figure 6.1), the outcomes of some measurements can determine the unitary operation that is applied before the remaining measurements are performed. The authors prove three theorems stating the necessary and sufficient conditions for perfect discrimination of two projective measurements within the simple and the M-M scheme. They use them to show that for general qubit projective measurements, the M-M scheme is needed and they derive the minimal number of uses of the tested apparatus. For general projective measurements on more dimensional systems they show that a finite number of uses of the apparatus within the M-U-M scheme suffices for perfect discrimination.

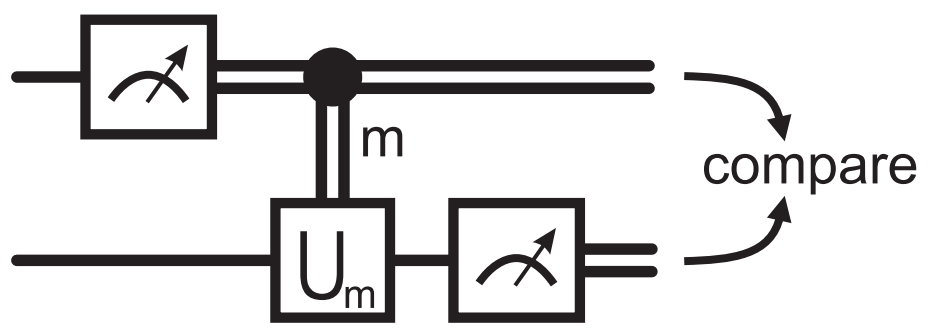

Figure 6.1. Illustration of the M-U-M scheme.

### 6.3 Unambiguous discrimination of unlabeled qubit observables

In contrast to the previous section, here we turn our attention to the discrimination of unlabeled observables. The first who considered this kind of problem were M. Ziman and T. Heinossari [69]. In their setting, the tested apparatus is modeled as a black box with leds. After a measurement is performed one of the leds is flashing to indicate the obtained outcome. The tested apparatus is known to be either the unlabeled measurement $\mathcal{A}$ or the unlabeled measurement $\mathcal{B}$. A single measurement outcome tells us that a support of the associated effect $E$ has overlap with the support of the measured state $\varrho$. However, in the unlabeled case this information does not help us to learn something about that particular effect. To illustrate this, we consider an unlabeled measurement described by effects $\{\mathcal{A}_1, \ldots, \mathcal{A}_n\}$, which form a particular POVM once the ordering is fixed. In fact, we have to label the leds, which is inevitably done in a random way. This causes that for each artificially named outcome the predicted probability is the same, i.e.

$$p_j(\mathcal{A}) = \frac{1}{n!} \sum_{\pi \in S(n)} Tr(\varrho \mathcal{A}_{\pi(j)}) = \frac{(n-1)!}{n!} \sum_{j'} Tr(\varrho \mathcal{A}_{j'}) = \frac{1}{n}\,, \tag{6.1}$$

where we used the fact that $n!$ is the total number of permutations on $J_n$, and that $(n-1)!$ is the number of them having a specific label $j'$ on the fixed ($j$th) position.

Using the apparatus once more, we can distinguish whether the observed outcomes coincide, or not. After fixing the labels $1, \ldots, n$ of the measurement device, the probability to observe a pair of outcomes $j, k$ reads

$$p_{jk}(\mathcal{A}) = \frac{1}{n!}\sum_{\pi\in S(n)} Tr(\varrho\mathcal{A}_{\pi(j)}\otimes\mathcal{A}_{\pi(k)}) \tag{6.2}$$
$$= \begin{cases} \frac{(n-1)!}{n!}\sum_{j'} Tr(\varrho\mathcal{A}_{j'}\otimes\mathcal{A}_{j'}) & \text{if} \quad j=k \\ \frac{(n-2)!}{n!}\sum_{j'\neq k'} Tr(\varrho\mathcal{A}_{j'}\otimes\mathcal{A}_{k'}) & \text{if} \quad j\neq k \end{cases},$$

where $(n-2)!$ is the number of permutations resulting in fixed operators $\mathcal{A}_{j'}, \mathcal{A}_{k'}$ for outcomes $j, k$. Let us note that the values of $p_{jk}$ do not depend on the particular values of $j, k$, but only on their relative relation (whether $j = k$, or $j \neq k$). Consequently, the probability to find the same/different outcomes in two shots reads

$$p_{\text{same}} = np_{jj} = \sum_j Tr(\varrho\mathcal{A}_j\otimes\mathcal{A}_j)\,,$$
$$p_{\text{diff}} = n(n-1)p_{jk} = \sum_{j\neq k} Tr(\varrho\mathcal{A}_j\otimes\mathcal{A}_k)\,.$$

We used the fact that for an $n$-valued measurement used twice, there are $n$ pairs of same outcomes and $n(n-1)$ pairs of different outcomes. In this two-shot scenario, the probabilities $p_{\text{same}}, p_{\text{diff}}$ depend on the particular properties of the effects $\mathcal{A}_1, \ldots, \mathcal{A}_n$, hence they contain some information about $\mathcal{A}$. Thus, with two uses of the unknown unlabeled apparatus we might be able to decide, whether it is apparatus $\mathcal{A}$ or $\mathcal{B}$. Ziman and Heinossari showed that for qubits, perfect discrimination is possible only if the measurements $\mathcal{A}$ and $\mathcal{B}$ correspond to Stern-Gerlach apparatuses oriented in mutually orthogonal directions. An example of this are the measurements of $\sigma_Z$ and $\sigma_X$ and all the other pairs can be obtained by their simultaneous rotation via a qubit unitary transformation.

Due to the freedom in the labeling of the outcomes the probabilities are equal for certain sequences of outcomes (see Eq. (6.2)). Hence, it is meaningful to distinguish only the symmetry of the sequence with respect to renumbering of the outcomes. Two outcomes can be either *same* $\leftrightarrow xx$ or *different* $\leftrightarrow xy$. For three measurement outcomes, there are five types of sequences: $xxx, xxy, xyx, xyy, xyz$. However, for more outcomes, the classification of the sequences becomes complicated.

Ziman and Heinossari concentrated on discrimination based on two measurement outcomes. If measurements $\mathcal{A}$ and $\mathcal{B}$ can not be perfectly discriminated, we need to allow the inconclusive result of the unambiguous discrimination. This means we need to drop one possible conclusion, because the relation of the two outcomes is only binary (same/different). Thus, the task Ziman and Heinossari studied might be called unambiguous detection of the unlabeled measurement $\mathcal{A}$ out of the measurements $\mathcal{A}, \mathcal{B}$. They showed that a qubit unlabeled measurement $\mathcal{A}$ can be unambiguously detected if and only if the measurement $\mathcal{B}$ is projective. Let us denote by $\eta_{\mathcal{A}}$, $\eta_{\mathcal{B}}$ the prior probability that the tested measurement is $\mathcal{A}$, $\mathcal{B}$, respectively. There are two optimal strategies for detecting the measurement $\mathcal{A}$. In one of them, *same* outcomes unambiguously indicate $\mathcal{A}$ and *diff* outcomes are inconclusive. The other optimal strategy uses a different probe state and the role of *same/diff* outcomes is exchanged. For sharp unlabeled measurements $\mathcal{A}$ and $\mathcal{B}$ defined by the unordered sets of effects $\{\frac{1}{2}(I+\vec{a}.\vec{\sigma}), \frac{1}{2}(I-\vec{a}.\vec{\sigma})\}$, $\{\frac{1}{2}(I+\vec{b}.\vec{\sigma}), \frac{1}{2}(I-\vec{b}.\vec{\sigma})\}$ ($\|a\| = \|b\| = 1$, $\vec{\sigma}$ - vector of sigma matrices) both optimal strategies succeed with probability $P_{succ} = \eta_{\mathcal{A}} \sin^2 \theta_{ab}$, where $\theta_{ab}$ is the angle between vectors $\vec{a}, \vec{b}$. The work of Ziman and Heinosaari defined the concept of unlabeled observable discrimination. They provided basic results for qubits, however there are many interesting open problems for qudits and for more than two uses of the tested apparatus.

### 6.4 Comparison of sharp qubit observables

Suppose that we are given a pair of experimental setups implementing qubit measurements, each of them designed by a different experimentalist. Is there a way to unambiguously compare their performance? Especially, are they same or different? As independent experimentalists we can think of these experimental setups as black boxes, producing outcomes after a qubit is inserted. Our conclusions then have to be based on the acquired measurement outcomes.

For quantum measurements, there are two natural variations of the comparison problem. First of all, we can ask whether the given black boxes are *identical*. This means that they produce the same measurement outcome statistics in any state. In particular, also the labeling of the outcomes is similar. For instance, two Stern-Gerlach apparatuses oriented in opposite directions are considered to be different in this strict sense. However, they can be made identical by simply re-labeling the outcomes in one of them. Thus, the other way to compare two black boxes is to ask whether they are *equivalent*, i.e., identical after suitable re-labeling of the outcomes.

As an example, suppose we are comparing whether two Stern-Gerlach apparatuses are identical. A singlet state of two qubits inserted into the measurements cannot lead to the same outcomes unless the measurement devices (including the labeling) are different. If labeling of the outcomes is not given or it is part of the comparison problem, then we can perform this singlet-based test for all possible labelings independently. Finding the same unambiguous conclusions in all of them leads to a conclusion also for measurements without apriori labels. Since for each of the Stern-Gerlach apparatuses we have two different choices of labels, we need to perform the singlet-based comparison four times, i.e. each of the apparatuses is used 4 times. We will show that there are also better strategies in which each of the unlabeled apparatuses is used only twice.

#### 6.4.1 Apriori information

From now on, we assume that the two compared measurement apparatuses are described by sharp non-degenerate observables, which we denote as $\mathcal{A}$ and $\mathcal{B}$. Otherwise the measurement apparatuses are completely unknown. This assumption represents a very important part of our apriori information. As such, these observables are in direct correspondence with orthonormal bases and have the same number of outcomes as the dimension of the Hilbert space ($n = d = 2$). Let us fix an orthonormal basis $|\psi_1\rangle, \ldots, |\psi_d\rangle$ and denote by $\mathcal{A}_j^U$ the projections onto vectors $U|\psi_j\rangle$, where $U$ is a unitary operator defined on $\mathcal{H}$. The projections $\mathcal{A}_1^U, \ldots, \mathcal{A}_d^U$ form a non-degenerate sharp observable $\mathcal{A}^U$. Moreover, every non-degenerate sharp observable is of the form $\mathcal{A}^U$ for some unitary operator $U$. As in any other comparison problem we shall assume that prior probabilities $\eta_{same}, \eta_{diff}$ of observables being same, different are both nonzero. Otherwise the comparison problem is meaningless and the conclusion is obvious from our prior knowledge. To properly define the problem we have to specify also the probability distribution in the subset of same observables and in the subset of different observables. The natural choice is to use to the Haar measure on the group of unitary operators that are in direct correspondence to sharp non-degenerate observables as we illustrated above. Thus, in the subset of same measurement apparatuses $\mathcal{A} = \mathcal{B} = \mathcal{A}^U$ the probability distribution equals $\eta_{same}dU$ and in the subset of different apparatuses $\mathcal{A} = \mathcal{A}^U \neq \mathcal{B} = \mathcal{A}^V$ the probability distribution reads $\eta_{diff}dUdV$. We shall use the following notation. For labeled observables the probability of observing outcomes $j, k$ on the two compared apparatuses will be denoted by $\overline{q}_{j,k}(\mathcal{A} = \mathcal{B})$, where the relation in the brackets indicates that the probability is conditioned on the measurement apparatuses being the same. According to our prior information we have

$$\begin{aligned}\overline{q}_{j,k}(\mathcal{A} = \mathcal{B}) &= \int Tr(\varrho \mathcal{A}_j^U \otimes \mathcal{A}_k^U)\, dU \\ &= \begin{cases} Tr(\varrho \int \psi \otimes \psi \;\, d\psi) & \text{if} \quad j = k \\ Tr(\varrho \int \psi \otimes \psi_\perp \; d\psi d\psi_\perp) & \text{if} \quad j \neq k \end{cases},\end{aligned} \tag{6.3}$$

where $d\psi_\perp$ denotes the integration over all vectors orthogonal to $\psi$. Similarly if the apparatuses are different the probability to observe outcomes $j, k$ is equal to

$$\begin{aligned}\overline{q}_{j,k}(\mathcal{A} \neq \mathcal{B}) &= \int Tr(\varrho \mathcal{A}_j^U \otimes \mathcal{A}_k^V)\, dUdV \\ &= Tr(\varrho \int \psi \, d\psi \otimes \int \varphi \, d\varphi) \\ &= Tr(\varrho \frac{1}{d} I \otimes \frac{1}{d} I) = \frac{1}{d^2}\end{aligned} \tag{6.4}$$

The overall average probability to observe outcomes $j, k$ on the two tested apparatuses is $\eta_{same}$ $\overline{q}_{j,k}(\mathcal{A} = \mathcal{B}) + \eta_{diff}\overline{q}_{j,k}(\mathcal{A} \neq \mathcal{B})$. The introduced notation can be easily extended to multiple outcomes of the compared apparatuses. For unlabeled observables we shall use analogous notation $\overline{p}_{j,k}(\mathcal{A} = \mathcal{B})$, $\overline{p}_{j,k}(\mathcal{A} \neq \mathcal{B})$, which differs only by using letter $p$ instead of $q$. However, as one might expect later we will show that for unlabeled observables we need to use each apparatus at least twice to be able to compare them.

### 6.4.2 Comparison of labeled observables

In this section we assume that the outcomes of the compared measurement apparatuses $\mathcal{A}$ and $\mathcal{B}$ are labeled by numbers $1, \ldots, d$. We study the simplest experimental scenario in which each of the apparatuses is used only once. Our goal is to find a test state $\varrho$ and divide the potential outcomes $(j, k)$ into three families associated with conclusions: i) observables are identical, ii) observables are different (not identical), iii) no conclusion (inconclusive result).

Using a pair of labeled measurements (each of them once) we distinguish $d^2$ different outcomes $(j, k)$ appearing with probabilities $\overline{q}_{jk}$ that depend on the equivalence of $\mathcal{A}$ and $\mathcal{B}$ (see equations (6.3),(6.4)). Our prior information causes that the probabilities $\overline{q}_{jk}(\mathcal{A} = \mathcal{B})$ and $\overline{q}_{jk}(\mathcal{A} \neq \mathcal{B})$ do not depend on particular values of $j, k$, but only on their mutual relation $j = k$, or $j \neq k$. That is, whatever test state is used, we can split the outcomes at most into two *outcome classes* $x \in \{\text{same}, \text{diff}\}$. Consequently, only two out of three conclusions can be made.

In general, conclusion $y \in \{identical, not\ identical\}$ based on observed outcome from the outcome class $x$ is unambiguous, if for all other alternatives $z$ , $z \neq y$ the conditional probability $p(y|x, z)$ of concluding $y$ vanishes. In order to conclude that the observables are different $(\mathcal{A} \neq \mathcal{B})$ the condition $\overline{q}_x(\mathcal{A} = \mathcal{B}) = 0$ must hold for some outcome class $x$. Similarly, if we can unambiguously conclude that $\mathcal{A} = \mathcal{B}$, then there must exist an outcome class $x'$ such that $\overline{q}_{x'}(\mathcal{A} \neq \mathcal{B}) = 0$. We refer to such conditions as the *no-error conditions*. Their validity is necessary to call a solution of the problem unambiguous. Outcomes that are not associated with unambiguous conclusions lead to an inconclusive result. Our task now is to show, which conclusion can be made unambiguously in an experiment involving single use of the compared measurement apparatuses. Let us note that

$$\int d\psi\, \psi^{\otimes k} = \frac{(d-1)!k!}{(d+k-1)!} P_{1\ldots k}^{sym} \equiv \frac{1}{d_k} P_{1\ldots k}^{sym}\,, \tag{6.5}$$

where $P_{1\ldots k}^{sym}$ is the projection onto the completely symmetric subspace of $\mathcal{H}^{\otimes k}$ and

$$d_k = Tr(P_{1\ldots k}^{sym}) = \frac{(d+k-1)!}{(d-1)!k!}$$

is the dimension of that subspace. For a fixed vector $\psi$

$$\begin{aligned} \int d\psi_\perp\, \psi_\perp^{\otimes k} &= \int_{\mathcal{H}_\psi^\perp} d\varphi\, \varphi^{\otimes k} \\ &= \frac{(d-2)!k!}{(d+k-2)!} (I-\psi)^{\otimes k} P_{1\ldots k}^{sym}\,, \end{aligned} \tag{6.6}$$

where we used $\mathcal{H}_\psi^\perp$ to denote the subspace of $\mathcal{H}$ orthogonal to $|\psi\rangle \in \mathcal{H}$.

We use these identities in the evaluation of the probabilities $\overline{q}_{jk}(\mathcal{A} = \mathcal{B})$ and $\overline{q}_{jk}(\mathcal{A} \neq \mathcal{B})$. In particular, from equations (6.3),(6.4) we have

$$\overline{q}_{jj}(\mathcal{A} = \mathcal{B}) = \frac{1}{d_2} Tr(\varrho P_{12}^{sym})\,, \tag{6.7}$$

$$\overline{q}_{jk}(\mathcal{A} = \mathcal{B}) = \frac{1}{d-1} Tr(\varrho(\frac{1}{d} I \otimes I - \frac{1}{d_2} P_{12}^{sym}))\,. \tag{6.8}$$

$$\overline{q}_{jj}(\mathcal{A} \neq \mathcal{B}) = \frac{1}{d^2} Tr(\varrho)\,, \tag{6.9}$$

$$\overline{q}_{jk}(\mathcal{A} \neq \mathcal{B}) = \frac{1}{d^2} Tr(\varrho)\,, \tag{6.10}$$

We see that if the measurement devices are different ($\mathcal{A} \neq \mathcal{B}$), then for all test states $\varrho$ the probabilities $\overline{q}_{jj}(\mathcal{A} \neq \mathcal{B})$ and $\overline{q}_{jk}(\mathcal{A} \neq \mathcal{B})$ do not vanish for any outcome. Because of that the identicality of the observables cannot be concluded unambiguously.

Using the relation $P_{12}^{sym} + P_{12}^{asym} = I \otimes I$ between the projectors onto the symmetric and antisymmetric subspace of $\mathcal{H} \otimes \mathcal{H}$ we can rewrite the operator

$$\frac{1}{d} I \otimes I - \frac{1}{d_2} P_{12}^{sym} = \frac{1}{d} P_{12}^{asym} + \frac{d-1}{d(d+1)} P_{12}^{sym}$$

in the spectral form. Since this is positive full-rank operator it follows that also $\overline{q}_{jk}(\mathcal{A} = \mathcal{B}) > 0$ for all test states. Therefore, the occurence of different outcomes cannot be used to unambiguously conclude that the measurements are different. However, $\overline{q}_{jj}(\mathcal{A} = \mathcal{B}) = 0$, if $\Pi_\varrho \leq P_{12}^{asym}$, where we denoted by $\Pi_X$ the projection onto the support of operator $X$. Hence, if we use test state $\varrho$ supported only on the antisymmetric subspace and observe the same outcomes then we can conclude with certainty that $\mathcal{A} \neq \mathcal{B}$.

In summary, the identicality of unknown sharp non-degenerate observables cannot be unambiguously confirmed if each of the labeled apparatuses is used only once. Using an antisymmetric test state $\varrho$ and observing the same outcomes on both apparatuses lead us to unambiguous conclusion that the apparatuses are different. For fixed observables $\mathcal{A} \neq \mathcal{B}$ the conditional probability of unambiguous conclusion reads

$$q_{\rm same}(\mathcal{A},\mathcal{B}) = \sum_j Tr(\varrho \mathcal{A}_j \otimes \mathcal{B}_j)\,. \tag{6.11}$$

On average

$$\overline{q}_{\rm same}(\mathcal{A} \neq \mathcal{B}) = \sum_{j=1}^{d} \overline{q}_{jj}(\mathcal{A} \neq \mathcal{B}) = d\frac{1}{d^2} = \frac{1}{d}\,.$$

This value gives the average conditional success probability for revealing the difference of the compared sharp labeled observables. It is independent of the used test state, however, the no-error conditions restrict the possible test states to antisymmetric states, i.e. to states supported only in the antisymmetric subspace spanned by $P_{12}^{asym}$. Let us stress that if we choose a test state $\varrho = \frac{1}{d_-} P^{asym}$, then $q_{\rm same}(\mathcal{A},\mathcal{B}) > 0$ whenever $\mathcal{A} \neq \mathcal{B}$.

### 6.4.3 Comparison of unlabeled measurements

In this section we assume that the outcomes of measurement apparatuses are not labeled. As previously, our goal is to design an experiment from which we are able to unambiguously conclude whether these apparatuses are same or not. But same now means that the observables are equivalent in the unlabeled sense.

Consider a fixed pair of unlabeled measurement apparatuses $\mathcal{A}$ and $\mathcal{B}$. A single usage of each of the apparatuses leads us to outcome $j$ on $\mathcal{A}$-apparatus and $a$ on $\mathcal{B}$-apparatus with probability

$$\begin{aligned} p_{j,a} &= \frac{1}{(d!)^2} \sum_{\pi,\pi' \in S(d)} Tr(\varrho \mathcal{A}_{\pi(j)} \otimes \mathcal{B}_{\pi'(a)}) \\ &= \frac{1}{d^2} \sum_{j',a'} Tr(\varrho \mathcal{A}_{j'} \otimes \mathcal{B}_{a'}) \\ &= \frac{1}{d^2} Tr(\varrho)\,, \end{aligned}$$

where we used that outcomes of each apparatus are labeled independently and we performed the summations in the same way as in equation (6.1). Since this probability is independent on whether $\mathcal{A} = \mathcal{B}$ or $\mathcal{A} \neq \mathcal{B}$ none of the outcomes can be used to make a conclusion. In fact $p_{j,a}$ is independent of particular observables at all. Hence, we need to use the unlabeled apparatuses more times. In particular, if each of them is used twice, then the independence of the labeling of the two apparatuses implies

$$
\begin{aligned}
p_{jk,ab} &= \frac{1}{d!d!}\sum_{\pi,\pi'} Tr(\varrho\mathcal{A}_{\pi(j)}\otimes\mathcal{A}_{\pi(k)}\otimes\mathcal{B}_{\pi'(a)}\otimes\mathcal{B}_{\pi'(b)}) \\
&= \begin{cases} \frac{1}{d^2}Tr(\varrho\mathcal{A}_{\text{same}}\otimes\mathcal{B}_{\text{same}}) & \text{if} \quad j=k, a=b \\ \frac{1}{d^2(d-1)}Tr(\varrho\mathcal{A}_{\text{same}}\otimes\mathcal{B}_{\text{diff}}) & \text{if} \quad j=k, a\neq b \\ \frac{1}{d^2(d-1)}Tr(\varrho\mathcal{A}_{\text{diff}}\otimes\mathcal{B}_{\text{same}}) & \text{if} \quad j\neq k, a=b \\ \frac{1}{d^2(d-1)^2}Tr(\varrho\mathcal{A}_{\text{diff}}\otimes\mathcal{B}_{\text{diff}}) & \text{if} \quad j\neq k, a\neq b\,, \end{cases}
\end{aligned}
\tag{6.12}
$$

where the summations are done analogously to equation (6.2) and we denoted

$$
\mathcal{A}_{\text{same}} \equiv \sum_j \mathcal{A}_j\otimes\mathcal{A}_j\,, \quad \mathcal{A}_{\text{diff}} \equiv \sum_{j\neq k}\mathcal{A}_j\otimes\mathcal{A}_k\,,
$$

similarly for $\mathcal{B}_{\text{same}}$ and $\mathcal{B}_{\text{diff}}$. We see that irrespectively whether $\mathcal{A}=\mathcal{B}$ or $\mathcal{A}\neq\mathcal{B}$ probability $p_{jk,ab}$ depends only on the mutual relation of the outcomes $j,k$ and $a,b$ of the two usages of the measurement $\mathcal{A}$ respectively $\mathcal{B}$. Hence, also for unknown $\mathcal{A}$ and $\mathcal{B}$ distributed according to our prior knowledge it is meaningful to distinguish at most four corresponding classes of outcomes.

Conditioned on measurement apparatuses being different ($\mathcal{A}\neq\mathcal{B}$) we shall calculate the average probability of observing outcomes from the four outcome classes. The probability to find the same outcomes on apparatus $\mathcal{A}$ and the same outcomes on apparatus $\mathcal{B}$, respectively, can be expressed as

$$
\begin{aligned}
\overline{p}_{\text{same,same}}(\mathcal{A}\neq\mathcal{B}) &= \sum_{j,a}\overline{p}_{jj,aa}(\mathcal{A}\neq\mathcal{B}) = \sum_{j,a}\int p_{jj,aa}(\mathcal{A}^U,\mathcal{A}^V)\,dUdV \\
&= d^2\int\frac{1}{d^2}Tr(\varrho\mathcal{A}^U_{\text{same}}\otimes\mathcal{A}^V_{\text{same}})\,dUdV \\
&= Tr(\varrho\mathcal{O}^{\mathcal{A}\neq\mathcal{B}}_{\text{same,same}})\,,
\end{aligned}
\tag{6.13}
$$

where by $p_{jj,aa}(\mathcal{A}^U,\mathcal{A}^V)$ we mean equation (6.12) with $\mathcal{A}=\mathcal{A}^U$,$\mathcal{B}=\mathcal{A}^V$. In equation (6.13) the factor $d^2$ stands for the number of same outcome pairs that can be observed on individual apparatuses. The operator $\mathcal{O}^{\mathcal{A}\neq\mathcal{B}}_{\text{same,same}}$ defined via this equation reads

$$
\begin{aligned}
\mathcal{O}^{\mathcal{A}\neq\mathcal{B}}_{\text{same,same}} &= \int dUdV\mathcal{A}^U_{\text{same}}\otimes\mathcal{A}^V_{\text{same}} \\
&= d^2\int d\psi d\varphi\,\psi\otimes\psi\otimes\varphi\otimes\varphi \\
&= d^2\overline{\mathcal{R}}_{\text{same}}\otimes\overline{\mathcal{R}}_{\text{same}}\,.
\end{aligned}
\tag{6.14}
$$

We used the definitions

$$
\begin{aligned}
\overline{\mathcal{R}}_{\text{same}} &= \int d\psi\psi\otimes\psi = \frac{1}{d_2}P^{sym}\,, \\
\overline{\mathcal{R}}_{\text{diff}} &= \int d\psi d\psi_\perp\psi\otimes\psi_\perp = \frac{1}{d}I - \frac{1}{d_2}P^{sym}\,.
\end{aligned}
$$

Similarly, for other outcomes we find that

$$\begin{aligned}
\mathcal{O}^{\mathcal{A}\neq\mathcal{B}}_{\text{diff,diff}} &= d^2(d-1)^2\overline{\mathcal{R}}_{\text{diff}}\otimes\overline{\mathcal{R}}_{\text{diff}} && (6.15)\\
\mathcal{O}^{\mathcal{A}\neq\mathcal{B}}_{\text{diff,same}} &= d^2(d-1)\overline{\mathcal{R}}_{\text{diff}}\otimes\overline{\mathcal{R}}_{\text{same}} && (6.16)\\
\mathcal{O}^{\mathcal{A}\neq\mathcal{B}}_{\text{same,diff}} &= d^2(d-1)\overline{\mathcal{R}}_{\text{same}}\otimes\overline{\mathcal{R}}_{\text{diff}}\ , && (6.17)
\end{aligned}$$

providing $\mathcal{A}\neq\mathcal{B}$. Let us define operators

$$\begin{aligned}
\Pi^{\mathcal{A}\neq\mathcal{B}}_{\text{same,same}} &= P^{sym}_{12}\otimes P^{sym}_{34} & \Pi^{\mathcal{A}\neq\mathcal{B}}_{\text{same,diff}} &= P^{sym}_{12}\otimes I_{34}\\
\Pi^{\mathcal{A}\neq\mathcal{B}}_{\text{diff,same}} &= I_{12}\otimes P^{sym}_{34} & \Pi^{\mathcal{A}\neq\mathcal{B}}_{\text{diff,diff}} &= I_{12}\otimes I_{34}\ ,
\end{aligned}$$

that project onto the supports of operators $\mathcal{O}^{\mathcal{A}\neq\mathcal{B}}_{\text{same,same}}$, $\mathcal{O}^{\mathcal{A}\neq\mathcal{B}}_{\text{same,diff}}$, $\mathcal{O}^{\mathcal{A}\neq\mathcal{B}}_{\text{diff,same}}$, $\mathcal{O}^{\mathcal{A}\neq\mathcal{B}}_{\text{diff,diff}}$, respectively.

Similarly, conditioned on measurement apparatuses being equivalent ($\mathcal{A}=\mathcal{B}$) we calculate the average probability of observing outcomes from the four outcome classes. For this purpose we define operators $\mathcal{O}^{\mathcal{A}=\mathcal{B}}_{\text{same,same}}$, $\mathcal{O}^{\mathcal{A}=\mathcal{B}}_{\text{same,diff}}$, $\mathcal{O}^{\mathcal{A}=\mathcal{B}}_{\text{diff,same}}$, $\mathcal{O}^{\mathcal{A}=\mathcal{B}}_{\text{diff,diff}}$ in analogous way to equation (6.13):

$$\overline{p}_{\text{same,same}}(\mathcal{A}=\mathcal{B}) = Tr(\varrho\mathcal{O}^{\mathcal{A}=\mathcal{B}}_{\text{same,same}}) = \sum_{j,a}\int p_{jj,aa}(\mathcal{A}^U,\mathcal{A}^U)\,dU\ ,$$

where

$$\begin{aligned}
\mathcal{O}^{\mathcal{A}=\mathcal{B}}_{\text{same,same}} &= d^2\int dU\frac{1}{d^2}\mathcal{A}^U_{\text{same}}\otimes\mathcal{A}^U_{\text{same}}\\
&= d\int d\psi\psi\otimes\psi\otimes\psi\otimes\psi\\
&\quad +d(d-1)\int d\psi d\psi_\perp\,\psi\otimes\psi\otimes\psi_\perp\otimes\psi_\perp && (6.18)
\end{aligned}$$

and in the second term of equation (6.18) the integration over $d\psi_\perp$ runs over all vectors orthog-onal to a fixed $\psi$.In a general case the operators $\mathcal{O}_{x,x'}^{\mathcal{A}=\mathcal{B}} = \int dU \mathcal{A}_x^U \otimes \mathcal{A}_{x'}^U$ read

$$\begin{aligned}
\mathcal{O}_{\text{same,diff}}^{\mathcal{A}=\mathcal{B}} &= d(d-1)\int \psi\otimes\psi\otimes[\psi\otimes\psi_\perp+\psi_\perp\otimes\psi] \\
&\quad+\frac{d!}{(d-3)!}\int \psi\otimes\psi\otimes\psi'\otimes\psi'_\perp\,, \\
\mathcal{O}_{\text{diff,same}}^{\mathcal{A}=\mathcal{B}} &= d(d-1)\int \psi\otimes\psi_\perp\otimes[\psi\otimes\psi+\psi_\perp\otimes\psi_\perp] \\
&\quad+\frac{d!}{(d-3)!}\int \psi'\otimes\psi'_\perp\otimes\psi\otimes\psi\,, \\
\mathcal{O}_{\text{diff,diff}}^{\mathcal{A}=\mathcal{B}} &= d(d-1)\int \psi\otimes\psi_\perp\otimes[\psi\otimes\psi_\perp+\psi_\perp\otimes\psi] \\
&\quad+\frac{d!}{(d-3)!}\int \psi\otimes\psi_\perp\otimes[\psi\otimes\psi'+\psi'\otimes\psi] \\
&\quad+\frac{d!}{(d-3)!}\int \psi\otimes\psi_\perp\otimes[\psi_\perp\otimes\psi'+\psi'\otimes\psi_\perp] \\
&\quad+\frac{d!}{(d-4)!}\int \psi\otimes\psi_\perp\otimes\psi'\otimes\psi'_\perp
\end{aligned} \tag{6.19}$$

where for simplicity we do not write explicitly the Haar measures $d\psi$, $d\psi'$, $d\psi_\perp$, $d\psi'_\perp$ and $\psi',\psi'_\perp$ are vectors orthogonal to $\psi$ and $\psi_\perp$. Of course, $\langle\psi|\psi_\perp\rangle = \langle\psi'|\psi'_\perp\rangle = 0$. Since for qubits the Hilbert space is two dimensional the terms containing $\psi'$ or $\psi'_\perp$ do not appear in these expressions. There are no two orthogonal vectors to a fixed $\psi$ in such case.

Let us note that the integration leading to $\mathcal{O}_{x,x'}^{\mathcal{A}\neq\mathcal{B}}$ includes the integration covered in $\mathcal{O}_{x,x'}^{\mathcal{A}=\mathcal{B}}$. Therefore,

$$\Pi_{x,x'}^{\mathcal{A}=\mathcal{B}} \leq \Pi_{x,x'}^{\mathcal{A}\neq\mathcal{B}}\,, \tag{6.20}$$

which implies that whenever $\overline{p}_{x,x'}(\mathcal{A}\neq\mathcal{B}) = Tr(\varrho\mathcal{O}_{x,x'}^{\mathcal{A}\neq B}) = 0$, then also $\overline{p}_{x,x'}(\mathcal{A}=\mathcal{B}) = Tr(\varrho\mathcal{O}_{x,x'}^{\mathcal{A}=\mathcal{B}}) = 0$, hence, in two shots we can not unambiguously conclude that the apparatuses are equivalent. We can only approve the difference of the measurement devices.

In what follows we are going to specify for which test states and for which outcomes classes $x,x' \in \{\text{same},\text{diff}\}$ the no-error conditions $Tr(\varrho\mathcal{O}_{x,x'}^{\mathcal{A}=\mathcal{B}}) = 0$ are satisfied and simultaneuously, whether the associated conditional success probability rates $\overline{p}_{\text{success}} = \overline{p}_{x,x'}(\mathcal{A}\neq\mathcal{B}) = Tr(\varrho\mathcal{O}_{x,x'}^{\mathcal{A}\neq\mathcal{B}}) > 0$ are nonvanishing. We shall show that for qubits ($d=2$)

$$\begin{aligned}
\Pi_{\text{same,same}}^{\mathcal{A}\neq\mathcal{B}} &= \Pi_{\text{same,same}}^{\mathcal{A}=\mathcal{B}} + Q_{\text{same,same}}\,, \\
\Pi_{\text{same,diff}}^{\mathcal{A}\neq\mathcal{B}} &= \Pi_{\text{same,diff}}^{\mathcal{A}=\mathcal{B}} + Q_{\text{same,diff}}\,, \\
\Pi_{\text{diff,same}}^{\mathcal{A}\neq\mathcal{B}} &= \Pi_{\text{diff,same}}^{\mathcal{A}=\mathcal{B}} + Q_{\text{diff,same}}\,, \\
\Pi_{\text{diff,diff}}^{\mathcal{A}\neq\mathcal{B}} &= \Pi_{\text{diff,diff}}^{\mathcal{A}=\mathcal{B}} + Q_{\text{diff,diff}}\,,
\end{aligned}$$

where $Q_{\text{same,same}} = O$, $Q_{\text{diff,diff}} \neq Q_{\text{same,diff}} = Q_{\text{diff,same}}$ are projections forming the relevant parts of the supports of potential test states $\varrho$ enabling us to conclude the difference. That is, we shall see that three out of four outcomes classes can be used to make the unambiguous conclusion.

*Support of operator* $\mathcal{O}^{\mathcal{A}=\mathcal{B}}_{\rm same,same}$

Evaluating the operator $\mathcal{O}^{\mathcal{A}=\mathcal{B}}_{\rm same,same}$ we obtain

$$\begin{aligned}\frac{1}{d}\mathcal{O}^{\mathcal{A}=\mathcal{B}}_{\rm same,same} &= \int \psi^{\otimes 4} + (d-1)\int \psi^{\otimes 2}\otimes\psi_\perp^{\otimes 2}\\ &= \frac{1}{d_4}P^{sym}_{1234} + \frac{2(d-1)}{d(d-1)}R_{12-34}P^{sym}_{34}\,,\end{aligned} \tag{6.21}$$

where

$$\begin{aligned}R_{12-34} &= \int \psi^{\otimes 2}\otimes(I-\psi)^{\otimes 2}\\ &= \frac{1}{d_2}P^{sym}_{12} + \frac{1}{d_4}P^{sym}_{1234} - \frac{1}{d_3}(P^{sym}_{123} + P^{sym}_{124})\,.\end{aligned}$$

Due to positivity of operators in Eq. (6.21) the unambiguous no-error conditions require that

$$Tr(\varrho P^{sym}_{1234}) = 0\,, \quad Tr(\varrho R_{12-34}P^{sym}_{34}) = 0\,,$$

hold simultaneously. Hence, the support of $R_{12-34}P^{sym}_{34}$ is of interest for us and in particular we should decide whether it is different from $\Pi^{\mathcal{A}\neq\mathcal{B}}_{\rm same,same} = P^{sym}_{12}\otimes P^{sym}_{34}$. If yes, then we can use this outcome for making the unambiguous conclusion.

Let us analyze properties of $R_{12-34}$ and its terms. First of all by definition $R_{12-34}P^{sym}_{34}$ is a positive operator, hence necessarily $[R_{12-34}, P^{sym}_{34}] = 0$ and also $[P^{sym}_{123} + P^{sym}_{124}, P^{sym}_{34}] = 0$. The support of the projections $P^{sym}_{12}$, $P^{sym}_{1234}$, $P^{sym}_{123}$, and $P^{sym}_{124}$ contains the completely symmetric subspace spanned by $P^{sym}_{1234}$. As it is shown in Appendix E it is their greatest joint subspace and since $\frac{1}{d_2} + \frac{1}{d_4} - \frac{2}{d_3} > 0$ the operator $R_{12-34}$ is indeed supported on the whole $P^{sym}_{1234}$.

It remains to analyze the properties of $R_{12-34}P^{sym}_{34}$ on the subspace $Q^+_{12} = P^{sym}_{12}\otimes P^{sym}_{34} - P^{sym}_{1234}$. In particular, we are interested whether

$$\langle\varphi|\frac{1}{d_2}Q^+_{12} - \frac{1}{d_3}(Q_{123} + Q_{124})|\varphi\rangle > 0 \tag{6.22}$$

for all $|\varphi\rangle$ from the support of $Q^+_{12}$, where $Q_{123} = P^{sym}_{123} - P^{sym}_{1234}$, $Q_{124} = P^{sym}_{124} - P^{sym}_{1234}$. For qubits these subspaces are described in details in Appendix E.1, where it is shown that the operator $Q_{123} + Q_{124}$ has two nonzero eigenvalues $4/3$ and $2/3$. However, the eigenvectors associated with $4/3$ are from the subspace spanned by $P^{sym}_{12}\otimes P^{asym}_{34}$, which is irrelevant due to multiplication of $R_{12-34}$ by $P^{sym}_{34}$. The eigenvectors associated with the eigenvalue $2/3$ are from $P^{sym}_{12}\otimes P^{sym}_{34}$, thus $\langle\varphi|Q_{123} + Q_{124}|\varphi\rangle \leq 2/3$ for all $|\varphi\rangle \in P^{sym}_{12}\otimes P^{sym}_{34} \geq Q^+_{12}$. Since $d_2 = 3$, $d_3 = 4$ (see equation (6.5))

$$\langle\varphi|\frac{1}{3}Q^+_{12} - \frac{1}{4}(Q_{123} + Q_{124})|\varphi\rangle \geq \frac{1}{3} - \frac{1}{6} > 0\,. \tag{6.23}$$

As a result we have shown that support of $R_{12-34}P^{sym}_{34}$ equals to support of $P^{sym}_{12}\otimes P^{sym}_{34}$, thus $\Pi^{\mathcal{A}=\mathcal{B}}_{\rm same,same} = P^{sym}_{12}\otimes P^{sym}_{34} = \Pi^{\mathcal{A}\neq\mathcal{B}}_{\rm same,same}$. In summary, an observation of pairs of same outcomes on both apparatuses cannot be used to make any unambiguous conclusion, because $Q_{\rm same,same} = O$.

*Support of operator* $\mathcal{O}^{\mathcal{A}=\mathcal{B}}_{\rm diff,diff}$

In this case our aim is to show that $Q_{\rm diff,diff} \neq O$. For qubits there are at most two mutually orthogonal vectors, hence

$$\mathcal{O}^{\mathcal{A}=\mathcal{B}}_{\rm diff,diff} = d(d-1)\int \psi\otimes\psi_\perp\otimes(\psi\otimes\psi_\perp+\psi_\perp\otimes\psi)\,.$$

Let us remind that for larger systems, this expression contains additional terms. Using the operators $R_{13-24}$, $R_{14-23}$ introduced in a similar way as $R_{12-34}$ defined in the previous section we obtain

$$\mathcal{O}^{\mathcal{A}=\mathcal{B}}_{\rm diff,diff} = 2(R_{13-24}P^{sym}_{24} + R_{14-23}P^{sym}_{23}). \tag{6.24}$$

Using the same arguments as for $R_{12-34}$ we find that $R_{13-24}P^{sym}_{24}$ is supported on $P^{sym}_{13}\otimes P^{sym}_{24}$ and $R_{14-23}P^{sym}_{23}$ is supported on $P^{sym}_{14}\otimes P^{sym}_{23}$. Therefore, for the test state $\varrho$ we can write the following no-error condition

$$0 = Tr(\varrho(P^{sym}_{13}\otimes P^{sym}_{24} + P^{sym}_{14}\otimes P^{sym}_{23}))\,. \tag{6.25}$$

The completely symmetric subspace $P^{sym}_{1234}$ is the greatest joint subspace of $P^{sym}_{13}\otimes P^{sym}_{24}$ and $P^{sym}_{14}\otimes P^{sym}_{23}$. According to Appendix E.2 the support of $P^{sym}_{13}\otimes P^{sym}_{24} + P^{sym}_{14}\otimes P^{sym}_{23}$ is 13 dimensional, because $d_4 = 5$ and $Q^+_{13} = P^{sym}_{13}\otimes P^{sym}_{24} - P^{sym}_{1234}$ and $Q^+_{14} = P^{sym}_{14}\otimes P^{sym}_{23} - P^{sym}_{1234}$ are both four dimensional. Since the total Hilbert space $\mathcal{H}^{\otimes 4}$ for qubits is 16-dimensional, it follows that test states satisfying the no-error conditions live in a three-dimensional subspace. In Appendix E.2 it is shown that this subspace is a linear span of vectors

$$\begin{aligned}
|\kappa_1\rangle &= \frac{1}{\sqrt{2}}(|00\rangle|\psi^+\rangle - |\psi^+\rangle|00\rangle)\,,\\
|\kappa_2\rangle &= \frac{1}{\sqrt{2}}(|0011\rangle - |1100\rangle)\,,\\
|\kappa_3\rangle &= \frac{1}{\sqrt{2}}(|11\rangle|\psi^+\rangle - |\psi^+\rangle|11\rangle)\,,
\end{aligned}$$

where $|\psi^+\rangle = \frac{1}{\sqrt{2}}(|01\rangle + |10\rangle)$. Thus, $Q_{\rm diff,diff} = \sum_j |\kappa_j\rangle\langle\kappa_j| \leq Q^+_{12} \leq P^{sym}_{12}\otimes P^{sym}_{34}$ and arbitrary test state $\varrho \leq Q_{\rm diff,diff}$ satisfies the no-error condition.

Let us optimize the conditional probability

$$\overline{p}_{\rm diff,diff}(\mathcal{A}\neq\mathcal{B}) = Tr(\varrho\mathcal{O}^{\mathcal{A}\neq\mathcal{B}}_{\rm diff,diff}) \tag{6.26}$$

where

$$\begin{aligned}
\mathcal{O}^{\mathcal{A}\neq\mathcal{B}}_{\rm diff,diff} &= 4(\frac{1}{2}I - \frac{1}{3}P^{sym}_{12})\otimes(\frac{1}{2}I - \frac{1}{3}P^{sym}_{34})\\
&= I - \frac{2}{3}(P^{sym}_{12} + P^{sym}_{34}) + \frac{4}{9}P^{sym}_{12}\otimes P^{sym}_{34}\,.
\end{aligned}$$

Arbitrary pure state $|\varphi\rangle \in Q_{\rm diff,diff}$ is an eigenvector of projections $P_{12}^{sym}$, $P_{34}^{sym}$ and $P_{12}^{sym} \otimes P_{34}^{sym}$. Therefore, the probability is independent of the test states $\varrho \leq Q_{\rm diff,diff}$ and reads

$$\overline{p}_{\rm diff,diff}(\mathcal{A} \neq \mathcal{B}) = 1 - \frac{4}{3} + \frac{4}{9} = \frac{1}{9}\,. \tag{6.27}$$

*Support of operator* $\mathcal{O}_{\rm same,diff}^{\mathcal{A}=\mathcal{B}}$

For qubits

$$\begin{aligned}\mathcal{O}_{\rm same,diff}^{\mathcal{A}=\mathcal{B}} &= d(d-1)\int \psi^{\otimes 2} \otimes (\psi \otimes \psi_\perp + \psi_\perp \otimes \psi) \\ &= d\left(\frac{1}{d_3}(P_{123}^{sym} + P_{124}^{sym}) - \frac{2}{d_4}P_{1234}^{sym}\right)\,,\end{aligned}$$

and since $P_{1234}^{sym} \leq P_{123}^{sym}, P_{124}^{sym}$; $1/d_3 > 1/d_4$ we can conclude that the no-error condition reads

$$Tr(\varrho(P_{123}^{sym} + P_{124}^{sym})) = 0\,. \tag{6.28}$$

Let us remind that $\Pi_{\rm same,diff}^{\mathcal{A}\neq\mathcal{B}} = P_{12}^{sym}$ and $P_{123}^{sym}, P_{124}^{sym} \leq P_{12}^{sym}$. The question is whether $\Pi_{\rm same,diff}^{\mathcal{A}=\mathcal{B}} = P_{12}^{sym}$, or not. We know (see Appendix E.1) that $P_{123}^{sym}, P_{124}^{sym}$ are not orthogonal, however, their greatest joint subspace is the completely symmetric one. The dimension of $P_{12}^{sym}$ is 12, whereas the total support of $P_{123}^{sym} + P_{124}^{sym}$ is 11 dimensional. It follows that there exist a unique vector such that $\Pi_{\rm same,diff}^{\mathcal{A}\neq\mathcal{B}}|\varphi_Q\rangle = |\varphi_Q\rangle$, and, simultaneuously, $\Pi_{\rm same,diff}^{\mathcal{A}=\mathcal{B}}|\varphi_Q\rangle = 0$, thus, $Q_{\rm same,diff} = |\varphi_Q\rangle\langle\varphi_Q|$. For such test state the observation of outcomes from $(same, diff)$ class leads to unambiguous confirmation of the difference of the measurement apparatuses.

*Support of operator* $\mathcal{O}_{\rm diff,same}^{\mathcal{A}=\mathcal{B}}$

There is no substantial difference in the analysis of this case and the previous one. We only need to exchange the role of pairs of indexes 12 and 34. Therefore, there exists a unique vector $|\varphi'_Q\rangle$ such that $\Pi_{\rm diff,same}^{\mathcal{A}=\mathcal{B}}|\varphi'_Q\rangle = 0$, but $\Pi_{\rm diff,same}^{\mathcal{A}\neq\mathcal{B}}|\varphi'_Q\rangle = P_{34}^{sym}|\varphi'_Q\rangle = |\varphi'_Q\rangle$. Surprisingly, we shall see that $|\varphi'_Q\rangle \equiv |\varphi_Q\rangle$, which means that the same test state $|\varrho_Q\rangle$ guarantees the unambiguity of both outcomes $\mathcal{O}_{\rm same,diff}, \mathcal{O}_{\rm diff,same}$.

On the systems $j$ and $k$ we define a singlet vector as $|\psi_{jk}^-\rangle = \frac{1}{\sqrt{2}}(|01\rangle_{jk} - |10\rangle_{jk})$. After a short calculation one can verify that the vector

$$|\varphi_Q\rangle = \frac{1}{\sqrt{3}}(|\psi_{13}^- \otimes \psi_{24}^-\rangle + |\psi_{14}^- \otimes \psi_{23}^-\rangle) \tag{6.29}$$

satisfies all the required properties, i.e. it is symmetric with respect to $1 \leftrightarrow 2$, $3 \leftrightarrow 4$ exchanges, i.e. $\Pi_{\rm same,diff}^{\mathcal{A}\neq\mathcal{B}}|\varphi_Q\rangle = \Pi_{\rm diff,same}^{\mathcal{A}\neq\mathcal{B}}|\varphi_Q\rangle = |\varphi_Q\rangle$, and $P_{123}^{sym}|\varphi_Q\rangle = P_{124}^{sym}|\varphi_Q\rangle = P_{134}^{sym}|\varphi_Q\rangle = P_{234}^{sym}|\varphi_Q\rangle = 0$, because both terms of $|\varphi_Q\rangle$ are antisymmetric exactly in one pair of all considered triples of indexes.

Using $|\varphi_Q\rangle$ as the test state we get

$$\begin{aligned}\overline{p}_{\rm same,diff}(\mathcal{A} \neq \mathcal{B}) &= \langle\varphi_Q|\mathcal{O}_{\rm same,diff}^{\mathcal{A}\neq\mathcal{B}}|\varphi_Q\rangle \\ &= \frac{4}{3}\langle\varphi_Q|\frac{1}{6}P_{12}^{sym} \otimes P_{34}^{sym} + \frac{1}{2}P_{12}^{sym} \otimes P_{34}^{asym}|\varphi_Q\rangle\,.\end{aligned}$$

Similarly, we find

$$p_{\text{diff,same}}(\mathcal{A}\neq\mathcal{B}) \;=\; \frac{4}{3}\langle\varphi_Q|\frac{1}{6}P_{12}^{sym}\otimes P_{34}^{sym}+\frac{1}{2}P_{12}^{asym}\otimes P_{34}^{sym}|\varphi_Q\rangle\,.$$

Since $P_{12}^{sym}|\varphi_Q\rangle = P_{34}^{sym}|\varphi_Q\rangle = |\varphi_Q\rangle$ implies $P_{12}^{sym}\otimes P_{34}^{sym}|\varphi_Q\rangle = |\varphi_Q\rangle$ and

$$\langle\varphi_Q|P_{12}^{asym}\otimes P_{34}^{sym}|\varphi_Q\rangle = \langle\varphi_Q|P_{12}^{sym}\otimes P_{34}^{asym}|\varphi_Q\rangle = 0\,,$$

we obtain

$$\overline{p}_{\text{same,diff}}(\mathcal{A}\neq\mathcal{B}) + \overline{p}_{\text{diff,same}}(\mathcal{A}\neq\mathcal{B}) = \frac{4}{9}\,. \tag{6.30}$$

This gives a better success rate than we achieved for the outcome class $(diff, diff)$. Unfortunately, $|\varphi_Q\rangle \notin Q_{\text{diff,diff}}$. In conclusion, $\overline{p} = 4/9$ is the optimal value of the conditional success probability $\overline{p}_{success}$ for unambiguous comparison of unlabeled sharp qubit observables in two shots.

Consider a fixed pair of unlabeled observables $\mathcal{A} = \{\psi,\psi_\perp\}$, $\mathcal{B} = \{\varphi,\varphi_\perp\}$ such that $\psi\neq\varphi$. Then the projections

$$\begin{aligned}\mathcal{O}_{\text{diff,same}} &= (\psi\otimes\psi_\perp+\psi_\perp\otimes\psi)\otimes(\varphi\otimes\varphi+\varphi_\perp\otimes\varphi_\perp)\,,\\ \mathcal{O}_{\text{same,diff}} &= (\psi\otimes\psi+\psi_\perp\otimes\psi_\perp)\otimes(\varphi\otimes\varphi_\perp+\varphi_\perp\otimes\varphi)\end{aligned}$$

determine the probability of observing outcomes from the corresponding measurement class via the relation $p_{\text{diff,same}} = Tr(\varrho\mathcal{O}_{\text{diff,same}})$. The success probability of revealing the difference of the observables using the test state $|\varphi_Q\rangle$ reads

$$p_{\text{success}}(\psi,\varphi) = \langle\varphi_Q|\mathcal{O}_{\text{same,diff}}+\mathcal{O}_{\text{diff,same}}|\varphi_Q\rangle\,. \tag{6.31}$$

Let us note that in a fixed orthonormal basis $|\psi\rangle, |\psi_\perp\rangle$ the test state $|\varphi_Q\rangle$ takes the form

$$|\varphi_Q\rangle = \frac{1}{\sqrt{3}}\left(|\psi^{\otimes 2}\otimes\psi_\perp^{\otimes 2}\rangle + |\psi_\perp^{\otimes 2}\otimes\psi^{\otimes 2}\rangle - |\psi^+\otimes\psi^+\rangle\right)\,,$$

where $|\psi^+\rangle = \frac{1}{\sqrt{2}}(|\psi\otimes\psi_\perp\rangle + |\psi_\perp\otimes\psi\rangle)$. Using the identities $|\langle\psi|\varphi\rangle| = |\langle\psi_\perp|\varphi_\perp\rangle| = \cos\theta$, $|\langle\psi|\varphi_\perp\rangle| = |\langle\psi_\perp|\varphi\rangle| = \sin\theta$ a direct calculation gives

$$\begin{aligned}\langle\varphi_Q|\mathcal{O}_{\text{same,diff}}|\varphi_Q\rangle &= \frac{1}{3}\langle\psi_\perp^{\otimes 2}|\varphi\otimes\varphi_\perp+\varphi_\perp\otimes\varphi|\psi_\perp^{\otimes 2}\rangle\\ &\quad+\frac{1}{3}\langle\psi^{\otimes 2}|\varphi\otimes\varphi_\perp+\varphi_\perp\otimes\varphi|\psi^{\otimes 2}\rangle\\ &= \frac{4}{3}|\langle\psi|\varphi\rangle|^2|\langle\psi_\perp|\varphi\rangle|^2\,.\end{aligned}$$

Since $\langle\varphi_Q|\mathcal{O}_{\text{same,diff}}|\varphi_Q\rangle = \langle\varphi_Q|\mathcal{O}_{\text{diff,same}}|\varphi_Q\rangle$ the success probability reads

$$p_{\text{success}}(\psi,\varphi) = \frac{2}{3}(\sin 2\theta)^2\,. \tag{6.32}$$

It vanishes only if $\theta = 0$, or $\theta = \pi/2$ when $\psi\equiv\varphi$, or $\psi\equiv\varphi_\perp$, respectively. As a result we get that the optimal test state detects unambiguously the difference for any pair of non-equivalent sharp qubit observables with strictly nonzero success probability. The actual probability depends on the angle between the observables. In fact, if sharp qubit observables are understood as ideal Stern-Gerlach apparatuses, then $\alpha = 2\theta$ is the angle between the measured spin directions. The probability achieves its maximum for orthogonal spin directions as one would expect.

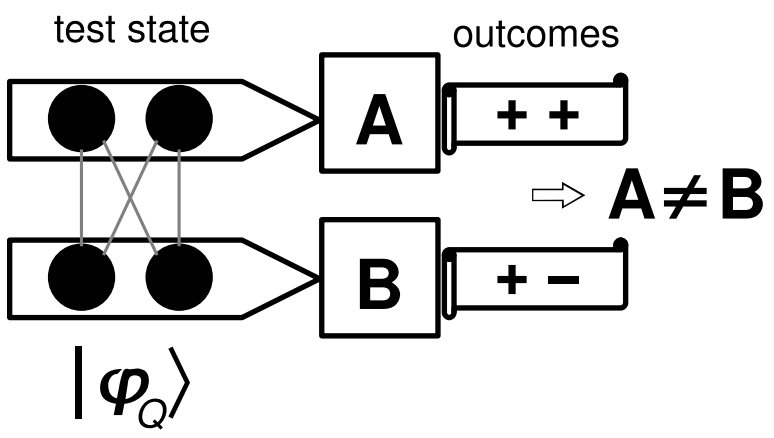

Figure 6.2. Illustration of the optimal scheme for unambiguous comparison of qubit apparatuses leading to unambiguous conclusion $A \neq B$ with average conditional probability $4/9$.

### 6.4.4 Summary

We have investigated the problem of unambiguous comparison of quantum measurements. We restricted our analysis to subset of sharp non-degenerate observables that can be associated with non-degenerate selfadjoint operators. Let us note that without any restriction the comparison problem has only a trivial solution.

We distinguished two different types of measurement apparatuses depending whether the labels of their outcomes are apriori given, or not. We give solution to single shot comparison of labeled measurements in arbitrary dimension. For unlabeled measurements the single usage of each of the apparatuses is not sufficient. In the two shots scenario we find solution for unlabeled qubit measurement apparatuses. In both cases, the unambiguous confirmation of the equivalence of measurements is not possible. Similarly, as in the case of pure states and unitary channels, also for sharp non-degenerate observables only the difference can be unambiguously concluded.

In summary, for the measurement apparatuses with labeled outcomes the optimal procedure exploits the so-called antisymmetric test states. For any such test state $\varrho$ the success is associated with the observation of the same outcomes. The difference of observables can be concluded with the average conditional probability

$$\overline{q}_{\text{success}}(\mathcal{A} \neq \mathcal{B}) = 1/d\,. \tag{6.33}$$

In the case of unlabeled measurements individual outcomes can be associated with an unambiguous conclusion only if the support of the test state belongs to at least one of the subspaces spanned by projections $I - \Pi^{\mathcal{A}=\mathcal{B}}_{x,x'}$, $x, x' \in \{\text{same}, \text{diff}\}$. We showed that only part of the test state acting on the support of the projections $Q_{\text{same,same}} = O$, $Q_{\text{diff,diff}}$ and $Q_{\text{same,diff}} = Q_{\text{diff,same}} = |\varphi_Q\rangle\langle\varphi_Q|$ may contribute to the success probability. Out of these possibilities, it turns out that the optimal test state is

$$|\varphi_Q\rangle = \frac{1}{\sqrt{3}}(|\psi^-_{13} \otimes \psi^-_{24}\rangle + |\psi^-_{14} \otimes \psi^-_{23}\rangle)\,, \tag{6.34}$$

for which the average conditional probability of the unambiguous conclusion equals

$$\overline{p}_{\text{success}}(\mathcal{A} \neq \mathcal{B}) = 4/9\,. \tag{6.35}$$

Using such test state and finding on one of the measurement apparatuses different outcomes, whereas on the second the same outcomes, we can conclude with certainty that the apparatuses are different. This scheme is illustrated on Fig. 6.2.

Let us compare these success probabilities with the comparison problem for pure states and unitary channels. In particular, for single shot comparisons

$$\overline{p}_{\rm state} = (d-1)/2d\,, \tag{6.36}$$

$$\overline{p}_{\rm unitary} = (d+1)/2d\,. \tag{6.37}$$

We see that unlike for states and channels the success rate for comparison of labeled measurements vanishes as the dimension is increasing. Unfortunately, for unlabeled measurements on systems of larger dimensions the situation is more complex and two shots are not sufficient to make any unambiguous conclusion. The problem is still open and will be analyzed elsewhere.

## 7 Conclusion

Quantum mechanics is a statistical theory and its predictions are conveyed by probabilities. However, in experiments we do not observe probabilities, but rather single experimental events called "clicks" or outcomes. Repeating the experiment many times we acquire frequencies, which express statistics of the observed clicks. When the number of repetitions of the experiment is really huge we intuitively expect that the predicted probabilities and the observed frequencies should coincide. From this point of view it may appear counterintuitive that single experiments may suffice to make reliable conclusions from the quantum mechanical predictions. In fact these conclusions can be even unambiguous if the probability distributions for all the alternatives are very well distinguishable. Mutual orthogonality of these probability distributions implies perfect discrimination among the alternatives. If a given click (outcome) can be observed for only one of the alternatives then we can use it to unambiguously conclude which alternative actually took place in the experiment. This means probability distributions must be orthogonal on the subset of outcomes usable for unambiguous conclusions. The remaining outcomes are inconclusive, as they may be a consequence of more than one alternative. In most experiments one can think of the subset of outcomes usable for unambiguous conclusions would be empty. However, in certain situations our prior knowledge allows us to design the experiment in such a way that it gives us unambiguous information about its constituents. The entire paper is devoted to study of this kind of unambiguous tasks. In Chapter 3 I formulated a framework which accommodates many problems of this type. I show that the prior information about any type of a constituent (state, channel, observable) allows us to reformulate the discrimination among finite number of alternatives as discrimination among finite number of average constituents.

Chapter 4 concentrates on tasks for quantum states. The relation of the quantum state comparison and the unambiguous identification to the discrimination of mixed states was recognized soon after the problem of unambiguous discrimination of mixed states was defined. My original results in this chapter cover two topics. One is comparison of ensembles of quantum states and the other one is unambiguous identification of coherent states.

In the comparison of two ensembles of $k$ and $l$ copies prepared by two preparators of unknown pure states I derived the conditional probability of revealing the difference of arbitrary pure states $|\psi_1\rangle, |\psi_2\rangle$. The conditional probability is a polynomial in $k, l, |\langle\psi_1|\psi_2\rangle|^2$ and does not depend on the dimension of the system. This is in contrast to the average success probability, which is essentially given by the ratio of the dimensions of the symmetric subspaces $\mathcal{H}_{sym}^{\otimes k+l}$ and $\mathcal{H}_{sym}^{\otimes k} \otimes \mathcal{H}_{sym}^{\otimes l}$. If the total number of available copies $N$ is fixed then the success probability is maximized for equal number of copies, i.e. $k = l = N/2$. The success probability can be slightly increased given the prior knowledge that the states are coherent. In this case the improvement is most significant for small number of copies. Moreover, I have proposed an optical implementation of the optimal quantum-state comparator of two finite ensembles of coherent states. This proposal is relatively easy to implement, since it consists only of $N - 1$ beam-splitters and a single photodetector.

Let me summarize the new results in the unambiguous identification (UI) part of Chapter 4. In UI we are given a set of identical quantum systems prepared in pure states, which are labeled as unknown and reference states. The promise is that one type of reference state is the same as the unknown state and the task is to find out unambiguously which one it is. My main focus was on the case where the set of possible reference states is formed only by coherent states of an electromagnetic field. I illustrated the relevance of this prior knowledge by showing that the specialized measurement outperforms the universal unambiguous identification, i.e. the UI measurements that can be applied for all pure states. The interesting qualitative difference between the specialized and the universal measurement for single copy of unknown and the reference states is in the probability of success for nearly orthogonal states. While the specialized measurement succeeds almost always the universal measurement produces conclusive result at most with probability $1/3$. Moreover, the specialized measurement can be easily experimentally realized, because it consists of beam splitters and photodetectors. The basic version of the setup was recently build and tested by L. Bartůšková et. al. [53] and we shortly summarize their experiment in Appendix C. The beamsplitter setup was motivated by an intuitive reduction of the unambiguous identification problem into specific "distribution" task and an unambiguous state comparison. The optical setup can be generalized to situations with more copies of the unknown and the reference states. The generalization is based on an idea of the "concentration" of the same type of states into strong coherent states that are subsequently identified by setups for the single-copy scenario. In the UI task it is assumed that the particular choice of the reference states is unknown to us, and only the probability distribution $\chi$ describing this choice is known. Nevertheless, even without having $\chi$ it is possible to derive the optimal choice of transmittivities in the beam-splitter setup for two types of reference states and an equal number of copies of each of the reference states ($n_B = n_C$). In that case the probability of identification for the reference states $|\alpha_1\rangle, |\alpha_2\rangle$ reads $P(|\alpha_1\rangle, |\alpha_2\rangle) = 1 - \exp[-\frac{n_A n_B}{n_A+2n_B}|\alpha_1 - \alpha_2|^2]$. Under the condition that the experimental setup consists only of linear optical elements and photodetectors I proved the optimality of the setup. In the limit of $n_B = n_C \to \infty$ the two reference states become known. Therefore, one needs to unambiguously discriminate the unknown state between two known pure states. The probability of success of our setup in this case coincides with the optimal value achieved by the Ivanovic-Dieks-Peres measurement [6–8]. I addressed also the question whether the coherent reference states can be recreated after our UI measurement. I showed that the reference states can be partially recovered only if the measurement yielded a conclusive outcome. The recovered reference states can be used in the next round of the UI if another unknown state is provided. This might be seen as a repeated search in a quantum database, where the data, i.e. the reference states, degrade with repeated use of the database. Another aspect I investigated was the influence of a particular type of noise on the reliability of the conclusions drawn by the UI setup. More precisely, I considered a communication scenario called the phase keying, with two coherent reference states of equal amplitude, but the opposite phases. I saw that the reliability of results depends only on the ratio of the amplitudes of the noise and the signal. However, for nonzero noise the unambiguity of the conclusions is lost.

The goal of Chapter 5 was to investigate unambiguous tasks for quantum channels. Many experiments probing channels are equivalent with respect to probability distribution on outcomes they generate. Without taking this equivalence into account optimization of any discrimination problem is very difficult, because we need to vary independently preparation and the measurement part of the experiment. *Process positive operator valued measure (PPOVM)* is a framework introduced by M. Ziman [31], which systematically takes this equivalence into account and works with the equivalence classes instead. Unambiguous discrimination of channels is much less studied than the discrimination of states. Using the PPOVM framework I defined the general problem and provided some insight for unitary channels. My main contribution in this chapter is the solution of comparison of two unknown unitary channels. Exploiting the framework of process POVM I have shown that the optimal strategy achieves the average conditional success probability $\overline{p}_{\text{success}} = (d+1)/(2d)$. Quantum channels are tested indirectly via their action on quantum states called also test states. I have shown that the optimal solution is achieved if and only if the test state is antisymmetric, i.e. its support is only in antisymmetric subspace. Let me note that such state is necessarily necessarily entangled. Hence, the entanglement is the key ingredient for comparison of unitary channels.

Unambiguous tasks for measurements were studied in Chapter 6. In contrast to states and channels quantum measurements act both quantumly and classically as they signalize the observed outcome. The prior knowledge about the signalling of outcomes motivates us two distinguish two types of equivalence among observables. We call observables *identical* if their outcomes are marked in the same way and they produce the same probability distribution for any measured state. We call observables *equivalent* if they can be made identical by suitable labeling of the outcomes. Suitable framework for description of experiments distinguishing observables was not established even for their single use. This is the reason why my studies were restricted to experiments, where only the tested measurements can be used and no feed forward of the outcomes is allowed. More precisely, I investigated the problem of unambiguous comparison of quantum measurements. I restricted the analysis to subset of sharp non-degenerate observables. It is important to note that without any restriction the unambiguous comparison of measurements has only a trivial solution. I distinguished two different types of measurement apparatuses depending whether the labels of their outcomes are apriori given, or not. I have presented solution to single shot comparison of labeled measurements in arbitrary dimension. For unlabeled measurements the single usage of each of the apparatuses is not sufficient. In the two shots scenario I give solution for unlabeled qubit measurement apparatuses. In both cases, the unambiguous confirmation of the equivalence of the measurements is not possible. Similarly, as in the case of pure states and unitary channels, also for sharp non-degenerate observables only the difference can be unambiguously concluded. For the measurement apparatuses with labeled outcomes the optimal procedure exploits the antisymmetric test states. For any such test state the success is associated with the observation of the same outcomes. The difference of observables can be concluded with the average conditional probability $\overline{q}_{\text{success}}(\mathcal{A} \neq \mathcal{B}) = 1/d$. In the case of unlabeled measurements the optimal test state also has some antisymmetry. The unambiguous conclusion for this test state is possible only if we find on one of the measurement apparatuses different outcomes, whereas on the second the same outcomes. The average conditional probability of revealing the difference of the apparatuses equals $\overline{p}_{\text{success}}(\mathcal{A} \neq \mathcal{B}) = 4/9$.

Let me compare these conditional success probabilities with the comparison problem for pure states and unitary channels. In particular, for single shot comparisons $\overline{p}_{\text{state}} = (d-1)/2d$, $\overline{p}_{\text{unitary}} = (d+1)/2d$. We see that unlike for states and channels the success rate for comparison of labeled measurements vanishes as the dimension is increasing. Unfortunately, for unlabeled measurements on systems of larger dimensions the situation is more complex and two shots are not sufficient to make any unambiguous conclusion.

Let me discuss my results also from a bit more general point of view. In classical physics we are used to have a direct relation of the measured property of the physical system to its implicitly assumed preexisting value before the measurement. In quantum physics it follows from Bells inequalities that for some sets of observables assuming preexisting values before the measurement is forbidden. On the other hand, there are well defined properties of the system before the measurement (e.g. states being same or different in the comparison problem), which are not directly observable in the experiment. Unambiguous approach to the discrimination tasks brings into the quantum world the direct relation of the measured outcomes to the preexisting properties of the quantum systems. This paper tries to understand what kind of tasks can be solved unambiguously and what kind of information we may acquire about the quantum system in an unambiguous way. Motivated by the results on unambiguous comparison of states, channels and observables, where equality of quantum objects can not be concluded, one might conjecture that equality of quantum objects can not be proved in general from finite statistics of outcomes. Although I do not prove this conjecture in the paper I consider it to be one of the the interesting problems that remain as an open question for further studies.

Recently, G. M. D'Ariano [70] and his coworkers from the group in Pavia, Italy developed a framework called *quantum combs*. Its aim is to describe most general transformations of constituents (states, channels, measurements) and represent them in a unified way. Similarly to PPOVM framework quantum combs are based on Choi-Jamiolkowski isomorphism. It seems that they are well suited also for experiments containing multiple uses of the tested constituent. Quantum combs were not yet applied to unambiguous discrimination tasks, however they were already applied in quantum algorithm learning [71], optimal tomography [72] or optimal cloning of a unitary transformation [73]. I think that application of quantum combs in unambiguous discrimination may advance our understanding of this kind of task. On the other hand, the actual calculations have to be the same as in the presently used calculus. Thus, one can not expect that complicated problems as a discrimination of mixed states will be solved by quantum combs easily. The more expectable advantage of better suited framework is in more clear and unified way of working with the quantum objects.

## Acknowledgments

I would like to express my gratitude to Vladimír Bužek, Mário Ziman and Mark Hillery for introducing me to the field of quantum information processing and quantum optics and for our fruitful collaborative work. I would also like to acknowledge L. Bartůšková, A. Černoch, J. Soubusta, and M. Dušek for permission to use their figures and text in the Appendix C, which describes experimental realization of unambiguous identifier of coherent states. This work was supported by the European Union research projects QAP 2004-IST-FETPI-15848 and HIP FP7-ICT-2007-C-221889, and via the projects APVV-0673-07 QIAM, and CE SAV QUTE.

# Appendices

## A Calculations for state comparison

### A.1 Proof of lemma

**Lemma**
Suppose we have a polynomial $Q_r(x) = \sum_{m=0}^{r} a_m x^m$ with the following properties:

1. $Q_r(1) = 0$
2. $a_m \geq 0$ for $m \leq r_0$ and $a_m \leq 0$ for $r_0 < m \leq r$

Then $Q_r(x) \geq 0$ for all $x \in [0,1]$.
*Proof:* For $x \in [0,1]$ and $a > b$ it follows that $x^a < x^b$. Therefore we can write

$$\begin{aligned} Q_r(x) &= \sum_{m=0}^{r_0} a_m x^m + \sum_{m=r_0+1}^{r} a_m x^m \\ &\geq x^{r_0} \sum_{m=0}^{r_0} a_m + x^{r_0+1} \sum_{m=r_0+1}^{r} a_m \qquad \text{(A.1)} \\ &= (1-x) x^{r_0} \sum_{m=0}^{r_0} a_m \qquad \text{(A.2)} \\ &\geq 0\,, \qquad \text{(A.3)} \end{aligned}$$

where we have used the fact that $0 = Q_r(1) = \sum_{m=0}^{r_0} a_m + \sum_{m=r_0+1}^{r} a_m$, i.e. $\sum_{m=r_0+1}^{r} a_m = -\sum_{0}^{r_0} a_m$.

### A.2 Projectors onto coherent states

Coherent states $|\alpha\rangle$ are intimately related to the group of phase-space displacements $G$ generated by the Glauber operator $D_\alpha = \exp(\alpha a^\dagger - \alpha^* a)$ via the following relation $D_\alpha|0\rangle = |\alpha\rangle$, where $|0\rangle$ is the vacuum (ground) state of a harmonic oscillator. Using the group invariant measure $dg$ (its support contains all coherent states) the operator $\Delta$ can be expressed as follows

$$\Delta = \int_G dg (D_g|0\rangle\langle 0|D_g^\dagger)^{\otimes N}. \qquad \text{(A.4)}$$

Applying the theorem proved in Ref. [74] to the representation of the group of displacements we find that

$$\Delta = \int_G dg (D_g|0\rangle\langle 0|D_g^\dagger)^{\otimes N} = \lambda \Delta_{\mathrm{coh}}^N\,, \qquad \text{(A.5)}$$

where $\lambda$ is a positive number ($\Delta$ is positive) and $\Delta_{\mathrm{coh}}^N$ is the projector onto the linear subspace spanned by the product states $|\alpha\rangle^{\otimes n}$. A particular choice of the group invariant measure $dg$ affects the value of the parameter $\lambda$. Our goal is to calculate the projector $\Delta_{\mathrm{coh}}^N$, hence we are looking for a measure $dg$ such that $\lambda = 1$. The canonical Lebesgue measure $d\alpha$ on the complex plane $\mathbb{C}$ is invariant under complex translations (displacements) and therefore the correct measure $dg$ is proportional to $d\alpha$, that is $dg = \mu d\alpha$ for some positive number $\mu$, i.e.

$$\Delta_{\rm coh}^N = \mu \int_{\mathbb{C}} d\alpha |\alpha\rangle\langle\alpha|^{\otimes N} . \tag{A.6}$$

Now, setting $\alpha = re^{i\theta}$, we have, expanding the coherent states in terms of number states,

$$\begin{aligned}\Delta_{\rm coh}^N|0\rangle^{\otimes N} &= \mu \int_{\mathbb{C}} d\alpha e^{-N|\alpha|^2/2} \times \\ &\times \sum_{l_1=0}^{\infty} \frac{\alpha^{l_1}}{\sqrt{l_1!}} \cdots \sum_{l_N=0}^{\infty} \frac{\alpha^{l_N}}{\sqrt{l_N!}} (\langle\alpha|0\rangle)^N |l_1,\dots l_N\rangle \\ &= 2\pi\mu \int_0^\infty dr\, re^{-Nr^2}|0\rangle^{\otimes N} \\ &= \mu\frac{\pi}{N}|0\rangle^{\otimes N} ,\end{aligned} \tag{A.7}$$

because $\int_0^{2\pi} e^{i\theta(l_1+\dots+l_N)}d\theta = 2\pi$ if $l_1 + \dots + l_N = 0$, and vanishes otherwise. The invariance of the canonical Lebesgue measure implies that

$$\begin{aligned}\Delta_{\rm coh}^N D_\beta^{\otimes N} &= D_\beta^{\otimes N} D_{-\beta}^{\otimes N} \Delta_{\rm coh}^N D_\beta^{\otimes N} \\ &= D_\beta^{\otimes N} \mu \int_{\mathbb{C}} d\alpha |\alpha-\beta\rangle\langle\alpha-\beta|^{\otimes N} \\ &= D_\beta^{\otimes N} \mu \int_{\mathbb{C}} d(\alpha-\beta) |\alpha-\beta\rangle\langle\alpha-\beta|^{\otimes N} \\ &= D_\beta^{\otimes N} \mu \int_{\mathbb{C}} d\alpha |\alpha\rangle\langle\alpha|^{\otimes N} \\ &= D_\beta^{\otimes N} \Delta_{\rm coh}^N\end{aligned} \tag{A.8}$$

The previous identity (A.8) implies

$$\Delta_{\rm coh}^N|\beta\rangle^{\otimes n} = \Delta_{\rm coh}^N D_\beta^{\otimes N}|0\rangle^{\otimes N} = D_\beta^{\otimes N}\Delta_{\rm coh}^N|0\rangle^{\otimes N} . \tag{A.9}$$

Consequently, for all $|\psi\rangle \in \mathcal{H}_{\rm coh} \equiv {\rm span}\{|\alpha\rangle^{\otimes N}\}$ it holds that

$$\Delta_{\rm coh}^N|\psi\rangle = \mu\frac{\pi}{N}|\psi\rangle , \tag{A.10}$$

and for all $|\psi_\perp\rangle \in \mathcal{H}_0^\perp$ we have $\Delta_{\rm coh}^N|\psi_\perp\rangle = 0$. The above equality fixes the invariant measure $dg$ to be $\frac{N}{\pi}d\alpha$, where $d\alpha$ is the Lebesgue measure on the complex plane.

## B Calculations for UI problems

### B.1 Eigenvalues of $E_0^{sb}$

The operator $\mathrm{E}_0^{sb}$ is defined in Eq.(4.94). As we have already mentioned there this operator is block diagonal consisting of three types of blocks. 1. Trivial $\langle iii|\mathrm{E}_0^{sb}|iii\rangle = 1$.

2. $3 \times 3$ matrix $\langle \sigma_1(iij)|\mathrm{E}_0^{sb}|\sigma_2(iij)\rangle$:

$$Q_3 = \begin{pmatrix} & iij & iji & jii \\ iij & 1-c_1/2 & 0 & c_1/2 \\ iji & 0 & 1-c_2/2 & c_2/2 \\ jii & c_1/2 & c_2/2 & 1-c_1/2-c_2/2 \end{pmatrix} \quad \text{(B.11)}$$

with eigenvalues

$$\begin{aligned} \lambda_1^{(3)} &= 1\,; \\ \lambda_{2,3}^{(3)} &= \frac{2-c_1-c_2 \pm \sqrt{c_1^2 - c_1c_2 + c_2^2}}{2} \end{aligned} \quad \text{(B.12)}$$

3. $6 \times 6$ matrix $\langle \sigma_1(ijk)|\mathrm{E}_0^{sb}|\sigma_2(ijk)\rangle$:

$$Q_6 = \begin{pmatrix} & ijk & kji & jki & ikj & kij & jik \\ ijk & X & c_1/2 & 0 & 0 & 0 & c_2/2 \\ kji & c_1/2 & X & c_2/2 & 0 & 0 & 0 \\ jki & 0 & c_2/2 & X & c_1/2 & 0 & 0 \\ ikj & 0 & 0 & c_1/2 & X & c_2/2 & 0 \\ kij & 0 & 0 & 0 & c_2/2 & X & c_1/2 \\ jik & c_2/2 & 0 & 0 & 0 & c_1/2 & X \end{pmatrix}, \quad \text{(B.13)}$$

where $X = 1 - \frac{c_1}{2} - \frac{c_2}{2}$. The corresponding eigenvalues read

$$\begin{aligned} &\lambda_1^{(6)} = 1 \\ &\lambda_2^{(6)} = 1 - c_1 - c_2 \\ &\lambda_{3,4}^{(6)} = \tfrac{2-c_1-c_2+\sqrt{c_1^2-c_1c_2+c_2^2}}{2} \\ &\lambda_{5,6}^{(6)} = \tfrac{2-c_1-c_2-\sqrt{c_1^2-c_1c_2+c_2^2}}{2}\,. \end{aligned} \quad \text{(B.14)}$$

For qubits we have $2^3$-dimensional Hilbert space and $\mathrm{E}_0^{sb}$ is represented by $8 \times 8$ matrix with two $3 \times 3$ blocks and two $1 \times 1$ blocks. So for qubits the sufficient condition for positivity of $\mathrm{E}_0^{sb}$ reads

$$2 - c_1 - c_2 \pm \sqrt{c_1^2 - c_1c_2 + c_2^2} \geq 0$$

For qudits, $d > 2$, at least one $6 \times 6$ block appears in the matrix of $\mathrm{E}_0^{sb}$. The eigenvalues of the $6 \times 6$ block satisfy the following inequality:

$$\lambda_1^{(6)} \geq \lambda_{3,4}^{(6)} \geq \lambda_{5,6}^{(6)} \geq \lambda_2^{(6)},$$

so the sufficient condition for positivity of $\mathrm{E}_0^{sb}$ when $d > 2$ is $\lambda_2^{(6)} = 1 - c_1 - c_2 \geq 0$.

### B.2 Evaluation of Gaussian type of integrals

As we have seen the following type of integrals

$$I_m = \frac{1}{(2\pi\sigma^2)^m}\int_{\mathbb{C}^m} d\alpha_1 \dots d\alpha_m e^{-\sum_{i=1}^m \frac{|\alpha_i|^2}{2\sigma^2} - \frac{a}{b}|x + \sum_{i=1}^m \alpha_i|^2}$$

emerge often in our calculation for the noise model. These integrals can be evaluated recursively using the relation

$$\frac{1}{(2\pi\sigma^2)}\int_{\mathbb{C}} d\alpha\, e^{-\frac{|\alpha|^2}{2\sigma^2} - \frac{a}{b}|x+\alpha|^2} = \frac{b}{b+2a\sigma^2} e^{-\frac{a}{b+2a\sigma^2}|x|^2} \tag{B.15}$$

we are going to derive now. Left hand side (LHS) of Eq. (B.15) can be rewritten using the following modification of the rectangular identity

$$\begin{aligned} k \quad & |\beta - \alpha_1|^2 + l|\beta - \alpha_2|^2 = \\ = \quad & \left|\sqrt{k+l}\beta - \frac{k\alpha_1 + l\alpha_2}{\sqrt{k+l}}\right|^2 + \frac{kl}{k+l}|\alpha_1 - \alpha_2|^2 \end{aligned} \tag{B.16}$$

as

$$\begin{aligned} LHS &= \frac{e^{-\frac{a}{b+2a\sigma^2}|x|^2}}{(2\pi\sigma^2)}\int_{\mathbb{C}} d\alpha\, e^{-\left|\sqrt{\frac{1}{2\sigma^2}+\frac{a}{b}}\alpha - \frac{2b\sigma^2}{b+2a\sigma^2}x\right|^2} \\ &= \frac{e^{-\frac{a}{b+2a\sigma^2}|x|^2}}{(2\pi\sigma^2)}\int_{\mathbb{C}} d\alpha\, e^{-\left|\sqrt{\frac{1}{2\sigma^2}+\frac{a}{b}}\alpha\right|^2} \\ &= \frac{b}{b+2a\sigma^2}e^{-\frac{a}{b+2a\sigma^2}|x|^2}\frac{1}{2\pi\sigma'^2}\int_{\mathbb{C}} d\alpha\, e^{-\frac{|\alpha|^2}{2\sigma'^2}} \\ &= \frac{b}{b+2a\sigma^2}e^{-\frac{a}{b+2a\sigma^2}|x|^2}, \end{aligned} \tag{B.17}$$

where we have used the fact that we are integrating over the whole complex plane. As a consequence, a constant shift of argument does not matter and the Gaussian distribution is normalized to unity. Hence we have proved Eq. (B.15), which we can be rewritten as

$$I_m(a,b) = \frac{b}{b+2a\sigma^2} I_{m-1}(a, b+2a\sigma^2).$$

From this recursive rule it follows that

$$I_m(a,b) = \frac{b}{b+2a\sigma^2 m} e^{-\frac{a}{b+2ma\sigma^2}|x|^2}, \tag{B.18}$$

which is the result we wanted to obtain.

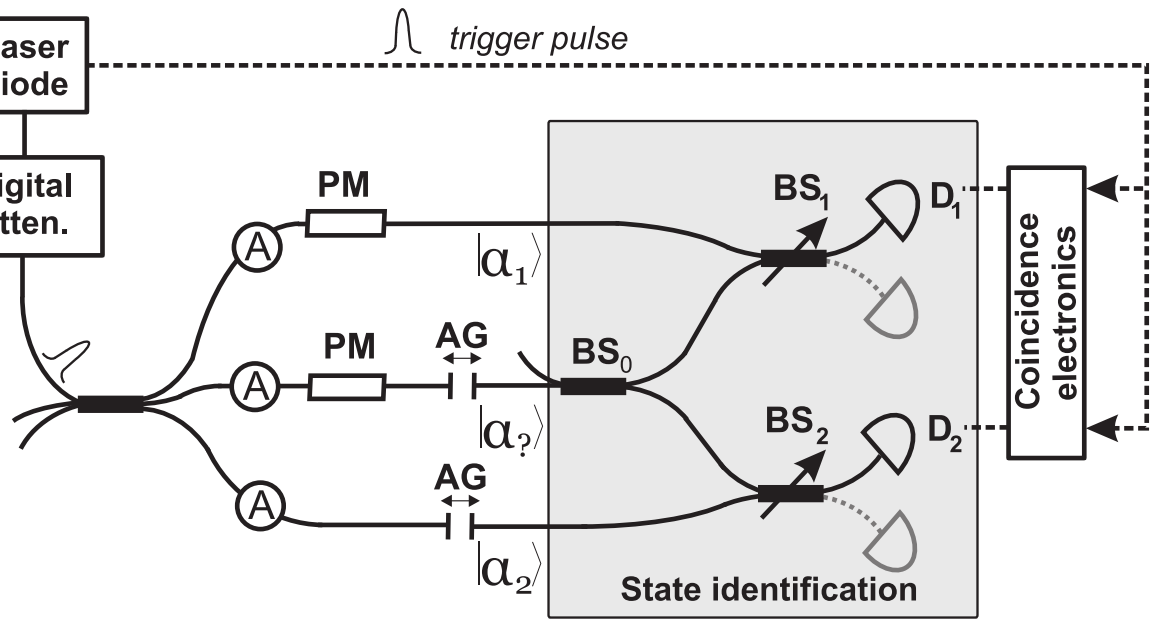

Figure C.1. The scheme of the experimental setup [53]. A - attenuators, PM - phase modulators, AG - adjustable air gaps, BS - beam splitters, D - detectors.

## C UI of coherent states – experimental realization

In this appendix we shortly summarize the experiment of Bartůšková, Černoch, Soubusta, and Dušek [53], who build and tested the unambiguous identificator for single copy of unknown and two reference states. The arrangement of the experiment is depicted in Fig. C.1 and it is based on the scheme that was derived in section 4.3.5. State $|\alpha_?\rangle$ is the unknown state that should be matched with one of the reference states $|\alpha_1\rangle, |\alpha_2\rangle$. As we shown in section 4.3.5 if $T_0 = 1/2, T_1 = 2/3, T_2 = 1/3$ hold for the transmittivities of beam splitters $BS_0$, $BS_1$, and $BS_2$, then one can unambiguously identify the unknown state using photodetectors $D_1$ and $D_2$. If $D_1$ clicks we conclude that $|\alpha_?\rangle = |\alpha_2\rangle$, if $D_2$ clicks it means that $|\alpha_?\rangle = |\alpha_1\rangle$. If neither of the detectors clicks we cannot make any conclusion about the state $|\alpha_?\rangle$ and this is an inconclusive result. In theory, both detectors can not click simultaneously. In practice, photodetectors have dark counts, dead times and less than 100% detection efficiency $\Gamma$. The non-ideal detection efficiency is for the measured coherent state $|\beta\rangle$ easily taken into account by replacing the probability of not registering photons $p_0(|\beta\rangle) = e^{-|\beta|^2}$ by $p_0(|\beta\rangle) = e^{-\Gamma|\beta|^2}$. Consequently, the probability of correct identification of state $|\alpha_1\rangle$ reads

$$p_1 = 1 - \exp\left(-\Gamma\frac{1-T_0}{2-T_0}|\alpha_1-\alpha_2|^2\right), \tag{C.19}$$

and of correct identification of state $|\alpha_2\rangle$

$$p_2 = 1 - \exp\left(-\Gamma\frac{T_0}{1+T_0}|\alpha_1-\alpha_2|^2\right), \tag{C.20}$$

where the same detection efficiency $\Gamma$ is assumed for both photodetectors. In a real setup it sometimes happens that detectors $D_1$ and $D_2$ click simultaneously (due to imprecisions and dark counts). Practically these double clicks, as well as no detections, correspond to inconclusive results because we cannot distinguish whether the unknown state was equal to state $|\alpha_1\rangle$, or to state $|\alpha_2\rangle$. The other two situations, when just one of detectors clicks, belong to conclusive results. They include both correct and erroneous identifications of the unknown state. Ideally, the erroneous identifications never occur and the probability of a conclusive result is equal to the probability of correct identification.

The experimental setup (see Fig. C.1) was built up on fiber optics. The coherent states were prepared by a laser diode, whose pulses were strongly attenuated and divided by a fiber coupler into three optical fibers. After passing through additional attenuators (A) and electro-optical phase modulators (PM) coherent states in the three fibers correspond to states $|\alpha_1\rangle$, $|\alpha_2\rangle$, $|\alpha_?\rangle$. The principle of the identification lies in the interference of light beams at beam splitters. As beam splitters $BS_1$ and $BS_2$ the two variable-ratio couplers were used, while beam splitter $BS_0$ was a $50:50$ fixed ratio fiber coupler. The whole setup works basically as two interconnected Mach-Zehnder (MZ) interferometers. To accomplish discriminating operation visibilities of both MZ interferometers had to be maximized. This was roughly done by aligning polarizations and setting the same optical paths using air gaps, for the arms corresponding to $|\alpha_1\rangle$ and $|\alpha_?\rangle$ and paths corresponding to $|\alpha_2\rangle$ and $|\alpha_?\rangle$. In order to to compensate the phase drift due to temperature fluctuations an active stabilization of the paths was done using phase modulators and another two photodetectors that regularly checked the visibility of the interference. The signal was detected by four single-photon counting avalanche photodiodes. Two of them, $D_1$ and $D_2$, served for both the discrimination and active stabilization while two others were used only for the stabilization. To minimize the influence of dark counts of detectors on a measurement only the coincidences between signals from detectors $D_1$, $D_2$ and pulses that triggered the laser diode were counted.

In the experiment the state identification for various combinations of states $|\alpha_1\rangle$ and $|\alpha_2\rangle$ was tested. Conclusive count rates $C_j^+$, when the state was correctly discriminated, and $C_j^-$ related to erroneous detections; $j=1,2$ were recorded. For example in cases when $j=1$ ($|\alpha_?\rangle = |\alpha_1\rangle$) the rate $C_1^+$ ($C_1^-$) was obtained by measuring coincidence rate between detector $D_2$ ($D_1$) and trigger pulse of the laser diode minus the coincidence rate between detectors $D_1$ and $D_2$ (related to double clicks). $C_2^+$ and $C_2^-$ were measured in a similar way. The fractions of correct and erroneous results read

$$P_j^+ = \frac{C_j^+}{C_{tot}}, \qquad P_j^- = \frac{C_j^-}{C_{tot}} \qquad (j=1,2) \tag{C.21}$$

respectively, where $C_{tot}$ is the total number of laser pulses per measurement period. The fraction of conclusive results is thus $P_j = P_j^+ + P_j^-$ (j=1, 2).

Experimental results are shown in Figs. C.2-C.5. Fig. C.2 and Fig. C.3 display the fraction of correct and erroneous results as a function of phase difference between coherent states $|\alpha_1\rangle$ and $|\alpha_2\rangle$. The theoretical curves of probabilities of a conclusive result (i.e. probabilities of correct identification) were calculated by equations (C.19), (C.20). They are identical for both program states due to assumed equality of the efficiencies of detectors $D_1$ and $D_2$. Measured data are presented as $P_j^+$ and $P_j^-$ according to equation (C.21). In the ideal case, when the visibility of interference is $100\%$ and there are no dark counts, the probability of a conclusive result is equal to the probability of correct identification. In the setup the effect of dark counts was minimized to be negligible and visibilities were around $98\%$. The imperfect interference affects the quality of discrimination mainly in situations when the overlap of coherent states $|\alpha_1\rangle$ and $|\alpha_2\rangle$ is relatively high. The probability of a conclusive result for the phase difference 180° between states rapidly grows with increasing intensities of states $|\alpha_1\rangle$ and $|\alpha_2\rangle$ (see Fig. C.4). Finally, Fig. C.5 shows the probability of a conclusive result as a function of intensity ratio $|\alpha_2|^2/|\alpha_1|^2$ whereas the intensity of the first state was fixed to 1.33 photons per pulse. The upper line is related to situations when the overlap of states is for given intensities minimal (phase difference 180° between the states) and the lower line corresponds to cases when the overlap is maximal (phase difference 0° between the states).

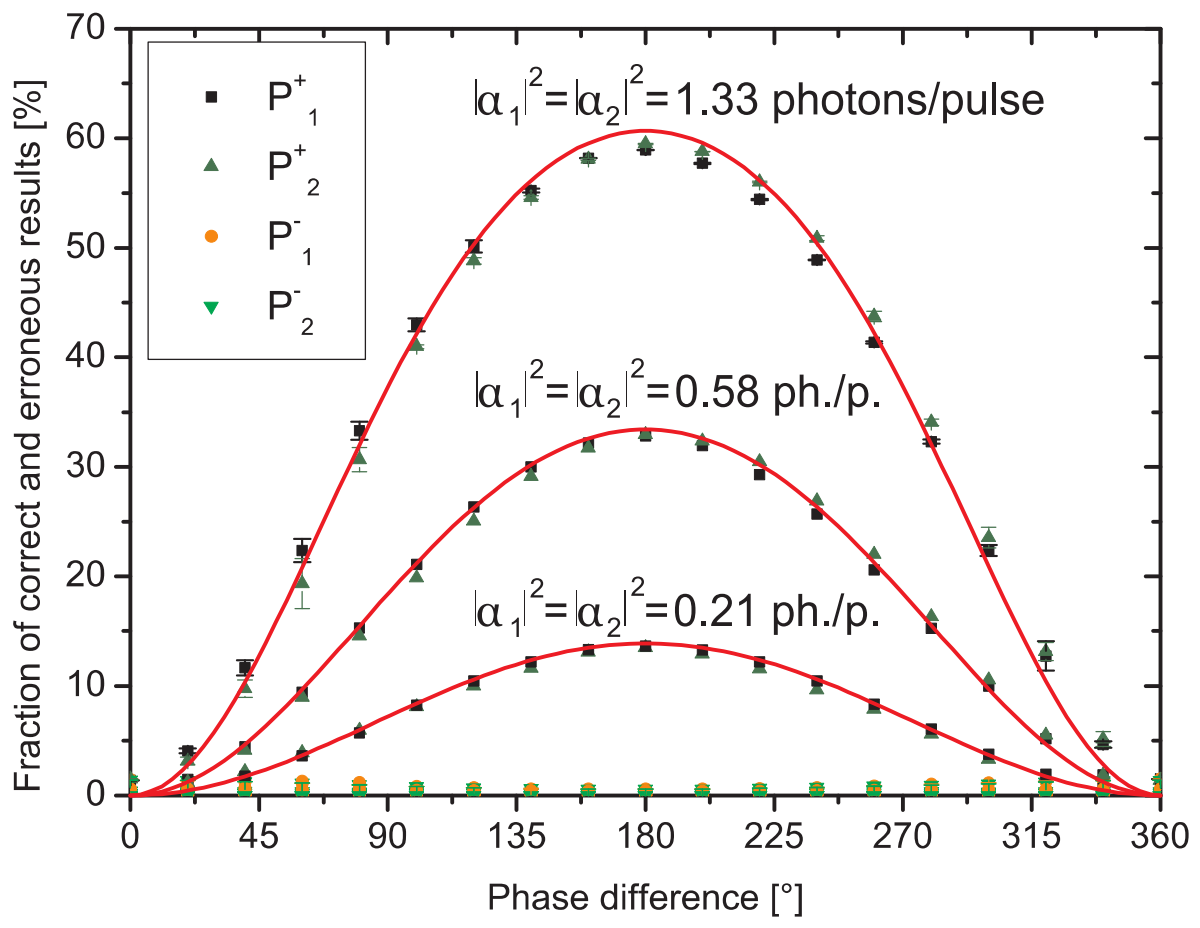

Figure C.2. Dependence of the fraction of correct and erroneous results on the phase difference between states $|\alpha_1\rangle$ and $|\alpha_2\rangle$ for three different intensities of states; $|\alpha_1|^2 = |\alpha_2|^2$. Solid lines represent theoretical predictions for the probability of a conclusive result. The graph originates from [53].

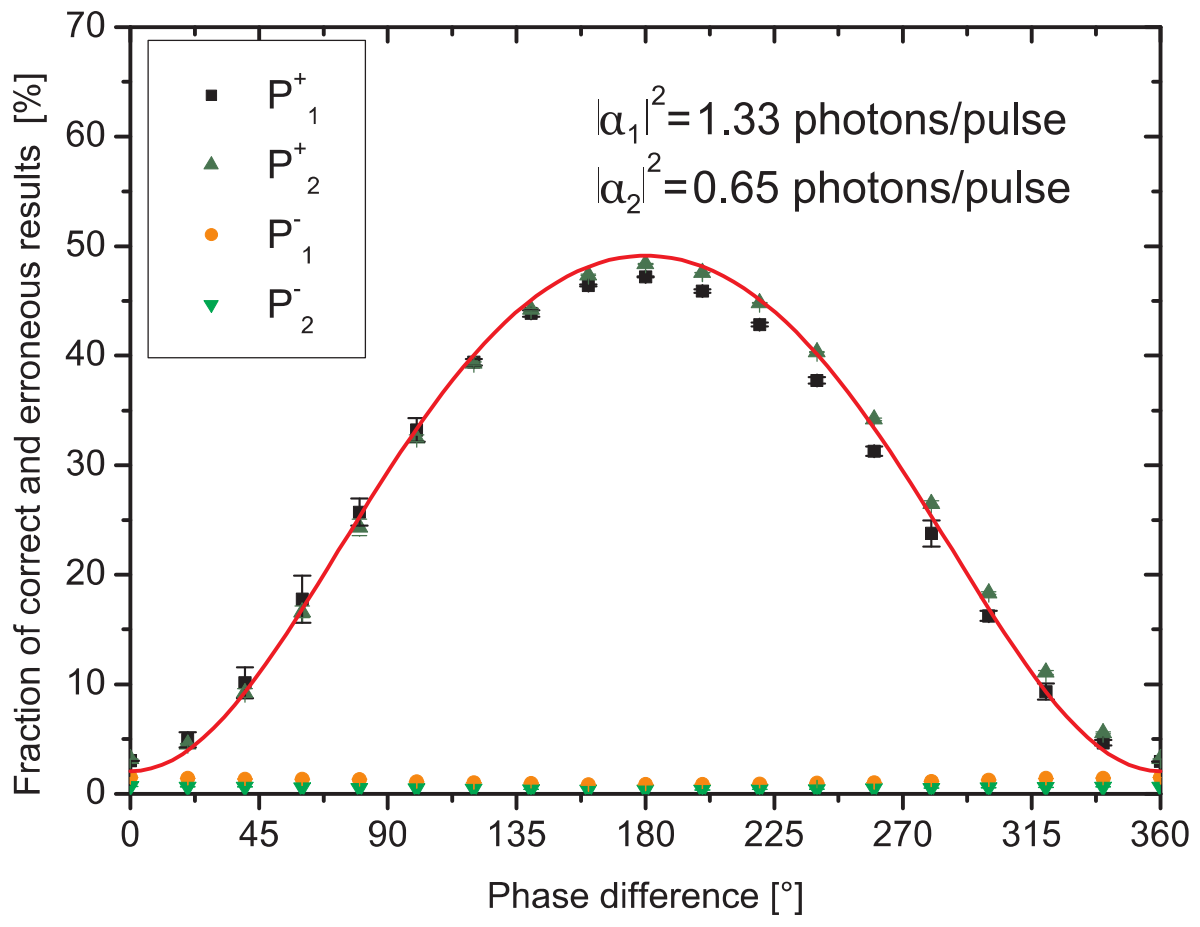

Figure C.3. Dependence of the fraction of correct and erroneous results on the phase difference between states $|\alpha_1\rangle$ and $|\alpha_2\rangle$; $|\alpha_1|^2 \neq |\alpha_2|^2$. Solid line represents a theoretical prediction for the probability of a conclusive result. The graph originates from [53].

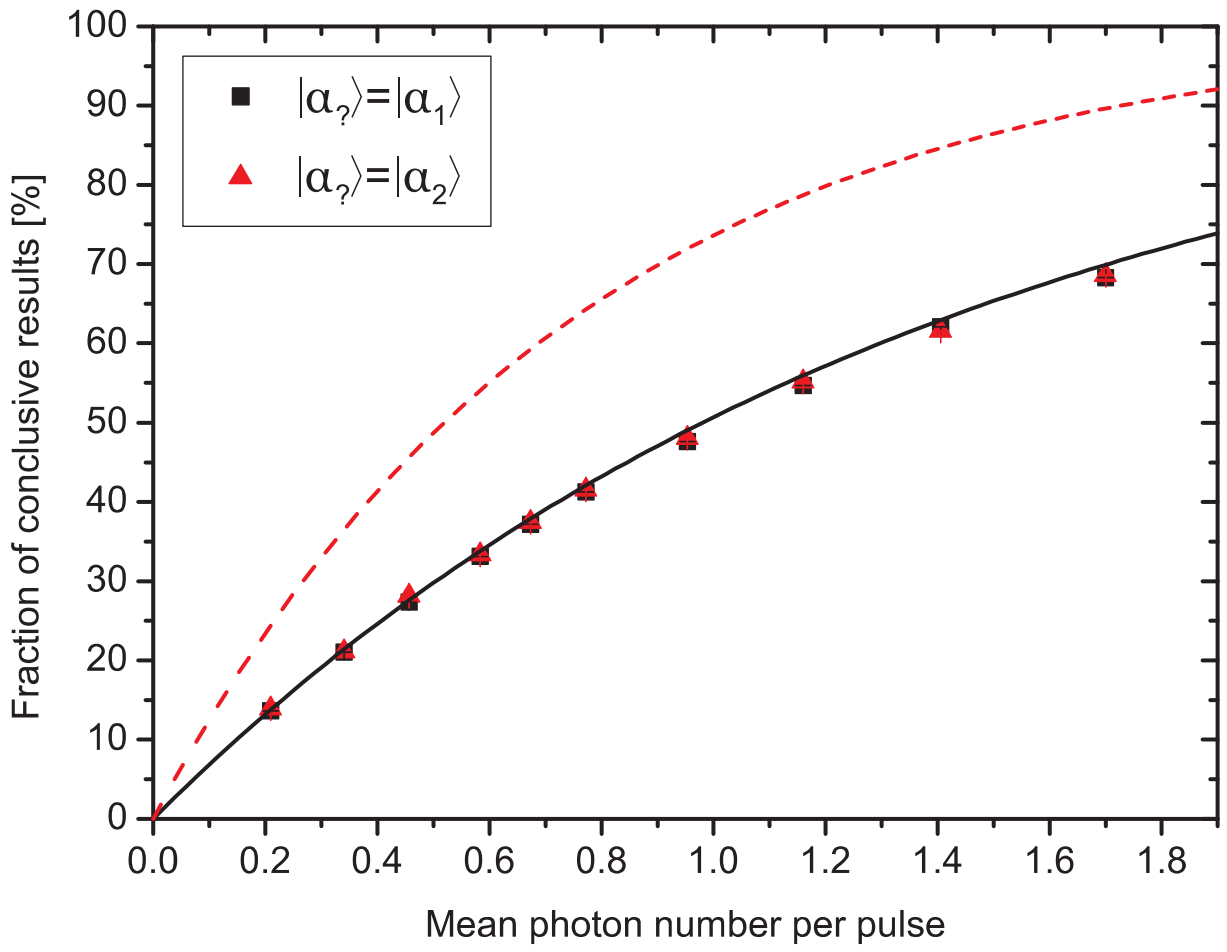

Figure C.4. Dependence of the probability of a conclusive result on the intensity of states ($|\alpha_1|^2 = |\alpha_2|^2$; phase difference between states was $180^\circ$). Solid line represents the theoretical prediction for detectors with $\eta = 53\%$. Dashed line is the theoretical limit for ideal detectors (quantum efficiency $\eta = 100\%$). The graph originates from [53].

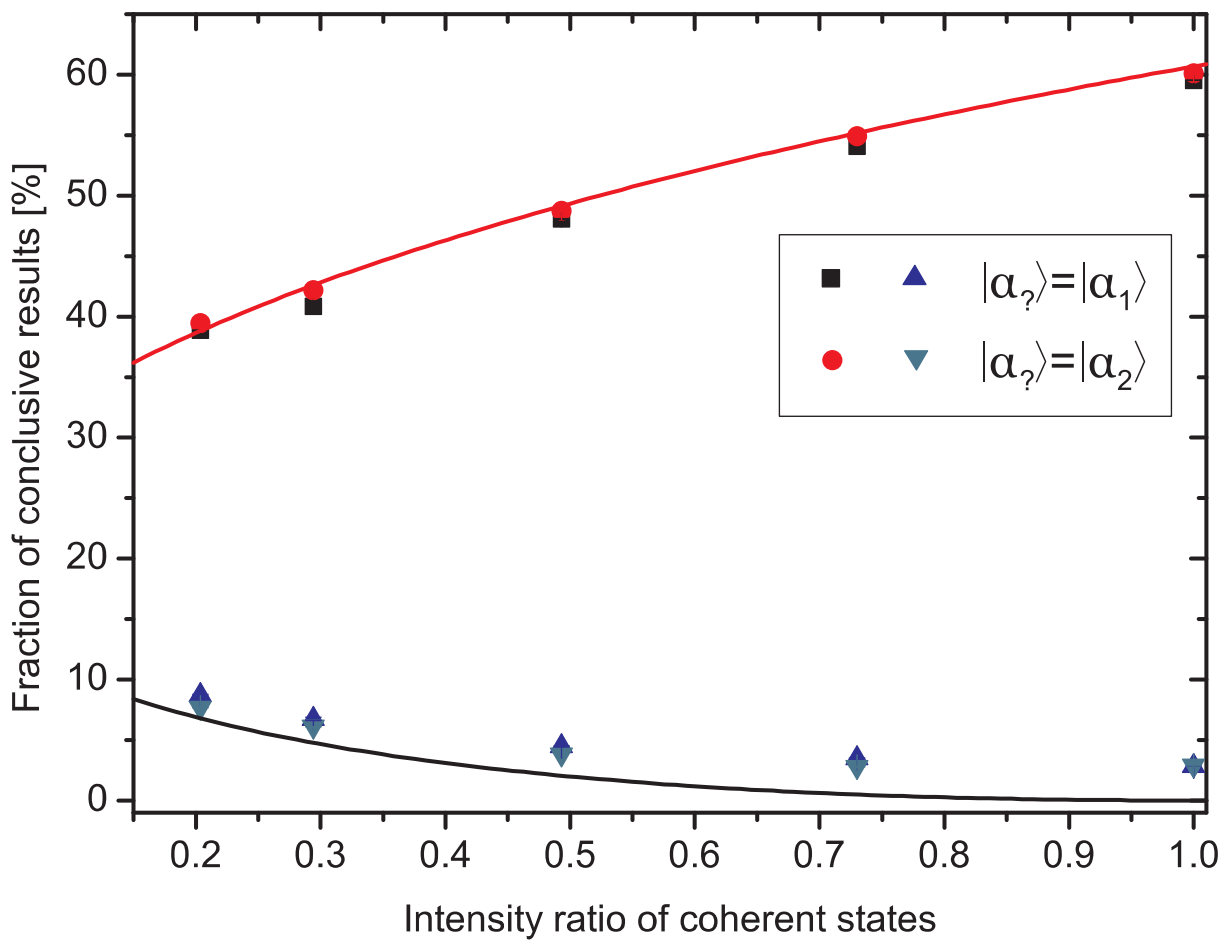

Figure C.5. Dependence of the probability of a conclusive result on the intensity ratio of states $|\alpha_2|^2/|\alpha_1|^2$ ($|\alpha_1|^2$=1.33 photons/pulse). The upper line represents the theoretical prediction for the phase difference $180^\circ$ between states $|\alpha_1\rangle$ and $|\alpha_2\rangle$ and the lower line corresponds to the phase difference $0^\circ$. The graph originates from [53].

## D Calculations for channels

### D.1 Average unitary channel

In this section we shall prove that the action of the average unitary channel can be expressed as

$$\mathcal{A}[X] = \int_{U(d)} dU\, UXU^\dagger = \frac{Tr(X)}{d} I\,, \tag{D.22}$$

where $dU$ is the unique *Haar invariant measure* defined on the group of unitary operators $U(d)$. By definition the image $\mathcal{A}[X]$ of any operator $X$ must commute with all unitary operators, i.e. $[\mathcal{A}[X], U] = 0$ for all $U \in U(d)$. The Schurr lemma implies that $\mathcal{A}[X] = c(X)I$. The transformation $\mathcal{A}$ is by definition trace-preserving. That is, $Tr(X) = c(X)Tr(I) = c(X)d$. It follows that $c(X) = Tr(X)/d$, hence the Eq.(D.22) holds.

### D.2 Twirling channel

We shall prove that the action of the twirling channel

$$\mathcal{T}[X] = \int_{U(d)} dU\, U \otimes UXU^\dagger \otimes U^\dagger\,, \tag{D.23}$$

on selfadjoint operators $X$ takes the form

$$\mathcal{T}[X] = \frac{Tr(XP^{sym})}{d_{sym}} P^{sym} + \frac{Tr(XP^{asym})}{d_{asym}} P^{asym}\,. \tag{D.24}$$

The properties of Haar invariant measure $dU$ implies that the operator $\mathcal{T}[X]$ commutes with all unitary operators of the type $U \otimes U$. If $X$ is selfadjoint, then $\mathcal{T}[X]$ is also selfadjoint and $\mathcal{T}[X] = \sum_j x_j P_j$, where $x_j$ are real eigenvalues and $P_j$ are the corresponding eigenprojectors. The commutation of $\mathcal{T}[X]$ with unitaries $U \otimes U$ implies that $[P_j, U \otimes U] = 0$ for all $U$. The subspaces $\mathcal{H}_j = P_j(\mathcal{H}_d \otimes \mathcal{H}_d) = \{\psi \in \mathcal{H}_d \otimes \mathcal{H}_d \text{ such that } P_j|\psi\rangle = |\psi\rangle\}$ are invariant under the action of operators $U \otimes U$.

It turns out there are only two invariant subspaces of $\mathcal{H}_d \otimes \mathcal{H}_d$ - *symmetric* and *antisymmetric* subspace. A vector $\psi \in \mathcal{H}_d \otimes \mathcal{H}_d$ is called symmetric (antisymmetric) if Swap$|\psi\rangle = \pm|\psi\rangle$, respectively, where we employed the *swap* operator. Let us denote by $P^{sym}$, $P^{asym}$ the projectors onto the symmetric and antisymmetric subspaces, respectively.

Consider an orthonormal basis $\{|j\rangle\}_{j=1}^d$ of $\mathcal{H}_d$. Defining the vectors $|\varphi_{jk}^\pm\rangle = \frac{1}{\sqrt{2}}(|j \otimes k\rangle \pm |k \otimes j\rangle)$ for $j < k$, $|\varphi_{jj}^+\rangle = |j \otimes j\rangle$ we can write

$$P^{sym} = \sum_{j \le k} |\varphi_{jk}^+\rangle\langle\varphi_{jk}^+|, \quad P^{asym} = \sum_{j<k} |\varphi_{jk}^-\rangle\langle\varphi_{jk}^-|. \tag{D.25}$$

Let us note that vectors $|\varphi_{jk}^\pm\rangle$ $(j, k = 1, \ldots, d)$ are forming an orthonormal basis of $\mathcal{H}_d \otimes \mathcal{H}_d$ and Swap$|\varphi_{jk}^\pm\rangle = \pm|\varphi_{jk}^\pm\rangle$. It follows that the dimensions of symmetric and antisymmetric subspaces are $d_\pm = d(d \pm 1)/2$, respectively. As a result we obtain that

$$\mathcal{T}[X] = a_+(X)P^{sym} + a_-(X)P^{asym} \tag{D.26}$$

is the spectral form of $\mathcal{T}[X]$. In order to verify that Eq.(D.23) and Eq.(D.24) define the same mapping, it is sufficient to verify their actions on elements of arbitrary operator basis. We shall use an orthonormal operator basis consisting of operators $E_{j\pm k,m\pm n} = |\varphi_{jk}^{\pm}\rangle\langle\varphi_{mn}^{\pm}|$.

According to Eq.(D.26) $Tr(Y^\dagger \mathcal{T}[X]) = 0$ for arbitrary operator $Y$ orthogonal to $P^{sym}$ and $P^{asym}$, i.e. if $Tr(Y^\dagger P^{sym}) = Tr(Y^\dagger P^{asym}) = 0$. This identity holds for both expressions of $\mathcal{T}$. Consequently, it is sufficient to verify that the values of $\Delta^{sym} = Tr(P^{sym}\mathcal{T}[E_{j\pm k,m\pm n}])$ and $\Delta^{asym} = Tr(P^{asym}\mathcal{T}[E_{j\pm k,m\pm n}])$ coincide for both expressions of the twirling channel given in Eq.(D.23) and in Eq.(D.24). Direct calculation gives

$$\begin{aligned}
\Delta^{sym} &= Tr(P^{sym}\int_{U(d)} dU\, U\otimes U E_{j\pm k,m\pm n} U^\dagger \otimes U^\dagger) \\
&= Tr(E_{j\pm k,m\pm n}\int_{U(d)} dU\, U\otimes U P^{sym} U^\dagger \otimes U^\dagger) \\
&= Tr(E_{j\pm k,m\pm n}P^{sym})\,; \\
\\
\Delta^{asym} &= Tr(E_{j\pm k,m\pm n}P^{asym})
\end{aligned}$$

and, simultaneuously,

$$\begin{aligned}
\Delta^{sym} &= \frac{Tr(E_{j\pm k,m\pm n}P^{sym})}{d_{sym}} Tr(P^{sym}P^{sym}) \\
&\quad + \frac{Tr(E_{j\pm k,m\pm n}P^{asym})}{d_{asym}} Tr(P^{sym}P^{asym}) \\
&= Tr(E_{j\pm k,m\pm n}P^{sym})\,; \\
\\
\Delta^{asym} &= Tr(E_{j\pm k,m\pm n}P^{asym})\,.
\end{aligned}$$

That is, the Eqs.(D.24) and (D.23) determine the same channel.

## E Calculations for Measurements

### E.1 Subspaces

In this appendix we shall analyze the subspaces of four quantum systems $\mathcal{H}^{\otimes 4}$, especially four qubits. Let us start with the simpler case of $\mathcal{H}\otimes\mathcal{H}$. Denote by $|j\rangle$ the basis of $\mathcal{H}$ and define

$$|\varphi_{jk}^{\pm}\rangle = \frac{1}{\sqrt{2}}(|j\otimes k\rangle \pm |k\otimes j\rangle)\,. \qquad \text{(E.27)}$$

for $j < k$. For $j = k$

$$|\varphi_{jj}^{+}\rangle = |j\otimes j\rangle\,. \qquad \text{(E.28)}$$

These vectors form an orthonormal bases of symmetric and antisymmetric subspaces of $\mathcal{H}\otimes\mathcal{H}$, i.e. they define the projections

$$P^{sym} = \sum_{j\le k} |\varphi_{jk}^{+}\rangle\langle\varphi_{jk}^{+}|, \quad P^{asym} = \sum_{j<k} |\varphi_{jk}^{-}\rangle\langle\varphi_{jk}^{-}|.$$

We shall use the notation $P_{12}^{sym} \equiv P_{12}^{sym} \otimes I_{34} = P_{12}^{sym} \otimes (P_{34}^{sym} + P_{34}^{asym})$. Let us stress that $P_{1234}^{sym} \leq P_{123}^{sym} \leq P_{12}^{sym}$. We shall be interested in properties of projections that are substracted from other projections to create the projections onto the completely symmetric subspace, for example, operators $Q_{12} = P_{12}^{sym} - P_{1234}^{sym}$ and $Q_{123} = P_{123}^{sym} - P_{1234}^{sym}$. Similar notations, definitions and relations hold also for other combination of indexes.

For qubits $\dim P_{12}^{sym} = d^2 \cdot d_2 = 12$, $\dim P_{12}^{sym} \otimes P_{34}^{sym} = d_2^2 = 9$, $\dim P_{123}^{sym} = \dim P_{124}^{sym} = d \cdot d_3 = 8$ and $\dim P_{1234}^{sym} = d_4 = 5$, thus, $\dim Q_{123} = \dim Q_{124} = 3$ and $Q_{12} = 7.$, etc.

## E.2 $P_{12}^{sym} \otimes P_{34}^{sym}$ and $P_{1234}^{sym}$

Let us start with the analysis of the subspace of $P_{12}^{sym}$ not contained in $P_{1234}^{sym}$, i.e. with $Q_{12}$. In the first step, let us split $Q_{12}$ into $Q_{12} = Q_{12}^{-} + Q_{12}^{+}$, where $Q_{12}^{+} = P_{12}^{sym} \otimes P_{34}^{sym} - P_{1234}^{sym}$, $Q_{12}^{-} = P_{12}^{sym} \otimes P_{34}^{asym}$. Due to asymmetry of $P_{12}^{sym} \otimes P_{34}^{asym}$ in $3 \leftrightarrow 4$ exchange the projections $P_{1234}^{sym}$ and $Q_{12}^{-}$ are orthogonal. For $Q_{12}^{+}$ the situation is more tricky. Our goal is to design a basis of the support of $Q_{12}^{+}$. The completely symmetric subspace $P_{1234}^{sym}$ is spanned by the following orthonormal basis

$$
\begin{aligned}
|\eta_0\rangle &= |\varphi_{00}^{+} \otimes \varphi_{00}^{+}\rangle \\
|\eta_1\rangle &= \frac{1}{\sqrt{2}}(|\varphi_{00}^{+} \otimes \varphi_{01}^{+}\rangle + |\varphi_{01}^{+} \otimes \varphi_{00}^{+}\rangle) \\
|\eta_2\rangle &= \sqrt{\frac{2}{3}}|\varphi_{01}^{+} \otimes \varphi_{01}^{+}\rangle + \sqrt{\frac{1}{6}}(|\varphi_{00}^{+} \otimes \varphi_{11}^{+}\rangle + |\varphi_{11}^{+} \otimes \varphi_{00}^{+}\rangle) \\
|\eta_3\rangle &= \frac{1}{\sqrt{2}}(|\varphi_{11}^{+} \otimes \varphi_{01}^{+}\rangle + |\varphi_{01}^{+} \otimes \varphi_{11}^{+}\rangle) \\
|\eta_4\rangle &= |\varphi_{11}^{+} \otimes \varphi_{11}^{+}\rangle\,.
\end{aligned}
$$

Our aim is to specify a basis spanning the support of $Q_{12}^{+}$. Since $\dim P_{1234}^{sym} = 5$ and $\dim P_{12}^{sym} \otimes P_{34}^{sym} = 9$ it follows we need to find four mutually orthogonal vectors in $P_{12}^{sym} \otimes P_{34}^{sym}$ that are also orthogonal to vectors $|\eta_j\rangle$. It is straightforward to verify that the following vectors

$$
\begin{aligned}
|\kappa_1\rangle &= \frac{1}{\sqrt{2}}(|\varphi_{00}^{+} \otimes \varphi_{01}^{+}\rangle - |\varphi_{01}^{+} \otimes \varphi_{00}^{+}\rangle) \\
|\kappa_2\rangle &= \frac{1}{\sqrt{2}}(|\varphi_{00}^{+} \otimes \varphi_{11}^{+}\rangle - |\varphi_{11}^{+} \otimes \varphi_{00}^{+}\rangle) \\
|\kappa_2'\rangle &= \sqrt{\frac{1}{3}}(|\varphi_{01}^{+} \otimes \varphi_{01}^{+}\rangle - |\varphi_{00}^{+} \otimes \varphi_{11}^{+}\rangle - |\varphi_{11}^{+} \otimes \varphi_{00}^{+}\rangle) \\
|\kappa_3\rangle &= \frac{1}{\sqrt{2}}(|\varphi_{11}^{+} \otimes \varphi_{01}^{+}\rangle - |\varphi_{01}^{+} \otimes \varphi_{11}^{+}\rangle)
\end{aligned}
$$

form such a basis.

Let us abbreviate the swap operator implementing the exchange of the subsystems $a, b$ by $S_{ab} \equiv \mathrm{Swap}_{\mathrm{ab}} = P_{ab}^{sym} - P_{ab}^{asym}$. This operation is unitary and arbitrary permutation can be written as a composition of swap operations. The following identities hold

$$P_{13}^{sym} \otimes P_{24}^{sym} = S_{23}(P_{12}^{sym} \otimes P_{34}^{sym})S_{23}\,,$$
$$P_{14}^{sym} \otimes P_{23}^{sym} = S_{34}(P_{13}^{sym} \otimes P_{24}^{sym})S_{34}\,,$$
$$P_{12}^{sym} \otimes P_{34}^{sym} = S_{24}(P_{14}^{sym} \otimes P_{23}^{sym})S_{24}\,.$$

The vectors $|\kappa_1\rangle, |\kappa_2\rangle, |\kappa_3\rangle$ defined with respect to division $P_{12}^{sym} \otimes P_{34}^{sym}$ are orthogonal to all vectors $|\kappa_j\rangle, |\kappa_2'\rangle$ defined with respect to splittings $P_{13}^{sym} \otimes P_{24}^{sym}$ and $P_{14}^{sym} \otimes P_{23}^{sym}$, i.e. $P_{13}^{sym} \otimes P_{24}^{sym}|\kappa_j\rangle = P_{14}^{sym} \otimes P_{23}^{sym}|\kappa_j\rangle = 0$. However, $\langle\kappa_2'|P_{13}^{sym} \otimes P_{24}^{sym}|\kappa_2'\rangle = \langle\kappa_2'|P_{14}^{sym} \otimes P_{23}^{sym}|\kappa_2'\rangle = 1/4$, because the vectors $|\kappa_2'\rangle$ defined with respect to different splittings are mutually nonorthogonal. This means that the 4 dimensional projections $Q_{12}^+, Q_{13}^+, Q_{14}^+$ are not orthogonal, however, there is a three-dimensional subspace of $Q_{12}^+$ (spanned by vectors $|\kappa_j\rangle$) orthogonal to both $Q_{13}^+$ and $Q_{14}^+$.

### E.3 $P_{123}^{sym} + P_{124}^{sym}$

For the purposes of this paper it is of interest to analyze the relation of the supports of projections $P_{123}^{sym}$ and $P_{124}^{sym}$. The swap operator $S_{34}$ can be written as a composition $S_{34} = S_{24}S_{23}S_{24}$. Consider a vector $|\varphi\rangle$ belonging to both subspaces, i.e. $P_{123}^{sym}\varphi = P_{124}^{sym}|\varphi\rangle = |\varphi\rangle$. For such vector $S_{12}|\varphi\rangle = S_{13}|\varphi\rangle = S_{14}|\varphi\rangle = S_{23}|\varphi\rangle = S_{24}|\varphi\rangle = |\varphi\rangle$ and therefore also $S_{34}|\varphi\rangle = S_{24}S_{23}S_{24}|\varphi\rangle = |\varphi\rangle$, hence the state $\varphi$ is symmetric also with respect to exchange $3 \leftrightarrow 4$. Consequently, it is invariant under the swap of arbitrary subsystems, i.e. it belongs to the completely symmetric subspace. Therefore, the greatest joint subspace of supports of $P_{123}^{sym}$ and $P_{124}^{sym}$ corresponds to the projection $P_{1234}^{sym}$.

Further we shall prove that the projections $Q_{123} = P_{123}^{sym} - P_{1234}^{sym}$ and $Q_{124} = P_{124}^{sym} - P_{1234}^{sym}$ are not mutually orthogonal and we shall specify the support of $Q_{123} + Q_{124}$. It is relatively stragihtforward to verify that the following unnormalized vectors

$$\begin{aligned}
|\omega_1\rangle &= |\varphi_{00}^+\rangle_{12}|\varphi_{01}^-\rangle_{34} + |\varphi_{00}^+\rangle_{13}|\varphi_{01}^-\rangle_{24} + |\varphi_{00}^+\rangle_{23}|\varphi_{01}^-\rangle_{14}\,,\\
|\omega_2\rangle &= |\varphi_{00}^+ \otimes \varphi_{11}^+\rangle - |\varphi_{11}^+ \otimes \varphi_{00}^+\rangle + 2|\varphi_{01}^+ \otimes \varphi_{01}^-\rangle\,,\\
|\omega_3\rangle &= |\varphi_{11}^+\rangle_{12}|\varphi_{01}^-\rangle_{34} + |\varphi_{11}^+\rangle_{13}|\varphi_{01}^-\rangle_{24} + |\varphi_{11}^+\rangle_{23}|\varphi_{01}^-\rangle_{14}\,,
\end{aligned}$$

form an orthogonal basis of the support of $Q_{123}$. These vectors are orthogonal to vectors $|\eta_j\rangle$ forming the completely symmetric subspace. In fact, they are completely symmetric only with respect to three indexes (123), but they not with respect to exchanges with the fourth qubit, hence, $P_{12}^{sym} \otimes P_{34}^{sym}|\omega_j\rangle$ is not proportional to $|\omega_j\rangle$. In the same way we can design a basis for each $Q_{jkl}$, in particular, for $Q_{124}$

$$\begin{aligned}
|\omega_1'\rangle &= -|\varphi_{00}^+\rangle_{12}|\varphi_{01}^-\rangle_{34} + |\varphi_{00}^+\rangle_{14}|\varphi_{01}^-\rangle_{23} + |\varphi_{00}^+\rangle_{24}|\varphi_{01}^-\rangle_{13}\,,\\
|\omega_2'\rangle &= |\varphi_{00}^+ \otimes \varphi_{11}^+\rangle - |\varphi_{11}^+ \otimes \varphi_{00}^+\rangle - 2|\varphi_{01}^+ \otimes \varphi_{01}^-\rangle\,,\\
|\omega_3'\rangle &= -|\varphi_{11}^+\rangle_{12}|\varphi_{01}^-\rangle_{34} + |\varphi_{11}^+\rangle_{14}|\varphi_{01}^-\rangle_{23} + |\varphi_{11}^+\rangle_{24}|\varphi_{01}^-\rangle_{13}\,.
\end{aligned}$$

Since $\langle\omega_j|\omega_k'\rangle = -2\delta_{jk}$ the pair of unnormalized vectors $|\omega_j\rangle, |\omega_j'\rangle$ forms a two-dimensional subspace orthogonal to remaining vectors. Equal superpositions $|\omega_j^+\rangle = |\omega_j\rangle + |\omega_j'\rangle$ are already symmetric in $3 \leftrightarrow 4$ exchange, hence $|\omega_j^+\rangle \in P_{12}^{sym} \otimes P_{34}^{sym}$. On the other hand, the vectors

$|\omega_j^-\rangle = |\omega_j\rangle - |\omega_j'\rangle$ are antisymmetric in $3 \leftrightarrow 4$, hence $|\omega_j^-\rangle \in P_{12}^{sym} \otimes P_{34}^{asym}$. It is easy to verify that they are orthogonal, i.e. $\langle\omega_j^+|\omega_j^-\rangle = 0$, because $\langle\omega_j|\omega_j\rangle = \langle\omega_j'|\omega_j'\rangle = 6$ and $\langle\omega_j|\omega_j'\rangle = \langle\omega_j'|\omega_j\rangle = -2$. Moreover, $\langle\omega_j^+|\omega_j^+\rangle = 8$ and $\langle\omega_j^-|\omega_j^-\rangle = 16$. Since $|\omega_j\rangle = \frac{1}{2}(|\omega_j^+\rangle + |\omega_j^-\rangle)$, $|\omega_j'\rangle = \frac{1}{2}(|\omega_j^+\rangle - |\omega_j^-\rangle)$ we have

$$\begin{aligned} Q_{123} + Q_{124} &= \frac{1}{6}\sum_j (|\omega_j\rangle\langle\omega_j| + |\omega_j'\rangle\langle\omega_j'|) \\ &= \sum_j \frac{1}{12}(|\omega_j^+\rangle\langle\omega_j^+| + |\omega_j^-\rangle\langle\omega_j^-|) \\ &= \sum_j \left(\frac{4}{3}\frac{1}{16}|\omega_j^-\rangle\langle\omega_j^-| + \frac{2}{3}\frac{1}{8}|\omega_j^+\rangle\langle\omega_j^+|\right) , \end{aligned}$$

where $\frac{1}{16}|\omega_j^-\rangle\langle\omega_j^-|$ and $\frac{1}{8}|\omega_j^+\rangle\langle\omega_j^+|$ are one-dimensional projections, hence, we get the spectral decomposition of $Q_{123} + Q_{124}$ with eigenvalues $2/3, 4/3$. For our purposes the relevant part is associated with vectors $|\omega_j^+\rangle$, because $|\omega_j^-\rangle$ are not from the support of $P_{12}^{sym} \otimes P_{34}^{sym}$.

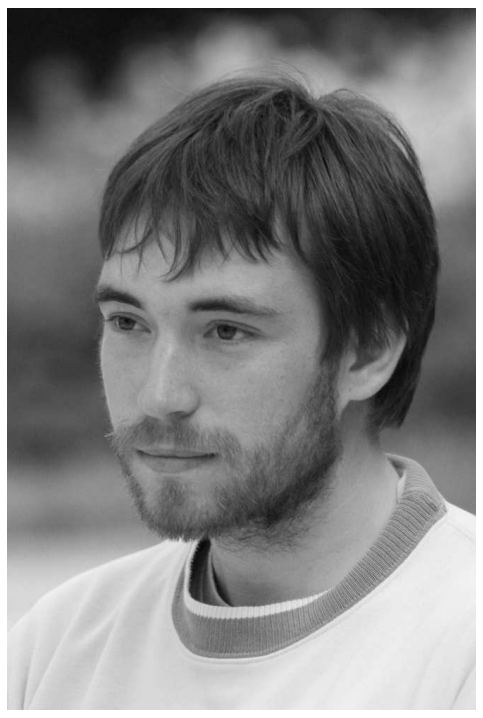

**Dr. Michal Sedlák** was born in 1982 in Piešťany in Slovakia. He graduated in 2005 at Comenius University in Bratislava, Faculty of mathematics, physics and informatics. He received his PhD in physics at Institute of Physics of Slovak Academy of Sciences in 2009 under the supervision of Prof. Vladimír Bužek. Since then, he worked as a researcher in Research Center for Quantum Information at Slovak Academy of Sciences. Currently, he has a research fellowship at Dipartimento di Fisica "A.Volta", Universita degli studi di Pavia in Italy. His research interests cover quantum discrimination problems, complexity of quantum circuits, and quantum information theory.